\pdfoutput=1
\documentclass[12pt,defended,oneside]{new_cit_thesis}
\usepackage{graphicx}
\usepackage{setspace}
\usepackage{amsmath}
\usepackage{color}
\usepackage{amssymb}
\usepackage{afterpage}
\usepackage{subfigure}

\begin{document}
\author{Lisa Maria Goggin}
\dedication{To Mum and Dad}    
\title{A Search For Gravitational Waves from Perturbed Black Hole Ringdowns in
LIGO Data}
\date{May 13th, 2008}
\maketitle
\begin{frontmatter}
    \makecopyright            
     \makededication           
     \begin{acknowledgements}  
The LIGO project has long been a source of wonder and amazement to me and I
feel privileged to have had the opportunity to contribute to it. My
involvement with the LSC has given me access to a brilliant and dedicated
group of scientists and engineers and for that I am very grateful. I believe I
have learned something from every member I have encountered, whether through
direct interaction or simply by your example of how good science is done. 

There are a number of people who have made a significant contribution to my
work that I would like to give credit to:

First and foremost I would like to thank my adviser Alan Weinstein for giving
me my first opportunity to work on LIGO as a SURF student at the 40m, for
supervising my senior thesis and for hiring me once more as a grad student.
Thank you for your guidance, patience and encouragement over the years and
allowing me the freedom to find how I work best. Your breadth of knowledge in
and enthusiasm for the project and field in general has been a great
inspiration.

Thank you to Yanbei Chen, Andrew Lange, and Frank Porter for serving on my
defense committee and to Albert Lazzarini for being part of my candidacy
committee.

Thank you to LIGO lab for funding me, enabling me to attend many conferences, and for the nicest office view I am ever likely to have!

Thanks to Duncan Brown for introducing me to the world of LIGO data analysis
and for your guidance over the last few years.  Your expertise was invaluable
in getting the search up and running.

Thanks to my office mates and friends Diego Fazi, Drew Keppel, and Pinkesh
Patel for your encouragement, the many interesting discussions and for
your canine friends.

Thanks to the members of the CBC group for your help with the analysis. In
particular thanks to co-chairs Patrick Brady and Gabriela Gonz\'alez for your
guidance and encouragement; to Jolien Creighton for writing the ringdown filtering code and patiently answering numerous questions; to Stephen Fairhurst for developing the IMR code and your helpful advice throughout the analysis; and to Sarah Caudill, Shourov Chatterji, Thomas Cokelaer, Alex Dietz, Nick Fotopoulos, Romain Gouaty, Chad Hanna, Siong Heng, Vicky Kalogera, Eirini Messaritaki, Richard O'Shaughnessy, Bangalore Sathyaprakash, Peter Shawhan, Xavier Siemens, Patrick Sutton, and John Whelan for your countless useful suggestions.

Thanks to Emanuele Berti explaining the finer points of black hole perturbation theory to me.

Thanks to the ringdown review committee, Laura Cadonati, Kipp Cannon, and
Stefan Ballmer who patiently and diligently reviewed the code and results.

Thanks to Stuart Anderson, Phil Ehrens, Dan Kozak, and Paul Armor for never
being too busy to help me out with computer crises. 

Thanks to Dennis Ugolini, Steve Vass, and Osamu Miyakawa for many fun and interesting days at the 40m.

To the friends who have kept me (somewhat) sane during my time in Pasadena: Guillaume Bres, Helen Gardarsdottir, Alan Hampton, Hannes Helgason, Vala Hjorleifsdottir, Gerdur Jonsdottir, Donal O'Connell, Tosin Otitoju, Jonathan Pritchard, Angel Ruiz, David Seal, Paul Skerritt, Allison Townsend, and Setthivoine You; thanks for all the good times.

Finally, to my wonderful family, without whose constant love and support this
would not have been possible; Julie, Amy, and Denise, thanks for making six thousand
miles feel so much shorter; Mum and Dad, thank you for always believing in me
and encouraging me to follow my dreams.

\end{acknowledgements}

     \begin{abstract}
According to General Relativity a perturbed black hole will return to a
stable configuration by the emission of gravitational radiation in a superposition of quasi-normal modes. Such a perturbation will occur due to the coalescence of a black hole binary, following their inspiral and subsequent merger. At late times the waveform, which we refer to as a ringdown, is expected to be dominated by a single mode. As the waveform is well-known the method of matched filtering can be implemented to search for this signal using LIGO data. 
LIGO is sensitive to the dominant mode of perturbed black holes with masses between 10 and 500 $M_\odot$, the regime of intermediate-mass black holes, to a distance of up to 300 Mpc. We present a search for gravitational waves from black hole ringdowns using data from the fourth LIGO science run. We implement a blind analysis of the data. We use Monte Carlo simulations of the expected waveform, and an estimation of the background from timeslides to tune the search. We present an analysis of the waveform parameter estimation and estimate the efficiency of the search. As there were no gravitational wave candidates found, we place an upper limit on the rate of black hole ringdowns in the local universe. 

     \end{abstract}
     \extrachapter{Preface}
     The work presented in this thesis was carried out within the LIGO
     Scientific Collaboration (LSC). The methods and results presented here
     are under review and are potentially subject to change. The opinions
     expressed here are those of the author and not necessarily those of the
     LSC. 
     
     The author gratefully acknowledges the support of the United States
     National
     Science Foundation for the construction and operation of the LIGO
     Laboratory,
     which provided support for this work.
     
     This thesis has LIGO document number LIGO-P080058-00-Z. 
     \clearpage
     \tableofcontents
     \listoffigures
     \listoftables

\end{frontmatter}

\chapter{Introduction}
\label{ch:intro}

Black hole ringdowns are amongst the most promising sources of gravitational waves,
detectable with current detectors out to very large distances, as far as 300
Mpc from the Earth. The ringdown is the final phase of a binary black hole
coalescence, following the inspiral and merger.

A central result of general relativity is that gravitational waves are emitted from an accelerating mass. It has been established using black hole perturbation theory that the waveform emitted by a perturbed black hole can be modeled as
a superposition of quasi-normal modes, with ``quasi'' referring to the fact
that the oscillation is damped. It is expected that at late times the
oscillation will be dominated by a single mode. Throughout this analysis we
will refer to a gravitational wave emitted from a perturbed black hole 
as a ``ringdown waveform'' or just ``ringdown''. 

This thesis presents the results of a search for gravitational waves from
perturbed black holes using data from the Laser Interferometer
Gravitational Wave Observatory (LIGO), a project dedicated to the detection of
gravitational waves. LIGO is run jointly between
Caltech and MIT, and funded by the National Science Foundation. The observatory
consists of three detectors at two sites; Hanford, WA hosts a 4 km
interferometer (H1) and a 2 km interferometer (H2), and Livingston, LA is home
to a second 4 km interferometer (L1). Construction of the interferometers
began in 1996, starting the Initial LIGO phase. After several engineering runs
the first LIGO science run (S1) began on August 23rd, 2002 and lasted a
little over two weeks. Between then and mid-2005 three more science runs took
place at a rate of about one per year with S2 and S4 lasting approximately one
month each and S3 collecting three months of data. By late 2005, LIGO reached
its initial design sensitivity, and on November 4th, 2005, S5 began. The goal
of S5 was to collect one year's worth of triple coincidence data; this was
achieved by September 30th, 2007.  At the time of writing, the LIGO detectors
are undergoing significant upgrades for the Enhanced LIGO phase
\cite{ELIGO}, which will implement many of the Advanced LIGO technologies
and see a factor of 2--3 increase in sensitivity. This will culminate in the S6
run, scheduled to start in the autumn of 2009.

We begin in chapter \ref{ch:GR} with an introduction to gravitational waves; we discuss how they may be detected through laser interferometry and outline some possible sources of gravitational waves. 
In chapter \ref{ch:astro} we provide the motivation for the search. We discuss theoretical and astrophysical black holes and introduce the waveform we are searching for. 
Chapter \ref{ch:matchedfilter} describes the method of matched filtering and the template bank used in the search. 
The pipeline that has been created to implement a search for ringdowns is detailed in chapter \ref{ch:search}. 
In chapter \ref{ch:s4} we describe some of the important details about the S4 science run. 
We describe the tools used to tune the search in chapter \ref{ch:tuning}, and explain how the final values of the constraints were arrived at. 
In chapter \ref{ch:paramest} we describe the results of a large scale Monte Carlo run. We evaluate the efficiency of the search, compare the expected and detected waveform parameters, and compare the recovered parameters between pairs of detectors. We also estimate the background and compare it to a subset of the data. 
In chapter \ref{ch:results} we describe the results of the search. We did not find any plausible gravitational wave candidates in the S4 data set. We place an upper limit on the rate of ringdowns and investigate some of the loudest candidate events. 
In chapter \ref{ch:imr} we investigate the effect that the presence of an inspiral and
merger preceding the ringdown would have on our ability to detect and estimate the parameters of ringdowns. 
In chapter \ref{ch:future} we make some recommendations for future ringdown
searches and document some issues we encountered in the course of the search.
We give a brief summary of our results and a final conclusion in chapter \ref{ch:concl}.

\chapter{Gravitational Waves}
\label{ch:GR}

In this chapter we introduce gravitational waves and outline how they may be detected using laser interferometry, in particular by the LIGO detectors. We also describe some of the likely sources of gravitational waves.

\section{Gravitational Waves in General Relativity}


``Spacetime grips mass, telling it how to move, and mass grips spacetime, telling it how to curve'',  a famous quote of John Wheeler's summarizing the mutual dependence of mass and spacetime in the theory of general relativity. This was a central result of Einstein's theory of general relativity, expressed mathematically by Einstein in what is now known as the Einstein equation, a set of ten nonlinear partial differential equations for ten metric coefficients, $g_{\alpha \beta}(x)$, relating the Einstein curvature tensor $G_{\alpha \beta}$ (a measure of local spacetime curvature) to the stress-energy tensor of matter $T_{\alpha \beta}$ (a measure of matter energy density),
\begin{equation}
G_{\alpha \beta} \left(g_{\alpha \beta}\right)= \frac{8 \pi G}{c^4} T_{\alpha \beta}, 
\label{eqn:Einsteineqn}
\end{equation}
where $G$ is Newton's constant and $c$ is the speed of light. 
A general solution for this equation has not been found, however various techniques exist for solving the equations under particular circumstances. One such case is weak time-varying fields producing ``ripples in spacetime'' or gravitational waves.

Under the assumption that the gravitational waves produced by the source are weak, the metric can be written as a small perturbation $h_{\alpha \beta}$ of the flat spacetime metric in Minkowski coordinates $\eta_{\alpha \beta}=\mathrm{diag}(-1, +1, +1, +1)$, where $|h_{\alpha \beta}| \ll 1$,
\begin{equation}
g_{\alpha \beta}(x) = \eta_{\alpha \beta} + h_{\alpha \beta}(x).
\label{eqn:metricpert}
\end{equation}
In the weak field limit the non-linear Einstein equation can be approximated
as linear, and with the choice of the transverse-traceless gauge, is given by
the wave equation
\begin{equation}
\left( -\frac{1}{c^2}\frac{\partial^2}{\partial t^2} + \overrightarrow{\nabla}^2\right) h_{\alpha \beta} (x) =0  \label{eqn:vacEinsteqn}.
\end{equation}
The solution to this equation is
\begin{equation}
h_{\alpha \beta}(x)=a_{\alpha \beta} e^{i \mathrm{\mathbf{k \cdot x}}},
\label{eqn:vacsoln}
\end{equation}
where $a_{\alpha \beta}$ is a symmetric $4 \times 4$ matrix of constants giving the amplitudes of the various components of the wave, and $\mathrm{\mathbf{k}}$ is the wave vector such that
\begin{equation}
\mathrm{\mathbf{k \cdot x}}=-k^t t+\overrightarrow{k}\cdot \overrightarrow{x}.
\end{equation}
Substituting equation (\ref{eqn:vacsoln}) into equation (\ref{eqn:vacEinsteqn}) gives the condition 
\begin{equation}
k_\alpha k^\alpha=0,
\end{equation}
showing that the wave propagates at the speed of light. Our choice of gauge gives us the following constraints:
\begin{eqnarray}
k_\alpha a^{\alpha \beta} &=& 0 \\
a^\alpha_\alpha &=& 0 \\ 
a_{\alpha \beta} u^\beta &=& 0,
\end{eqnarray}
where $u^\beta$ is some fixed four-velocity.
The first constraint restricts $a^{\alpha \beta}$ to be orthogonal or transverse to $k_\alpha$, the second requires that the matrix is traceless and the third, if we orient the coordinate axes such that the direction of propagation is along the z-axis, implies that $a_{\alpha  z}=0$.  These conditions reduce the number of components of $a_{\alpha \beta}$ from
ten to just four,
\begin{equation}
a_{\alpha \beta}= \left( \begin{array}{cccc} 
                        0 & 0 & 0 & 0 \\  
                        0 & a_{xx} & a_{xy} & 0 \\
                        0 & a_{xy} & -a_{xx} & 0 \\
                        0 & 0 & 0 & 0
                  \end{array} \right).
\end{equation}
We write the final form of the solution to the source-free, linearized
Einstein equation as
\begin{equation}
h_{\alpha \beta}= \left( \begin{array}{cccc}
                        0 & 0 & 0 & 0 \\
                        0 & h_+ & h_\times & 0 \\
                        0 & h_\times & -h_+ & 0 \\
                        0 & 0 & 0 & 0
                  \end{array} \right)
                  e^{i\omega \left( z-t \right) }.
\end{equation}
Thus the transverse traceless gravitational wave travels at the speed of light and is composed of two independent polarizations; $h_+$ is known as the plus polarization and $h_\times$ is the cross-polarization.

The energy density is given by the stress-energy tensor 
\begin{eqnarray}
T^{GW} &=& \frac{1}{32 \pi} \frac{c^2}{G} \sum_{i,j} \left\langle h_{i,j,0}^{TT},h_{i,j,0}^{TT}\right\rangle \nonumber \\ 
&=& \frac{1}{16 \pi} \frac{c^2}{G} \left\langle |h_{+,0}|^2+|h_{\times,0}|^2 \right\rangle,
\label{eqn:GWenergy}
\label{eqn:energydensity}
\end{eqnarray}
where $\left\langle  ... \right\rangle $ denotes an average over several wavelengths \cite{Hart03}.


\section{Detection of Gravitational Waves}

\subsection{Effects of Gravitational Waves on Test Masses}
\label{sec:gweffects}

The effects of a gravitational wave cannot be seen in isolated bodies, but only by observing the change in separation between pairs of masses. Take as an example a pair of test masses separated by a distance $L$ (as measured in the unperturbed flat spacetime) along the x-axis, and a gravitational wave propagating along the z-axis. In the perturbed spacetime the distance between the test masses $L'$ is
\begin{equation}
L'(t) = \int_0^L  \left[ 1+h_{xx}\left( t,x\right)  \right]^{1/2} \textrm{d}x 
\end{equation}
which, in the long wavelength approximation can be expressed as
\begin{equation}
L'(t) \approx  L \left[ 1+\frac{1}{2} h_{xx} \left(t,0 \right) \right]
\end{equation}
and thus for a change in distance between the two test masses $\delta L'=L'-L$, the strain produced by the gravitational wave is
\begin{equation}
\frac{ \delta L' \left(t,0 \right) }{ L } = \frac{1}{2} h_{xx} \left( t,0 \right). \label{eqn:dLL}
\end{equation}

\begin{figure}[h]
\centering
\begin{center}
\includegraphics[scale=0.6]{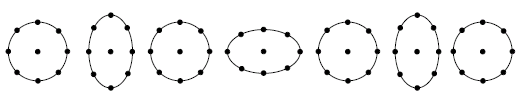}
\caption{Illustration of the effect of a gravitational wave with plus polarization on a ring of test particles when the direction of propagation of the gravitational wave is orthogonal to the plane of the particles.}
\label{fig:testmasses_plus}
\end{center}
\end{figure}

\begin{figure}[h]
\centering
\begin{center}
\includegraphics[scale=0.6]{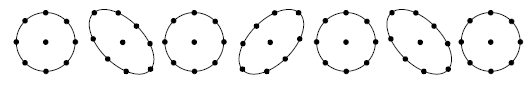}
\caption{Illustration of the effect of a gravitational wave with cross polarization on a ring of test particles when the direction of propagation of the gravitational wave is orthogonal to the plane of the particles.}
\label{fig:testmasses_cross}
\end{center}
\end{figure}

An illustration of the effects of a gravitational wave is shown in figures \ref{fig:testmasses_plus} and \ref{fig:testmasses_cross}. These show a circular configuration of free test masses in the $z=0$ plane. From the view point of the central test mass the gravitational wave manifests itself by stretching space between it and the other test masses in one direction transverse to the direction of propagation and contracting in the orthogonal direction in the same plane, changing the circular pattern of the test masses to an elliptical configuration.  Half a period later the effect is reversed; those masses which were displaced furthest from the central test mass are now brought closest and vice versa. The gravitational wave polarization which causes maximal stretching along the x and y axes is known as the plus polarization. Rotating the coordinate axes by 45$^{\circ}$ in the $z=0$ plane demonstrates the cross polarization. The most general gravitational wave traveling in the $z$ direction is a linear superposition of these two polarizations.


It is evident from figures \ref{fig:testmasses_plus} and \ref{fig:testmasses_cross}, the gravitational wave is invariant under a rotation of 180$^{\circ}$ about its direction of propagation, $\theta_{inv}=180^{\circ}$. This is related to the spin $S$ of the zero-mass particle associated with the field, which in the case of gravity is the graviton, by the relation $S=360^{\circ}/\theta_{inv}$, and thus is a consequence of the fact that the graviton is spin-2 (the quantum analogue of a classical rank-2 tensor field) \cite{MTW}.


\subsection{Detecting Gravitational Waves with Laser Interferometry}

The detection of gravitational waves through laser interferometry takes advantage of the effects just described, and uses laser light as a displacement measuring device. We replace the central test mass in the configuration above with a 50\% reflecting mirror known as a beam splitter (BS), and replace the ring of test masses with two highly reflecting mirrors placed at an equal distance from, but in orthogonal directions to the BS. These mirrors are referred to as end test masses in the x and y direction, ETMX and ETMY. The BS directs an input beam of laser light towards the ETMs. If the distance between the ETMs and the beamsplitter is the same in the two arms, the phase of the light reflected from the ETMs is the same. A gravitational wave traveling in a direction orthogonal to the plane of the detector will increase the distance between the BS and the ETM in one arm and decrease the distance in the other arm. This will produce a phase difference on the light received by a photodiode at the output of the interferometer. The longer the distance the light has to travel the greater the phase shift will be. This optical configuration describes a Michelson interferometer \cite{Saul94}.

\subsection{The Laser Interferometer Gravitational Wave Observatory, LIGO}

\begin{figure}[h]
\centering
\begin{center}
\includegraphics[scale=0.6]{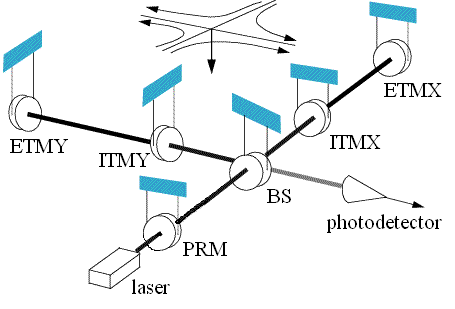}
\caption{Schematic of the LIGO interferometer. Picture courtesy of LIGO lab.}
\label{fig:FPMifo}
\end{center}
\end{figure}

The LIGO detectors are Michelson interferometers with the additional feature of Fabry-Perot arms. These are resonant cavities, formed by placing an additional mirror just after the laser in both arms at an integral number of wavelengths from the ETM. These mirrors are known as input test masses, ITMX and ITMY. These resonant cavities allow the light to circulate many times, effectively increasing the length of the arms. The ITMs and ETMs are separated by 4 km in H1 and L1, and by 2 km in H2. The light circulates in the resonant cavities approximately 200 times. LIGO also employs power recycling, in which an additional mirror (the power recycling mirror PRM) is placed between the laser and the BS, resonantly enhancing the light stored in the interferometer. A schematic of the LIGO detectors is shown in figure \ref{fig:FPMifo}.

\begin{figure}[h]
\centering
\begin{center}
\includegraphics[scale=0.6]{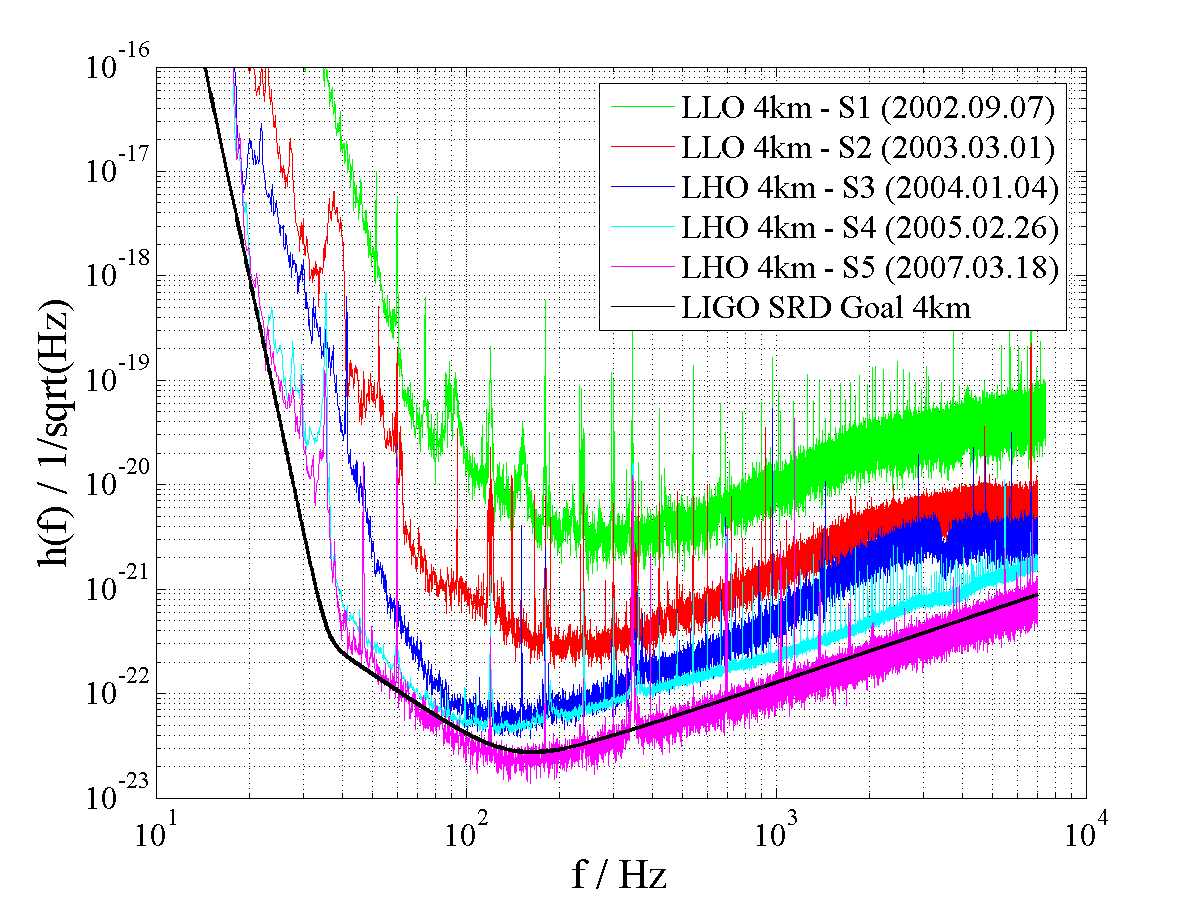}
\caption{Best LIGO strain sensitivity curves for the science runs S1 to S5.
The LIGO design sensitivity curve is shown in black.}
\label{fig:S1S5noise}
\end{center}
\end{figure}

As mentioned in chapter \ref{ch:intro} LIGO has successfully completed six science runs since 2002. In figure \ref{fig:S1S5noise} we show the best strain sensitivity curves from each of the five science runs, along with the design sensitivity curve. The plot demonstrates the large increases in sensitivity achieved between runs to the point where, in S5, the LIGO detectors achieved design sensitivity. The plot also shows that the LIGO detectors are most sensitive to gravitational waves between $\sim40$ Hz and 2 kHz. 

\subsubsection{Calibration of the LIGO Detectors} \label{sec:calibration}

Calibration of the data is essential to determining the sensitivity to distant sources of gravitational radiation \cite{Diet05,Siem04}. Here we describe how it is achieved.

The data to be analysed is taken from the gravitational wave channel
DARM\_ERR, $q(t)$. This is the error signal on the feedback loop which is used
to control the differential motion of the interferometer arms (DARM). It is
related to the gravitational wave strain in the Fourier domain $h(f)$ by the
response function $R(f)$
\begin{equation}
h(f)=R(f)q(f).  \label{eqn:response}
\end{equation}
Accurate reconstruction of the strain from the error signal, i.e., determination of the response function $R(f)$, is essential and is done
through the process of calibration \cite{Diet05,Siem04}.

\begin{figure}[h]
\centering
\begin{center}
\includegraphics[scale=0.4]{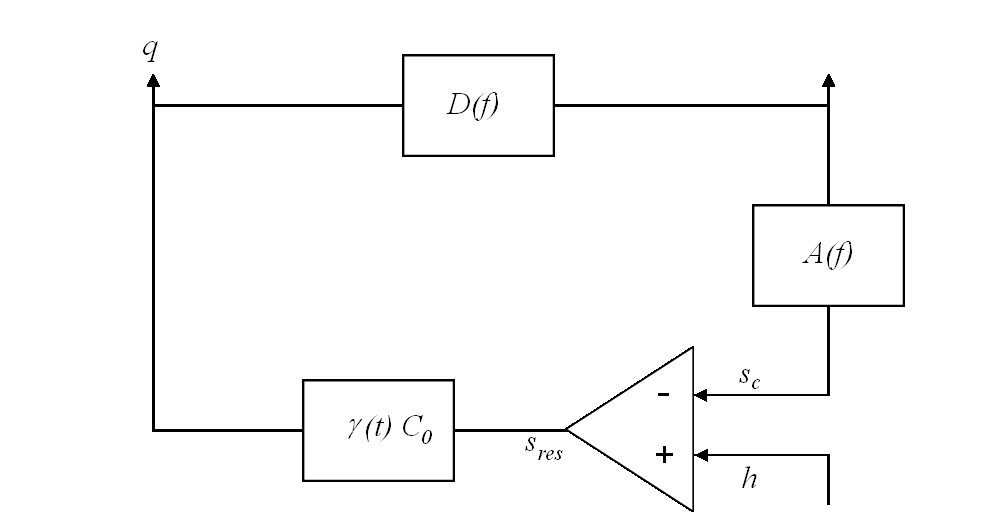}
\caption{Detector control loop}
\label{fig:controlloop}
\end{center}
\end{figure}

The response of the interferometer to a gravitational wave strain can be
characterized by a loop gain function $G(f)$, which is parameterized by three
functions, a sensing function C(f), an actuation function A(f), and a digital
filter function D(f), by
\begin{equation}
G(f)=C(f)D(f)A(f).
\end{equation}
The control loop is shown in figure \ref{fig:controlloop}. A control strain
$s_c$ subtracts from the gravitational wave strain plus noise, $h=s_{GW}+n$,
leaving a residual strain $s_{res}$. The sensing function
$C(t,f)=\gamma(t)C_0(f)$ consists of a reference sensing function $C_0$ and a
loop gain $\gamma(t)$, which depends on the light power stored in the arms.
$C_0$ is dominated by the cavity pole frequency response $(1+f/f_p)^{-1}$,
with $f_p \approx 90$ Hz. As described in the next section, $\gamma(t)$ is a relative measurement which changes over time as the alignment of the mirrors varies. The sensing function converts the residual strain into a digital error signal $q$ which is
read out by the channel DARM\_ERR in arbitrary units of counts at a rate of 16384 Hz.
The digital filter $D(f)$ converts the error signal to a control signal $d$
that is sent to the mirrors as an actuation. This quantity is known precisely. The actuation function converts
the control signal to strain by sending a current to coils surrounding magnets
which are attached to the mirrors. This produces a force and hence a
displacement of the mirrors, adjusting the lengths of the cavities. 
From the figure we can see that
\begin{eqnarray}
q(f) &=& \gamma(t)C_0(f)s_{res}, \\
h &=& s_{res}+s_c,
\end{eqnarray}
and
\begin{equation}
s_c=s_{res}C_0 \gamma(t) D(f)A(f).
\end{equation}
Substituting these quantities in equation (\ref{eqn:response}) and solving for
$R$ gives the response function
\begin{equation}
R(f)=\frac{1+\gamma(t)G_0(f)}{\gamma(t)C_0(f)},
\end{equation}
where  $G_0=C_0(f) D(f) A(f)$ is the reference open loop gain.

\subsubsection{Calibration Lines}
As was mentioned above, the calibration coefficient $\gamma$ is a  
function of time. In practice it changes on the order of minutes. In order
to track these changes, sinusoidal signals of known frequency are added to the
control signal. For the S4 run these were at 46.7, 393.1, and 1144.3 Hz for
H1; 54.1, 407.3, and 1159.7 Hz for H2; and 54.7, 396.7, and 1151.5 Hz for L1. By
digitally heterodyning the error signal, the control signals, and the
excitation signal with the injected sinusoid, the calibration coefficients can
be found.


\subsection{Antenna Response and Effective Distance}
\label{sec:antresp}
\begin{figure}[h]
\centering
\begin{center}
\includegraphics[scale=0.4]{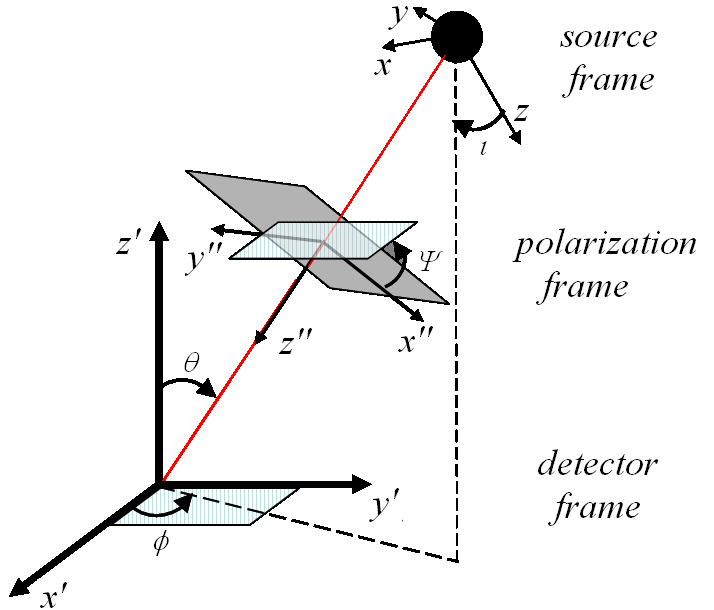}
\caption{Illustration of source ($x,y,z$), radiation ($x'',y'',z''$) and
detector ($x',y',z'$) frames.}
\label{fig:coordinateframes}
\end{center}
\end{figure}

Up until now we have been concerned with gravitational waves propagating from
directly above (or below) the interferometer, so-called ``optimally
positioned and oriented'' sources that give the maximum response in the interferometer.
However, in reality not only can a gravitational wave come from any
direction in the sky but its orientation may be such that the detectors can
only capture some portion of it. In calculating the strain produced by a
given source these considerations need to be accounted for. Figure \ref{fig:coordinateframes} displays coordinates for the emission, propagation, and reception of a gravitational
wave. The source has axes $(x,y,z)$, with the $z$-axis in the direction of the
angular momentum. The line between the source and the detector $r$ makes an
angle $\iota$ with the $z$-axis. This is the angle of inclination. At the detector the
local coordinate axes are $(x',y',z')$, with $x'$ and $y'$ along the arms of
the interferometer. $r$ makes an angle $\theta$ with the $z'$ axis and an
angle $\phi$ with the $x'$ axis. In between we have the propagation
coordinates, $(x'',y'',z'')$ such that $z''$ lies along $r$ and the $x''$ and
$y''$ axis make an angle $\Psi$ with the $x'$ and $y'$ axes. This is the
polarization angle. 

\begin{figure}[h]
\centering
\begin{center}
\includegraphics[scale=0.6]{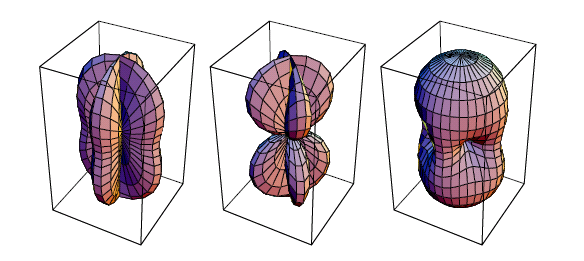}
\caption{The plus, cross, and unpolarized combination $\sqrt{F_+^2 +
F_\times^2}$ antenna patterns for the LIGO detectors. (This figure was taken
from \cite{Sigg98}.)}
\label{fig:antennapattern}
\end{center}
\end{figure}

Thus the strain produced at the detector is given by
\begin{equation}
h(t)=h_+(t)F_+(\theta,\phi,\Psi)+h_\times(t)F_\times(\theta,\phi,\Psi),
\end{equation}
where the plus and cross detector beam functions $F_+$ and $F_\times$ \cite{Schu87} are given by
\begin{eqnarray}
F_+&=&-\frac{1}{2}(1+\cos^2\theta) \cos2\phi \cos 2 \Psi -\cos \theta \sin
2\phi \sin 2\Psi
\label{eqn:Fplus} \\
F_\times &=& \frac{1}{2}(1+\cos^2\theta) \cos2\phi \sin 2 \Psi -\cos \theta
\sin 2\phi \cos 2\Psi.
\label{eqn:Fcross}
\end{eqnarray}
The detector plus, cross, and unpolarized combination $\sqrt{F_+^2 + F_\times^2}$ are shown in figure \ref{fig:antennapattern}. The figures show that there is a null point in the antenna pattern. If a gravitational wave is traveling in a direction orthogonal to the plane of the detector and the polarization angle is at 45 degrees to the $x'$ and $y'$ axes the effect of the gravitational wave will be the same in both arms and no phase shift will be produced.

We define the effective distance $D_{eff}$ as the distance to an optimally positioned and
oriented source that produces the same strain in the detector as a source at a
given position, polarization, and inclination at a distance $D$,
\begin{equation}
D_{eff} = \frac{D}{ \sqrt{ F_+^2 \left(1+\cos^2 \iota \right)^2/4
              +       F_\times^2 \cos^2 \iota}}. 
\label{eqn:Deff}
\end{equation}


\section{Sources of Gravitational Waves}

Even though gravitational waves have not yet been directly observed, their existence has been inferred through careful monitoring of the orbital period of the binary pulsar PSR 1913+16, discovered by Hulse and Taylor in 1974 \cite{Huls75}. They observed that the orbital period of the binary system was decreasing in a manner precisely consistent with the loss of energy and angular momentum due to gravitational radiation. For this they were awarded the Nobel Prize in 1993. 

Below we briefly outline the main sources of gravitational waves, categorized by waveform type. The LIGO Scientific Collaboration (LSC) data analysis efforts are structured around searches for these different waveform morphologies.

\subsection{Binary Coalescence}
\label{sec:bincoals}
A system composed of either two neutron stars, two black holes, or one of each bound together by gravity forms a binary system. According to general relativity the objects will lose energy through the emission of gravitational radiation. As a result their orbits shrink and the two stars spiral in towards one another eventually combining to form a single star, most likely a black hole. This process is called binary coalescence. The coalescence can be divided into three phases according to how well we can model the waveform at different times. The ``inspiral phase'' is defined as that time while the two stars are distinct objects orbiting around one another and the gravitational waveform emitted can be well approximated by the post-Newtonian model (i.e., the velocities are low). The post-Newtonian approximation breaks down as the stars begin their final few orbits and plunge in towards one another. We refer to this as the ``merger'' phase. Although numerical simulations are telling us more about the waveform produced at this stage (see chapter \ref{ch:imr}) it is still not represented by an analytic waveform. We refer to this waveform as an unmodeled burst. (Searches for this type of waveform will be discussed in the next section.) After the plunge, the resulting star tries to return to a stable configuration by emitting gravitational waves in a series of quasi-normal modes. These are also well modeled and this phase is known as the ``ringdown phase''. 

The search for gravitational waves from the ringdown phase is the subject of
this thesis, thus we dedicate chapter \ref{ch:astro} to a discussion of this
waveform and black holes in general.

The waveform produced during the inspiral phase is colloquially known as a chirp waveform, because the frequency and amplitude of the signal increases rapidly with time. For a binary of total mass $M$, separation $a$, and orbital period $T$ at a distance $r$, the characteristic strain expected from the inspiral can be approximated as \cite{Thor87}
\begin{equation}
h  \sim  \frac{G}{c^4} \frac{E_k}{r},
\end{equation}
where $E_k=M(\pi a/T)^2$ is the kinetic energy of an equal mass binary due to non-spherical
motion. Employing Kepler's third law $T^2=4\pi^2a^3/GM$ we can estimate the strain as
\begin{equation}
h \sim 10^{-20} \left(\frac{6.3 \textrm{ kpc}}{r}\right) \left(\frac{M}{2.8\ M_\odot}\right)^{5/3}  \left(\frac{T}{1\textrm{ s}}\right)^{-2/3}.
\end{equation}
As the signal is well known it can be searched for using the method of matched filtering (introduced in chapter \ref{ch:matchedfilter}). Inspiral searches on LIGO data over the past five data runs have targeted binaries containing neutron stars, stellar mass black holes, and primordial black holes. Details of these analyses may be found in the following papers \cite{S1BNS,S2BNS,S2PBH,S2BBH,S3SBBH,S3S4,GRB070201}.

\subsection{Unmodeled Bursts}
There are many astrophysical sources which are likely to emit what is best described as a burst of gravitational waves whose exact form is not well known. This includes gravitational waves from the merger of two stars described above, supernova explosions, gamma ray burst (GRB) engines, and possibly sources we are not even aware of. Data analysis algorithms capable of identifying short-duration excesses of strain power and correlating these between detectors are employed to search for unmodeled sources. A selection of papers describing results of LIGO burst analyses are \cite{BurstS1,BurstS2,BurstS3,BurstS4,GRBS2S3S4}.

\subsection{Periodic Sources}
The mechanism by which a rapidly spinning neutron star is most likely to emit gravitational waves occurs if its shape deviates from axisymmetry. This deviation is expressed as the ellipticity $\varepsilon$ of the neutron star, $\varepsilon=\left(I_{xx}-I_{yy}\right)/I_{zz}$, where $I_{jj}$ represent the moments of inertia about the principle axes. The resulting gravitational wave has a frequency twice the rotational frequency $f_{rot}$. The expected strain for a neutron star at a distance $r$ is 
\begin{eqnarray}
h &\sim& \frac{4 \pi^2 G}{c^4} \frac{I_{zz} f_{rot} \varepsilon}{r} \\
&=& 2\times 10^{-26} \left( \frac{f_{rot}}{1\textrm{ kHz}} \right)^2 
     \left(\frac{10\textrm{ kpc}}{r} \right) \left( \frac{\varepsilon}{10^{-6}} \right)
\end{eqnarray}
\cite{Thor87}. Both all-sky and targeted searches have been undertaken within the LSC, details may be found in the following papers: \cite{PULS1,PULS2,PULS2b,PULS2c,PULS3S4,PULS4}.

\subsection{Stochastic Background}
Analogous to the cosmic microwave background of electromagnetic radiation is the stochastic background of gravitational radiation. This may be composed of gravitational waves of cosmological origin as well as of astrophysical origin. The latter is a random superposition of weak signals from supernovae, binary coalescences, and rotating neutron stars. Detection of gravitational waves of a cosmological origin would provide a unique opportunity to explore the early universe, as other forms of radiation, such as electromagnetic or neutrino, cannot probe such early times. Several searches for a stochastic gravitational background with LIGO data have been completed; see \cite{StocS1,StocS3,StocS4,StocS4b,StocS4c} for more details.

\chapter{Black Holes}
\label{ch:astro}

\section{Introduction}
This chapter is concerned with theoretical and astrophysical black holes. We discuss the solution to the Einstein equation for perturbed black holes and the analytic waveform of the emitted gravitational radiation far from the source, the -- ringdown waveform. This motivates the search for ringdowns in LIGO data described in later chapters. We discuss astrophysical black holes and outline previous searches for ringdowns.


\section{Theoretical Black Holes}

The first reference to objects now known as black holes came from the British
geologist John Michell in 1784 \cite{Michell}. In a letter to Henry Cavendish
describing a method of determining a star's distance, magnitude, and mass, he
discusses the possibility of a star with such a large gravitational force that
light would be prevented from escaping its surface:
\begin{quote}
``If the semi-diameter of a sphere of the same density as the Sun were to
exceed that of the Sun in the proportion of 500 to 1, a body falling from an
infinite height towards it would have acquired at its surface greater velocity
than that of light, and consequently supposing light to be attracted by the
same force in proportion to its vis inertiae (inertial mass), with other
bodies, all light emitted from such a body would be made to return towards it
by its own proper gravity.''
\end{quote}
It took another hundred and thirty two years for this notion to be revisited.
Soon after Einstein published his theory of general relativity in 1915 Karl
Schwarzschild found an exact solution to the Einstein equation for the geometry
outside of a non-spinning spherically symmetric star \cite{Schw16b}. He found that if a star of mass $M$ is confined to a radius $R=2GM/c^2$, electromagnetic radiation is infinitely red-shifted and the star appears dark. This radius later
became known as the Schwarzschild radius. In 1939, Oppenheimer and Snyder performed the first rigorous calculation demonstrating the formation of a black hole from the implosion of an idealized star using the formalism of general relativity \cite{Oppe39}. The name black hole itself was coined by John Wheeler in 1968 \cite{Whee68}.

General relativity tells us that a black hole is a region of spacetime where the
gravitational field is so powerful that nothing, not even light can escape. At
the center is the singularity, a point of zero volume and infinite density
where all of the black hole's mass is located. Spacetime is infinitely curved
at this point. The singularity is enclosed by the event horizon. A black hole
can be completely specified by three parameters: its mass, spin and charge.
All observable properties of the black hole depend only on those three
parameters; this is the so-called ``no hair'' theorem \cite{Whee71}.

The geometry outside of a non-spinning black hole is given by the
Schwarzschild metric,
\begin{equation}
ds^2=- \left( 1-\frac{2GM}{c^2r }\right) c^2 dt^2 +\left(1-\frac{2GM}{c^2r}
\right)^{-1}dr^2+r^2d\theta ^2 + r^2 \sin ^2 \theta d\phi ^2
\end{equation}
and the geometry outside of an uncharged spinning black hole with angular momentum $J$ is described, in terms of the Boyer-Lindquist coordinates $\left(t,r,\theta,\phi \right)$, by the Kerr metric,
\begin{eqnarray}
ds^2&=& -\left(1-\frac{2GMr}{c^2\rho^2}\right) c^2 dt^2 - \frac{4GMar \sin^2
\theta}{c\rho^2} d\phi dt +\frac{\rho^2}{\triangle}dr^2 \nonumber \\
& & +\rho ^2 d\theta ^2 +\left(r^2+a^2+ \frac{2GMra^2 \sin^2
\theta}{c^2\rho^2} \right) \sin ^2 \theta d\phi ^2,
\end{eqnarray}
where the spin $a \equiv J/cM$, $\rho^2 \equiv r^2+a^2\cos^2 \theta$ and $\triangle
\equiv r^2-2GMr/c^2+a^2$.

From this point forth when talking about spin we refer to the dimensionless spin parameter $\hat{a}=Jc/GM^2$. This ranges between 0 for a Schwarzschild black hole and 1 for an extreme Kerr black hole. It is related to the spin $a$ defined above by $\hat{a}=ac^2/GM$.


\section{Quasi-Normal Modes of Black Hole Oscillation}
\label{sec:QNM}

An astrophysical black hole can become perturbed by a number of processes, for example by a massive object falling into it, by the merger of two black holes, or in its formation through the asymmetric core collapse of a massive star. In this section we discuss the emitted gravitational waveform.

\subsection{Schwarzschild Black Holes}  
In 1957 Regge and Wheeler \cite{Regg57} investigated the stability of the Schwarzschild black hole to small perturbations. Their study found that a disturbance of the black hole from sphericity would not grow with time, but would oscillate about the equilibrium configuration in a superposition of quasi-normal modes. They found that the solution to the linearized Einstein equation could be expressed in terms of spherical harmonics $Y_{lm}$. Each mode has a characteristic complex angular frequency $\omega_{lm}$; the real part is the angular frequency and the imaginary part is the inverse of the damping time $\tau$. In subsequent sections we will express $\omega_{lm}$ in terms of the oscillation frequency $f_{lm}$ and the quality factor $Q_{lm}$:
\begin{equation}
\omega_{lm} =2 \pi f_{lm}- i \tau_{lm}^{-1} = 2 \pi f_{lm} - i \frac{\pi f_{lm}}{Q_{lm}}.
\end{equation}

\begin{figure}[h]
\centering
\begin{center}
\includegraphics[scale=0.6]{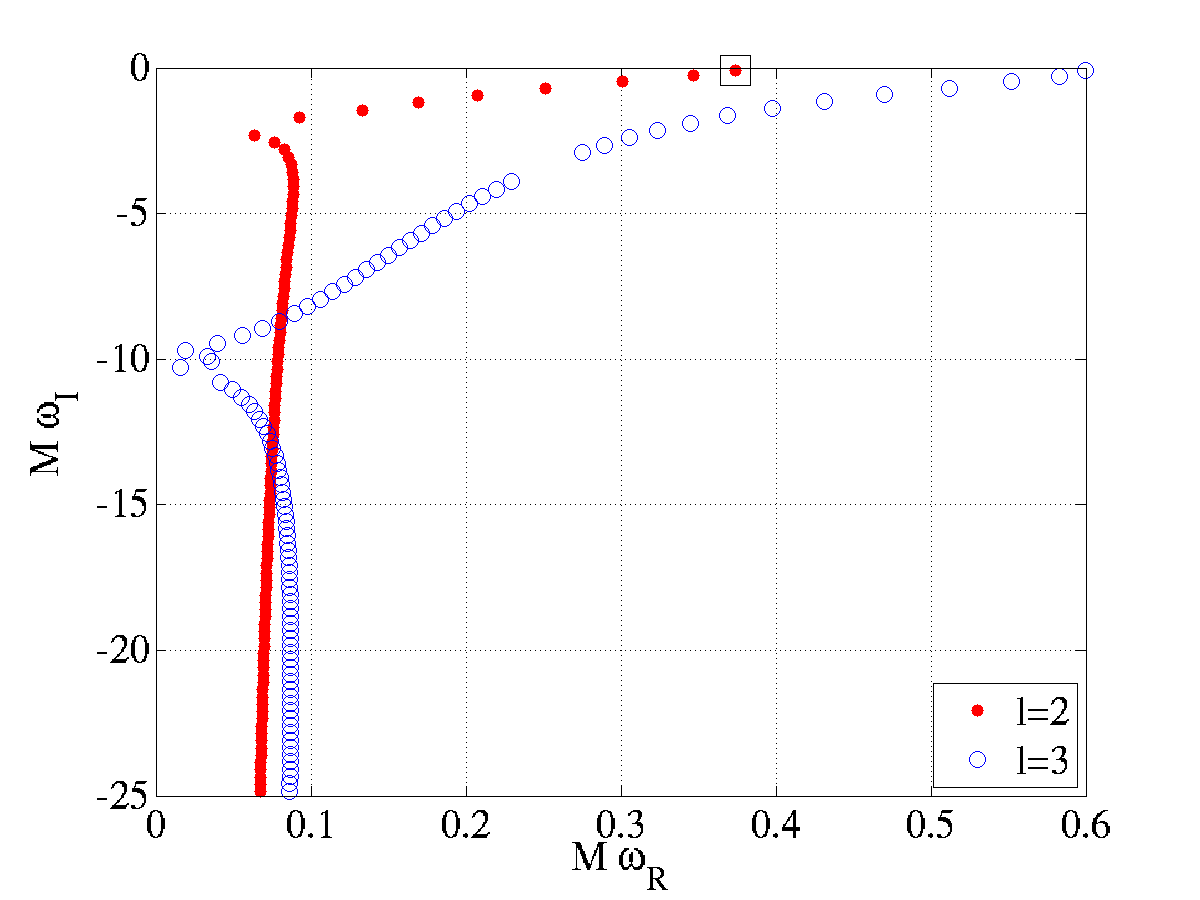}
\caption{The imaginary part of the frequency $\omega_I$ versus the real part
$\omega_R$ for overtones of the $l=2$ and $l=3$ modes of a Schwarzschild
black hole. The most slowly damped mode has $l=2$, $n=0$ is marked by a black
box. (The data in the plot is from \cite{Leav85} and \cite{Bert04}.)}
\label{fig:Schwarzmodes}
\end{center}
\end{figure}

In 1985 Leaver \cite{Leav85} determined the fundamental $l=2$ and $l=3$ modes, as well as the first 62 overtones, indexed by $n$. The first of these are listed in table \ref{tab:Leavermodes} \cite{Leav85}. The most slowly damped mode (i.e., that with the lowest value of the imaginary part of the frequency) was found to be the $l=2$, $n=0$ mode. Figure \ref{fig:Schwarzmodes} \cite{Leav85,Bert04} shows the real part versus the imaginary part of the frequency, for a selection of the $l=2$ and $l=3$ modes. The figure demonstrates that the imaginary part of the frequency grows very quickly with $n$ indicating that higher-order modes do not contribute significantly to the emitted gravitational radiation. In contrast, the real part of the frequency asymptotes to a constant value. The $l=2$, $n=0$ mode is marked with a box in the figure. It was verified in 1993 that an infinity of these modes exist \cite{Bach93}.  

\begin{center}
\begin{small}
\begin{table}[htdp]
\centering
\caption{Quasi-normal modes of oscillation for a non-spinning black hole \cite{Leav85}.}
\label{tab:Leavermodes}
\begin{tabular}{cccc}
\hline \hline
$l$ & $n$ & $M\omega$  \\
\hline
2 & 0 & $0.3737 - \imath 0.0890$ \\
  & 1 & $0.3467 - \imath 0.2739$ \\
  & 2 & $0.3011 - \imath 0.4783$ \\
3 & 0 & $0.5994 - \imath 0.0927$ \\
  & 1 & $0.5826 - \imath 0.2813$ \\
  & 2 & $0.5517 - \imath 0.4791$ \\
\hline \hline
\end{tabular}
\end{table}
\end{small}
\end{center}

\subsection{Kerr Black Holes}
In 1973 Teukolsky {\cite{Teuk73} addressed the problem of perturbations of a rotating black hole. In this case the linear equations describing the gravitational perturbations were decoupled into spin-weighted spheroidal  harmonics ${}_sS_{lm}$, where the spin weight $s$ is -2 for gravitational  perturbations. In the same study as referenced above, Leaver presented the  $l=2$ modes for different $m$ and spin, showing that spin removed the $2l+1$  degeneracy in $m$. This is demonstrated in figure \ref{fig:Kerrmodes} \cite{Bert05} where the spin of the $l=2$,  $n=0$ mode (shown as a single point in figure \ref{fig:Schwarzmodes}) is allowed to vary from $0\leq  \hat{a} <1$ resulting in five different quasi-normal frequencies for each value of $\hat{a}$. The point where the five lines converge has $\hat{a}=0$. In figure \ref{fig:KerrREMOva} we plot the real part of the frequency as a function of the spin for the $l=3$ and $l=2$ modes. Note that for the $l=m=2$ mode $M\omega_R$ ranges from 0.37 for a non-spinning black hole to 0.9 for a maximally spinning black hole. In figure \ref{fig:KerrREMOva} we plot the real part of the frequency as a
function of the spin for the $l=3$ and $l=2$ modes.

\afterpage{\clearpage}
\begin{figure}[h]
\centering
\begin{center}
\includegraphics[scale=0.55]{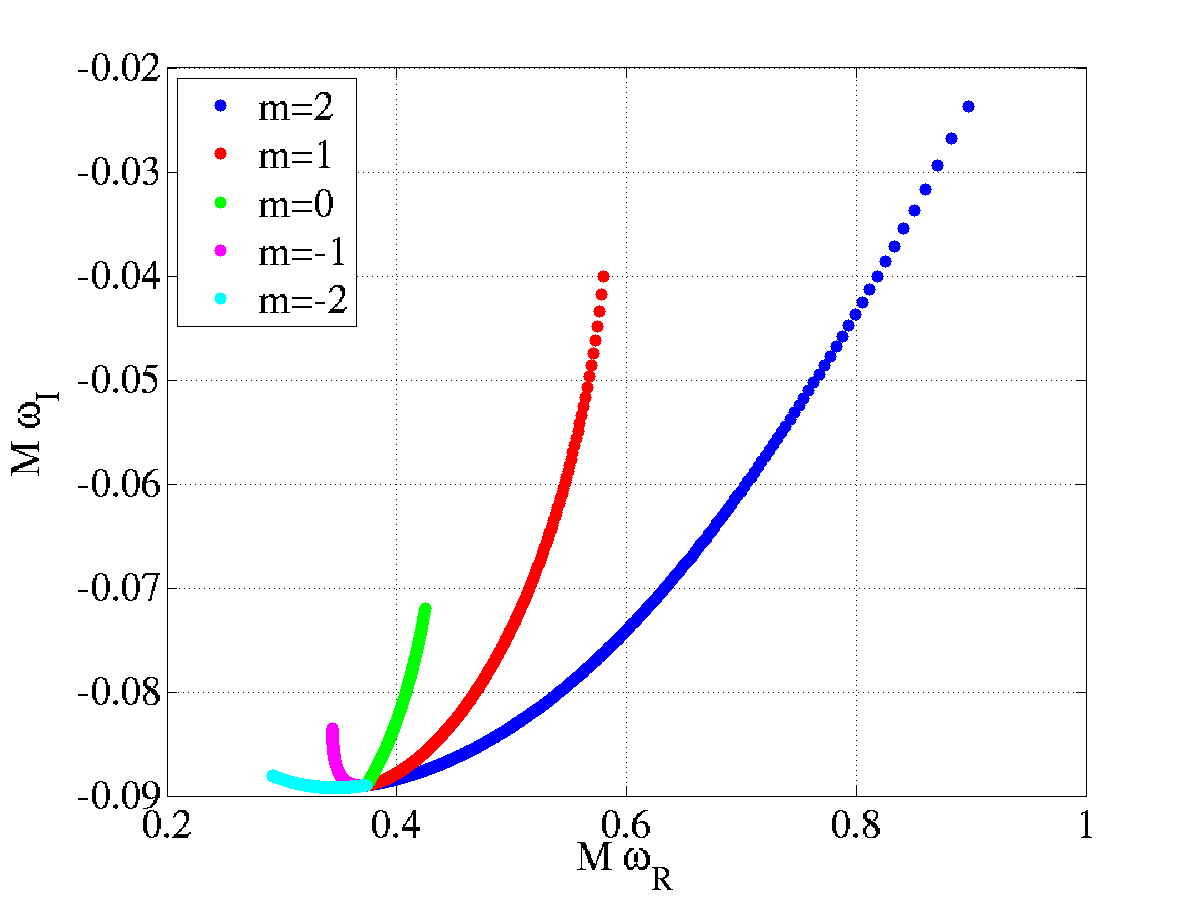}
\caption{A demonstration of how spin removes the degeneracy in $m$ for the $l=2$ mode. The cyan, magenta, green, red, and blue sets of points correspond to $m=-2$, $m=-1$, $m=0$, $m=1$, and $m=2$, respectively. The frequency where the points converge corresponds to $\hat{a}=0$. (The data in this plot is from \cite{Bert05}.)}
\label{fig:Kerrmodes}
\end{center}
\end{figure}
\begin{figure}[h]
\centering
\begin{center}
\includegraphics[scale=0.55]{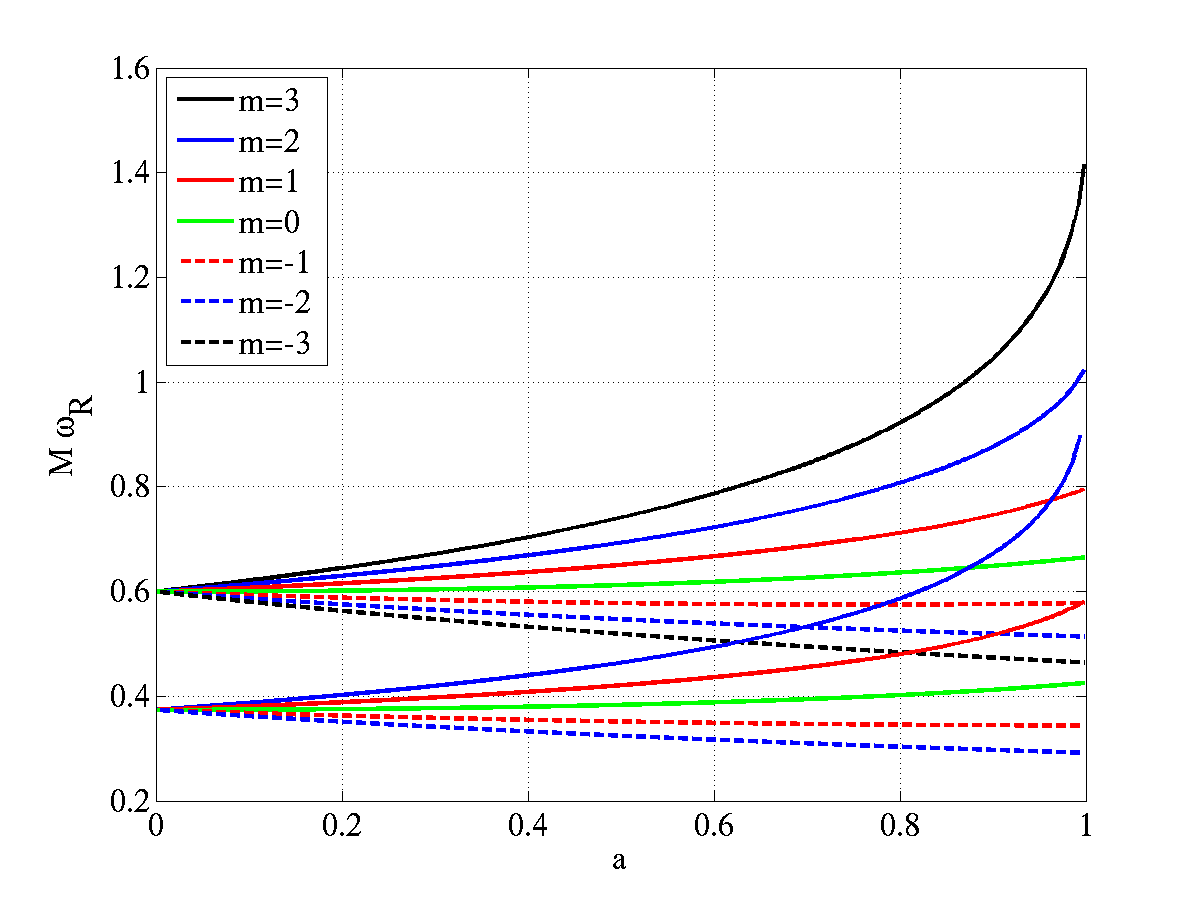}
\caption{The real part of M$\omega$ as a function of spin for the $l=3$ (upper
group) and $l=2$ (lower group) modes. (The data in this plot is from \cite{Bert05}.)}
\label{fig:KerrREMOva}
\end{center}
\end{figure}

Other major contributions to the understanding of the quasi-normal modes of a black hole came from Vishveshwara \cite{Vish70}, Zerilli \cite{Zeri70}, Press \cite{Pres71}, Price \cite{Pric72}, Chandrasekhar and Detweiler \cite{Detw75}, Ferrari and Mashhoon \cite{Ferr84}. A nice review of quasi-normal modes can be found in \cite{Kokk99}.


\subsection{The $l=m=2$ Mode}

\subsubsection{The Black Hole Physical Parameters}

Echeverria \cite{Eche89} found an analytic fit to Leaver's calculations, relating the complex frequency of the $l=m=2$ mode to the black hole's physical parameters mass $M$ and dimensionless spin parameter $\hat{a}$,
\begin{eqnarray}
f_{220}&=&\frac{1}{2 \pi} \frac{c^3}{GM}\left[ 1-0.63\left( 1-\hat{a}\right)
^{\frac{3}{10}}\right]  \label{eqn:Echeverria_fofMa} \\
Q_{220}&=&2\left( 1-\hat{a}\right) ^{-\frac{9}{20}}.
\label{eqn:Echeverria_Qofa}
\end{eqnarray} 
The inverse of these equations is given by
\begin{eqnarray}
M&=&\frac{1}{2\pi} \frac{c^3}{G f_{220}} \left[ 1-0.63 \left(
\frac{2}{Q_{220}}\right)
^{\frac{2}{3}}\right] \label{eqn:Echeverria_MoffQ} \\
\hat{a}&=&1-\left( \frac{2}{Q_{220}}\right) ^{\frac{20}{9}}.
\label{eqn:Echeverria_aofQ}
\end{eqnarray}
Note that the spin of the black hole depends only on the quality factor, as
shown in figure \ref{fig:EcheQva}, whereas the mass depends on both
quality and frequency. Figure \ref{fig:EchefvM} shows the frequency as a
function of mass for three spin values, $\hat{a}={0,0.5,0.98}$. The mass range
reflects the sources that LIGO is most sensitive to.

\afterpage{\clearpage}

\begin{figure}[h]
\centering
\begin{center}
\includegraphics[scale=0.6]{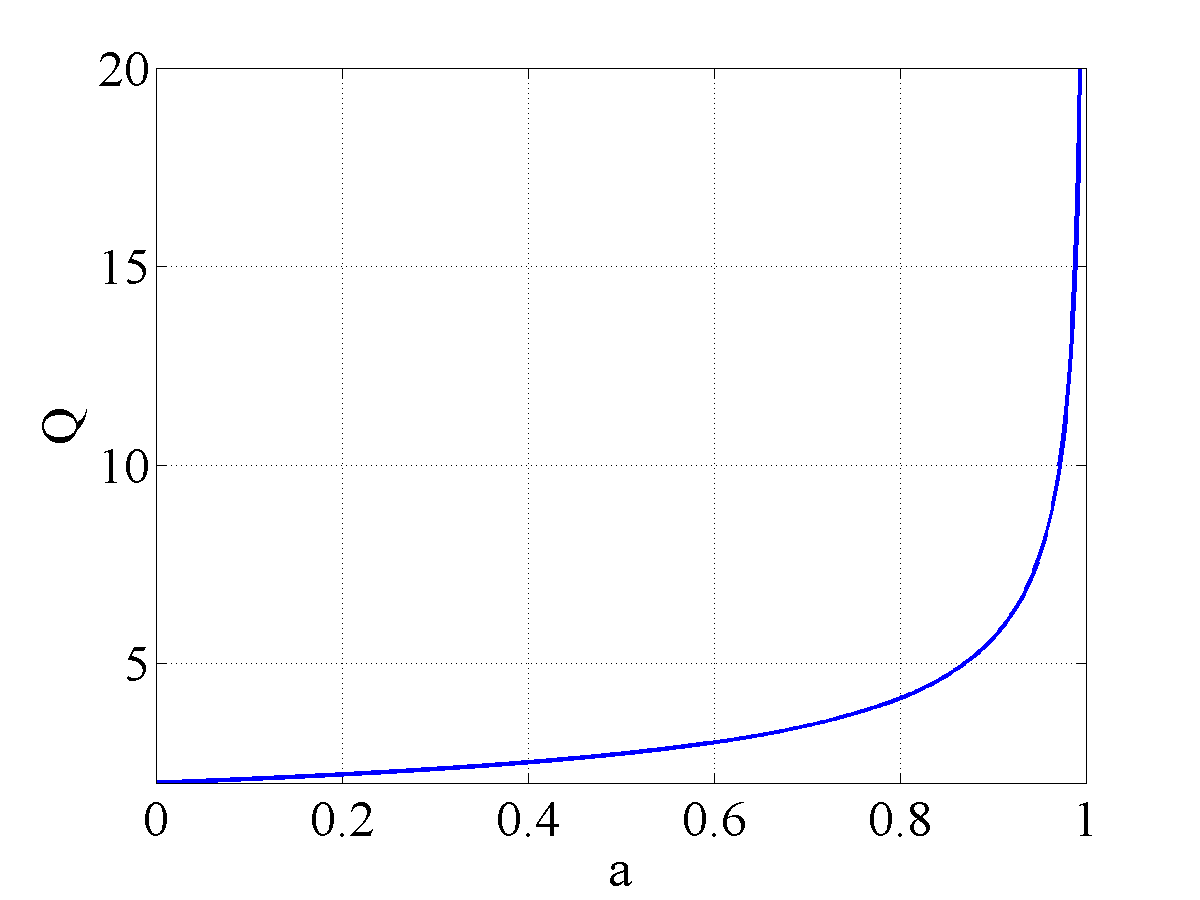}
\caption{Quality factor versus dimensionless spin factor for the $l=m=2$ mode.}
\label{fig:EcheQva}
\end{center}
\end{figure}

\begin{figure}[h]
\centering
\begin{center}
\includegraphics[scale=0.6]{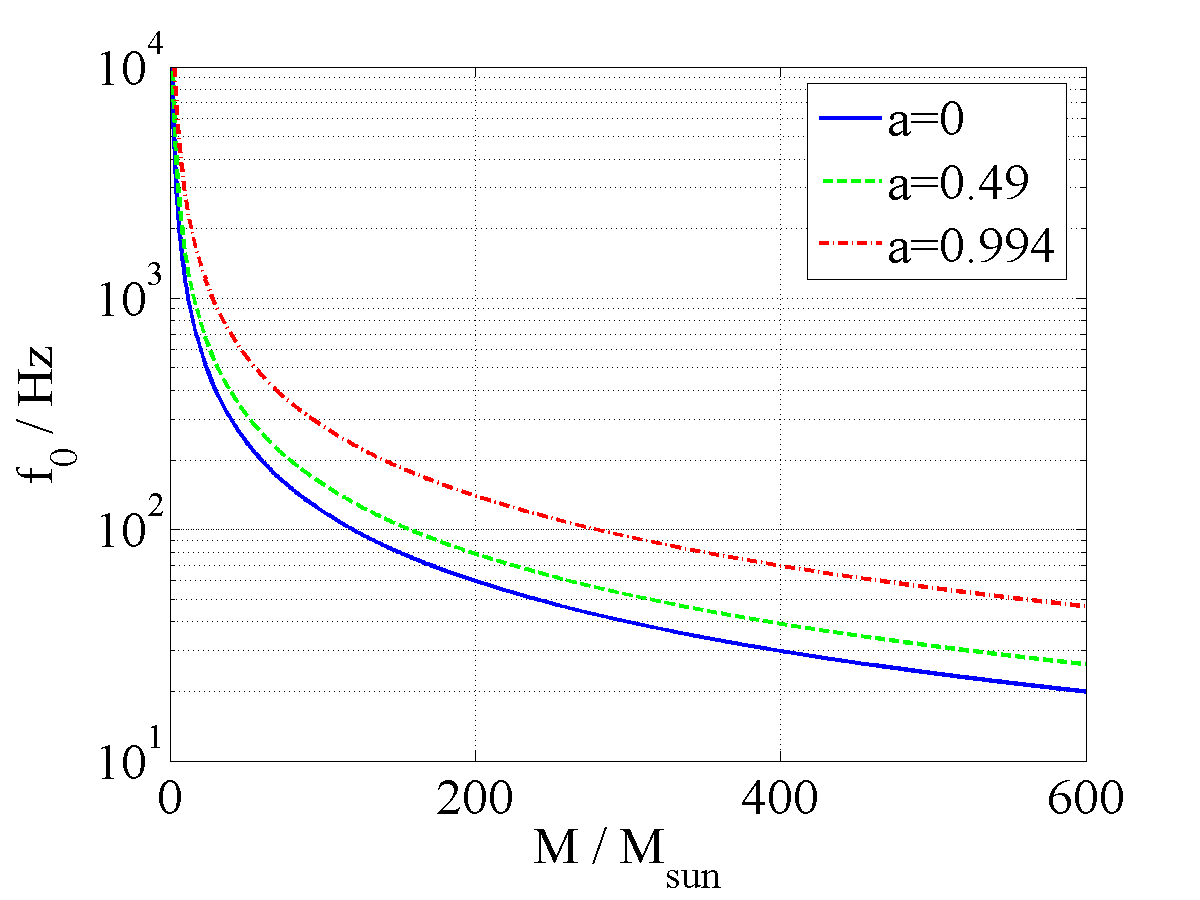} 
\caption{Frequency versus mass for $\hat{a}=0$ (blue line), $\hat{a}=0.5$ (green dashes), and $\hat{a}=0.98$ (red dash-dot) for the $l=m=2, n=0$ mode.} 
\label{fig:EchefvM}
\end{center}
\end{figure} 

We can use a simple model from continuum wave mechanics to predict the frequency for a given mass. Taking the wavelength $\lambda$ to be the circumference of the black hole at the Schwarzschild radius, $\lambda=2\pi\left(2GM/c^2\right)$, the frequency is given by
\begin{equation}
f_0=\frac{c}{\lambda}= \frac{c^3}{4\pi G M},
\end{equation}
which is the same order of magnitude as equation
(\ref{eqn:Echeverria_fofMa}). The quality factor is related to the rate of
dissipation of energy,
\begin{equation}
\frac{2\pi f_0}{Q} =\frac{\textrm{d}\epsilon /\textrm{d}t}{\epsilon}.
\end{equation}
Energy is lost from the perturbation due to gravitational waves
escaping to infinity or falling in to the hole. We see from equation (\ref{eqn:Echeverria_Qofa}) that the quality factor grows with the spin $\hat{a}$.
One explanation for this is that as the black hole spins, the spin energy couples to the perturbation, amplifying it and decreasing the damping time. This is analogous to the r-mode instability in rotating neutron stars \cite{Ande01}.

From this point on we assume that the gravitational wave far from the source is dominated by the most slowly damped mode, $l=m=2$ and neglect any contributions from higher-order modes. We write the central frequency of the waveform as $f_0$ and the quality as $Q$.

\subsubsection{The Ringdown Waveform}
Far from the source the waveform can be approximated by
\begin{equation}
h_0(t)= \Re \left\{ \frac{\mathcal{A}}{r} e^{-\imath \omega t}\right\} = \Re\left\{\frac{\mathcal{A}}{r} e^{-\imath \left(2 \pi f_0 - \imath \pi f_0/Q \right)t} \right\}
\end{equation}
where $\mathcal{A}$ is the amplitude of the $l=m=2$ mode and $r$ is the distance from the source. This is usually expressed as
\begin{equation}
h_0(t)=\frac{\mathcal{A}}{r} e^{- \frac{\pi f_0}{Q} t} \cos \left( 2 \pi f_0 t\right)
\end{equation}
and this is the form we will use in subsequent chapters. An example of three
ringdown waveforms with $f_0=100$ Hz and $Q=2,10,20$ for a source at a
distance of 100 Mpc is shown in figure \ref{fig:waveform}.
The plus and cross polarizations of the wave are
\begin{eqnarray}
h_+(t) &=& \left(1+\cos^2 \iota \right) h_0(t) \\ \label{eqn:hplus}
h_\times (t) &=& 2 \cos \iota \; h_0(t) \label{eqn:hcross}
\end{eqnarray}
where $\iota$ is the inclination angle of the source. The strain produced in the detector is then
\begin{equation}
h(t)=h_+(t)F_+(\theta,\phi,\Psi)+h_\times(t)F_\times(\theta,\phi,\Psi)
\label{eqn:hpFphcFc}
\end{equation}
(as described in section \ref{sec:antresp}). 

\begin{figure}[h]
\centering
\begin{center}
\includegraphics[scale=0.6]{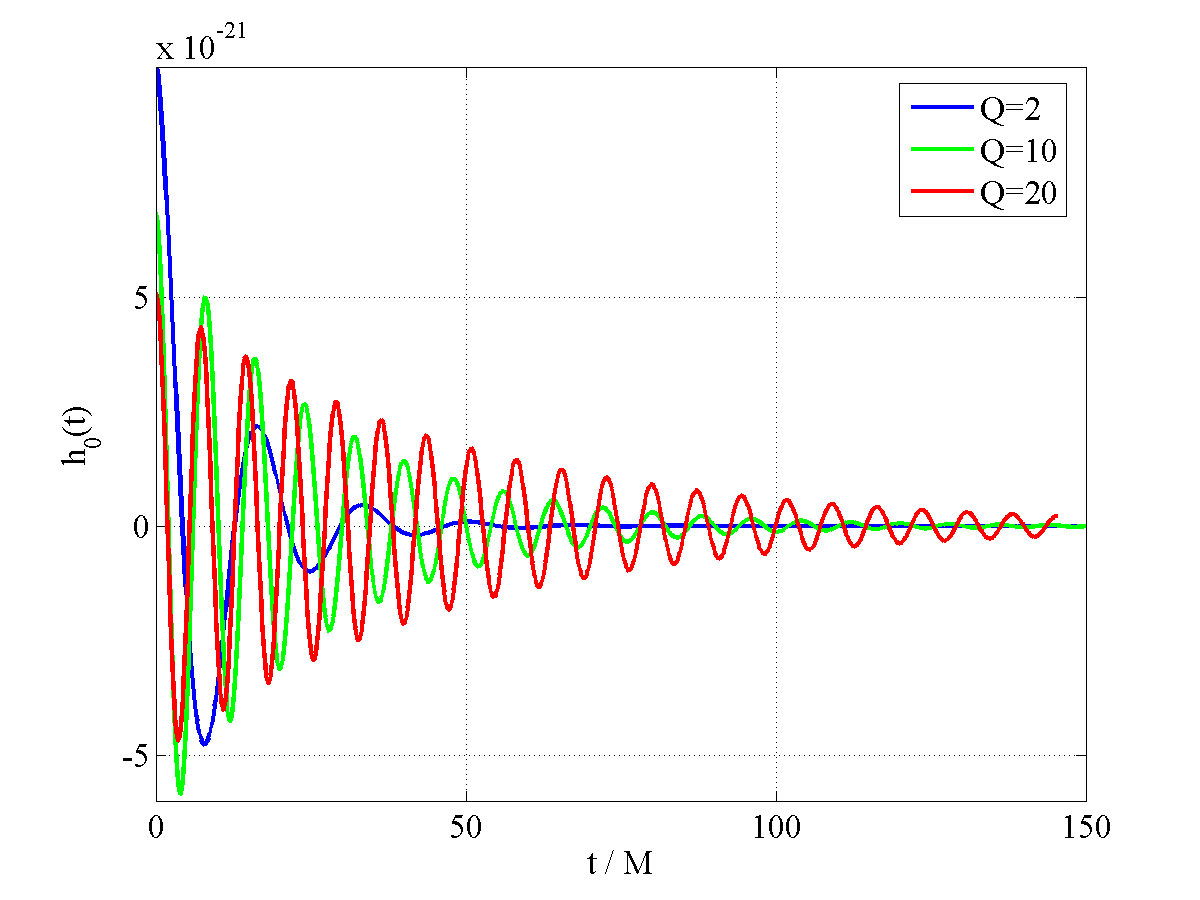}
\caption{The ringdown waveform produced by a source radiating 1\% of its mass
in gravitational wave located at a distance of 100 Mpc with frequency
of 100 Hz and quality factor of 2 (blue), 10 (green), and 20 (red).}
\label{fig:waveform}
\end{center}
\end{figure}

\subsubsection{The Ringdown Peak Amplitude}

We can evaluate $\mathcal{A}$ from the stress-energy tensor, equation (\ref{eqn:energydensity}). If $\epsilon$ is the fraction of the black hole's mass radiated as gravitational waves, then 
\begin{equation}
\epsilon Mc^2=\int_V T_{00}\; \textrm{d}V = 
                  \frac{1}{16 \pi} \frac{c^2}{G} \int_V \;\textrm{d}V \, 
                  (\dot{h}_{+}^{2}(t)+\dot{h}_{\times}^{2}(t)).
\end{equation}
Solving this equation for $\mathcal{A}$ gives
\begin{equation}
\mathcal{A}=\sqrt{\frac{5}{2}\epsilon}\left(\frac{GM}{c^2}\right)
Q^{-\frac{1}{2}}F(Q)^{-\frac{1}{2}} g(a)^{-\frac{1}{2}}
\label{eqn:amplitude}
\end{equation}
where
\begin{eqnarray}
F(Q)&=& 1+\frac{7}{24 Q^2} \nonumber \\
g(a) &=& 1-0.63 \left( 1-a \right) ^{3/10}. \nonumber
\end{eqnarray}


\subsection{Energy Emitted as Gravitational Waves}

The amount of energy emitted as gravitational waves during the ringdown phase, $\epsilon$, depends on the magnitude of the perturbation. For the example of a mass $m$ falling into a black hole of mass $M$ it is reasonable to expect that the energy released is proportional to some function of the ratio $m/M$. This was first calculated by Davis, Ruffini, Press, and Price in 1971 \cite{Davi71}, for the case of a mass $m\ll M$ falling into a non-spinning black hole. They found that the energy emitted was given by
\cite{Davi71},
\begin{equation}
\epsilon \approx 0.0104  \Big(\frac{m}{M}\Big)^2,
\end{equation}
with $\sim90\%$ of the radiation emitted in the $l=2$ mode and $\sim8\%$ in the $l=3$ mode.

Flanagan and Hughes \cite{Flan98} estimate an upper limit of 3\% on the energy emitted in the $l=m=2$ mode for the binary coalescence of equal-mass black holes by considering the mode's amplitude when the distortion of the horizon of the black hole is of order unity. For an unequal-mass binary they assume that the amount of energy emitted is reduced by the factor $\left(4\mu/M_T\right)^2$, where $\mu$ is the reduced mass of the binary and $M_T$ is the total mass.

Numerical simulations of binary coalescence can tell us how much energy is radiated at various stages of the evolution. Figure \ref{fig:Buonenergy} \cite{Buon06} shows a plot of $M\omega$ as a function of time for an equal-mass non-spinning binary. The plot is annotated with the amount of energy emitted. From this we can clearly see that the value of $\epsilon$ depends on how we define the start point of the ringdown. For our purposes, an estimate of  $\epsilon=1\%$ is reasonable.
\begin{figure}[h]
\centering
\begin{center}
\includegraphics[scale=0.7]{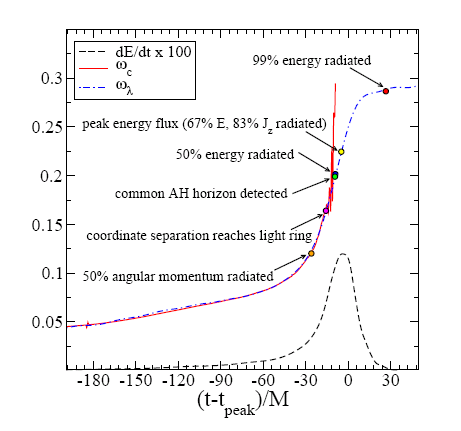}
\caption{$M\omega$ as a function of time for the gravitational wave emitted
as an equal-mass black hole binary undergoes coalescence. The energy emitted is indicated by the annotations. $\omega_c$ and $\omega_\lambda$ correspond to different methods for extracting the frequency from the numerical waveforms. This figure was taken from Buonanno et al. \cite{Buon06}.}
\label{fig:Buonenergy}
\end{center}
\end{figure}


\section{Astrophysical Black Holes}
\label{sec:astro}

Black holes do not emit electromagnetic radiation, and thus cannot be observed with a telescope. However the influence of their strong gravitational field on nearby matter can be observed electromagnetically and it is by this indirect means that astronomers can infer the presence of a dark compact object. If there is sufficient evidence to rule out alternative sources such as a cluster of neutron stars or brown dwarfs it is called a black hole candidate. Only the detection of gravitational waves will provide unambiguous evidence for the presence of a black hole. However the gravitational wave community can benefit from astronomers' observations of the electromagnetic signature of an event that could be accompanied by gravitational waves such as a supernova or gamma ray burst.

Astrophysical black holes have been divided by mass into three categories. Stellar mass black holes, which are believed to form as the end points of stellar evolution, lie in the range $3\leq M/M_\odot \leq 20$; 3 $M_\odot$ is the upper limit on the mass of the neutron star \cite{Kalo96}. Supermassive black holes, the engines behind radio galaxies and quasars, are observed to have masses in excess of $10^6\ M_\odot$ and little is known about their formation \cite{Ferr05}. Recent claims of evidence of black holes with masses in between these categories prompted the creation of a third category, intermediate-mass black holes, with a mass interval roughly defined as $10^2\leq M/M_\odot\leq 10^5$. It remains to be seen whether these are the high (low) mass end of the stellar mass (supermassive) black hole population or distinct population.  An excellent review of intermediate-mass black holes may be found in \cite{Mill04}.


\section{Previous Searches for Gravitational Waves from Perturbed Black Holes}
The first LSC ringdown search was carried out by Creighton \cite{Crei99} on
data from the LIGO 40 m prototype in 1994 using the GRASP software
\cite{Alle97}. A single filter was used to demonstrate the implementation of
matched filtering in a gravitational wave search. Although the poor detector sensitivity made detection extremely unlikely, the methods used laid the foundations for
subsequent searches, including the analysis described in later chapters. The
method of coincidence analysis was demonstrated by dividing the data set in
two and treating one half as if it originated from a second detector located
3000 km from the first.

In 2004 Adhikari \cite{Adhi04} performed a matched filter ringdown search with
coincidence analysis from the 300-hour-long second LIGO science run (S2) using
the LIGO Algorithm Library (LAL) \cite{LAL} software package. In this analysis
simulated signals were injected (in software) into the data stream and a
detection efficiency was calculated as a function of strain. For the most
sensitive band of the detector at Livingston, 150--450 Hz, the 50\% efficiency
was located at a peak strain of $5\times 10^{-20}$.

The TAMA collaboration also carried out a search for ringdowns using data from
their 300 m interferometer \cite{Tsun04}. In total they analysed 1000 hours of
data from their sixth and eighth science runs in 2001 and 2003,
respectively. The detectors were maximally sensitive to ringdowns  at a
frequency of $\sim 1$ kHz at a strain of $\sim5\times10^{-21}$. This was
approximately a factor of 2 more sensitive at 1 kHz than the S2 LIGO data, but
at least an order of magnitude less sensitive than LIGO at 200 Hz for the S2 run.

The search described in this thesis was carried out on data from the fourth
LIGO science run (S4) which amounted to $\sim360$ hours of triple coincident
data at significantly improved strain sensitivity than any of the previous
searches. Matched filtering was implemented using the LAL software. We performed a coincidence analysis between each of the three LIGO detectors and, using a
study of simulated signals, present an upper limit on the rate of ringdowns in
the local universe.



\chapter{Matched Filtering}
\label{ch:matchedfilter}

\section{Introduction}

In chapter \ref{ch:GR} we described the generation and propagation of gravitational waves from perturbed black holes and the general form of the wave far from the source. We discussed the output of the detectors which may or may not contain a signal buried in the noise. This chapter is concerned with how to uncover such a signal. When the signal is known, the optimal method of extracting the signal from Gaussian noise is matched filtering \cite{Turi60,Thor08,Kuma05}; we will demonstrate this in section \ref{sec:optfilter}.

The convention used for the Fourier transform of a signal $w(t)$ in this analysis is 
\begin{equation}
\tilde{w}(f) = \int_{-\infty}^{\infty} w(t) e^{-2 \pi \imath f t} dt
\label{eqn:fourier}
\end{equation}
and for the inverse Fourier transform is
\begin{equation}
w(t) = \int_{-\infty}^{\infty} \tilde{w}(f) e^{2 \pi \imath f t} df.
\label{eqn:invfourier}
\end{equation}


\section{Wiener Optimal Filtering}
\label{sec:optfilter}

Consider a detector output $s(t)$ which may or may not contain a weak signal of known form $h(t)$ superimposed on the noise $n(t)$  
\begin{equation}
s(t) = \left\{ \begin{array}{ll} n(t) & \textrm{signal absent}   \\ 
                                 n(t)+h(t) & \textrm{signal present.} 
       \end{array} \right.
\end{equation}
We assume without loss of generality that the signal, if present, occurs at $t=0$. We also assume that the detector output is a stationary random process with zero mean, Gaussian probability distribution, and a one-sided power spectrum $S_n(f)$ defined by 
\begin{equation}
\frac{1}{2} \delta(f-f')S_n(f)=\left< \tilde{n}(f) \tilde{n}(f') \right>.
\end{equation}
As we know the form of the signal we are looking for, the best way to ascertain whether or not it is present in the data is to pass the detector output through a filter $K(t)$. The output $Z$ of the filter is a number given by
\begin{equation}
Z \equiv \int_{-\infty}^{\infty}K(t)s(t) dt. \label{eqn:filter}
\end{equation}
Our aim is to choose $K(t)$ such that $Z$ will have a large value if the signal is present and a small value if it is not. We define
\begin{equation}
H\equiv \int_{-\infty}^{\infty}K(t)h(t)dt \qquad \textrm{and} \qquad N\equiv \int_{-\infty}^{\infty}K(t)n(t)dt
\end{equation}
where 
$H$ is the filtered signal, $N$ is the filtered noise and
\begin{equation}
Z=H+N
\end{equation}
if a signal is present. Note however that whereas $h(t)$ is a well-defined signal with finite duration, $n(t)$ is a random process
\begin{equation}
N(t)\equiv \int_{-\infty}^{\infty} K(t-t')n(t')dt'
\end{equation}
and thus we average over an ensemble of instantiations of the noise,
\begin{equation}
\left< N^2 \right>=\int_0^\infty |\tilde{K}(f)|^2 S_n(f) df.
\end{equation}
Using the convolution theorem and the fact that $K(t)$ and $h(t)$ are real (and so $\tilde{K}(-f)=\tilde{K}^*(f)$ and $\tilde{h}(-f)=\tilde{h}^*(f)$) we can write the filtered signal as
\begin{equation}
H=2 \int_0^\infty \tilde{K}^*(f)\tilde{h}(f) df.
\end{equation}
Next we define a statistic, the signal-to-noise ratio (SNR) $\xi$, and aim to find a filter $K(t)$ that maximizes this quantity:
\begin{eqnarray}
\xi& \equiv &\frac{H^2}{\left<N^2\right>} \\ 
&=&\frac{4 \left| \int_{0}^{\infty} \tilde{K}^*(f)\tilde{h}(f) df \right|^2}
      { \int_{0}^{\infty} \left|\tilde{K}(f)\right|^2 S_n(f) \; df } \\
&=&\frac{4 \left| \int_{0}^{\infty} \left[\tilde{K}^*(f) \sqrt{S_n(f)}\right]
     \left[\tilde{h}(f) / \sqrt{S_n(f)}\right] df \right|^2}
     { \int_{0}^{\infty} \left|\tilde{K}(f)\right|^2 S_n(f) \; df }.
\label{eqn:stat}
\end{eqnarray}
The Cauchy-Schwarz inequality tells us that for two arbitrary functions $A(f)$ and $B(f)$
\begin{equation}
\left| \int A(f) B(f) df \right|^2 \leq \int \left|A(f)\right|^2 df \int \left|B(f)\right|^2 df.
\label{eqn:schwarz}
\end{equation}
Identifying $\tilde{K}(f) \sqrt{S_n(f)}$ with $A(f)$ and $\tilde{h}(f) /
\sqrt{S_n(f)} $ with $B(f)$ we can see that to attain the maximum value of $\xi$, 
we need the equality in equation \ref{eqn:schwarz} to hold, which only occurs
when $A$ and $B$ are equal up to a constant $C$, thus,
\begin{equation}
\tilde{K}(f) \sqrt{S_n(f)} = C \frac{ \tilde{h}(f) }{ \sqrt{S_n(f)} }
\end{equation}
or
\begin{equation}
\tilde{K}(f) = C \frac{ \tilde{h}(f) }{ S_n(f) }.  \label{eqn:optfilter}
\end{equation}
Thus the optimal filter for detecting signals of known form in coloured Gaussian noise is the Fourier transform of the signal $\tilde{h}(f)$ weighted by the inverse of the power spectrum. 


\section{Detection Statistic for Gravitational Wave Searches}
\label{sec:filterstat}
We can employ the method of matched filtering in the search for gravitational waves from perturbed black holes, as the waveform is known; it is a damped sinusoid,
\begin{equation}
h(t)=\cos\left(2\pi f_0 t\right) \;e^{-\frac{\pi f_0}{Q}t}.
\end{equation}
In this section we derive the statistic to be employed in the ringdown search.

Returning to our initial equation for the filter output equation (\ref{eqn:filter}), and allowing the signal to occur at some unknown time, the filter output is
\begin{equation}
Z(t) = C \int_{-\infty}^{\infty}K(t-t')s(t') dt'.
\end{equation}
Using the convolution theorem this can be expressed as
\begin{equation}
Z(t) = C \int_{-\infty}^{\infty} \tilde{s}(f) \tilde{K}^*(f) e^{\imath 2 \pi f t} df
\end{equation}
and substituting in the expression for the optimal filter, equation (\ref{eqn:optfilter}), we get
\begin{equation}
Z(t) = C \int_{-\infty}^{\infty} \frac{ \tilde{s}(f) \tilde{h}^*(f) e^{ \imath 2 \pi f t}}{ S_n(f) } df.
\end{equation}
At this stage we could choose a threshold value of $|Z|$ above which a signal would be
defined as being present and below which the signal is absent. However rather
than thresholding directly on the filter output we first normalise by the
variance of the optimal filter $\sigma^2$,
\begin{equation}
\sigma^2 = 2 \int_0^{\infty}\frac{ \tilde{h}(f) \tilde{h}^*(f) }{ S_n(f) } df,
\end{equation}
for which $C=2$.
We define a statistic $\rho$, the SNR of the normalize output of the optimal filter, as
\begin{equation}
\rho(t) = \frac{\left|Z(t)\right|}{\sigma}  
\label{eqn:rho}
\end{equation}
and choose a value $\rho^*$ on which to threshold. Thus 
\begin{equation}
\textrm{if} \qquad \rho  \left\{ \begin{array}{ll} < \rho^* & \textrm{the signal is absent}   \\
                           \geq \rho^* & \textrm{the signal is present.}
      \end{array} \right.
\end{equation}
With this comes the possibility of false alarm and false dismissal; the former occurs when $\rho \geq \rho^*$ and no signal is present and the latter occurs when $\rho < \rho^*$ and a signal is present. Thus $\rho^*$ must be chosen carefully so as to minimize the rate of false alarms and false dismissals. As will be described in chapter \ref{ch:s4}, the data we are dealing with in this search is non-Gaussian and non-stationary and so we need to apply further measures to minimize the rate of false alarms and false dismissals.

It is convenient to define the inner product of $a$ and $b$ as
\begin{equation}
(a|b)=2 \int_{-\infty}^\infty \frac{ \tilde{a}(f) \tilde{b}^*(f) }{ S_n(f) } df = 4 \Re \left[ \int_{0}^\infty \frac{ \tilde{a}(f) \tilde{b}^*(f) }{ S_n(f) } df \right],
\label{eqn:innerprod}
\end{equation} 
where $\Re \left[X\right]$ denotes the real part of $X$. This allows us to express $\sigma^2$ as
\begin{equation}
\sigma^2=(h|h)
\end{equation} 
and the SNR as
\begin{equation}
\rho(t)= \frac{1}{\sigma} (s|h(t)) = \frac{(s|h(t))}{\sqrt{(h|h)}}. \label{rhoinnerprod}
\end{equation}


\section{Discrete Quantities}
\label{sec:discretize}
As described in chapter \ref{sec:calibration} the output of the detector is not continuous but a discrete time series sampled every $\Delta t$ seconds. Thus, in order to filter the data we need to modify the expressions described above. First consider time; we have $N$ data points (where we assume $N$ is even) sampled over a time $T$. The discretized time series can be expressed as
\begin{equation}
t_{j}=j \Delta t  \qquad \qquad j=0,1,\ldots,N-1
\end{equation}
and a function of time $w(t)$ when discretized is denoted by $w(t_{j})$. When we Fourier transform a function we seek a discrete frequency array
\begin{equation}
f_{k}=\frac{k}{N \Delta t}, \qquad \qquad k=-\frac{N}{2},\ldots,\frac{N}{2}
\end{equation}
where $k$ is an integer. A continuous function of frequency $w(f)$, once discretized, is written as $w(f_{k})$. Thus the Fourier transform in equation (\ref{eqn:fourier}) can be approximated by the sum
\begin{eqnarray}
\tilde{w}(f) &\approx& \sum_{j=0}^{N-1} w(t_{j}) e^{- \imath 2 \pi f_{k} t_{j}} \Delta t
\nonumber \\ \nonumber\\ 
&=& \sum_{j=0}^{N-1} w_{j}e^{-\imath 2 \pi  \left(k/N \Delta t \right) \Delta t j}
\Delta t \nonumber \\ \nonumber\\
&=& \Delta t \sum_{j=0}^{N-1}w_{j} e^{-\imath 2 \pi jk/N} \nonumber \\
\nonumber\\
&=& \tilde{w}_{k} \Delta t,
\end{eqnarray}
where the quantity $ \tilde{w}_{k}$ is the discrete Fourier transform. Similarly the inverse Fourier transform,  equation (\ref{eqn:invfourier}), can be approximated by the sum
\begin{eqnarray}
w(t) &\approx&\sum_{k=0}^{N-1}\tilde{w}(f_{k})e^{\imath 2 \pi f_{k} t_{j}}  \Delta f
\nonumber \\ \nonumber\\
&=& \sum_{k=0}^{N-1}\Delta t\tilde{w}_{k} e^{\imath 2 \pi \left(k/N \Delta t\right) \Delta t_{j}} \frac{1}{N \Delta t} \nonumber \\ \nonumber\\
&=& \frac{1}{N} \sum_{k=0}^{N-1} \tilde{w}_{k} e^{\imath 2 \pi j k/N},
\end{eqnarray}
where we have used that 
\begin{equation}
\Delta f=f_{k+1}-f_{k}= \frac{k+1}{N \Delta t} -\frac{k}{N \Delta t}
=\frac{1}{N \Delta t}.
\end{equation}
The discrete form of the waveform we are searching for is
\begin{equation}
h(t_j) = \cos \left( 2\pi f_0 j \Delta t \right) e^{-\pi f_0 j \Delta t/Q}.
\label{eqn:distmplt}
\end{equation}
In a similar manner we can express the filter output and the estimated variance as a sum of discrete quantities;
\begin{eqnarray}
Z(t_j) &=&  4 \sum_{k=0}^{\frac{N}{2}-1} \frac{\tilde{s}^{*}(f_{k})\tilde{h}(f_{k})}
                  {S_{n}(f_{k})} e^{- \imath 2 \pi f_{k} t_{j}} \Delta f \nonumber \\
&=&  4 \sum_{k=0}^{\frac{N}{2}-1}\frac{ \Delta t \tilde{s}_{k}^{*}\Delta t
         \tilde{h}_{k}}{S_{n}(f_{k})} e^{- \imath 2 \pi \left(k/N \Delta t\right) \Delta t j}
         \frac{1}{N \Delta t} \nonumber \\
&=&  \frac{4 \Delta t}{N} \sum_{k=0}^{\frac{N}{2}-1} \frac{\tilde{s}_{k}^{*}
          \tilde{h}_{k}}{S_{n}(f_{k})} e^{- \imath 2 \pi j k/N}
          \label{eqn:discfilter}
\end{eqnarray}
and
\begin{eqnarray}
\sigma^{2}&=& 4 \sum_{k=0}^{\frac{N}{2}-1} \frac{ \tilde{h}^{*}(f_{k})\tilde{h}(f_{k})}
          {S_{n}(f_{k})} \Delta f \nonumber \\
&=& 4 \sum_{k=0}^{\frac{N}{2}-1}\frac{\Delta t \tilde{h}_{k}^{*}\Delta t \tilde{h}_{k}}
          {S_{n}(f_{k})}\frac{1}{N \Delta t} \nonumber \\
&=&  \frac{4 \Delta t}{N} \sum_{k=0}^{\frac{N}{2}-1}
      \frac{\tilde{h}_{k}^{*}\tilde{h}_{k}}{S_{n}(f_{k})}.
      \label{eqn:discsigma}
\end{eqnarray}
We will use these equations in chapter \ref{ch:search} when describing the
pipeline.


\section{Templated Matched Filtering Searches}
\label{sec:metric}

In section \ref{sec:optfilter} we demonstrated that the optimum filter to use
in extracting a signal from noise when the signal is known is the matched
filter. However it is often the case, particularly in the gravitational wave
searches that we are concerned with here, that although the form of the signal
is known, the exact values of the intrinsic parameters $\lambda_i$ (these
parameters are central frequency and quality in the ringdown search) are unknown. To
overcome this we can create an array of filters, a template bank, such that
each template has a different value of the intrinsic parameters covering the
space of parameters of interest, and filter the data with each one. Of course,
given that these are discretely placed over the parameter space, it is not
likely that one of the filters will have the exact parameters of the waveform
we are looking for, but if the parameters are close enough, then the SNR will be high and may exceed the threshold.  In practice, templates are
laid out in a very specific way, covering the entire space with as few
templates as possible. A nice discussion of template spacing for gravitational
wave searches can be found in \cite{Owen96}. Some of the main points are
illustrated here. We define the match $M$ between two templates
$\tilde{u}(f;\mu,\lambda)$ and $\tilde{u}(f;\mu+\Delta \mu,\lambda+\Delta
\lambda)$ as the inner product between the two templates maximized over the
extrinsic parameters $\mu$ (such as time of arrival and phase),
\begin{equation} 
M(\lambda,\Delta \lambda) \equiv \mathrm{max}_{\mu,\Delta \mu}
       \left(u(\mu,\lambda)| u(\mu+\Delta \mu,\lambda+\Delta \lambda)\right).
\end{equation}
This is the fraction of the maximum SNR achieved by filtering a signal with a template with the same form but slightly different parameters. Expanding $M$ in a power series about $\Delta \lambda=0$ gives 
\begin{equation}
M(\lambda,\Delta \lambda) \approx 1+\frac{1}{2}\left(\frac{\partial^2 M}
       {\partial \Delta \lambda^i \partial \; \Delta \lambda^j} \right)_{\Delta \lambda^k=0}
       \Delta \lambda^i \; \Delta \lambda^j
\end{equation}
from which we can define the metric
\begin{equation}
g_{ij}(\lambda)=-\frac{1}{2} \left(\frac{\partial^2 M}{\partial
       \Delta \lambda^i \; \partial \Delta \lambda^j} \right)_{\Delta \lambda^k=0}
\end{equation}
so that the mismatch $(1-M)$ between two nearby templates is equal to the square of the proper distance between them
\begin{equation}
ds^2_{ij}=g_{ij} \; \Delta \lambda^i \; \Delta \lambda^j.
\end{equation}
For the ringdown templates the mismatch between two templates differing in frequency by d$f_0$ and in quality by d$Q$ is given by \cite{Alle97}
\begin{eqnarray} ds^2 &=& \frac{1}{8} \Bigg[ \frac{3+16Q^4}{Q^2(1+4Q^2)^2} \:
                        \mathrm{d} Q^2 - 2 \frac{3+4Q^2}{f_0 Q(1+4Q^2)} \:
                        \mathrm{d} Q \:\mathrm{d} f_0 \nonumber \\
                      & & \qquad \qquad +\frac{3+8Q^2}{f_0^2} \: \mathrm{d} f_0^2 \Bigg]. 
\label{eqn:metric}
\end{eqnarray}
We will find it useful when laying out the template bank to define $\phi=\log(f_0)$, as then the metric coefficients no longer depend on $f_0$,
\begin{eqnarray}
ds^2 &=& \frac{1}{8} \Bigg[ \frac{3+16Q^4}{Q^2(1+4Q^2)^2} \:\mathrm{d} Q^2 - 2
\frac{3+4Q^2}{Q(1+4Q^2)} \:
\mathrm{d} Q \: \mathrm{d} \phi + \left(3+8Q^2\right) \: \mathrm{d} \phi^2
\Bigg] \\
&= & g_{QQ} \:\mathrm{d} Q^2 + g_{Q\phi} \:\mathrm{d} Q \:\mathrm{d} \phi
+g_{\phi\phi} \:\mathrm{d} \phi^2.
\label{eqn:metricphi}
\end{eqnarray}
The templates form a two-dimensional lattice whose unit cell has sides of proper length $dl^2$. The highest mismatch will occur for a signal whose parameters lie in the middle of the cell, that is for $ds^2=dl^2/2$.


\chapter{The Search Pipeline}
\label{ch:search}

\section{Overview}
The set of steps we take in analysing data output from the interferometers in order to detect gravitational waves is known as a search pipeline. As discussed in chapter \ref{ch:matchedfilter}, for the case where the waveform is known we implement the method of matched filtering. However, as the noise in the data stream is non-stationary and non-Gaussian, matched filtering alone is not enough to extract a gravitational wave from the noise. Noise can often mimic the signal we are searching for, and so a large effort goes into characterizing the noise to best separate it from a potential gravitational wave signal. We implement several consistency checks on any candidate events to increase our detection confidence.

The ringdown search pipeline is summarized in figure \ref{fig:pipeline}. Each step will be explained in detail in this chapter, but in brief the main steps are as follows: 
\begin{itemize}
\item{Data from each detector is read in from frame files and conditioned.}
\item{The template bank is generated and the data is filtered, yielding a set of trigger files for each detector.}
\item{The triggers from each detector are then brought together and compared (i.e., put through a coincidence test).}
\item{Those triggers failing the coincidence test are discarded and those that pass are followed up on as candidate events.}
\end{itemize}

\begin{figure}[h] 
\centering
\begin{center}
\includegraphics[scale=0.3]{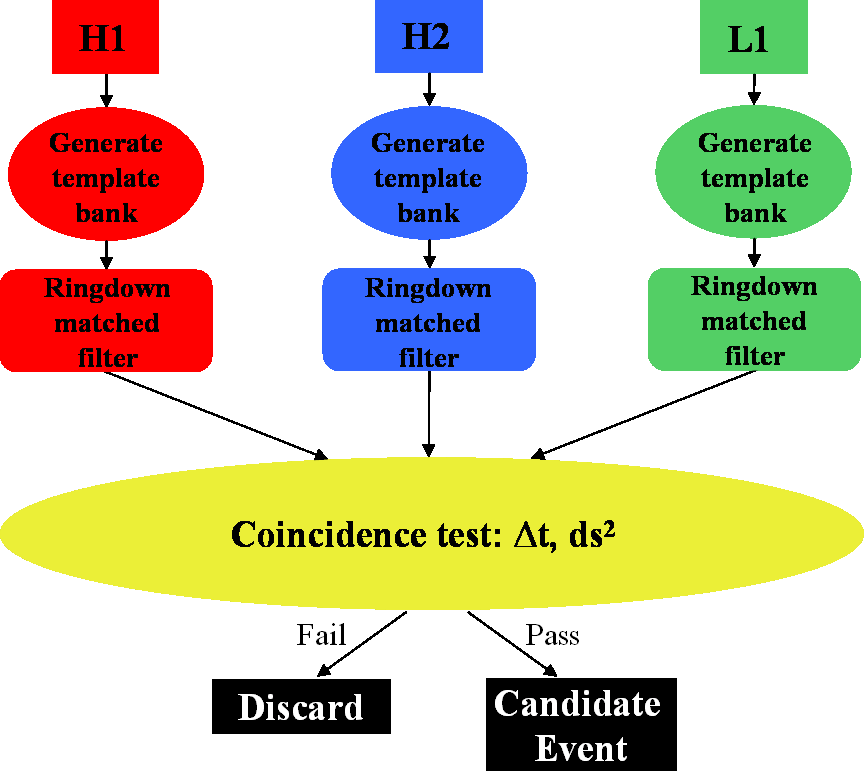}
\caption{The ringdown search pipeline.}
\label{fig:pipeline}
\end{center}
\end{figure}

The pipeline is run on large computer clusters using the Condor high-throughput computing system \cite{Cond}, a specialized workload management system for compute-intensive jobs. The steps in the pipeline can be broken into individual jobs which are scheduled by DAG Man (Directed Acyclic Graph Manager) and submitted to Condor to run in an order specified by a directed acyclic graph (DAG). Condor places the jobs in a queue, chooses what cluster node to run them on and allows the user to monitor their progress.


\section{Preliminaries}

Prior to launching the pipeline three files are required: list of times to be analysed, a list of times to be vetoed, and a configuration file containing arguments required by the search pipeline.

\subsection{Segment Lists}
We begin by generating a list of times for which data is available for each detector. This segment file contains the GPS start time, GPS end time, and duration of each interval of time for which the detector was taking science quality data. A sample of a segment file is shown in table \ref{tab:H1segs}. In generating the list we apply category 1 vetoes which preclude times when the quality of the data was unacceptable from the segment list (see section \ref{sec:dataquality} for more details). 

\begin{center}
\begin{small}
\begin{table}[htdp] 
\centering
\caption{A sample of the H1 S4 segment list}
\label{tab:H1segs}
\begin{tabular}{cccc}
\hline \hline
segment number & GPS start time & GPS end time & Duration  \\
\hline
   1 & 793154953 & 793155973 & 1020 \\
   2 & 793162453 & 793162693 &  240 \\
   3 & 793166413 & 793170673 & 4260 \\
   4 & 793171813 & 793175893 & 4080 \\
   5 & 793176613 & 793179853 & 3240 \\
\hline \hline
\end{tabular}
\end{table}
\end{small}
\end{center}

\subsection{Create the Veto Lists} 
We also create a list of times when the quality of the data was suboptimal due to a known source of noise making the detection of gravitational waves impossible. Triggers during these times are removed before the coincidence stage of the pipeline. These are known as category 2 and 3 vetoes (and are described further in section \ref{sec:dataquality}). 

\subsection{Create the Configuration File}
The configuration file lists all of the arguments needed by the pipeline. All of the parameters mentioned in the following sections, such as thresholds, coincidence windows, clustering windows, and template bank boundaries are specified in this file.
An example of the configuration file used in the S4 search can be found in appendix \ref{app:ini}. 


\section{Launch the Pipeline}
\label{sec:segmentation}
The pipeline is launched by running the python script {\fontfamily{pcr}\selectfont lalapps\_inspiral\_hipe}. This reads in the parameters listed in the configuration file, and generates a set of files with instructions on how the different parts of the pipeline are to run.

The first step is the segmentation of the data. The input lists of variable-length science segments are subdivided into contiguous 2176-s-long analysis segments overlapping each other by 128 s on each end, and written to a file. If there is a non-integer number of 2176 s analysis segments in the science segment, with the remainder $n$ s in length, then the final analysis segment begins (2176-$n$) s earlier, overlapping the previous analysis segment, but only the previously unanalysed data is analysed (see appendix \ref{app:segmentation} for an example of segmentation). 

Files containing instructions on how each of the main jobs, ``datafind'', ``ringdown'', and ``rinca'' is to run are also created;
\begin{itemize}
\item{The datafind job runs {\fontfamily{pcr}\selectfont LSCdataFind} to get the location of the frame files on disk.}
\item{The ringdown job runs the main data conditioning and filtering code \\ {\fontfamily{pcr}\selectfont lalapps\_ring}.}
\item{The rinca job runs the coincidence step of the analysis {\fontfamily{pcr}\selectfont lalapps\_rinca}.}
\end{itemize}
The DAG file is submitted to the Condor pool with the {\fontfamily{pcr}\selectfont condor\_submit\_dag} command. The DAG specifies that the datafind jobs run first, followed by the ringdown jobs and then the rinca jobs.


\section{The Filtering Section of the Pipeline}

Each of the steps described in this section are run on an individual analysis segment basis.

\subsection{Read In and Condition the Data}
\label{sec:readindata}
The uncalibrated data read in from frame files is sampled at 16384 s$^{-1}$; however to reduce the computational cost we re-sample to 8192 s$^{-1}$, and a Butterworth low-pass filter is used to remove any power above the new Nyquist frequency, 4096 s$^{-1}$. The data is then high-pass filtered to remove power below the frequency range of interest, 40 Hz. Although our basic analysis segment is 2176 s in length, an additional 8 s of science data is read in before the start time of the segment and after the end time defining each analysis segment. This padding data is used for these data conditioning steps in order to avoid any corruption of the data in the analysis segment. Once these data conditioning steps have been completed, the 16 s of padding data is removed. No data between contiguous segments is lost, however the first 72 s at the start of a science segment and the last 72 s at the end of the science segment are not used in the search. (An example of segmentation is given in appendix \ref{app:segmentation}.)

\subsection{Calculate the Response Function}

Next the response function $R(f)$ for the segment is calculated. As discussed in section \ref{sec:calibration} the output of the gravitational wave channel is converted to strain via the response function using the calibration coefficients read in for that particular epoch. The numerical value of the response function is very small, $\sim10^{-15}$, and so to save the computational cost of extra precision we scale this quantity by the dynamical range factor $dyn=10^{20}$. The scaled response function, shown in figure \ref{fig:respfn} has units of strain counts$^{-1}$. 

\begin{figure}[h] 
\centering 
\begin{center}
\includegraphics[scale=0.6]{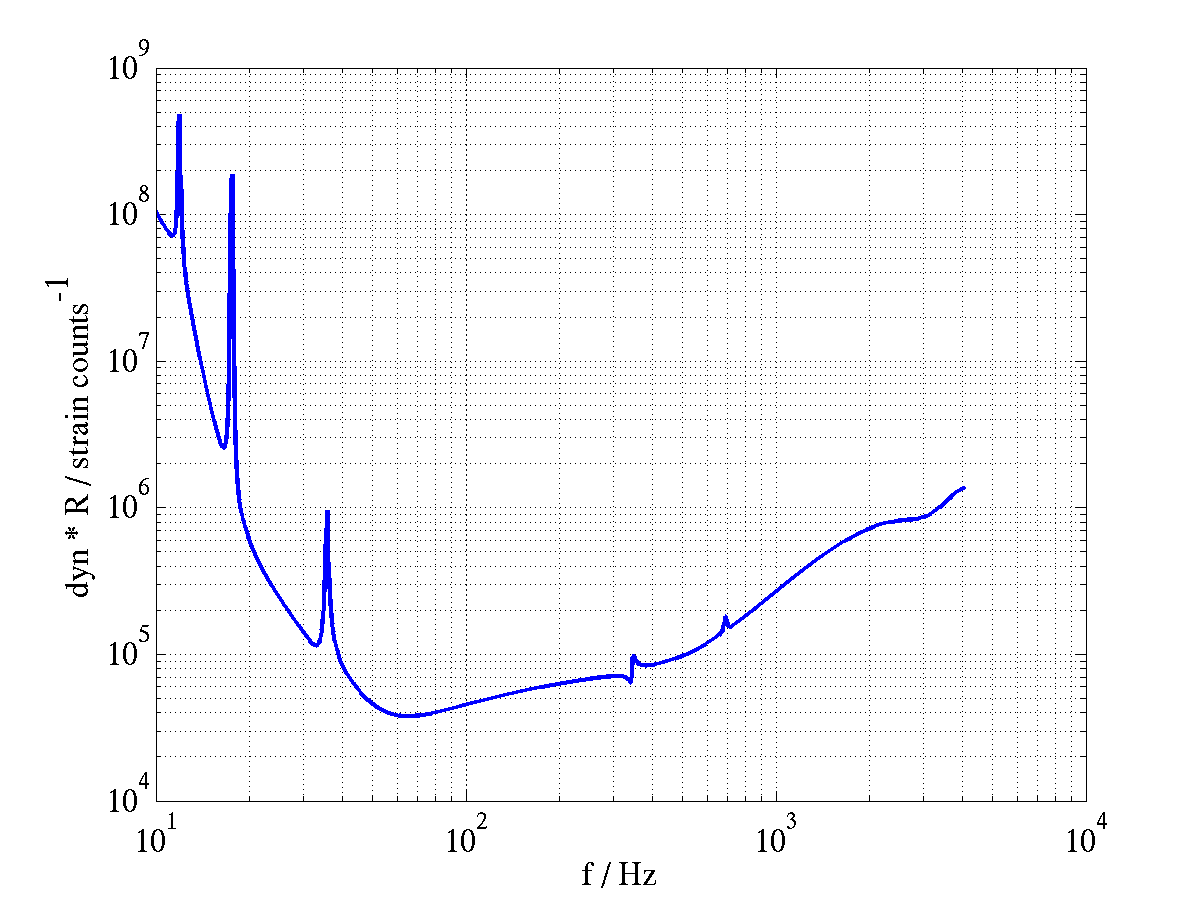}
\caption{The response function (scaled by $dyn$) for H1 between GPS times 793165201 and 793167377.}
\label{fig:respfn}
\end{center}
\end{figure}

\subsection{Calculate the Power Spectrum}
The interferometer noise is characterized by the one-sided power spectral density $S_n(f)$ introduced in chapter \ref{ch:matchedfilter}. For this search it is calculated using the median-mean method. In each 2176 s segment the first and last 64 s are discarded and the remaining 2048 s is split into sixteen 256 s blocks which overlap each other by 128 s. (This segmentation is discussed in more detail in appendix \ref{app:segmentation}.) These are divided (by block number) into even and odd groups and are transformed to the Fourier domain. The bin-by-bin median of the even blocks is calculated, as is the bin-by-bin median of the odd blocks. The uncalibrated spectral density $S_v(f)$  in units of s counts$^2$ is then the bin-by-bin mean of these two medians. In practice it is the inverse of this quantity that is required and so we invert it to get the inverse uncalibrated power spectrum $S_v^{-1}$, with units of s$^{-1}$ counts$^{-2}$. We calibrate the spectrum by dividing by the square of the (scaled) response function. Finally, the inverse calibrated power spectrum shown in figure \ref{fig:invcalspec} in units of s$^{-1}$ strain$^{-2}$ can be written as
\begin{equation}
\frac{1}{S_n(f)}=\frac{1}{dyn^2R^2} \: S_v^{-1}.
\end{equation}

\begin{figure}[h] 
\centering 
\begin{center}
\includegraphics[scale=0.6]{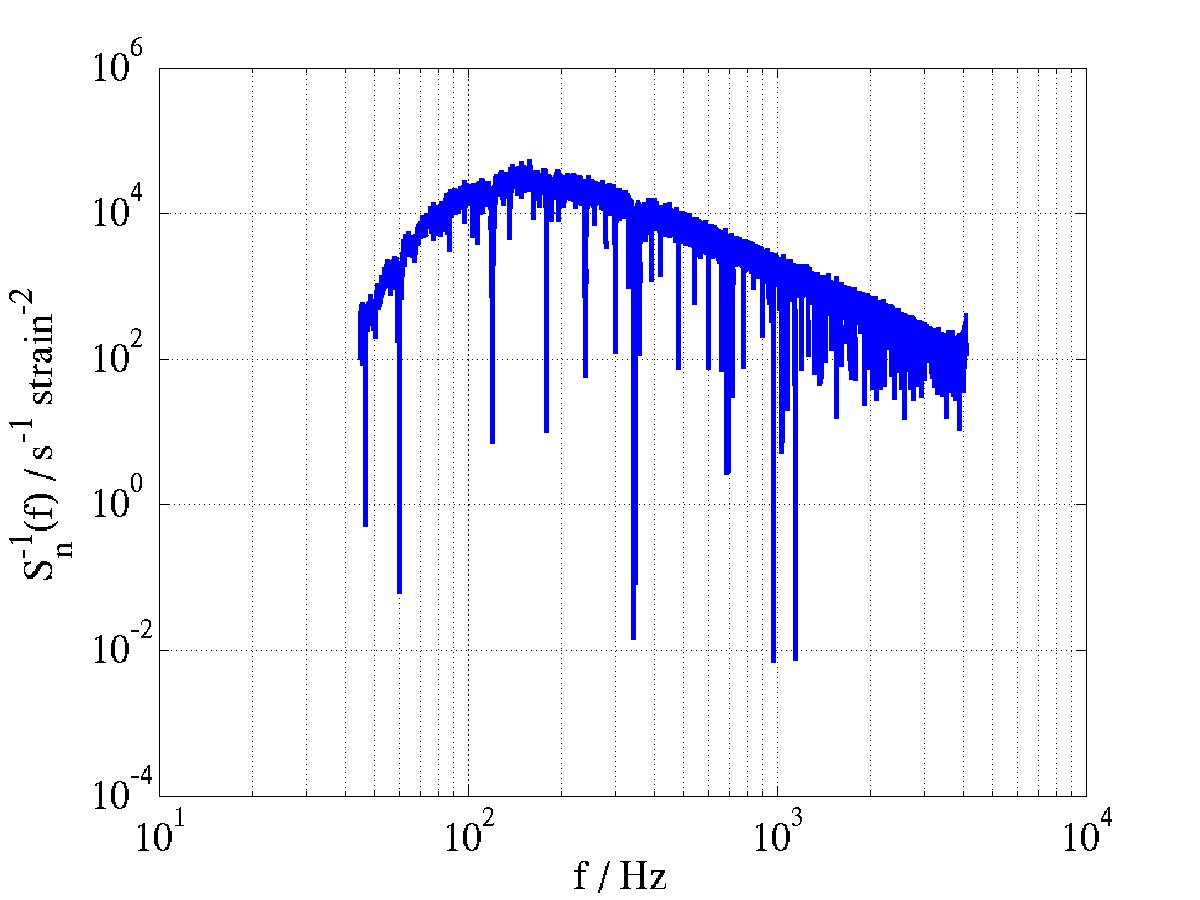}
\caption{The inverse calibrated spectrum for H1 between GPS times 793165201 and 793167377}
\label{fig:invcalspec}
\end{center}
\end{figure}

\subsection{Calibrate and Fourier Transform the Data}
Next, 2176 s of data $s(t)$ is read in in units of ADC counts, and divided into sixteen 256 s overlapping blocks. Each block is Fourier transformed and calibrated, converting it into units of s strain. 
Finally we multiply the data by the inverse calibrated spectrum, giving the frequency series
\begin{equation}
\frac{S_v^{-1}}{dyn\: R} \: \Delta t \: \tilde{s}_k,
\label{eqn:filtdata}
\end{equation}
with units of strain$^{-1}$.


\subsection{Generate the Template Bank} \label{sec:tmpltbnk}

Five user specified parameters are required to lay out the template bank; the maximum mismatch $ds_{\textrm{max}}^2$ (the maximum value of the mismatch between a signal and the nearest template that will be tolerated in the search, see section \ref{sec:metric}), and the frequency and quality boundaries, $f_{\textrm{min}}$, $f_{\textrm{max}}$, $Q_{\textrm{min}}$, and $Q_{\textrm{max}}$ (the tuning of these constraints is discussed in section \ref{sec:tunbank}). Recall from section \ref{sec:metric} that if we define $\phi=\log(f_0)$ the metric coefficients no longer depend on $f_0$ and so we use $\phi_{\textrm{min}}=\log(f_{\textrm{min}})$ and $\phi_{\textrm{max}}=\log(f_{\textrm{max}})$. Then, starting at the point $(Q_{\textrm{min}},\phi_{\textrm{min}})$ one moves across the $Q_{\textrm{min}}$ line incrementing $\phi$ in steps of $\sqrt{2ds^2_{\textrm{max}}/g_{\phi \phi}}$ until $\phi_{\textrm{max}}$ is  reached. Then $Q$ is incremented by $\sqrt{2ds^2_{\textrm{max}}/g_{QQ}}$ and the process is repeated until the point ($\phi_{\textrm{max}},Q_{\textrm{max}}$) is reached. The tuning of these parameters is discussed in section \ref{sec:tunbank} and the final template bank is shown in figure \ref{fig:tpltbank}.


\subsection{Create the Template}

As discussed in section \ref{sec:discretize}, the template used in this search
is given in its discrete form by
\begin{eqnarray}
h_{j}=e^{-\frac{\pi f_0 j \Delta t}{Q}}\cos(2\pi f_{0} j \Delta t),
\label{eqn:ringCTD}
\end{eqnarray}
where, as mentioned earlier $\Delta t=1/8192$ s. The length of the template was set to  ten e-folding times, $t_{\textrm{max}}=10\:\tau$ where $\tau=Q/\pi f_0$. The template time series is Fourier transformed to give
\begin{equation}
\tilde{h}(f_k)=\Delta t \; \tilde{h}_k \label{eqn:filttmplt}
\end{equation}
with units of strain/Hz.

\subsection{Filter the Data}

The filtering step is done template by template for each of the sixteen 256 s overlapping blocks of data in a 2176 s analysis segment. We multiply the template and the data weighted by the power spectrum and inverse Fourier transform to get the time series
\begin{equation}
z(t_j)= \frac{2}{\Delta t\; N} \Re \left\{  \sum_{k=0}^{\frac{N}{2}-1}\frac{S_v^{-1}}{dyn \:R} ( \Delta t \; \tilde{s}_k^*) \; (\Delta t \; \tilde{h}_k) e^{\imath 2 \pi jk/N} \right\}.
\label{eqn:zoft}
\end{equation}
The variance of the template $\sigma_T$ is evaluated according to equation (\ref{eqn:discsigma}) as
\begin{equation}
\sigma_T^2=\frac{4}{N\Delta t} \sum_{k=0}^{\frac{N}{2}-1} \frac{S_v^{-1}}{dyn^2 \: |R|^2} 
               \Big[\Re (\Delta t \: \tilde{h}_k)^2+\Im (\Delta t \: \tilde{h}_k)^2\Big] dyn^2.
\label{eqn:variance}
\end{equation}
Next, using the user-defined threshold on the SNR, $\rho^*$, we calculate the
equivalent threshold on the filter output, $z^*$, for a given template,
$T$,
\begin{equation}
z^*=\rho^* \frac{\sigma_T}{2\; dyn}
\end{equation}
and compare the  series $z(t)$ to $z^*$. Note that this is done on a template
by template basis for each 256 s block of data.

\subsection{Cluster the Filter Output}
\label{sec:clustfilter}
For data handling purposes it is preferable at this stage to cluster those triggers above threshold in time. This is achieved with a sliding window of 1 s in width; starting at the beginning of the time series, the loudest trigger is temporarily assigned to $z_{\textrm{max}}$ and the window is moved so that the left edge coincides with the time of $z_{\textrm{max}}$, $t(z_{\textrm{max}})$. If a trigger within this new window exceeds $z_{\textrm{max}}$, then this becomes the new $z_{\textrm{max}}$ and the window moves once more. In order for this to be a symmetric window, the sliding is continued until a window is reached where there are no further triggers above $z_{\textrm{max}}$. When this occurs, $t(z_{\textrm{max}})$ is deemed to be the time of the trigger and written out to a file. The window is shifted by 1 s and the process starts again, continuing until the end of the block is reached. This is illustrated by figure \ref{fig:ringfiltclust}; the threshold $z^*$ is denoted by the green horizontal lines, and it is clear that there are four groups of data points with $z>z^*$ which are separated from one another by more than 1 s. The loudest data point in each of these clusters, denoted by a red circle, is recorded as a trigger. This process is repeated for each template. The effectiveness of this method of clustering in reducing the level of background is discussed in section \ref{sec:tunclustfilter}. 
 
\begin{figure}[htb]
\centering
\begin{center}
\includegraphics[scale=0.6]{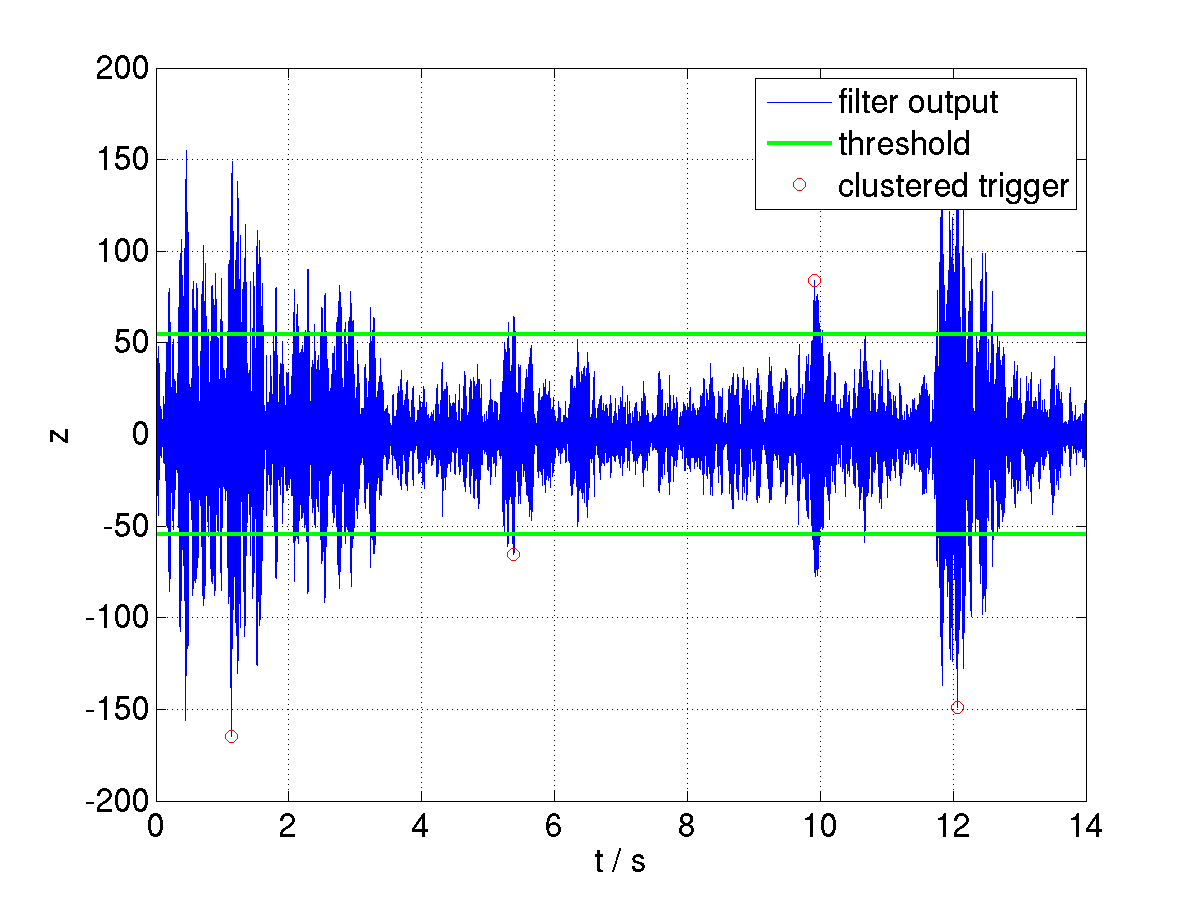}
\caption{An example of single-template clustering with a window of 1 s. The
threshold is marked by a horizontal green line; triggers between the two lines are
below threshold. The red circles mark the loudest trigger in the cluster.}
\label{fig:ringfiltclust}
\end{center}
\end{figure}


\subsection{Calculate and Record the Trigger Parameters}

The SNR of the clustered triggers is calculated as
\begin{equation}
\rho(t_j)=z(t_j) \frac{2\; dyn}{\sigma_T}.
\end{equation}
The amplitude $\mathcal{A}$ is calculated from the template parameters $f_0$ and $Q$ according to equation (\ref{eqn:amplitude}) with $\epsilon=0.01$.  The sensitivity $\sigma$ of the detector to a signal at 1 Mpc is evaluated as
\begin{equation}
\sigma^2= \left( 1 \: \textrm{Mpc}\right) \: \sigma_T^2 \:  \mathcal{A}^2,
\end{equation}
from which the effective distance of the trigger is calculated
\begin{equation}
D_{eff}=\left(1 \textrm{ Mpc}\right) \; \frac{\sigma}{\rho}.
\end{equation}
All of the parameters mentioned here are recorded for each trigger. The
triggers from all of the templates in each of the 256 s blocks are written out
to a file.


\section{Coincidence Analysis}
\label{sec:coinc}

Prior to comparing triggers from multiple detectors, category 2 and 3 vetoes are applied, and the triggers are clustered across all templates in the bank. Then, the remaining triggers from all three detectors are brought together and compared. Those that fail the coincidence test are discarded and those that pass are written out to a file to undergo further examination. There are four possible types of coincidence: a triple coincidence consisting of triggers from H1, H2, and L1 and three types of double coincidence, H1H2, H1L1, and H2L1.

\subsection{Apply Category 2 and 3 Vetoes}
Contiguous trigger files written out from the filtering stage are read in together by the coincidence code  {\fontfamily{pcr}\selectfont lalapps\_rinca} for each of the detectors.
Category 2 and 3 vetoes are applied to the data at this stage. The reason these times were not vetoed at the segment selection stage is because the science segments would be interrupted further and the
likelihood of data being lost increases (recall a minimum science segment
length of 2176 s is required in the analysis). The data during times flagged as category 2 and 3 is not so bad that including it adversely effects the calculation of the power spectrum. (This is not true for category 1 times and they are removed at the segment selection stage.) Thus, as the detection of gravitational waves during category 2 or 3 times would be very difficult, these times are removed before we compare data from multiple detectors.

\subsection{Cluster the Single Detector Triggers Across the Template Bank}
The most information that we need at the end of the coincidence stage is whether or not triggers from different detectors were coincident in $f_0$ and $Q$ at a particular time, i.e., coincidence of a single pair of templates is sufficient to draw our attention to a particular time; we do not need to know about every pair  of templates that were found in coincidence at that time. In theory the template closest in $f_0$ and $Q$ to the actual signal will ring off the loudest; in practice noise will change this somewhat, however we expect the values of these parameters to be close in different detectors. Thus, recording the loudest trigger in a time interval shorter than the duration of the gravitational waves we are sensitive to is sufficient. Therefore, before we compare triggers, we first cluster over all templates in a (fixed) time window of 1 ms, retaining the loudest trigger in that interval. 

\subsection{Implement the Time Coincidence Test}
The time coincidence test requires that triggers be seen within a given time window $\delta t$ of each other for co-located detectors, and $\delta t+10$ ms for the Hanford-Livingston pairs, to allow for the extreme case where gravitational waves are emitted from a distant source along the line connecting the two detectors. As will be explained in section \ref{sec:tunecoinc}, the value of $\delta t$ is determined by evaluating how wide this window needs to be in order to recover as many simulated signals added to the data as possible, while keeping the rate of accidental coincidences low. The optimal value for this search was found to be $\delta t=2$ ms. (Note that the window is applied to each trigger in a pair; thus triggers can be a total of $2\delta t$ s apart.)

\begin{figure}[h] 
\centering
\begin{center}
\includegraphics[scale=0.6]{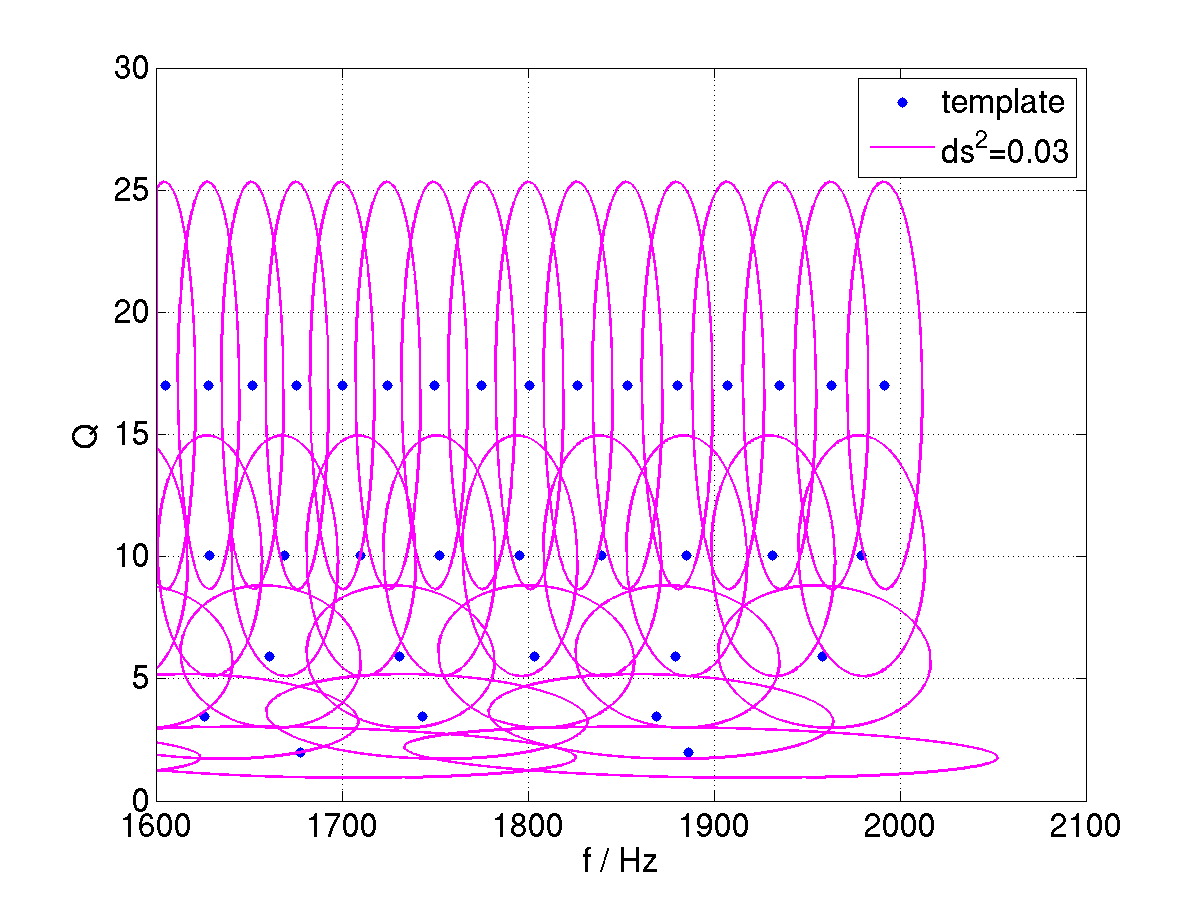}
\caption{Contours of $ds^2=0.03$ for the portion of the template bank between $f=1600$ Hz and $2100$ Hz and $Q=2$ to $30$.}
\label{fig:fQbankz}
\end{center}
\end{figure}

\subsection{Implement the Parameter Coincidence Test}
For triggers surviving the time coincidence test the next consideration is the waveform parameters. As demonstrated in section (\ref{sec:tmpltbnk}) the templates are not distributed uniformly throughout the bank. Thus the use of fixed windows  $\delta f_0$ and $\delta Q$ is not a suitable coincidence test. A more appropriate test is based on the metric distance $ds^2$ as this quantity depends on both $f_0$ and $Q$. Thus it may be used to define a window that essentially varies the size of $\delta f_0$ and $\delta Q$ depending on the region of the bank under investigation. This is illustrated in figure \ref{fig:fQbankz}. This window can be described as a contour of a constant $ds^2$ about each template.


\subsection{Implement a H1H2 Distance Cut}
\label{sec:distancecut}
If a gravitational wave ringdown is detected in both H1 and H2, then given that H1 is twice as long as H2 and they have correlated displacement noise, we expect that $\rho_{\textrm{H1}} \approx 2\rho_{\textrm{H2}}$. Furthermore, $\sigma_{\textrm{H1}}^2 \approx 2\sigma_{\textrm{H2}}^2$ and thus we would expect the effective distance measured by each instrument to be consistent. Inconsistent effective distances from co-located detectors would suggest that the coincident pair was not due to a gravitational wave signal but caused by noise. Thus a coincidence including a H1H2 trigger is retained only if
\begin{equation}
\frac{d_{eff_{H1}}}{d_{eff_{H2}}}>\kappa,
\end{equation}
where $\kappa$ has a user-specified value. As will be discussed in section \ref{sec:distcuttuning}, the conservative value of $\kappa=2$ was used in the search. 

This test is not possible to implement in L1 primarily because the arms of this detector are not aligned with those of the Hanford detectors, nor do the arms lie in the same plane. Thus, one could imagine a situation where a source was optimally aligned and oriented for L1 producing a loud trigger, whereas in the Hanford detectors the signal would be weaker by a factor $\gamma$. Assuming comparable sensitivities of the two 4 km interferometers, this would mean that the effective distance calculated for H1 would be a factor of $\gamma$ higher than that calculated for L1.

After the time coincidence, parameter coincidence, and distance cut are imposed, the resulting list of coincident groups (i.e., H1H2L1 triple coincidences, H1L1 double coincidences, H1H2 double coincidences, and H2L1 double coincidences) are time sorted, with the members of each group listed alphabetically by interferometer name.
This list of coincidences (which contains all the information from the original trigger files) is written out to a table.

\subsection{Cluster the Coincidences}
\label{sec:postproc}
The output coincidence files generally have multiple groups of coincident triggers lying in a short time window, as a noise event (or indeed a gravitational wave signal) will ring up several templates, a number of which may be found in coincidence with one or more templates in the second and possibly the third interferometer. As just one group is sufficient for drawing our attention to a particular time we can cluster the coincident groups of triggers within a short time window, retaining the most ``significant'' group, using the LAL program {\fontfamily{pcr}\selectfont lalapps\_coincringread}. A (fixed) window of 10 s was used in this analysis. Ranking the significance of the groups is achieved by defining a detection statistic $\rho_{DS}$, discussed below. The final clustered groups were once more written out to a {\fontfamily{pcr}\selectfont sngl\_ringdown} table and followed up as detection candidates.

\subsection{The Detection Statistic}
\label{sec:detstat}
Defining a detection statistic is a mechanism for ranking groups of coincidences. It is a process whereby a single number is assigned to a coincident group of triggers describing the collective significance of the group. In choosing a detection statistic the aim is to find one which will best discriminate between signal and background. We use the detection statistic in a number of stages throughout the analysis: for clustering, for comparison of foreground and background, and in deciding which coincidences to follow up at the end of the pipeline.

There are a number of valid criteria that one could impose in the choice of detection statistic, for example one could choose the group of coincident triggers whose elements are closest in time, or which have the smallest $ds^2$. The parameter we choose to use for the selection process is SNR, i.e., we measure the loudness of the group. For triple coincidences a good statistic to discriminate between signal and background is the sum of the squares of the SNRs,
\begin{equation}
\rho_{DS}=\rho_{\textrm{H1}}^2+\rho_{\textrm{H2}}^2+\rho_{\textrm{L1}}^2.
\end{equation}
For doubles, the above statistic was impractical because of the high level of non-Gaussian noise. Instead, we required a $\rho_{DS}$ that prevents a high-SNR glitch in one interferometer from (unfairly) influencing which group is chosen as the most significant. Hence for a double coincidence in interferometers 1 and 2 the detection statistic used is the chopped-L;\footnote{This is similar to the bitten-L used in the inspiral analysis, but with a flat ``bite''.}
\begin{equation}
\rho_{DS}=\textrm{min}\{\rho_{\textrm{ifo1}}+\rho_{\textrm{ifo2}}, \: a \rho_{\textrm{ifo1}}+b, \: a \rho_{\textrm{ifo2}}+b\},
\end{equation}
where $a$ and $b$ are parameters that are tuned for the particular search. A
discusssion of the detection statistics and the tuning of $a$ and $b$ is given in section \ref{sec:distcuttuning}. 


\chapter{The S4 Data Set}
\label{ch:s4}
\section{Introduction}
In this chapter we discuss general aspects of the S4 run.

\section{The Fourth LIGO Science Run }
\label{sec:s4}
The fourth LIGO Science (S4) run took place between February 22$^{\mathrm{nd}}$ and March 24$^{\mathrm{th}}$, 2005, --- GPS times 793130413 to 795679213, a total of 708 hours. This yielded 567.4 hours of analysable data from H1, 571.3 hours from H2, and 514.7 hours from L1. In this analysis we require that data be available from at least two detectors at any given time. This results in approximately 364 hours of triple coincidence and 210 hours of double coincidence as shown in figure \ref{fig:s4vennhr}.

At their best during S4, H1 and L1 had comparable sensitivities at high frequencies, and H1 was more sensitive below $\sim 200$ Hz. H2 was about a factor of two less sensitive than the 4 km interferometers. Figure \ref{fig:s4strain} displays the best noise curves for each of the interferometers during the S4 run.  

\afterpage{\clearpage}
\begin{figure}[h]
\centering
\begin{center}
\includegraphics[scale=0.4]{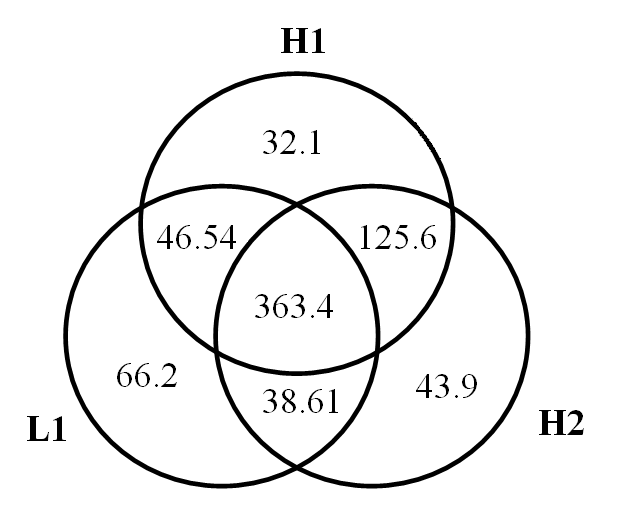}
\caption{Time in hours of analysable S4 data} \label{fig:s4vennhr}
\end{center}
\end{figure}

\begin{figure}[h] 
\centering
\begin{center}
\includegraphics[scale=0.6]{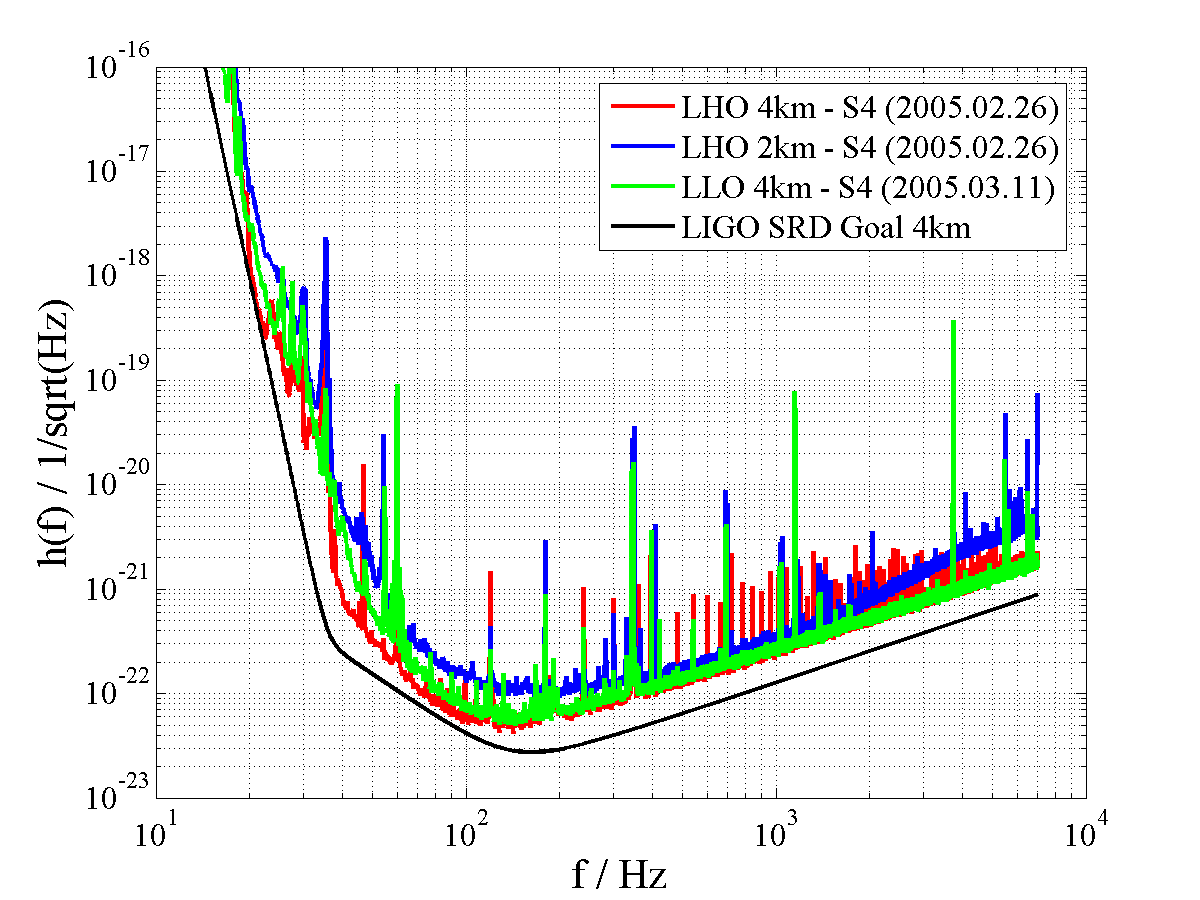}
\caption{S4 strain sensitivity}
\label{fig:s4strain}
\end{center}
\end{figure}


\section{Data Quality and Veto Categorization}
\label{sec:dataquality}

There are times during each science run when some component of the instrument
malfunctions or an external disturbance couples to the gravitational wave
channel introducing short bursts of noise into the data stream. This is
troublesome for data analysis as these glitches can often match many templates
in a matched filter search, producing high SNR triggers. This is of particular
concern to the ringdown search as the templates are short in duration, just
like the glitches. This increases the false alarm rate making it more
difficult to detect a gravitational wave.
In many cases glitches can also be seen in auxiliary channels, for example, a
glitch in the gravitational wave channel due to a seismic disturbance can also
be seen in the seismometer channels. During and after a science run, times
during which it was known that the quality of the data was compromised because
of noise are flagged. We refer to these as data quality flags.

Members of the inspiral and burst analysis groups within the LIGO Scientific
Collaboration have used information from auxiliary channels to flag times when
the level of noise in the detector was unacceptably high or the quality was
questionable, and grouped these times into four categories depending on the
severity of the disturbance. The categorization of data quality flags depends
primarily on two quantities: (i) the efficiency, that is the number of noise
triggers above a particular SNR that it vetoes, and (ii) the dead-time, the
total science time cut by applying the veto. Ideally we want to veto times
with maximum efficiency and minimum dead-time.

The first category of data quality flags vetoes time when the data is of very
poor quality and should not be analysed. These are applied at the segment
selection stages. We try to keep this set to a minimum because, as described
in section \ref{sec:segmentation}, a minimum analysis segment length of 2176 s
is required for filtering. Thus we want to avoid unnecessarily breaking up the
data.

Data during which detection of a gravitational wave is not possible because of
a known correlation between an environmental or instrumental disturbance and
the gravitational wave channel is marked by category 2 data quality flags.
Examples include times during which hardware injections (an actuation on one
of the test masses simulating a gravitational wave) are being performed
(accounting for approximately $1\%$ of the data) and overflows in any of the
digital signals used in feedback loops controlling the detector lengths.
Category 2 vetoes effect $4 \%$ of the triple coincidence data, $56 \%$ of the
H1L1 data, $3 \%$ of the H1H2 data, and $27 \%$ of the L1H2 data
\cite{tuning}.

The third category of data quality flags includes those times during which
there was a statistical correlation between the gravitational wave channel and
an external disturbance. A gravitational wave may be searched for, but caution
must be taken when determining confidence levels. Data quality flags under
this category include elevated levels of dust reported by a particle counter
in the vicinity of photo-detectors. These times may be associated with
glitches in the gravitational wave channel due to a dust particle passing
across the beam.

For S4, triggers from the binary neutron star inspiral search were used to
evaluate efficiency and dead-times which led to the data quality flag
categorization. This categorization was then implemented in the binary black
hole inspiral search (BBH) \cite{S3S4} and the ringdown search. In the
inspiral analysis, data with category 2 data quality flags is searched over
for gravitational wave candidates but not included in the upper limit
calculation, while category three data is included.
The inspiral analyses implement various tests to check the consistency of the
data around a given trigger with nearby templates, so-called signal-based
vetoes. In the ringdown analysis we do not implement signal-based vetoes and
for this reason category 2 and 3 vetoes were combined and no data lying in
either category was used in the calculation of the upper limit.  This affected
12\% of triple time, 8\% H1-H2 time, 62\% of H1-L1 time, and 29\% of H2-L1
time.

Times flagged as category four showed a weak but positive correlation with
false alarms. Data from these times were not vetoed, but the presence of such
a flag on a candidate event decrease our confidence in it being caused by a
gravitational wave.


\section{Horizon Distance}
\label{sec:horizondistance}
Using the noise curves discussed in section \ref{sec:s4} we can get a sense of how sensitive we are to gravitational waves from a particular black hole. The horizon distance $D_H$ is the distance at which we can detect a ringdown from an optimally oriented and located black hole with signal-to-noise ratio (SNR) of 8, and is given by
\begin{equation}
D_H(f)=\frac{1}{8}  \left( \frac{2}{S_n(f)} \right)^\frac{1}{2} h_{rss}
\; \left(1 \textrm{ Mpc}\right)
\end{equation}
where $S_n(f)$ is the average power spectral density of the detector noise and $h_{rss}$ is the root sum squared of the strain for a signal with optimal orientation at 1 Mpc,
\begin{eqnarray}
h_{rss}^{2} = \left(  \frac{\mathcal{A}}{\textrm{1 Mpc}} \right) ^2 \left( \frac{2}{\pi f_0} \right) Q \left( \frac{1+2Q^2}{1+4Q^2}\right) 
\end{eqnarray}
and $\mathcal{A}$ is given by equation (\ref{eqn:amplitude}).
The physical distance of the source is always less than or equal to the
horizon distance. Figure \ref{fig:range} shows that in S4 a gravitational wave from an optimally oriented black hole of mass $250 \ M_\odot$ at a distance of 350 Mpc directly above (or below) the detector, will produce an SNR of 8 in H1. The source would need to be at 140 Mpc and 270 Mpc to produce the same SNR in H2 and L1 respectively. In this calculation we have assumed that the black hole is spinning with $\hat{a}=0.9$ and that 1\% of the mass is radiated as gravitational waves. To put this into an astrophysical context consider figure \ref{fig:atlas}. This shows the universe out to $\sim300$ Mpc. According to \cite{Powe}, there are approximately three million large galaxies in this region of the sky. From \cite{Kopp08} we can estimate the number of sources as $2.2\times 10^6 \textrm{ L}_{10}$ or $1.3 \times 10^6$ MWEG.

\afterpage{\clearpage} 
\begin{figure}[h]
\centering
\begin{center}
\includegraphics[scale=0.55]{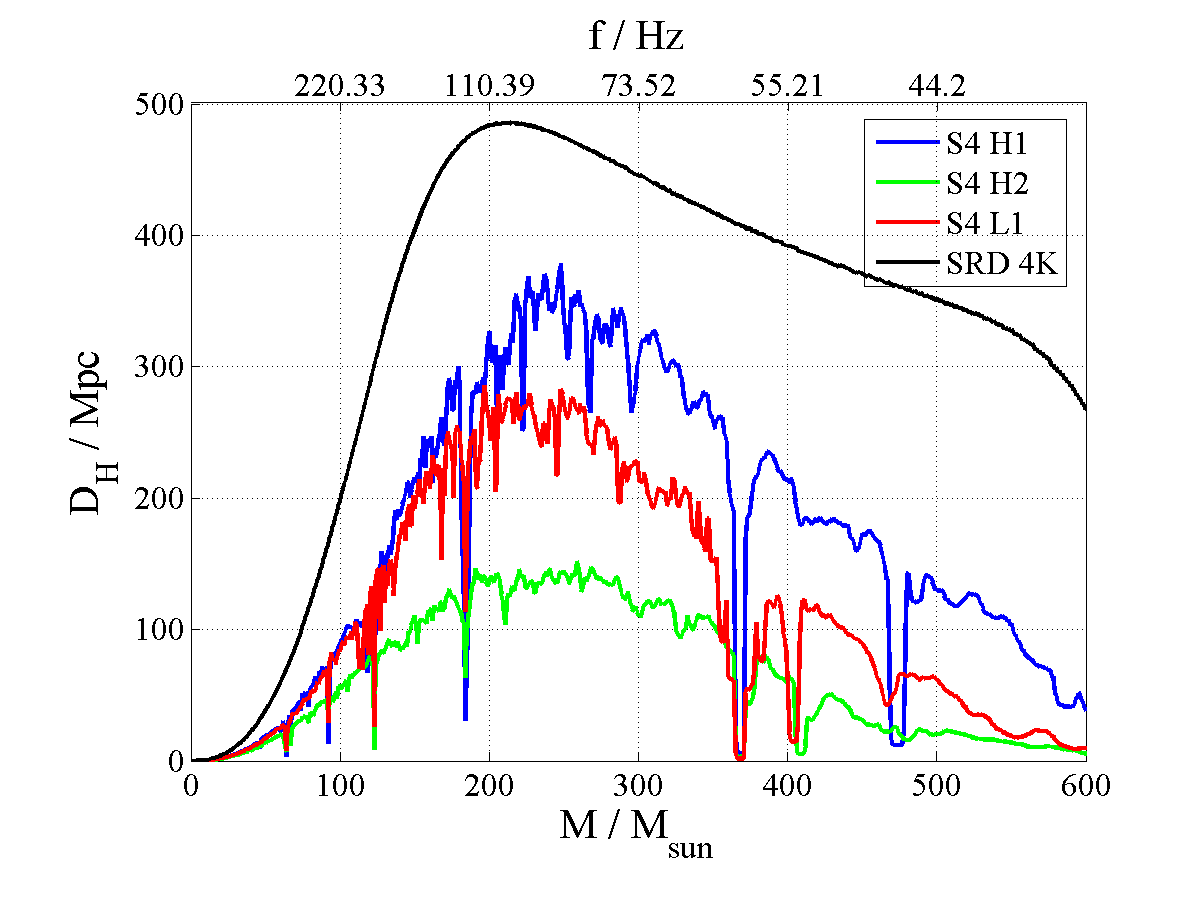}
\caption{S4 ringdown horizon distance versus mass and frequency for $\hat{a}=0.9$ and $\epsilon=1\%$.}
\label{fig:range}
\end{center}
\end{figure}

\begin{figure}[h]
\centering
\begin{center}
\includegraphics[scale=0.45]{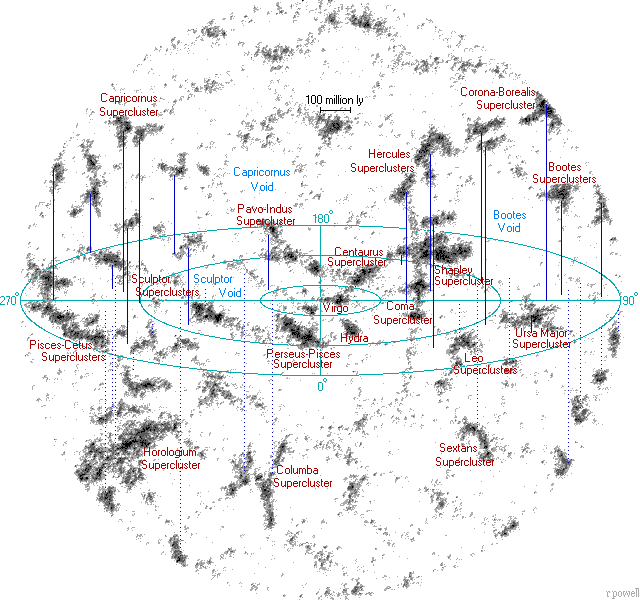}
\caption{Illustration of the galaxies with $\sim 300$ Mpc of the Earth. {\it
Picture Credit}: Richard Powell \cite{Powe}.}
\label{fig:atlas}
\end{center}
\end{figure}

A standard figure of merit used in the control room is the ``inspiral horizon distance''; this is the distance out to which an optimally oriented and located $1.4\ M_\odot - 1.4\ M_\odot$ binary neutron star system can be seen with an SNR of 8. This quantity has been used throughout the five science runs as a means of evaluating the sensitivity of the detector during a run and as a comparison between different science runs. We can make an analogous plot for ringdowns; we choose as our standard candle a source that gives the maximum horizon distance in the above plot. This corresponds to an optimally located and oriented black hole with mass $250\ M_\odot$ and spin $\hat{a}=0.9$ with $\epsilon=1\%$. Using the value of $\sigma$ output from the filter with these parameters we can plot the horizon distance as a function of time for the duration of the run. As can be seen from figure \ref{fig:HorzDist}, H2 was the most stable of the three detectors, particularly during the first half of the run, with an average horizon distance of $\sim 150$ Mpc. During the last 15 days the horizon distance fluctuated about this value. H1 had the largest horizon distance for this source, averaging $\sim 320$ Mpc. The small-scale fluctuations can be attributed to the diurnal activities in the area with a minimum in the horizon distance usually occurring around 10am local time, and the quietest time of day occurring around 11pm. The larger dips were attributed instrumental problems such as the beam drifting in the interferometer causing mis-alignment, or prolonged environmental disturbances such as high wind (in excess of 15 mph) causing increased seismic noise. The L1 range was consistently lower than H1 for the run, reaching at best $\sim 320$ Mpc. This dropped to 120 Mpc for days 6--15 because of lower light power in the interferometer.

\begin{figure}[h]
\centering
\begin{center}
\includegraphics[scale=0.6]{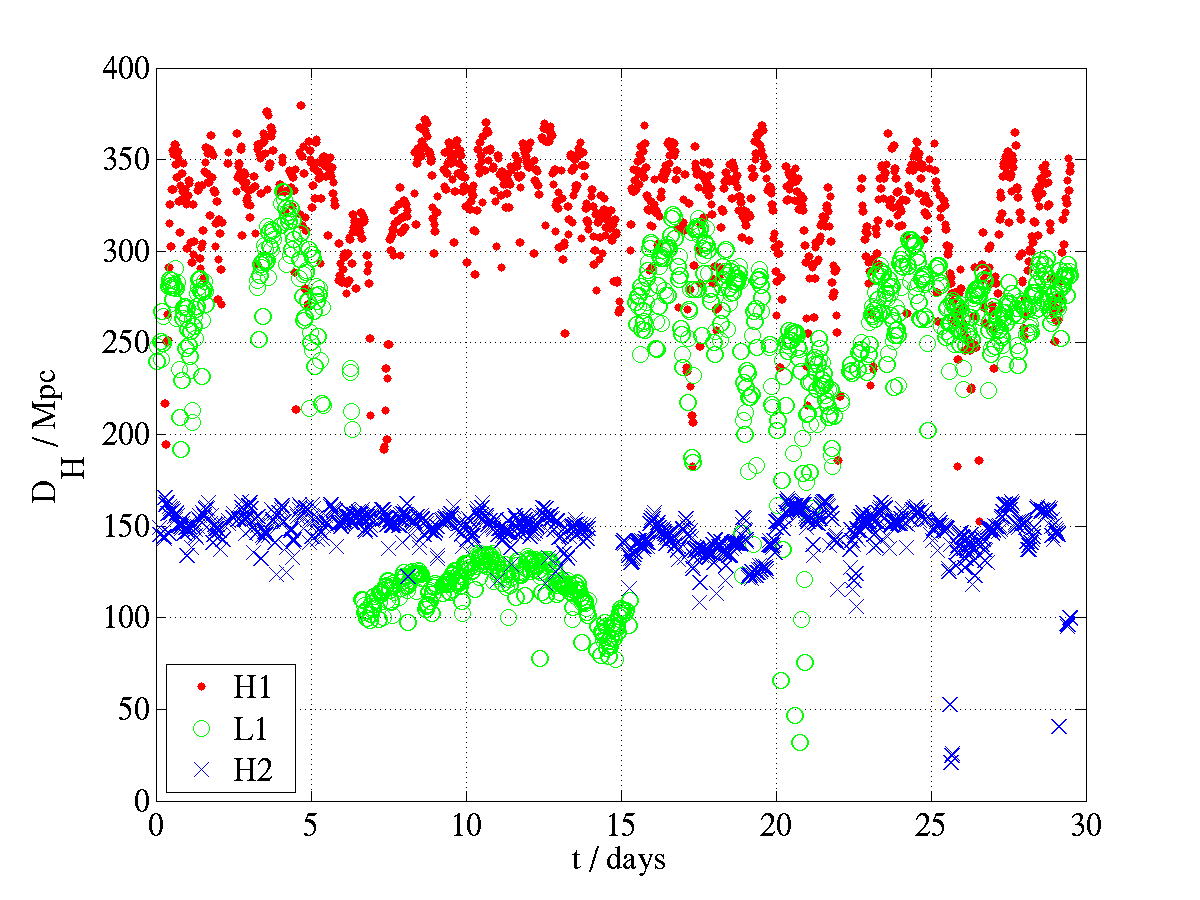}
\caption{S4 ringdown horizon distance versus time for an optimally located and
oriented black hole of mass $250\ M_\odot$, spin of 0.9 and $\epsilon=1\%$
producing an SNR of 8 in the detector.}
\label{fig:HorzDist}
\end{center}
\end{figure}

\section{Detection of Binary Compact Coalescence}

As mentioned in section \ref{sec:bincoals}, a promising source of
gravitational waves is expected to be from compact binary coalescences, in which case the
ringdown will be preceded by an inspiral and merger. In this section we
examine which of the searches, the inspiral or ringdown, is more sensitive to
gravitational waves from a given binary.

Compact binaries can be composed of two neutron stars (NS-NS), two black holes
(BH-BS), or a neutron star and a black hole (NS-BH). As discussed in section
\ref{sec:astro}, neutron stars have a maximum mass of $3\ M_\odot$ and stellar
mass black holes are believed to lie in the range 3 $M_\odot < $ M $ < 20 \
M_\odot$. Black holes with masses above 20 $M_\odot$ are referred to as
intermediate-mass black holes, however, their existence is still in question.

In figure \ref{fig:inspringsources} we plot the strain of a number of compact
binaries at an arbitrary distance as the frequency of their emitted radiation passes through the Initial LIGO band. The binaries considered are the NS-NS pair, 1.4 $M_\odot$--1.4 $M_\odot$, a NS-BH binary with component masses 1.4 $M_\odot$--3 $M_\odot$, and BH-BH pairs with equal mass components of 10 $M_\odot$, 20 $M_\odot$, 40 $M_\odot$, and 100 $M_\odot$.  The blue line describes the inspiral, as it sweeps through a range of frequencies. The end-point of the inspiral for the four least massive binaries is at the inner-most stable circular orbit (ISCO), at which point the frequency is $f_{ISCO}=c^3/(6\sqrt{6}\pi GM)$. The plot shows that for these binaries a large proportion of the inspiral is in the LIGO band. In order to detect the inspiral from the higher mass binaries with a matched filter however, it is necessary to evolve the waveform further, to the light-ring, at which point the frequency of the gravitational waves is $f_{LR}=c^3/(3\sqrt{3}\pi GM)$. It is clear from the figure that as the mass of the binaries increases, less and less of the inspiral is in-band.  The ringdown on the other hand, which appears as a dot marking the single frequency of the gravitational radiation emitted, is out of the LIGO band for the low mass binaries, but as the masses increase, the ringdown frequency decreases, bringing it into the LIGO band. For the 100 $M_\odot$ pair the ringdown is at the most sensitive frequency of the LIGO detectors. In calculating the ringdown strain, we assume that the final black hole has a spin of 0.7 (this is the expected spin for a the black hole formed following the merger of two equal-mass, non-spinning compact objects) and that 1\% of the mass is radiated.

\begin{figure}[h]
\centering
\begin{center}
\includegraphics[scale=0.6]{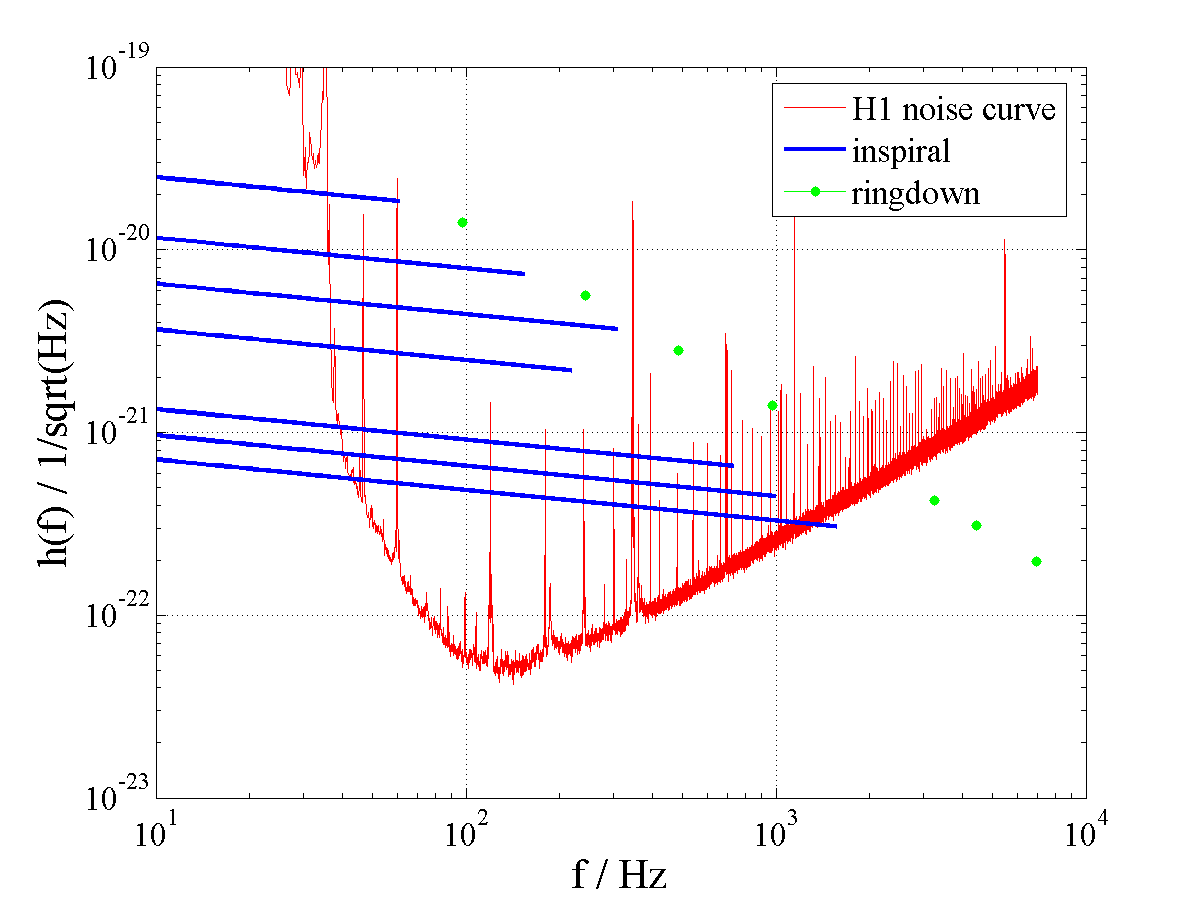}
\caption{A plot of S4 strain sensitivity versus frequency (red). The blue lines
represent the inspiral phase of a binary coalescence for binaries with
masses (from bottom to top): 1.4-1.4, 1.4-3, 3-3, 10-10, 20-20, 40-40, and 100-100, in units of $M_\odot$. The green dots represent the ringdown (for the same list of binaries) for a final black hole with a spin of 0.7 and assuming 1\% of the mass has been radiated. }
\label{fig:inspringsources}
\end{center}
\end{figure}

We can estimate how sensitive each of the searches is to a given source by calculating the horizon distance. As before, the horizon distance is the maximum distance to which an optimally oriented and located black hole will produce an SNR of 8 in the detector. In this section we calculate the ringdown horizon distance for a black hole with spin $\hat{a}=0.7$. Figure \ref{fig:inspringDhorz} shows the H1 inspiral and ringdown horizon distances for binaries with (total) mass of up to 100 $M_\odot$, using the S4 noise curve.  The plot shows that the maximum inspiral horizon distance, $\sim 95$ Mpc is attained for a binary with a total mass of $\sim 30\ M_\odot$. For this mass, the ringdown search is only able to see to $\sim10$ Mpc, and therefore, we are much less sensitive to the ringdown than we are to the inspiral of a black hole binary with that mass. As the mass increases however, the ringdown search becomes more sensitive and the inspiral search becomes less sensitive, with the cross-over point occurring at $\sim62\ M_\odot$.  Figure \ref{fig:inspringDhor} shows the ringdown horizon distance for the entire ringdown mass range along with the inspiral horizon distance, demonstrating how much wider the ringdown mass range and further the distance is for ringdowns.

\begin{figure}[h]
\centering
\begin{center}
\includegraphics[scale=0.6]{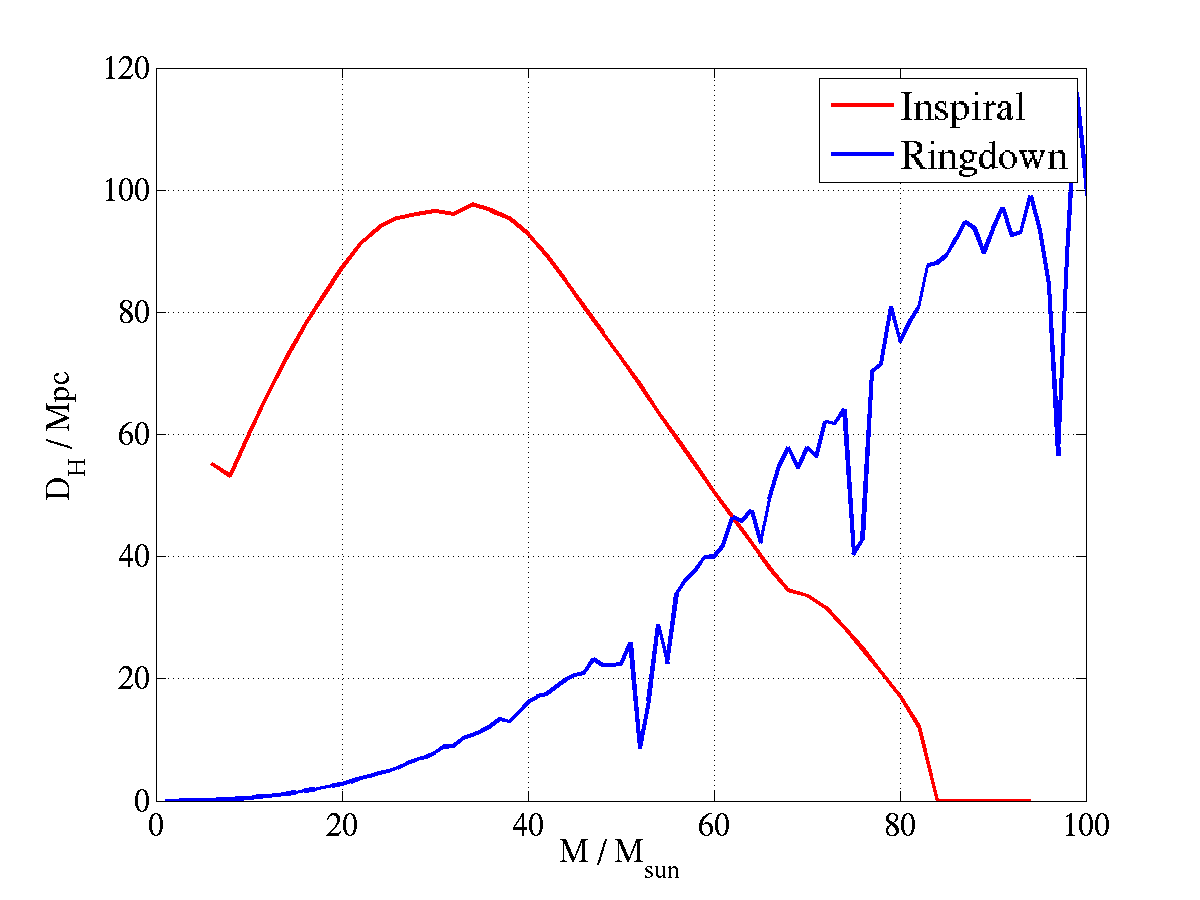}
\caption{The horizon distance as a function of mass for the S4 inspiral and
ringdown searches up to 100 $M_\odot$. A final spin of 0.7 and $\epsilon=1\%$
are assumed in the calculation of the ringdown horizon distance.}
\label{fig:inspringDhorz}
\end{center}
\end{figure}

\begin{figure}[h]
\centering
\begin{center}
\includegraphics[scale=0.6]{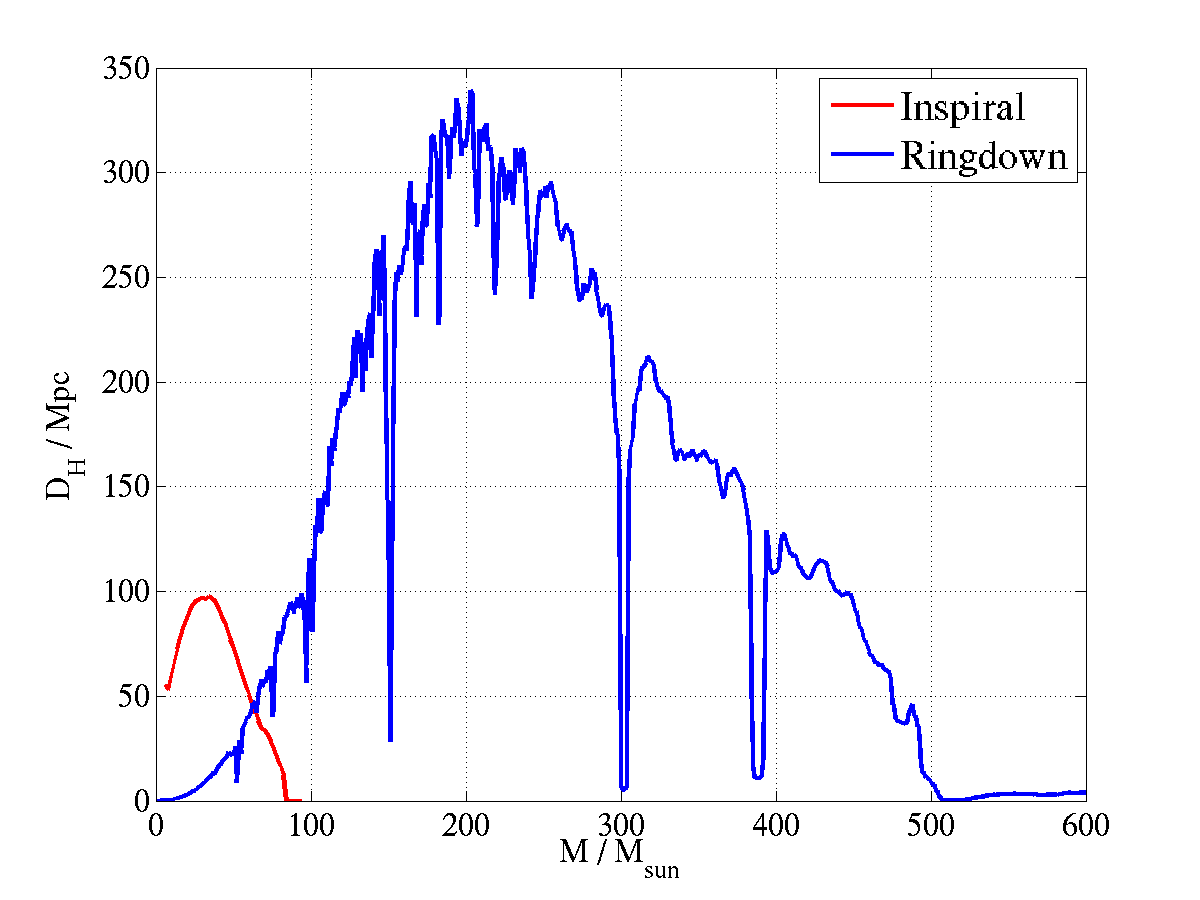}
\caption{The horizon distance as a function of mass for the S4 inspiral and ringdown searches, covering the full mass range of the ringdown search. A final spin of 0.7 and $\epsilon=1\%$ are assumed in the calculation of the ringdown horizon distance.}
\label{fig:inspringDhor}
\end{center}
\end{figure}

\section{Predicted Ringdown Rates}
Here we discuss the predicted rate of ringdowns for the S4 search.

\subsection{Stellar Mass Black Holes}
Blue light luminosity is a tracer of massive star formation, and therefore is also expected to scale linearly with the binary coalescence of massive stars \cite{Phin91}. We measure luminosity in terms of the blue light luminosity, in units of L$_{10}$ which is defined as L$_{10}=10^{10}L_{B,\odot}$, and $L_{B,\odot}=2.16 \times 10^{33}$ erg s$^{-1}$ is the solar blue light luminosity. Population synthesis models predict that the merger rate, $R$, of 10 $M_\odot$-10 $M_\odot$ black hole binaries is $R=0.4$ L$_{10}^{-1}$ Myr$^{-1}$ \cite{Osha05}. Figure \ref{fig:inspringDhorz} showed that for a 20 $M_\odot$ black  hole the ringdown search can see to approximately 3 Mpc.  We can see how many sources we are sensitive to at that distance with figure \ref{fig:galaxy}, the cumulative blue light luminosity $C_L$ as a function of distance. This plot shows that at 3 Mpc we can see approximately 5 L$_{10}$. Therefore the expected rate of stellar mass black hole coalescences detectable at S4 sensitivity, $R_{S4}$, is $R_{S4}=R C_L = 0.4$ L$_{10}$ Myr$^{-1}\; \times$ 5 L$_{10}=2$ Myr$^{-1}$. Thus, given the sensitivity (and thus the distance reach) of S4, the detection of a stellar mass black hole ringdowns is unlikely. However at these low masses the S4 ringdown search is sensitive to the local group of galaxies, and thus should an event have occurred there during the run, the ringdown search would be capable of detecting it.

\begin{figure}[h]
\centering
\begin{center}
\includegraphics[scale=0.6]{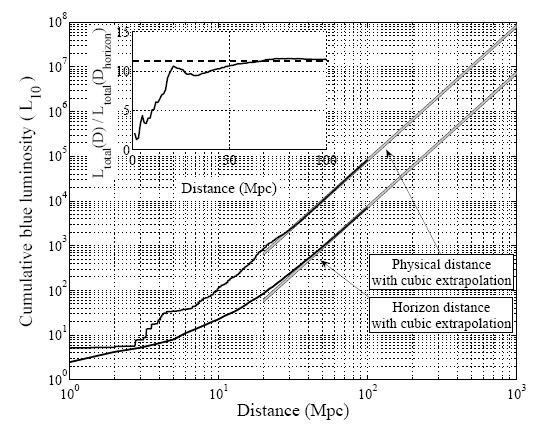}
\caption{The cumulative blue light luminosity as a function of horizon
distance. This
figure was taken from \cite{Kopp08}.}
\label{fig:galaxy}
\end{center}
\end{figure}

\subsection{Intermediate-Mass Black Holes}
The most likely formation scenario for an intermediate-mass black hole (IMBH) is
 by core collapse of a very massive star \cite{Port99}. Recent simulations have indicated that star clusters with binary
fractions larger than 10\% will produce two IMBHs which will form a binary (IMBHB), which will eventually merge \cite{Gurk06}. A model to predict the rate
of ringdowns from IMBHB merger for a given detector sensitivity \cite{Freg06}
is
\begin{equation}
R_{S4}=R \; \frac{4\pi}{3} \left(\frac{D_H}{2.26}\right)^3 \rho_{GC} \; g,
\end{equation}
where $R$ is the rate of IMBHB merger per cluster, taken to be the age of the universe (as
this is expected to occur just once in each cluster) (13.7)$^{-1}$ Gyr$^{-1}$; $D_H$
is the horizon distance, which, from figure \ref{fig:inspringDhor} is 300 Mpc
for a final black hole mass of 200 $M_\odot$; $\rho_{GC}$ is the number
density of star clusters sufficiently massive to produce an IMBH, taken to be
the current density of globular clusters, $\rho_{GC}=8h^3$ Mpc$^{-3}$; and $g$,
the fraction of globular clusters with a binary fraction high enough to
produce an IMBHB, is the most uncertain term in the model and is taken to be
10\%. With these values the predicted rate is $R_{S4}=1\times10^{-4}$
yr$^{-1}$. Therefore, once again we see that it is not expected that this type
of event would occur within the range of the S4 search. However, our knowledge of IMBHs from electromagnetic observations is very poor, and hence there is a large uncertainty associated with the assumptions made in the calculation just described. Gravitational wave searches could provide the evidence needed to affirm the existence of this population of black holes.


\chapter{Tuning the Search}
\label{ch:tuning}

\section{Overview}

In chapter \ref{ch:search} we outlined the analysis pipeline, the set of steps that are followed to analyse data with the aim of detecting any gravitational wave ringdown signals that may be present. We described a number of thresholds, cuts, coincidence windows, and clustering techniques that are implemented to isolate the most likely gravitational wave candidates. 
In LIGO matched filtering searches we adopt the ``blind search'' philosophy,
whereby the constraints are decided upon prior to looking at the
full data set. 
In this chapter we discuss methods used to determine optimal values for each of these constraints in the S4 search, i.e., we tune the search.
To achieve this we employ the following three tools, each of which will be described in more detail in this chapter:
\begin{itemize}
\item{Monte Carlo simulations of the ringdown waveform (injections),}
\item{time-shifted data sets (timeslides),}
\item{a representative subset of the data (playground).}
\end{itemize}
To tune the search we run the pipeline several times on injections,
timeslides and playground each time modifying the constraints on the pipeline
to get the desired result.
Once the tuning has been finalized the data set is unblinded, or in LIGO language, the box is opened. Implementing a blind search prevents any bias on the part of the analyst when tuning the search from influencing the result.


\section{Implementing an Injection Analysis}
\label{sec:injections}

Injections are simulations of the signals we are trying to detect whose parameters are randomized within a given distribution. These simulated signals are added to the data stream and the pipeline is used to recover them from the noise. We employ injections for many purposes in the course of the analysis; in this section we discuss the use of injections to tune the search. We utilize injections and timeslides in tandem to find a balance between recovering as many simulated signals in coincidence between multiple detectors as possible while keeping the rate of false coincidences to a minimum. (Background estimation via timeslides will be discussed in the next section.)

\subsection{Creating the Injection File}
A table of injection parameters is created using the LAL program {\fontfamily{pcr}\selectfont lalapps\_rinj} based on a set of input arguments.  

\begin{itemize}
\item{Time: Injections are placed at a random time within an interval of 250 s every $7000/\pi$ seconds. This ensures that (i) there is not more than one injection per analysis segment, (ii) the injection does always occur at the same number of seconds after the start of the analysis segment, and (iii) the injection does not occur at the same fraction of a second each time.}

\item{Frequency and Quality: For the purpose of tuning we want to cover the parameter space available to the search (this is dependent on the detector sensitivity). Injections are made uniformly in quality factor $Q$ between values of 2 and 22 and in logarithmic frequency $\log_{10}(f_0)$ with $45 \textrm{ Hz}\leq f_0 \leq 2500\textrm{ Hz}$. The distribution of injections in frequency is shown in figure \ref{fig:injf}.} 

\item{Sky location and source orientation: The injections are placed uniformly in logarithmic distance between 1 kpc and 200 Mpc and uniformly in sky position (right ascension $\alpha$ and declination $\delta$), as shown in figure \ref{fig:injsphere}. The inclination $\iota$ and polarization $\Psi$ angles and initial phase $\phi_0$ of the injection are also uniformly distributed.}

\item{From these parameters the amplitude $\mathcal{A}$, mass, spin, and the arrival time and effective distance at each site are calculated using equations (\ref{eqn:amplitude}), (\ref{eqn:Echeverria_MoffQ}), (\ref{eqn:Echeverria_aofQ}), and (\ref{eqn:Deff}) . The percentage of mass radiated as gravitational waves $\epsilon $ is fixed at 1$\%$.}
\end{itemize}

\begin{figure}[ht] 
\centering
\begin{center}
\includegraphics[scale=0.6]{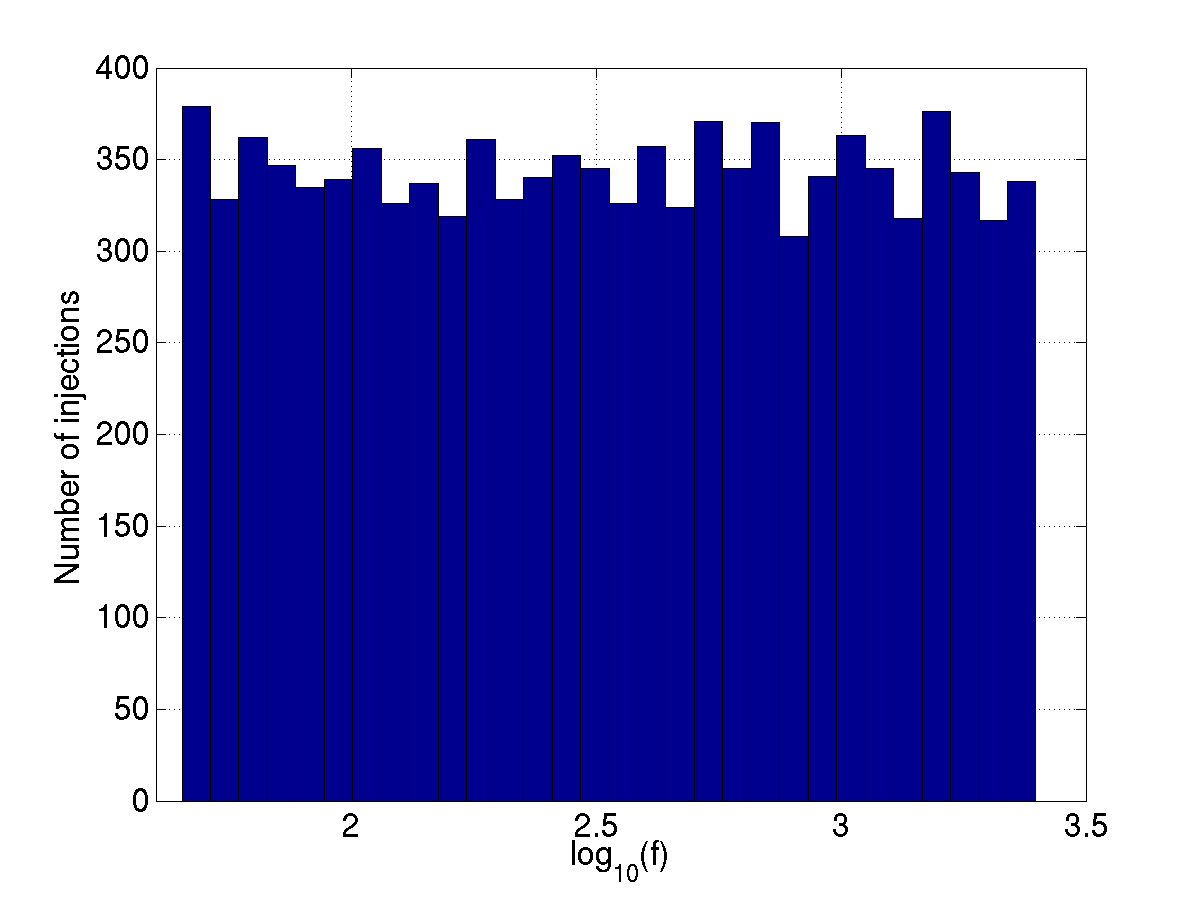}
\caption{Distribution of injections in frequency over nine injection runs.}
\label{fig:injf}
\end{center}
\end{figure}

\begin{figure}[ht] 
\centering
\begin{center}
\includegraphics[scale=0.6]{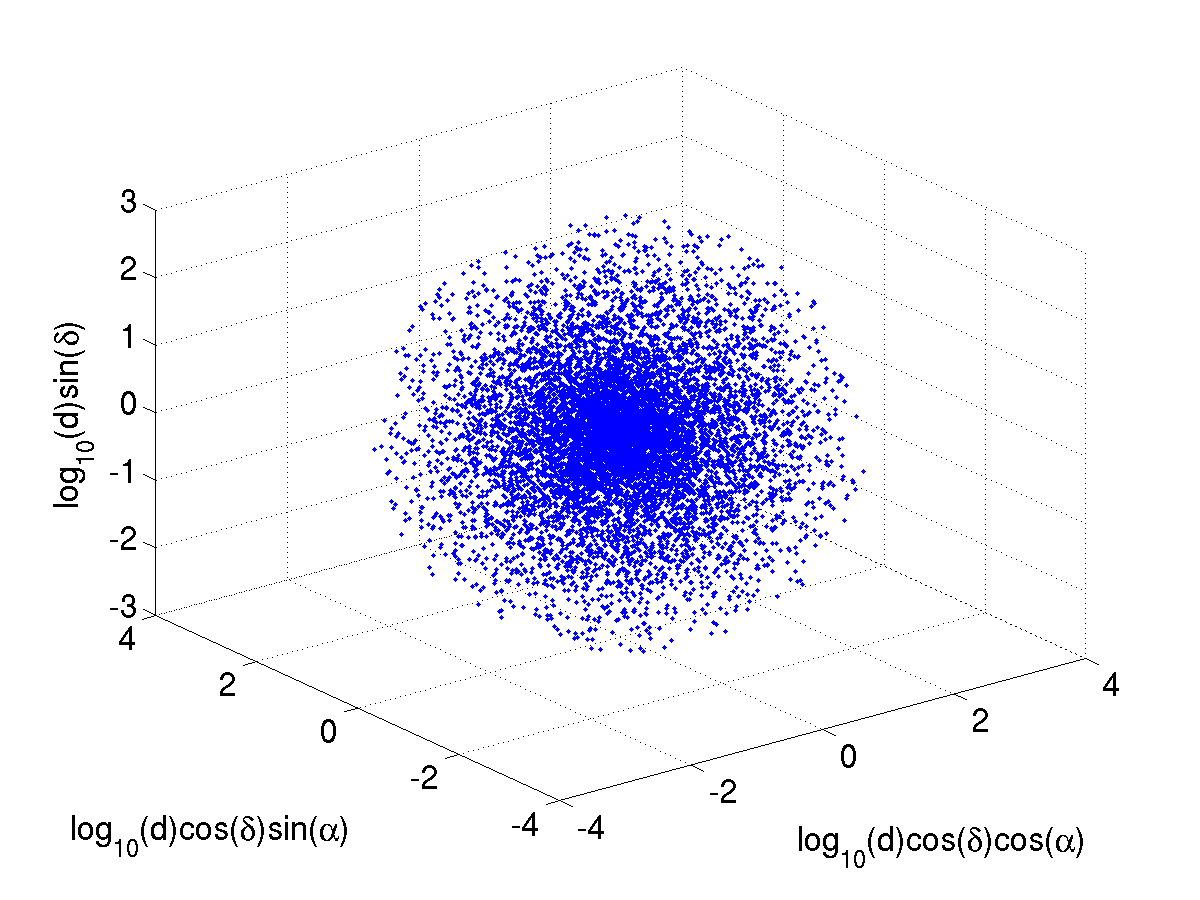}
\caption{Distribution of injections' sky location over nine injection runs.}
\label{fig:injsphere}
\end{center}
\end{figure}


\subsection{Adding the Injection to the Data}
In the context of the pipeline described in chapter \ref{ch:search}, the injection list is read in directly after the raw data. The injection parameters are passed to the ringdown waveform generation code which creates a time series array for the plus and cross amplitudes, $a_+$ and $a_\times$, the frequency $f$ and phase $\phi$, where
\begin{eqnarray}
a_+(t)=  \mathcal{A} (1+\cos^2 \iota) \: e^{-\frac{\pi f_0}{Q} t}, \label{eqn:aplus} \\
a_\times(t) =  \mathcal{A} (2 \sin \iota) \: e^{-\frac{\pi f_0}{Q} t}, \label{eqn:across} \\
f(t)=f_0, \\
\phi(t)=2 \pi f_0 t + \phi_0.
\end{eqnarray}
From these the plus and cross polarizations of the gravitational wave are
created:
\begin{eqnarray}
h_{+_{inj}}(t)=a_+ \cos(2\pi f_0 t+\phi_0), \\
h_{\times_{inj}}(t)=a_\times \sin(2\pi f_0 t+\phi_0).
\end{eqnarray}
Using information about the source and detector positions, the detector antenna patterns $F_+(\alpha,\delta,\Psi)$ and $F_\times(\alpha,\delta,\Psi)$ are calculated and the injection waveform is then given by 
\begin{equation}
h_{inj}(t)=h_{+_{inj}}(t) F_+(\alpha,\delta,\Psi)+h_{\times_{inj}}(t)F_\times(\alpha,\delta,\Psi).
\end{equation}

The transfer function (the inverse of the response function) is applied to the waveform so that it has the same units as the raw data, counts. The waveform is then added data point by data point to the raw data as shown in figure \ref{fig:datainj1plot}, and the pipeline continues as normal. (As can be seen from figure \ref{fig:datainj1plot} there is a discontinuity where the injection starts. This did not have an impact on our ability to recover the injections or on the accuracy of the parameter estimation.)

\begin{figure}[ht] 
\centering
\begin{center}
\includegraphics[scale=0.6]{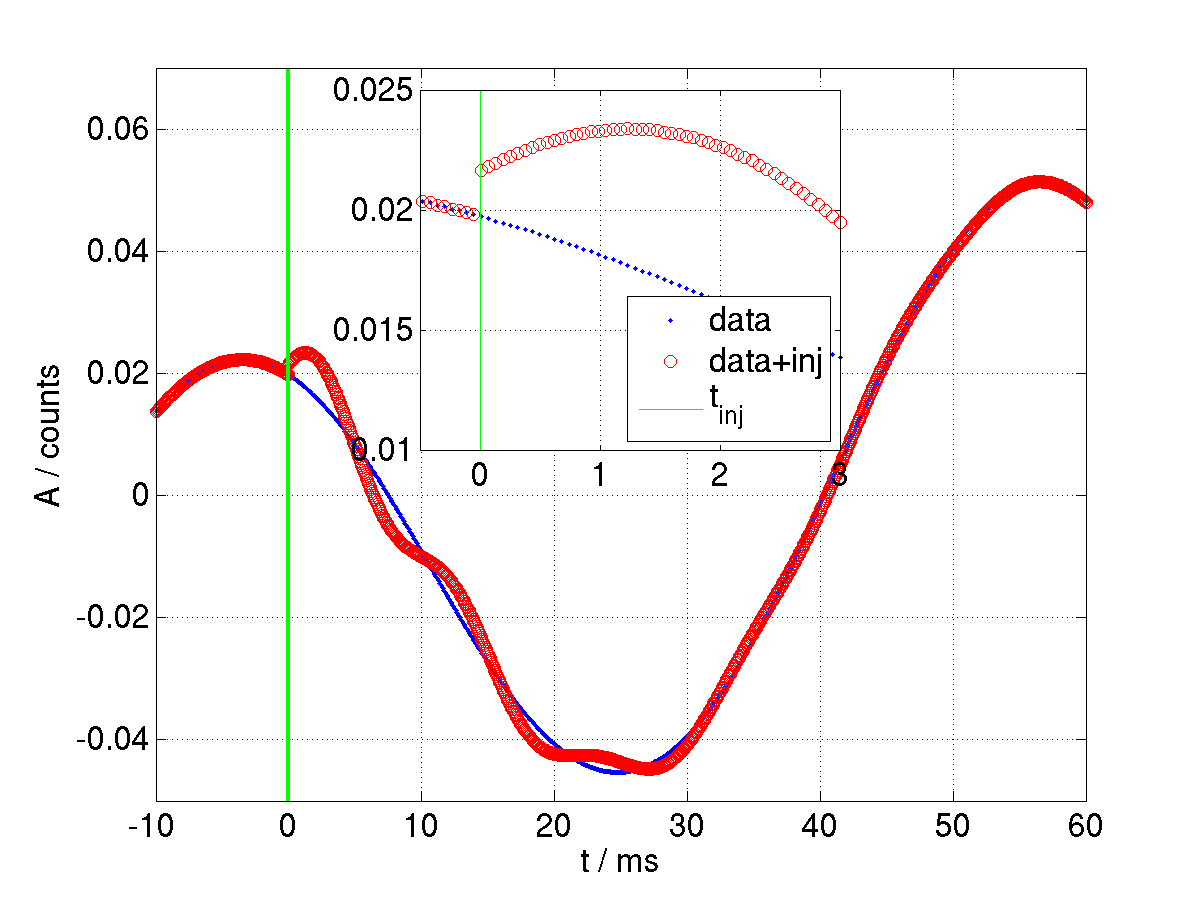}
\caption{Demonstration of the addition of an injection to the raw data; amplitude (in arbitrary units of countes) versus time for 70 ms of raw data (blue) and raw data plus an injection (red). The green line marks the time of the injection according to the injection file. The inset zooms in on the injection, $t_{inj}=0$.}
\label{fig:datainj1plot}
\end{center}
\end{figure}


\subsection{Identifying the Injection Triggers}

The output of the pipeline is a list of coincident triggers consisting of the signals that were injected and coincidences in the data. In order to keep the search blind, at this stage we only want to examine the times at which we added a simulated signal.  We do this by invoking the injection options in {\fontfamily{pcr}\selectfont lalapps\_coincringread}. The input list of injections is read in along with the coincident triggers, and only those groups of coincidences within a 100 ms time window of each injection are examined further. As before, the coincident groups are clustered according to the detection statistic. It is important to note that, even though additional information about the waveform such as the waveform parameters, we do not utilize this as we want to apply a method that can also be applied to the zero-lag data where this additional information is not available. We can, however, use this knowledge after the fact to evaluate how well the injections are recovered (that is the subject of chapter \ref{ch:paramest}). The parameters of the successfully recovered injections were written to a ``found'' file, while those that were not recovered were written to a ``missed'' file. Injections in the list for times where data was not available are discarded, reserving the missed category for injections that were added to the data but were not recovered.


\section{Background Estimation via Timeslides}
\label{sec:timeslides}
As with any search it is important to get an estimate of the background or false alarm rate. Unlike a particle detector we cannot simply turn off the source, or change the orientation of the instrument like one would do in astronomy. Because we require coincidence in time between triggers in multiple detectors, an alternative is to take the trigger files from the filtering stage of two instruments and shift one set in time with respect with the other by a time much longer than the expected length of the signal (and light travel time for separated detectors) and look for coincidences. This is illustrated by the cartoon in figure \ref{fig:cartoonts}, where we can see that the timeslides are done on a ring ensuring that each timeslide contains the same duration of data. In pairs of detectors with uncorrelated noise (H1L1 and H2L1) this provides a good estimate rate of accidental coincidences of noise triggers as the noise sources are completely independent. In the case of H1H2 this is not such a good estimate because by sliding the data sets we are removing known noise correlations from the background and actually under-estimating the rate of false coincidences. We slide L1 fifty times in steps of 5 s and fifty times in steps of $-5$ s, H2 in steps of $\pm10$ s, while H1 stays in place. Timeslides are implemented at the coincidence stage of the pipeline and are initiated by including the ``timeslide'' option in the configuration file. Just as with the zero-lag and injection runs we cluster the coincidences and write them to a {\fontfamily{pcr}\selectfont sngl\_ringdown table}.
\begin{figure}[ht]
\centering
\begin{center}
\includegraphics[scale=0.4]{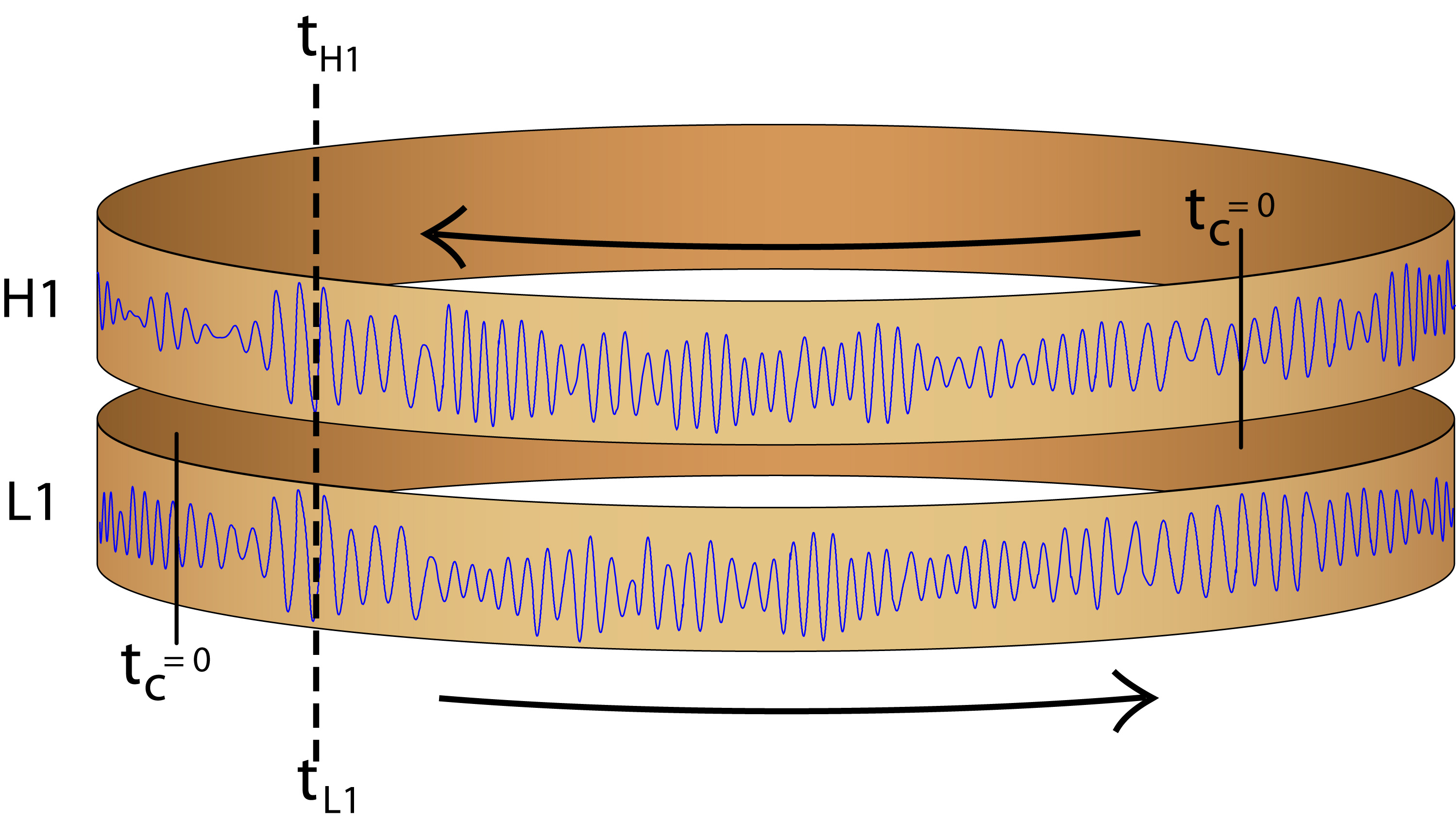}
\caption{Cartoon illustrating timeslides; data is time-shifted such that any
coincidences are accidental, providing an estimate of the rate of false
alarms. (Picture  from R. Tucker.)}
\label{fig:cartoonts}
\end{center}
\end{figure}


\section{Tuning the Constraints}
Next we describe the process involved in tuning the various constraints using
injections and timeslides and the final values chosen.

\subsection{SNR Threshold, Template Bank Limits, and Sampling Rate}
\label{sec:tunbank}
On the first run through the data the SNR threshold was set to 6 in each detector. The template bank frequency limits reflected the sensitivity range of the interferometers, $f_{\textrm{min}}=50$ Hz, $f_{\textrm{max}}=2000$ Hz, and the quality factor range $Q_{\textrm{min}}=2$ and $Q_{\textrm{max}}=20$, was chosen based on the likely range of spins from black holes $0 \leq \hat{a} \leq 0.98$. The maximum mismatch of the template bank was set to 3\%. We downsampled from 16384 s$^{-1}$ to a new sampling rate of 4096 s$^{-1}$. As this yielded a very low rate of background events it was decided to lower the threshold to 5.5 and extend the template bank to search between 40 Hz and 4 kHz. This new $f_{\textrm{max}}$ necessitated a higher sampling rate and thus 8192 s$^{-1}$ was used. The result was a factor of two increase in the number of triggers in H1 and L1 and  a factor of four increase in the number of H2 triggers, giving approximately 10$^6$ triggers from each interferometer in the playground. However, for reasons outlined in section \ref{sec:wings} the increased scope of the template bank was not feasible and it so was returned to its previous frequency range. The threshold and sampling rate remained at 5.5 and 8192 s$^{-1}$, respectively. This resulted in a drop of about 10\% in the number of triggers in H1 and L1 and about 20\% from H2. To summarize, the final values were:
\begin{itemize}
\item{threshold: $\rho^*=5.5$,}
\item{sampling rate: 8192 s$^{-1}$,}
\item{maximum mismatch: $ds^2_{\textrm{max}}=0.03$,}
\item{template bank boundaries: $f_{\textrm{min}}=50$ Hz, $f_{\textrm{max}}=2000$ Hz,
$Q_{\textrm{min}}=2$, $Q_{\textrm{max}}=20$.}
\end{itemize}
With these values the final bank, shown in figure \ref{fig:tpltbank} consisted of 583 templates. The same template bank was used for each detector throughout the run. 

\begin{figure}[h] 
\centering 
\begin{center}
\includegraphics[scale=0.6]{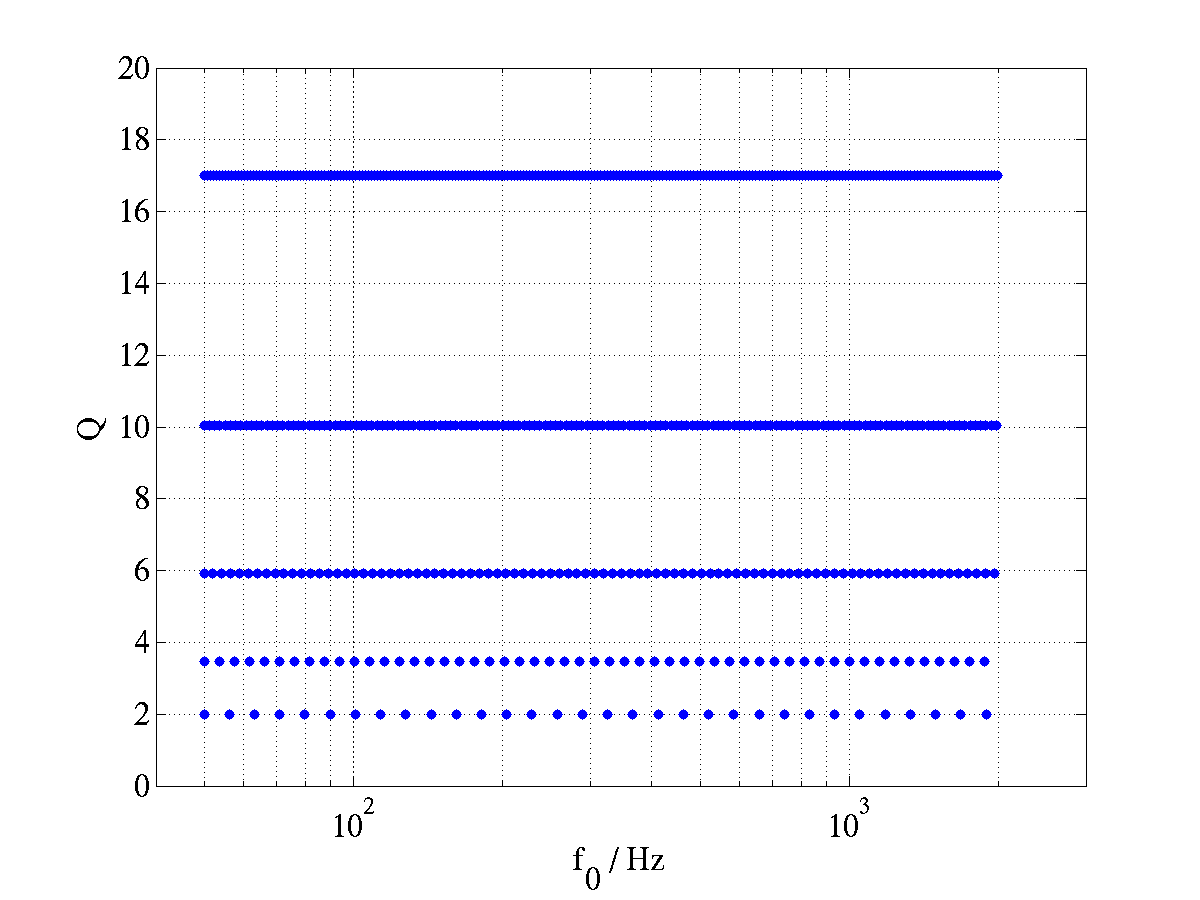}
\caption{The template bank for the S4 ringdown search which, with the parameters $Q_{min}=2$, $Q_{\textrm{max}}=20$, $f_{min}=50 \textrm{ Hz}$, $f_{\textrm{max}}=2000 \textrm{ Hz, and } ds^2_{\textrm{max}}=0.03$ contained 584 templates.}
\label{fig:tpltbank}
\end{center}
\end{figure}


\subsection{Clustering the Filter Output}
\label{sec:tunclustfilter}
In section \ref{sec:clustfilter} we described a method of clustering the matched filter output using a sliding window. This method is implemented to reduce the number of output triggers. It assumes that any triggers occurring within 1 s of each other are correlated. Of course there is the possibility that a noisy block of data could be clustered entirely, giving one trigger for a particular template out of the 256 s of data, but our threshold is sufficiently high that this does not occur. We call the length of time clustered over the dead-time. The upper panel of figure \ref{fig:deadtimebefore} displays the dead-time before each of the triggers in H1 during S4 and the upper panel of figure \ref{fig:deadtimeafter} displays the dead-time after the trigger. The plots show that a high proportion of triggers have dead-times longer than 5 s associated with time. However looking just at the triggers that survive the category 2 and 3 vetoes (the lower panels in figures \ref{fig:deadtimebefore} and \ref{fig:deadtimeafter}), it is apparent that many of these long dead-times occurred during noisy times, and thus the clustering successfully reduces the number of noise triggers output from each filter.

\begin{figure}[htb]
\centering
\begin{center}
\includegraphics[scale=0.6]{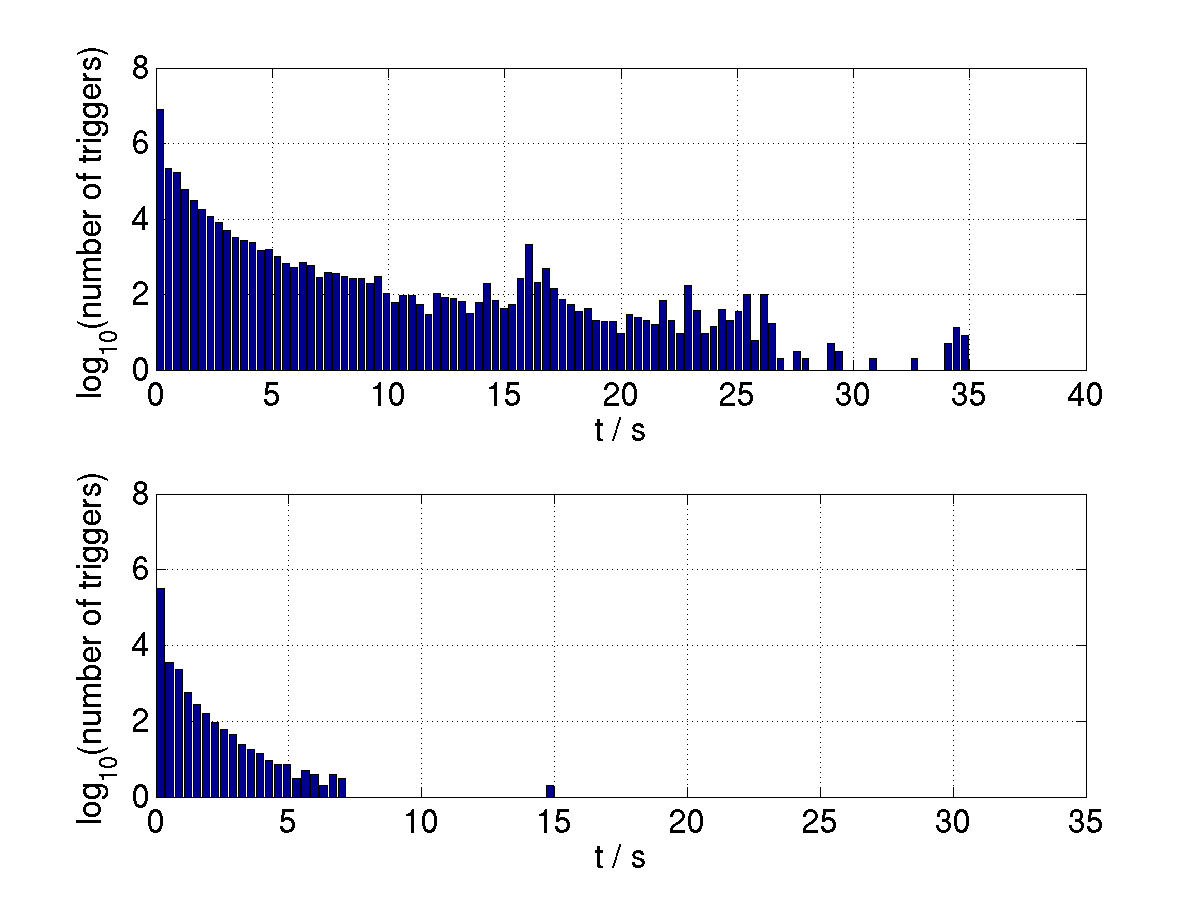}
\caption{Dead-time before triggers in H1 due to clustering of the filter
output. The upper panel shows all the triggers while the lower panel only
shows those that are not cut by category 2 or 3 vetoes.}
\label{fig:deadtimebefore} 
\end{center} 
\end{figure}  

\begin{figure}[htb]
\centering 
\begin{center}
\includegraphics[scale=0.6]{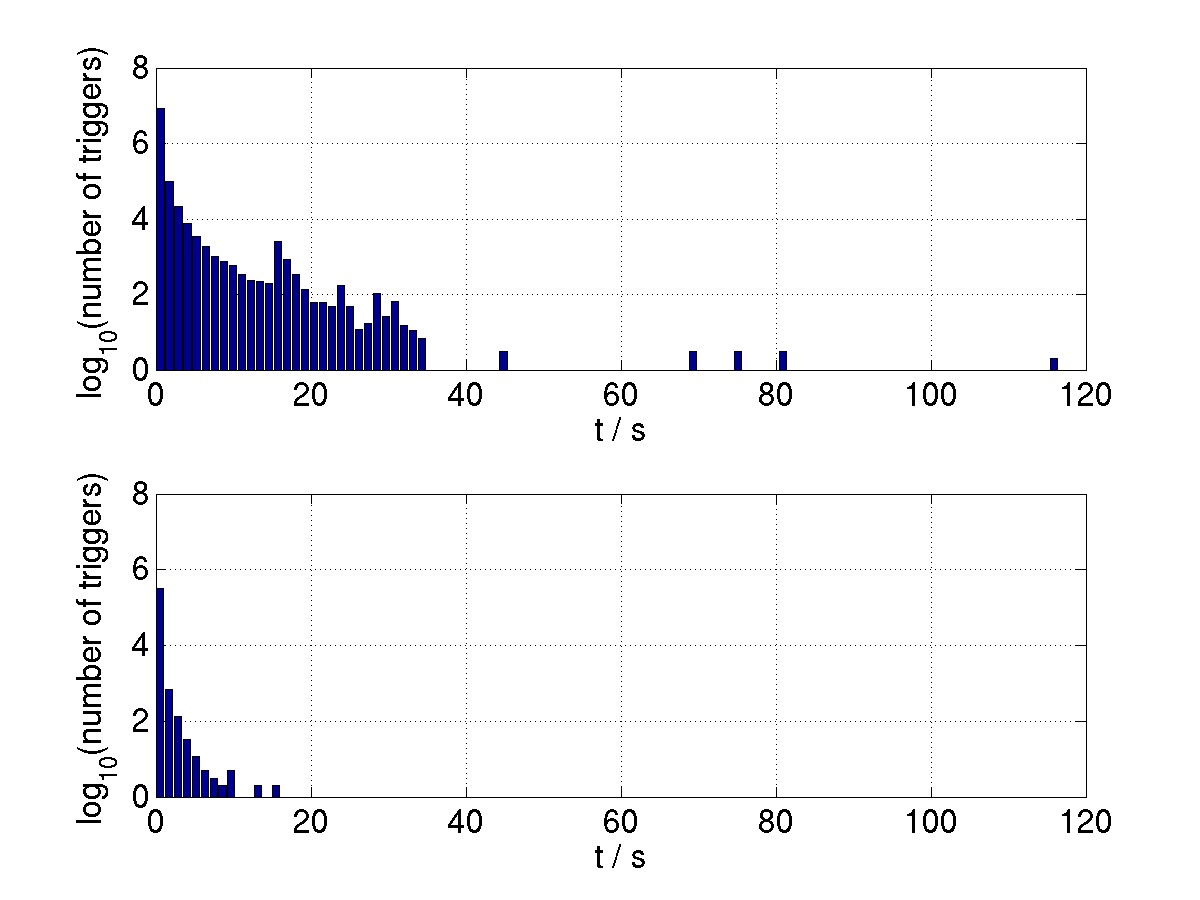} 
\caption{Dead-time after triggers in H1 due to clustering of the filter output. The upper panel shows  all the triggers while the lower panel only shows those that are not cut by category 2 or 3 vetoes.}
\label{fig:deadtimeafter}
\end{center}
\end{figure}


\subsection{Coincidence Windows} 
\label{sec:tunecoinc}
To tune the coincidence windows described in section \ref{sec:coinc} we ran the pipeline on injections and timeslides several times, starting with very wide windows and each time narrowing the windows and comparing the results. Starting with $ds_{coinc}^2=1$ (recall that in addition to being used for template placement the mismatch is also used to define the parameter coincidence test) and time window $\delta t=5$ ms (on either side of the trigger) these parameters were decreased until we reached the point where we started to lose injections. It was observed throughout this iterative process that as one window was tightened the accuracy of the other generally increased. This demonstrates that the clustering techniques described in chapter \ref{ch:search} were not too stringent. The final tuned values were $ds_{coinc}^2=0.05$ and $\delta t=2$ ms for H1H2, and $\delta t=12$ ms for Hanford-Livingston pairs to allow for the maximum gravitational wave travel time between the sites.

\subsection{H1H2 Distance Cut} 
\label{sec:distcuttuning}

When the coincidence windows had been fixed, the effective distance distributions of injections and timeslides were considered. As discussed in section \ref{sec:distancecut} we expect the values of effective distance found by H1 and H2 to be similar for real signals whereas for false coincidences they should be more randomly distributed. Figure \ref{fig:dcut}, a plot of H1-recovered effective distance for injections and timeslides, shows that even though the distributions overlap, there is some portion of the timeslides that can be isolated from the injections. We choose a value of $\kappa=d_{eff_{H1}}/d_{eff_{H2}}=2$, denoted in the plot by a green line. With this value we easily retained all of our injections and were able to reduce the background by 15\%. Figure \ref{fig:histdistcut} displays a normalized histogram of $\kappa=2$ for injections and timeslides. 

\begin{figure}[h]
\centering
\begin{center}
\includegraphics[scale=0.6]{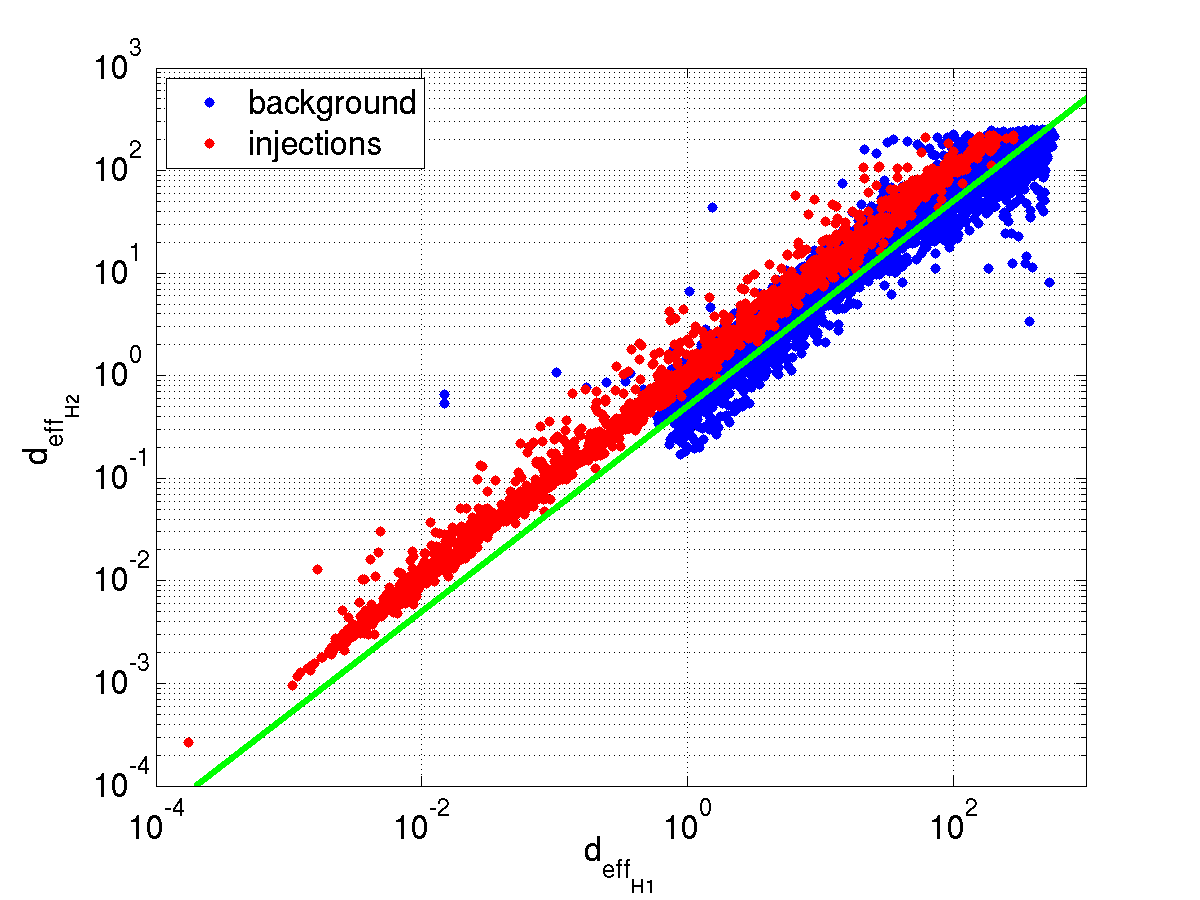}
\caption{The effective distance of H2 versus H1 for injections (red) and timeslides (blue). The green line marks the distance cut; all triggers below this line are discarded.}
\label{fig:dcut}
\end{center}
\end{figure}

\begin{figure}[h]
\centering 
\begin{center}
\includegraphics[scale=0.6]{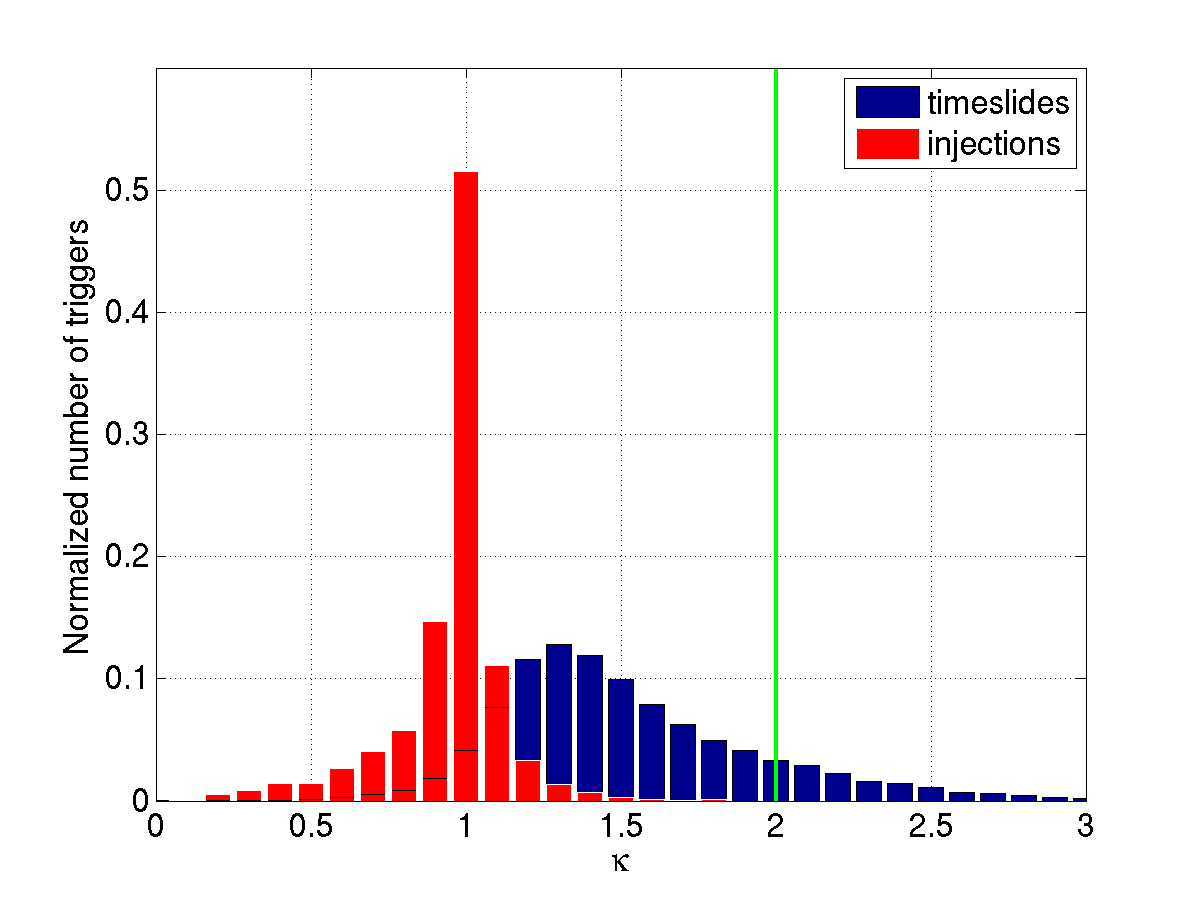}
\caption{The normalized $\kappa=d_{eff_{H1}}/d_{eff_{H2}}$ histograms of injections (red) and timeslides (blue). The green line marks the distance cut; all triggers to the right of this line are discarded.}
\label{fig:histdistcut}
\end{center}
\end{figure}

\subsection{Detection Statistic} 

As discussed in section \ref{sec:detstat}, the detection statistic is a ranking mechanism using the SNRs of coincident triggers. The exact form of the detection statistic for a given population (doubles or triples) depends on the properties of the SNR distributions of simulated signals and background signals.
For triple coincidences the level of background is very low and the SNRs of
the false coincidences all tended to have low values of SNR. As distant simulated signals have a similar distribution the most logical detection statistic is the sum of the squares of the individual triggers.
This, however, is not so for double coincidence background events; these tend to have long ``tails'', that is, coincidences with a very loud SNR in one detector and a much lower SNR in the other. The contour plot of the H1L1 background SNR distribution shown in figure \ref{fig:bgcont} illustrates this. Injections on the other hand generally lie on the diagonal, --- the component SNRs are comparable. If we implemented the triples' detection statistic for doubles then the background trigger at the point (5.5,1000) in figure \ref{fig:bgcont} would be given the same significance as a trigger at (250,250) ($250=10^{2.4}$). Clearly this does not make sense. Thus, while we cannot rule out the possibility that a real signal lies in the tails, we do not want to give it a high significance. Our choice of detection statistic should reflect this.

\begin{figure}[ht]
\centering
\begin{center}
\includegraphics[scale=0.6]{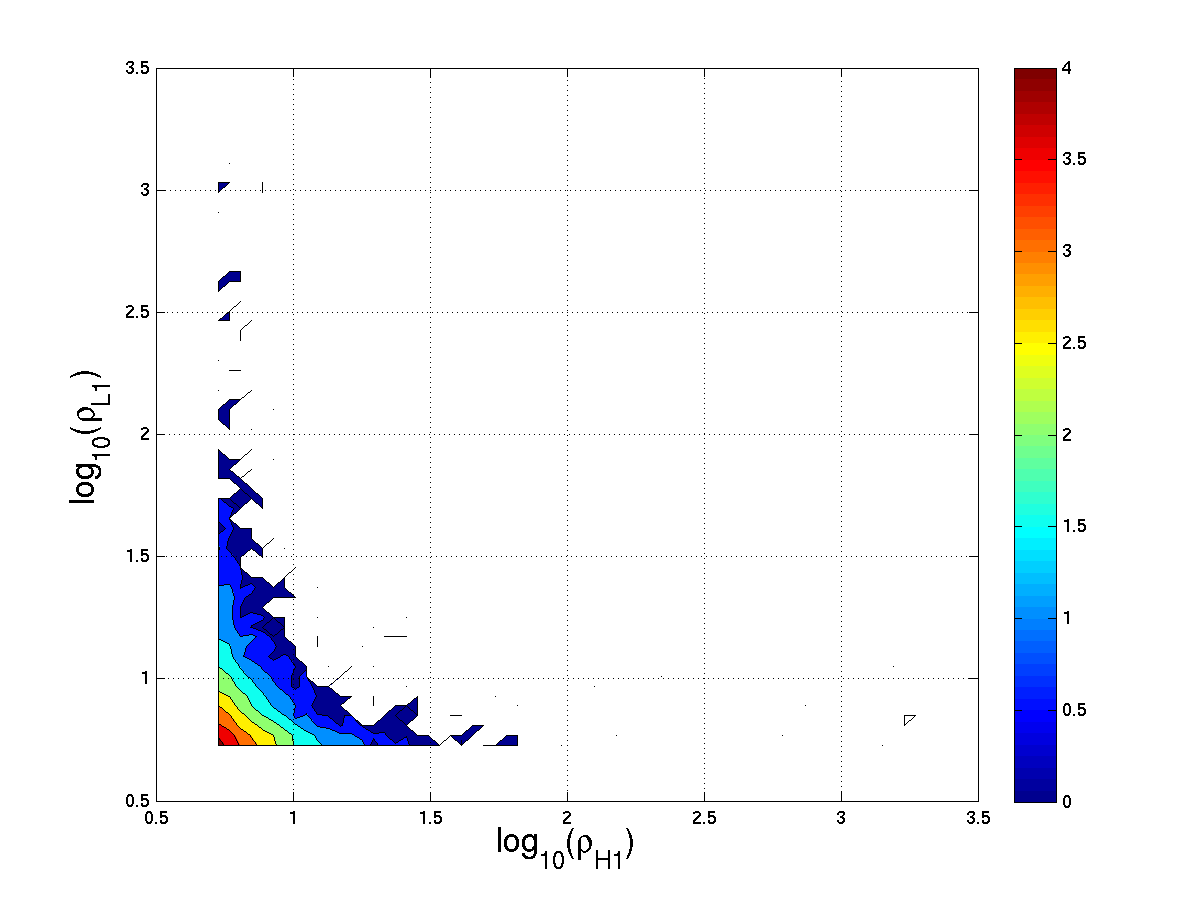}
\caption{A contour plot of the L1 signal-to-noise ratio versus the H1 signal-to noise-ratio for double coincident timeslide event. The colour-bar represents $\log_{10}(N)$ where $N$ is the number of triggers.}
\label{fig:bgcont}
\end{center}
\end{figure}

A simple statistic to do this is the ``chopped-L'' statistic; for a coincidence with signal-to-noise ratio (SNR) $\rho_{\textrm{ifo}1}$ in detector 1 and $\rho_{\textrm{ifo}2}$ in detector 2 the ranking in significance is
\begin{equation}
\rho_{DS}=\textrm{min}\{\rho_{\textrm{ifo1}}+\rho_{\textrm{ifo2}}, \: a_1 \rho_{\textrm{ifo1}}+b_1, \: a_2 \rho_{\textrm{ifo2}}+b_2\}.
\end{equation}
The values of $a$ and $b$ were chosen by considering both the injections and the timeslide distributions. For simplicity we choose a symmetric statistic $a_1=a_2=a$, $b_1=b_2=b$. Figure \ref{fig:detstatcont} shows the SNR distribution of H1L1 timeslides and injections with detection statistic contours of constant values of the detection statistic marked. Tuning the values of $a$ and $b$ essentially amounts to varying where the horizontal and vertical lines cross the diagonal for a given contour. The aim is to find a balance between having as many injections lie on the diagonal while keeping the horizontal and vertical lines forward of the tails. Suitable values for this search were found to be $a=2$, $b=2.2$. 
\begin{figure}[ht]
\centering
\begin{center}
\includegraphics[scale=0.6]{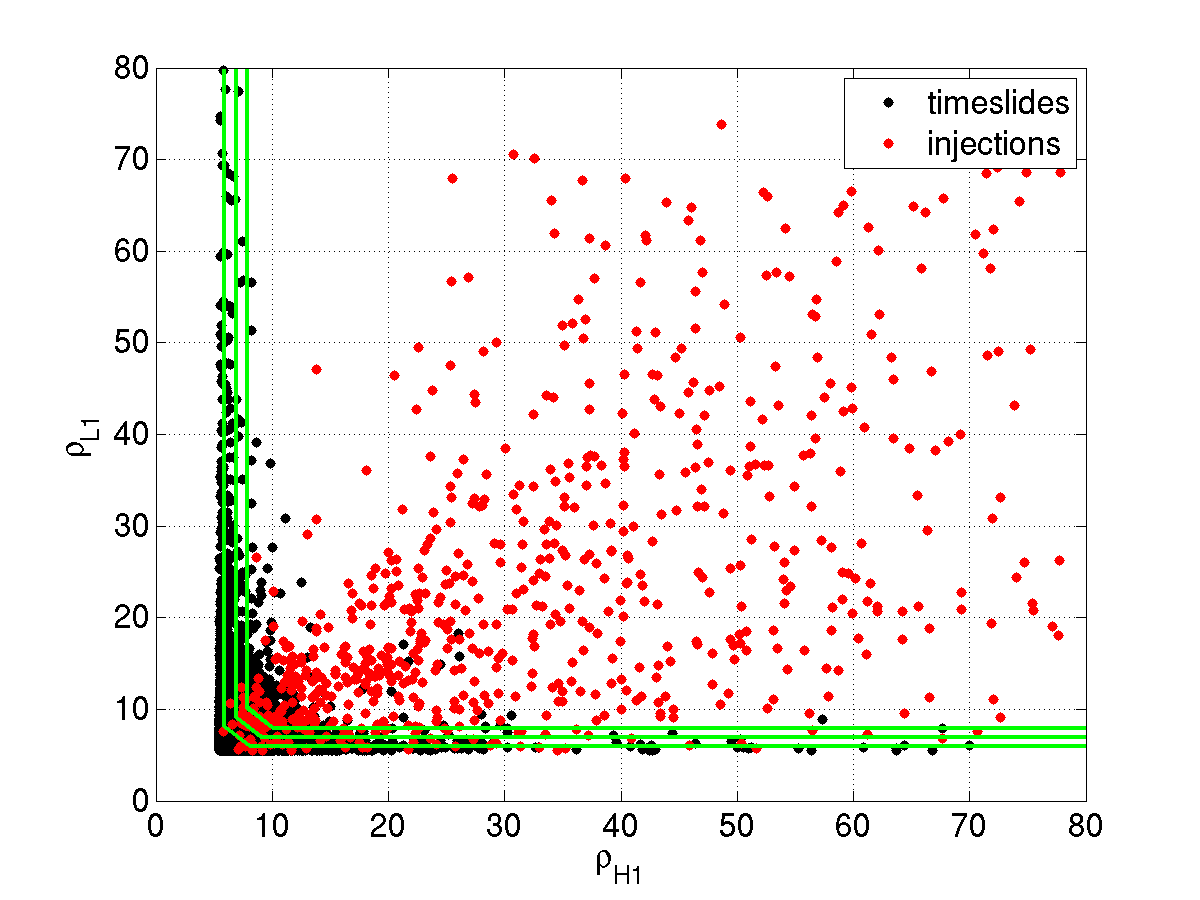}
\caption{The H1L1 SNR distribution for timeslides (black) and injections (red). Contours of constant values of the detection statistic with $a=2$ and $b=2.2$ are marked in green.}
\label{fig:detstatcont}
\end{center}
\end{figure}


\section{Playground}
\label{sec:playground}

A trigger at GPS time $t$ is in the playground if it satisfies 
\begin{equation}
\textrm{mod}(t-t_{\textrm{S2}},6370)<600
\end{equation}
where $t_{\textrm{S2}}$ is the GPS start time of the second LIGO science run 729273613. This constitutes approximately 9.5\% of the data set. In the development of the pipeline and in the tuning process we are allowed to look at this representative subset and leave the remaining 90.5\% blind until all the parameters have been decided upon. It is, of course, possible that a gravitational wave may lie in the playground; this by no means invalidates the detection. However information from the playground is not used in determining an upper limit, in order to avoid bias on the part of the analyst from influencing cuts that determine the upper limit. 

In this analysis we did not use the playground directly for tuning, however it
was used in the development of the pipeline for testing code. The distribution
of coincident events found in the playground is discussed in section
\ref{sec:pgbg}.

\chapter{Parameter Estimation}
\label{ch:paramest}

\section{Introduction}
In the previous chapter we discussed how the search was tuned, with our goal being to recover as many injections as possible while keeping the rate of false alarms to a minimum. Once all of the constraints were decided upon we ran a large scale injection run; the findings of the injection run are described in this chapter. In particular we discuss the efficiency of the search and the accuracy with which we recover the injected parameters. We also compare recovered parameters between detectors and the parameters of accidental coincidences. In addition we unblind a fraction of the data, the playground, and compare this to our estimated false alarm rate.


\section{Recovery of Simulated Signals}
\label{sec:mfinjs}
Nine injection runs were made into times when all three interferometers were in science mode (hereafter referred to as triple time). As discussed in section \ref{sec:injections}, injections were made at a maximum rate of one per 2176 s analysis segment, totaling 5142 injections over the nine runs. The primary purpose for such a large scale Monte Carlo simulation is to evaluate the efficiency, which is needed to calculate an upper limit. As we intend to place an upper limit on the rate of triple coincidences, we do not perform injections into times when data was available from only two instruments (double time).

\subsection{Single Interferometer Injections}
Before delving into the results of the coincidence analysis it is interesting to consider the single detector results; we choose H2 as an example. We can compare the triggers from the output of the filtering stage to the list of injections using {\fontfamily{pcr}\selectfont lalapps\_ringread} (in much the same way as we use {\fontfamily{pcr}\selectfont lalapps\_coincringread} for the coincidence analysis, as described in section \ref{sec:postproc}) to ascertain how many injections were missed and how many were found. As we discussed in relation to the coincidence analysis, this process does not make use of any information regarding the injections, apart from the time interval into which they were injected. As a consequence, a trigger due to noise occurring within that time interval with a higher signal-to-noise ratio (SNR) than the injection would be erroneously identified with the injection. In such a situation the recovered waveform parameters (frequency and quality) will most likely not be close to those injected. 

In figure \ref{fig:mfH2} we plot the effective distance versus frequency of the injections missed (in red) and found (in blue) in the nine injection runs. The plot shows that, for a given frequency, there is a distance at which we begin to miss injections. As we move across in frequency from 50 Hz this distance increases to a maximum of 200 Mpc at approximately 90 Hz and then falls off again to 300 kpc at the high-frequency end of the template bank. Superimposed is the expected horizon distance (discussed in section \ref{sec:horizondistance}) for a ringdown with an SNR of 5.5 and spin of 0.98  created using the best S4 H2 noise curve. The boundary between missed and found injections follows the horizon distance curve nicely, illustrating how features in the noise curve effect the distance out to which an injection can be recovered. (Such features include the calibration line at 54 Hz, the power line at 60 Hz and its harmonic at 120 Hz, and the violin mode resonances of the test mass suspensions at 340 Hz.)  This plot shows that we find a trigger within 100 ms of every injection  that we would expect to recover (i.e., below the horizon distance line). In addition we find some distant injections, however in most of these cases, it is spurious noise in the detector that we are finding and not the injection itself. We will see in the next section that this effect is dramatically reduced by requiring coincidence between detectors, illustrating of the power of the coincidence test.

\begin{figure}[htb]
\centering
\begin{center}
\includegraphics[scale=0.6]{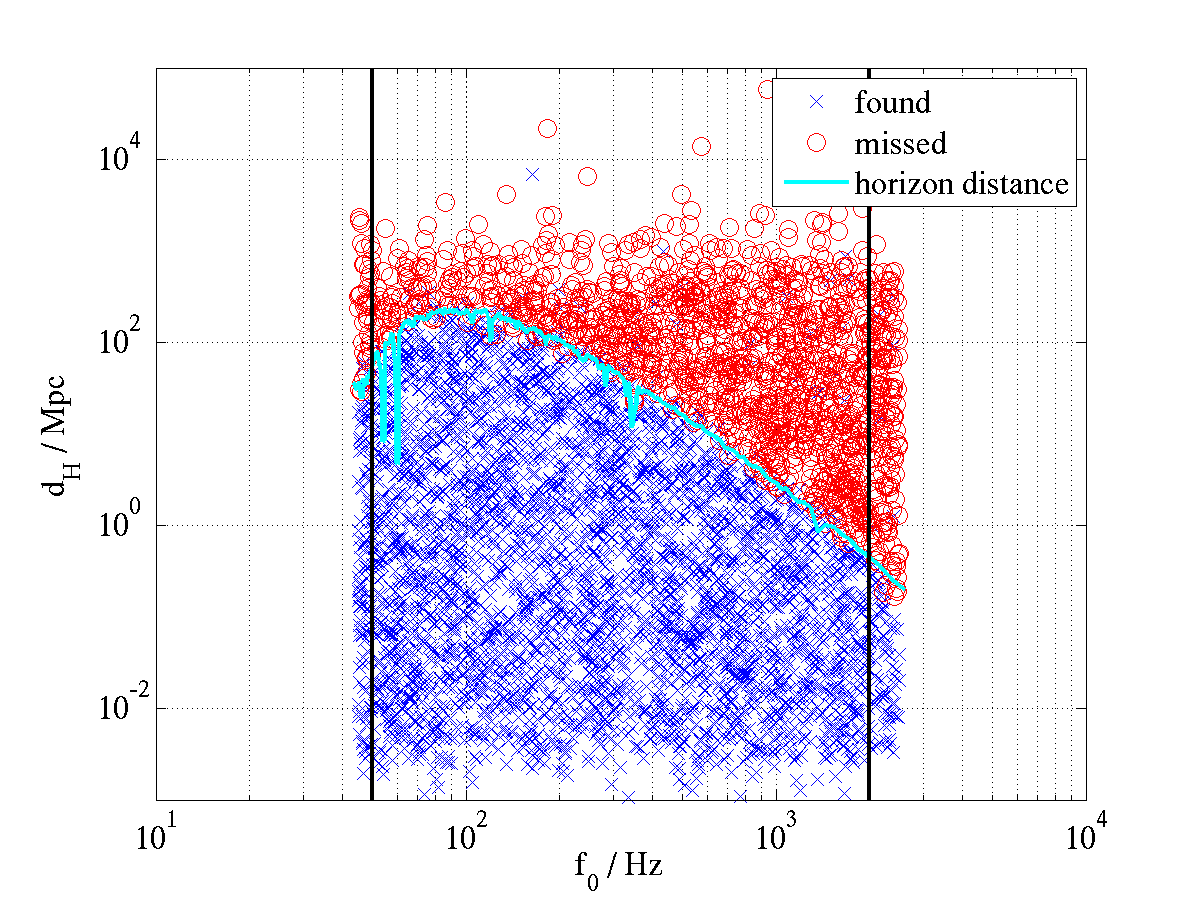}
\caption{Hanford effective distance versus frequency for missed (red) and found (blue) injections in the H2 single interferometer analysis. The cyan line is the horizon distance for a source with a spin of 0.98 producing an SNR of 5.5 in the detector, assuming that 1\% of its mass is radiated as gravitational waves. }
\label{fig:mfH2}
\end{center}
\end{figure}

\subsection{Coincident Injections}
Running {\fontfamily{pcr}\selectfont lalapps\_coincringread} on the output of the coincidence step gives us a list of the triggers identified with the injections, one coincidence for each injection (as discussed in section \ref{sec:postproc}). These can be categorized as injections found in triple coincidence (referred to as triples in triple time, or simply triples) and injections found in coincidence in two detectors (doubles in triple time, or doubles). The doubles are further divided into those that were missed in the third detector because that time was vetoed and those that were simply not seen in the third detector. Note that intervals within triple time when one detector was vetoed are still regarded as triple time; we consider this a loss of efficiency (of detecting triples) rather than a loss of live time. One could, in theory, account for the lost analysis time, however the extra complexity this entails in terms of bookkeeping is not justified given the small difference this makes.

\subsubsection{Missed and Found Injections}

Figure \ref{fig:mf_allvH} shows a plot similar to that discussed above, the effective distance versus frequency of missed and found injections from the coincidence analysis. The injections found in triple coincidence are marked in blue and those injections found in two detectors are denoted by green, cyan, and magenta stars for H1H2, H1L1, and H2L1 doubles, respectively. The injections vetoed in one detector and found in the other two have a black circle surrounding the star to emphasize the fact that technically they were not missed in the third detector (although they are treated that way in the calculation of the efficiency). The doubles are shown on their own in figure \ref{fig:mf_doubsH}. As before, the missed injections are marked in red. Our sensitivity to triples is limited by the least-sensitive instrument, H2, and thus the distance out to which we see triples depends on how far H2 can see. For that reason the distance at which we no longer find injections in triple coincidence is approximately the same as the distance that we start missing H2 injections in figure \ref{fig:mfH2}. Beyond this limit H1 and L1 are still sensitive enough to detect ringdowns, and so we see a thin line of H1L1 double coincidences beyond the distance at which the triples end. At high frequencies there is also a band of H1H2 injections mixed in with the triples. This is because during the S4 run the sensitivity in L1 decreased as the laser power was lower (as discussed in section \ref{sec:s4}), and thus injections made at large distances during this time were missed in L1 while those made when L1 was running with full power were found. The remaining uncircled doubles (i.e., those missed in the third detector) scattered throughout the predominantly blue area should have been found in the third detector and were followed up on an injection-by-injection basis. Further investigation showed that these were predominantly due to excess noise in the third detector, causing the SNR to peak at a frequency other than the injected frequency, and as a result this detector failed the coincidence test. However, with these exceptions, this plot shows that we are recovering the all injections we would expect to recover, and this reassures us that the tools we are employing in the analysis are sound.

\begin{figure}[htb]
\centering
\begin{center}
\includegraphics[scale=0.7]{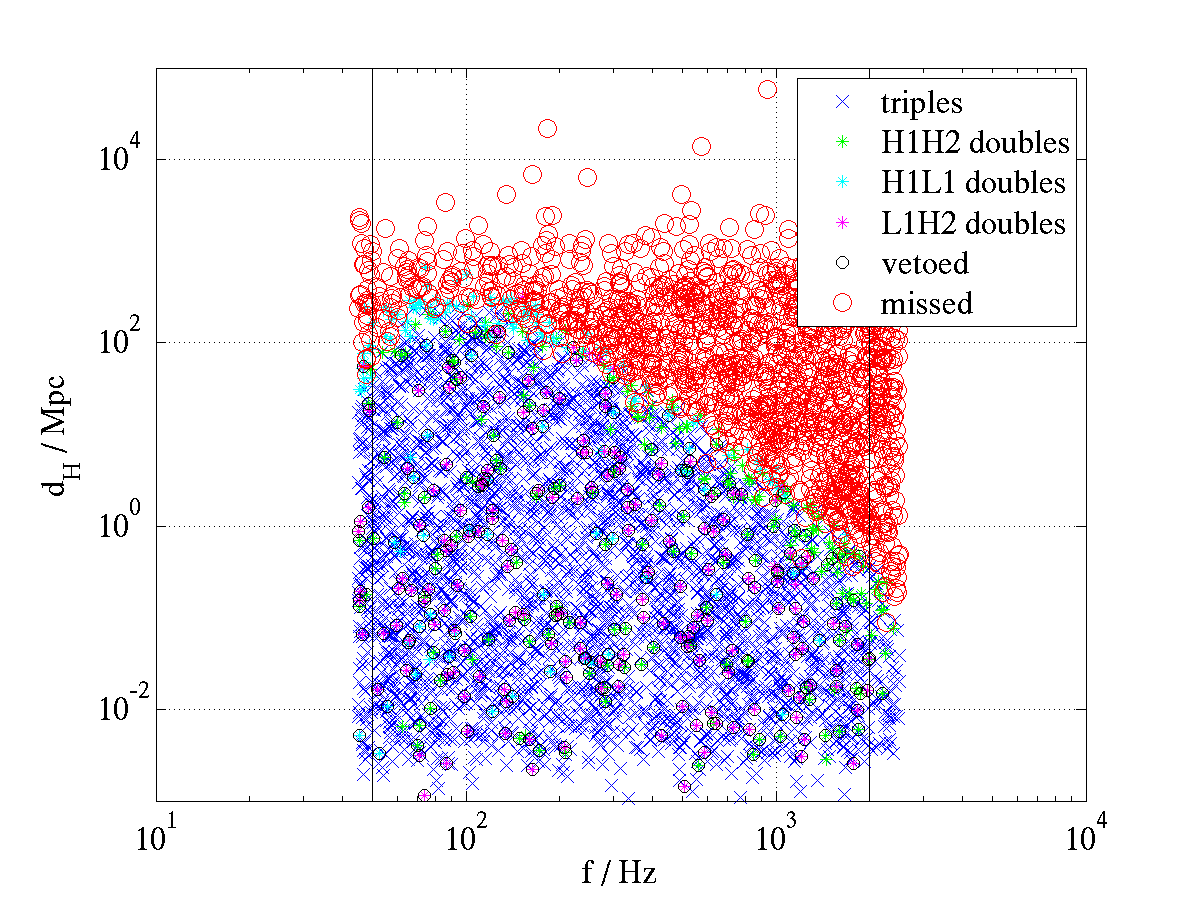}
\caption{Hanford effective distance versus frequency for injections missed (red circles) and found in coincidence. Injections found in triple coincidence are marked as blue crosses, injections found in double coincidence are shown as green (H1H2) cyan (H1L1) and magenta (L1H2) stars and those that were vetoed are also marked with a black circle.}
\label{fig:mf_allvH}
\end{center}
\end{figure}

\begin{figure}[htb]
\centering
\begin{center}
\includegraphics[scale=0.7]{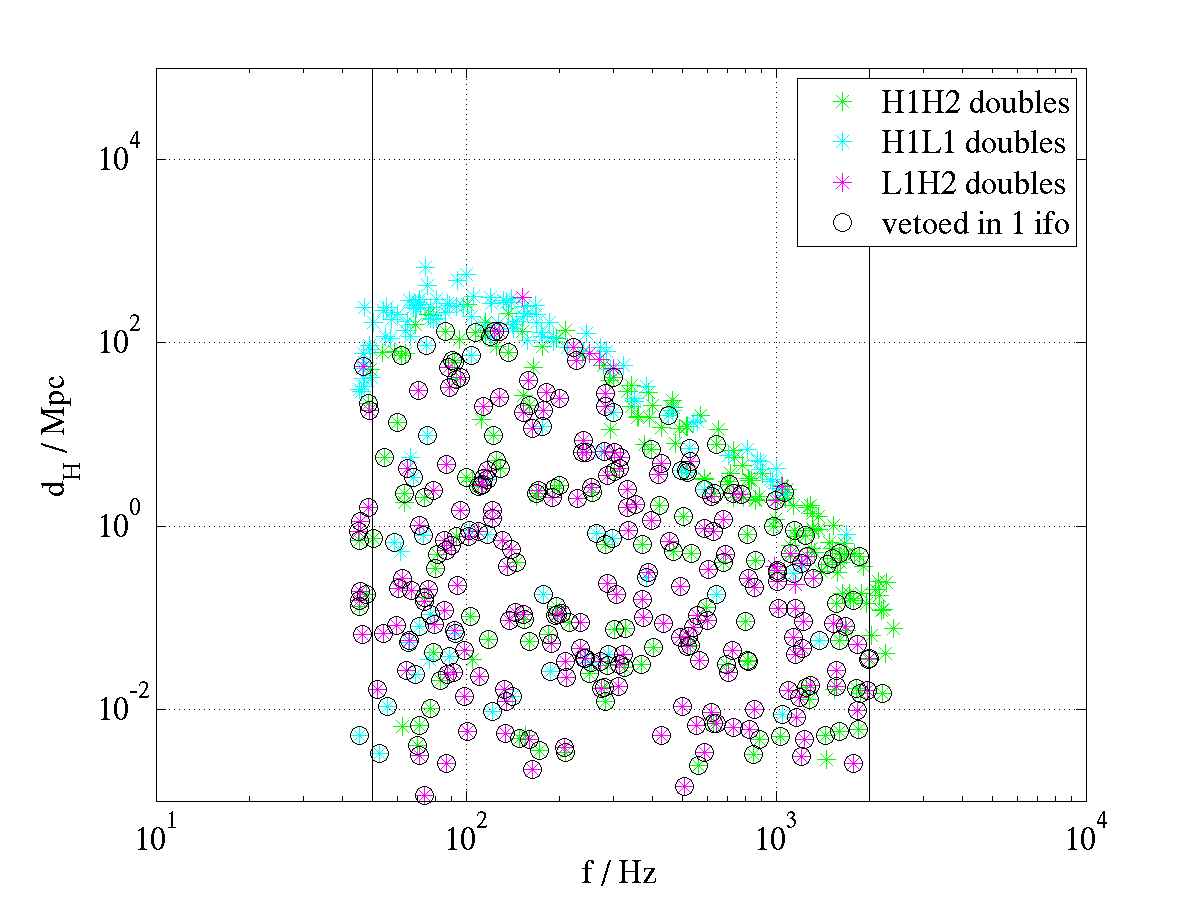}
\caption{Hanford effective distance versus frequency for injections found in double coincidence. The coloured stars represent each of the detector pairs (H1H2 doubles are marked in green, H1L1 in cyan and L1H2 in magenta). The black circles mark those doubles that were vetoed in the third detector.}
\label{fig:mf_doubsH}
\end{center}
\end{figure}

\subsubsection{Efficiency of Finding Triple Coincidences}

We evaluate the efficiency $\varepsilon$ of finding triples, that is the fraction of injections found in triple coincidence, as a function of injected (physical) distance. This is implemented by binning the injections in logarithmic distance and calculating the efficiency in each bin. A plot of efficiency versus distance is shown in  figure \ref{fig:effvd_tintt_45_2000}. The uncertainty in the efficiency is assumed to be binomial,
\begin{equation}
\sigma_\varepsilon^2=\frac{ \varepsilon (1-\varepsilon) }{N},
\end{equation}
where $N$ is the total number of injections made.
\begin{figure}[htb]
\centering
\begin{center}
\includegraphics[scale=0.6]{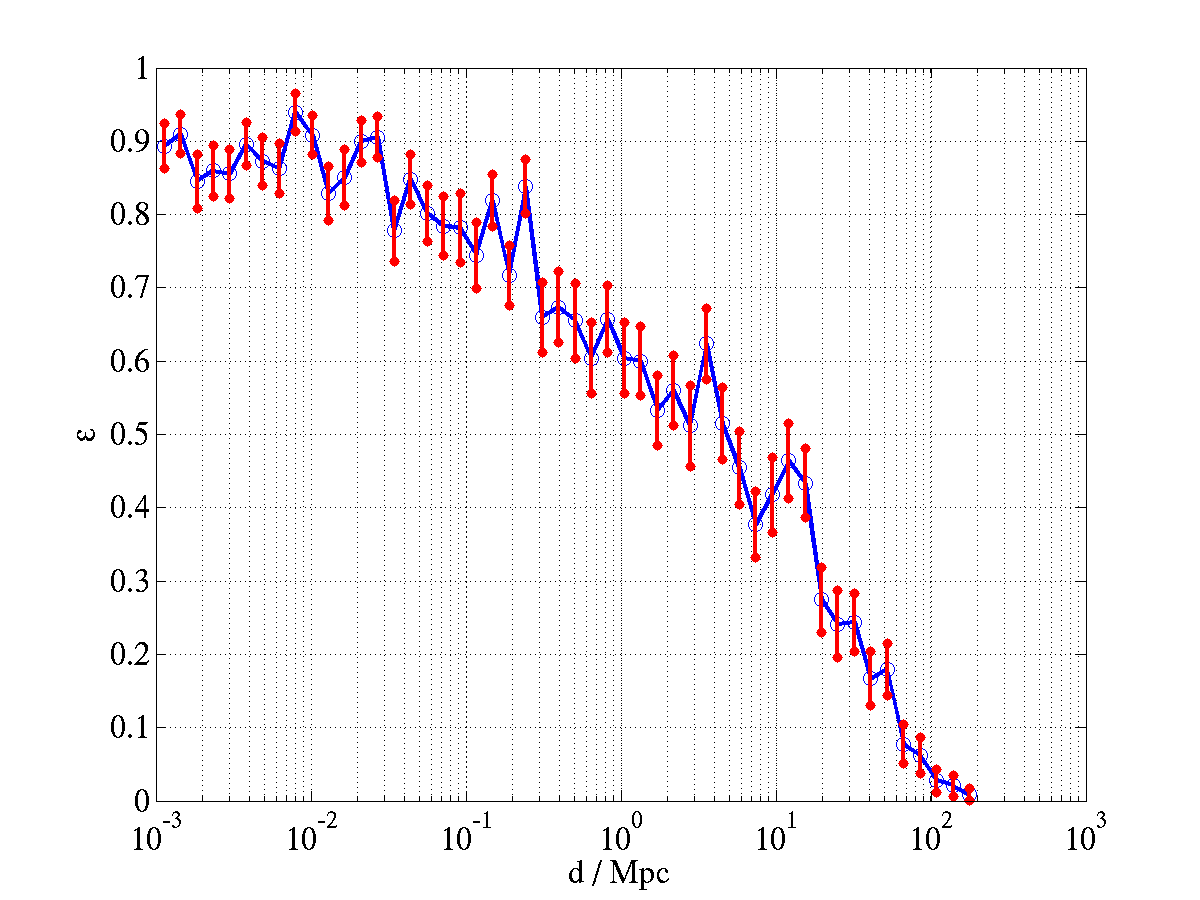}
\caption{The efficiency of finding injections in triple coincidences as a function of physical distance for injections made between 45 Hz and 2.5 kHz.}
\label{fig:effvd_tintt_45_2000}
\end{center}
\end{figure}

The first thing to note is that the efficiency is never unity, even at small distances, because, as mentioned in section \ref{sec:coinc} we apply category 2 and 3 vetoes, and thus some injections that otherwise may have been found as triples are found as doubles or not at all. The second feature of note is the gradual slope. This is because the plot encompasses all frequencies and, as is obvious from figure \ref{fig:mf_allvH}, the efficiency is a strong function of frequency. In chapter \ref{ch:results} we present analogous plots for smaller frequency bands. Considering injections at all values of the central frequency, we see from figure \ref{fig:effvd_tintt_45_2000} that the 50\% efficiency point lies at a distance of approximately 4 Mpc.  


\section{Comparison of Injected and Detected Parameters for Injections Found in Triple Coincidence}

Comparing the recovered parameters to the injected parameters gives us a sense of the accuracy with which we can expect to recover the parameters of a gravitational wave ringdown. Here we compare the injected and detected time, metric distance, and effective distance of injections found in triple coincidence.

\subsection{Time of Arrival} 
\label{subsec:injdettime}

Figure \ref{fig:histH1H2L1dt} shows a histogram of the difference between the detected and injected times $\delta t$ for H1, H2, and L1. This distribution is highly asymmetric for all three detectors with a peak close to zero and a tail extending to negative $\delta t$. Figure \ref{fig:L1injdettvf} shows that the accuracy of the time of the injection is a strong function of frequency (L1 is shown, H1 and H2 displayed a similar trend), with largest $\delta t$ occurring at low frequencies and decreasing as the frequency increases. Further investigation showed that this is because the injections turn on suddenly at $t=0$ with a random phase as demonstrated in figure \ref{fig:datainj1plot}. The templates have a phase of zero, and thus the maximum time difference is inversely proportional to the frequency. The spread in $\delta t$ represents the randomness of the initial phase of the injection. 

\afterpage{\clearpage}
\begin{figure}[htb]
\centering
\begin{center}
\includegraphics[scale=0.6]{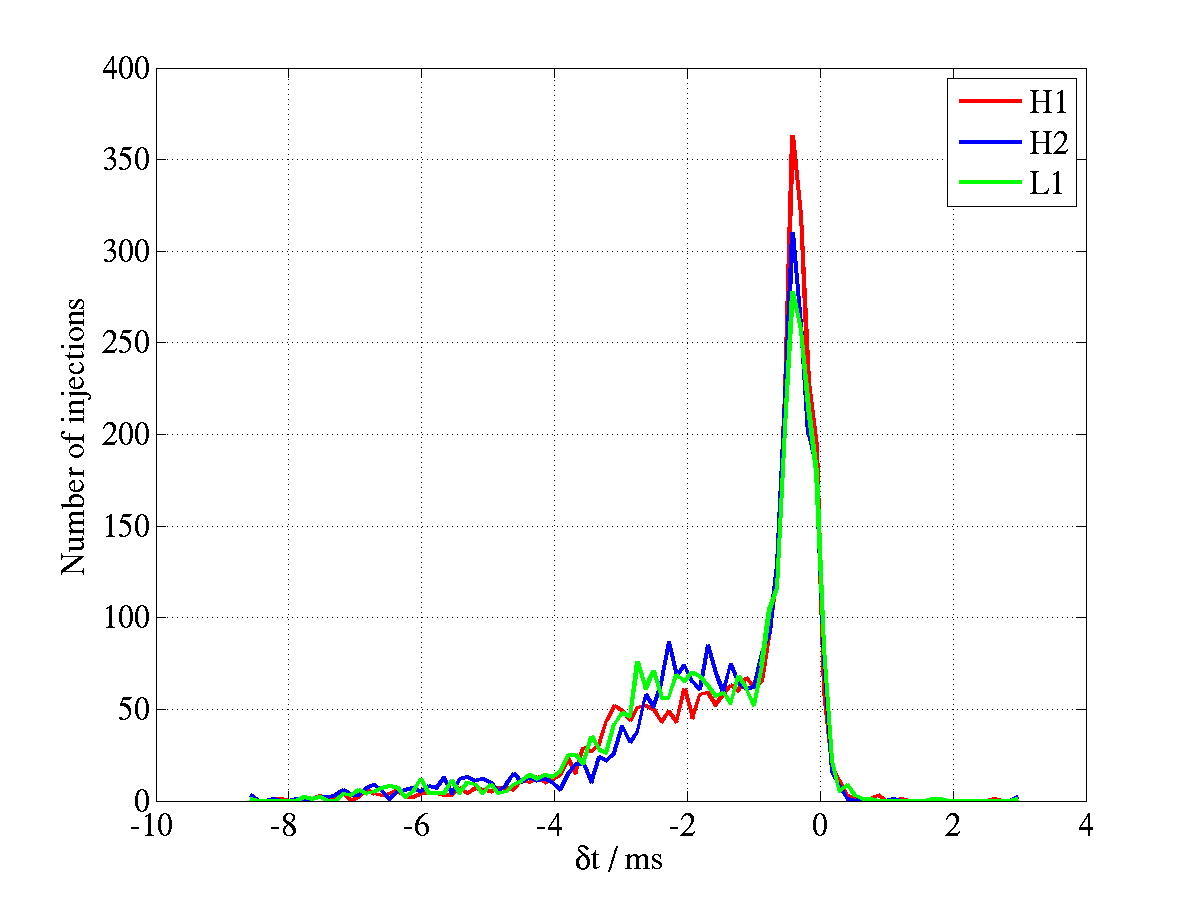}
\caption{Histogram of difference between detected and injected time of arrival for H1, H2, and L1 injections found in triple coincidence.}
\label{fig:histH1H2L1dt}
\end{center}
\end{figure}

\begin{figure}[htb]
\centering \begin{center}
\includegraphics[scale=0.6]{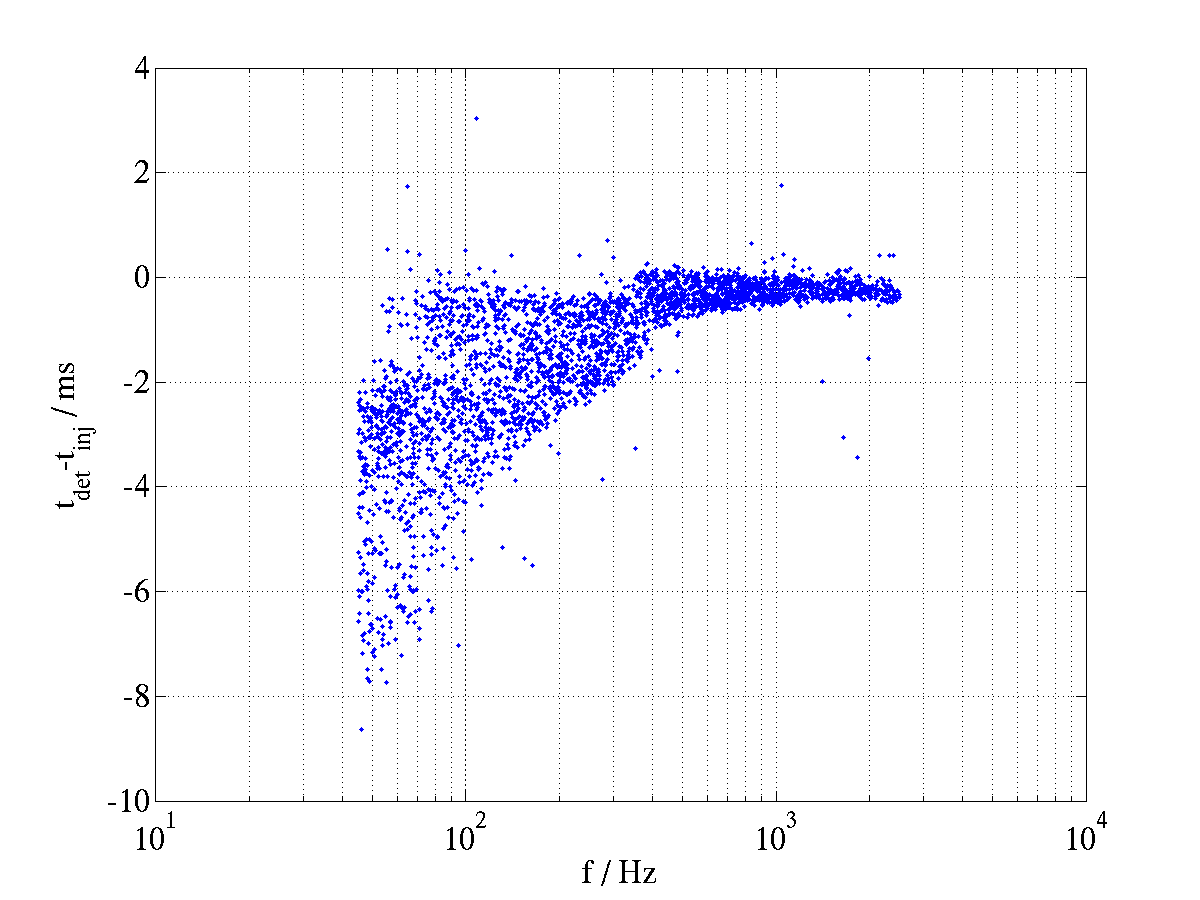}
\caption{Difference between L1 detected and injected time versus injected frequency for injections found in triple coincidence.}
\label{fig:L1injdettvf}
\end{center}
\end{figure}


\subsection{Metric Distance}

We can calculate the metric distance between the injection and the template that rang off with the loudest SNR. Recall from section \ref{sec:tmpltbnk} that the template bank is laid out such that the maximum metric distance between any point within the $f_0$ and $Q$ boundaries and the nearest template is less that $ds^2_{\textrm{max}}$, where in this analysis we have set $ds^2_{\textrm{max}}=0.03$. We can calculate this distance between an injection and the template that recovered it using equation 
(\ref{eqn:metric}). In the implementation of the coincidence test an error was
made with the result that the metric distance is under-estimated by a factor
of 4. Therefore, in the context of the following discussion the value of the maximum mismatch in the layout of the template bank is $ds^2=0.03/4=0.0075$.
A histogram of $ds^2$ calculated between the injected and detected quantities for H1, H2, and L1 triple coincident injections is shown in figure \ref{fig:histH1H2L1ds2}. The maximum distance between a template and any point in the space is marked by a black vertical line at $ds^2=0.0075$; to the left of this line are all the injections that were found by the correct (i.e., the closest) template. This accounts for approximately $\sim70\%$ of the injections and gives us a high level of confidence in our ability to recover the parameters of ringdowns.

Figure \ref{fig:L1injdetds2vf} shows how $ds^2$ varies with frequency in L1 (H1 and H2 displayed similar structure). Not surprisingly, the points follow the general trend of the noise curve (see figure \ref{fig:s4strain}) with the lower values of $ds^2$ lying in the most sensitive frequency band of the detector and higher values of $ds^2$ at the highest and lowest frequencies. The injections made outside the template bank obviously have the highest values of $ds^2$, as the nearest template is always going to have $ds^2>0.0075$. The spread in $ds^2$ for a given frequency can be attributed to the coarseness of the $Q$ parameter in the template bank. This is demonstrated by figure \ref{fig:L1injdetds2vQ}, a plot of $ds^2$ versus quality factor. Here we see that as the injected value of $Q$ deviates from one of its five templates, the mismatch increases until such point as the next $Q$ template is closer than the previous.  Thus we see a series of arches, each with the same maximum height. 

We can examine both $f_0$ and $Q$ on a scatter plot with $ds^2$ as the colour scale; this is shown in figure \ref{fig:L1fvQvds2}. From this we can see that the lowest value of the mismatch between a template and an injection occurs along the lines of $Q$ templates, indicated by black lines. The highest values of $ds^2$ occur, as expected, along the template boundaries. However it is interesting to note that the frequency dependence that we observed in figure \ref{fig:L1injdetds2vf} is most pronounced at low values of $Q$. Injections on the $Q=17$ line are mostly found by the closest template (dark blue to cyan on the colour scale). In contrast, on the $Q=3.6$ line, injections are rarely found by the closest template above an injected central frequency of $\sim300$ Hz.

\afterpage{\clearpage}

\begin{figure}[htb]
\centering
\begin{center}
\includegraphics[scale=0.55]{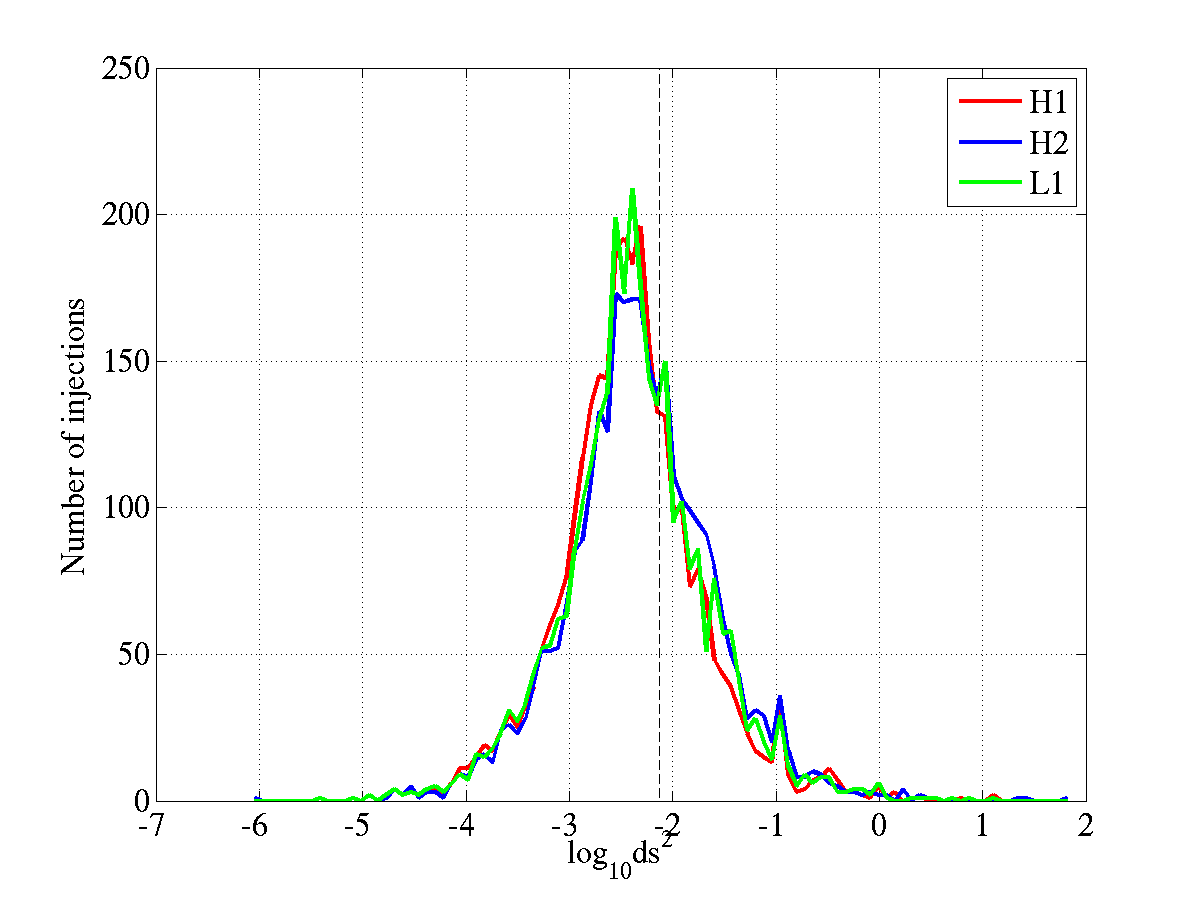}
\caption{Histogram of the mismatch $ds^2$ between injections and the templates
they were found with for H1, H2, and L1 triple coincidence injections. The
black vertical line marks $ds^2=0.0075$, the template bank maximum
mismatch. The plot shows that $\sim 70\%$ of the injections found in triple
coincidence were found with the correct template.}
\label{fig:histH1H2L1ds2}
\end{center}
\end{figure}

\begin{figure}[htb]
\centering \begin{center}
\includegraphics[scale=0.55]{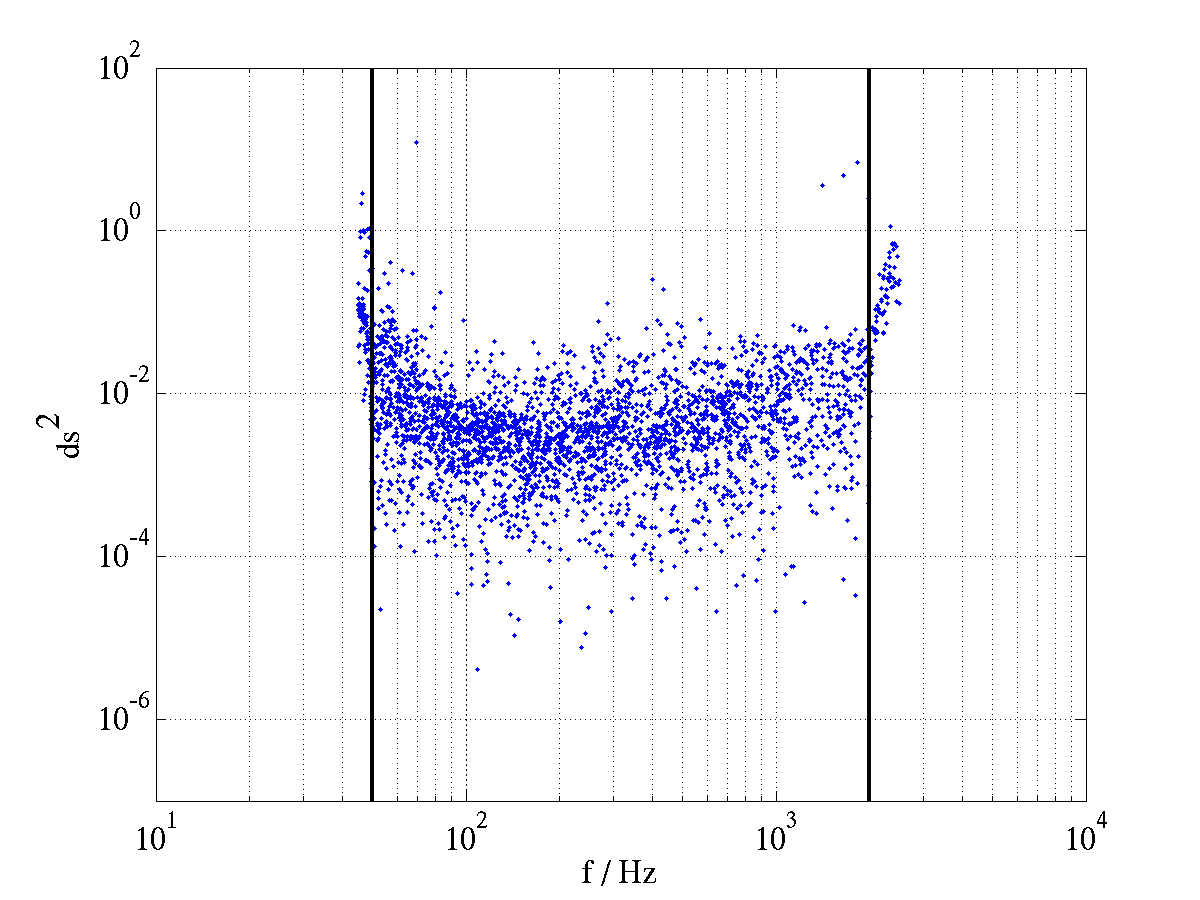}
\caption{Plot of $ds^2$ versus central frequency for the L1 component of injections found in triple coincidence}
\label{fig:L1injdetds2vf}
\end{center}
\end{figure}

\begin{figure}[htb]
\centering \begin{center}
\includegraphics[scale=0.55]{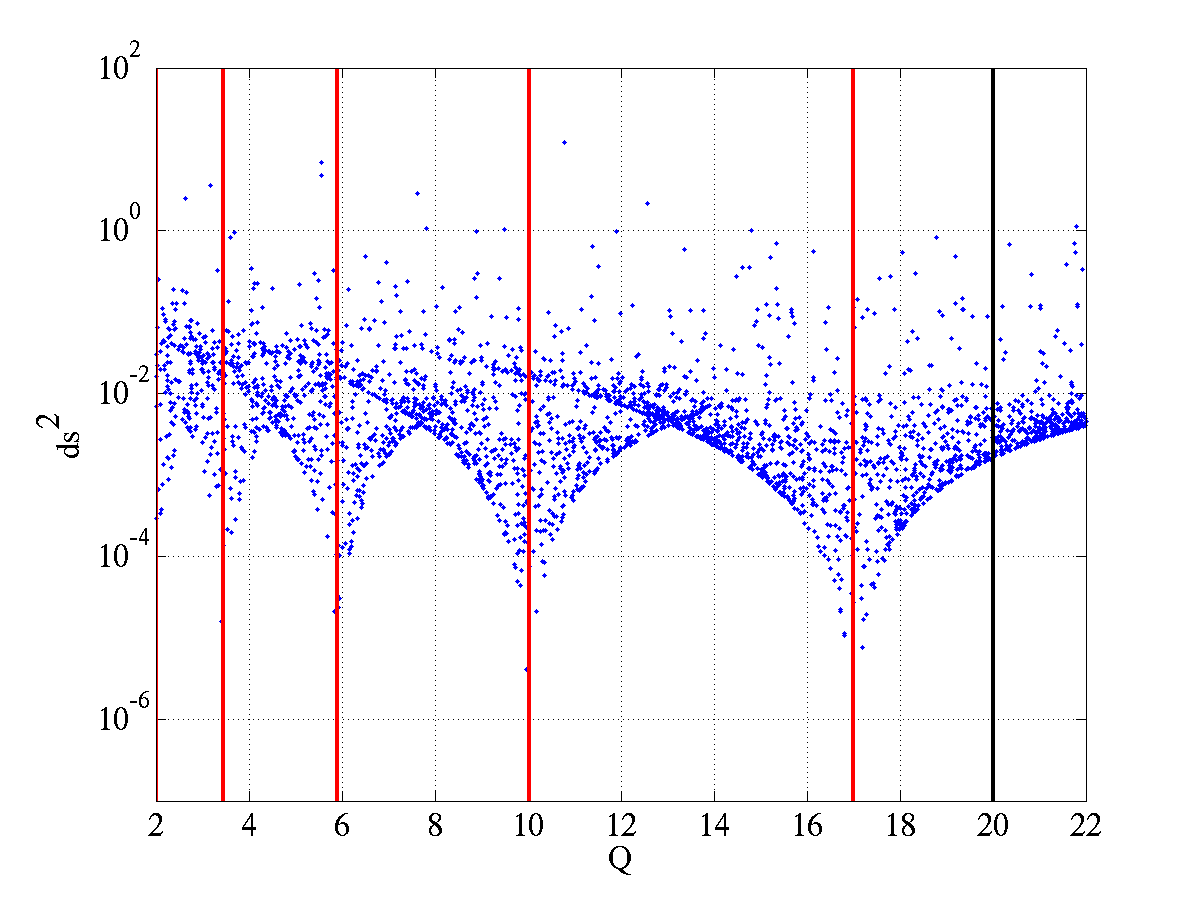}
\caption{Plot of $ds^2$ versus quality for the L1 component of injections found in triple coincidence. The red lines mark the five values of $Q$ in the template bank and the black line marks the upper $Q$ boundary of the template bank.}
\label{fig:L1injdetds2vQ}
\end{center}
\end{figure}

\begin{figure}[htb]
\centering \begin{center}
\includegraphics[scale=0.55]{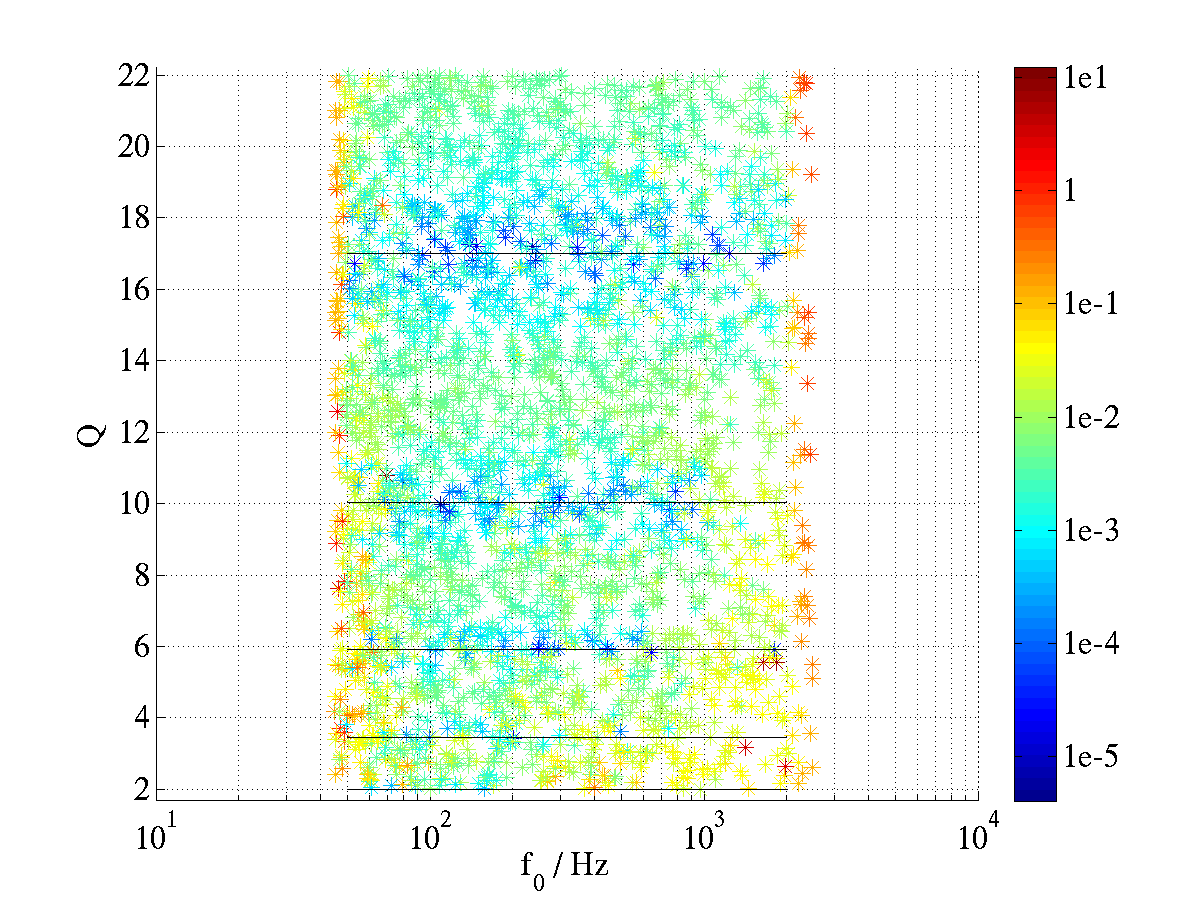}
\caption{Plot of quality versus frequency for the L1 component of injections found in triple coincidence. The colour scale is the mismatch between the injected and recovered parameters $ds^2$. Each horizontal line denotes the frequency range of the templates for each of the five values of $Q$.}
\label{fig:L1fvQvds2}
\end{center}
\end{figure}

\subsection{Effective Distance} 
Another check we can make is between injected and detected effective distance. We find the best way to evaluate this is by the fractional difference
\begin{equation}
\frac{\delta d_{eff}}{\left<d_{eff}\right>} = 
\frac{ 2 \left[ d_{eff}(det)-d_{eff}(inj)\right] }{ \left[d_{eff}(det)+d_{eff}(inj)\right]}
\label{eqn:fracdeltad}
\end{equation}
where $det$ stands for detected and $inj$ for injected. A histogram of this quantity evaluated for H1, H2, and L1 detected effective distances is shown in figure \ref{fig:histH1H2L1ddd}. From this we see a distribution which is sharply peaked at zero with a slight asymmetry in the tails for all three detectors. Figure \ref{fig:H2injdetdvf} shows that the accuracy with which the effective distance is recovered is frequency dependent in H2. Similar plots for H1 (figure \ref{fig:H1injdetdvf}) and L1 (not shown) display a similar behaviour. Below 100 Hz and above 1 kHz the noise increases and we become less accurate. It should be noted that injections made below 50 Hz and above 2 kHz are outside the template bank and thus are recovered by the ``wrong'' template. Therefore the SNR is less than the SNR of an exactly matched template, and hence the effective distance is over estimated. This explains the asymmetric tails seen in the histograms. The feature at approximately 340 Hz is believed to be due to the test mass suspension violin mode resonances. 

\afterpage{\clearpage}
\begin{figure}[h]
\centering \begin{center}
\includegraphics[scale=0.6]{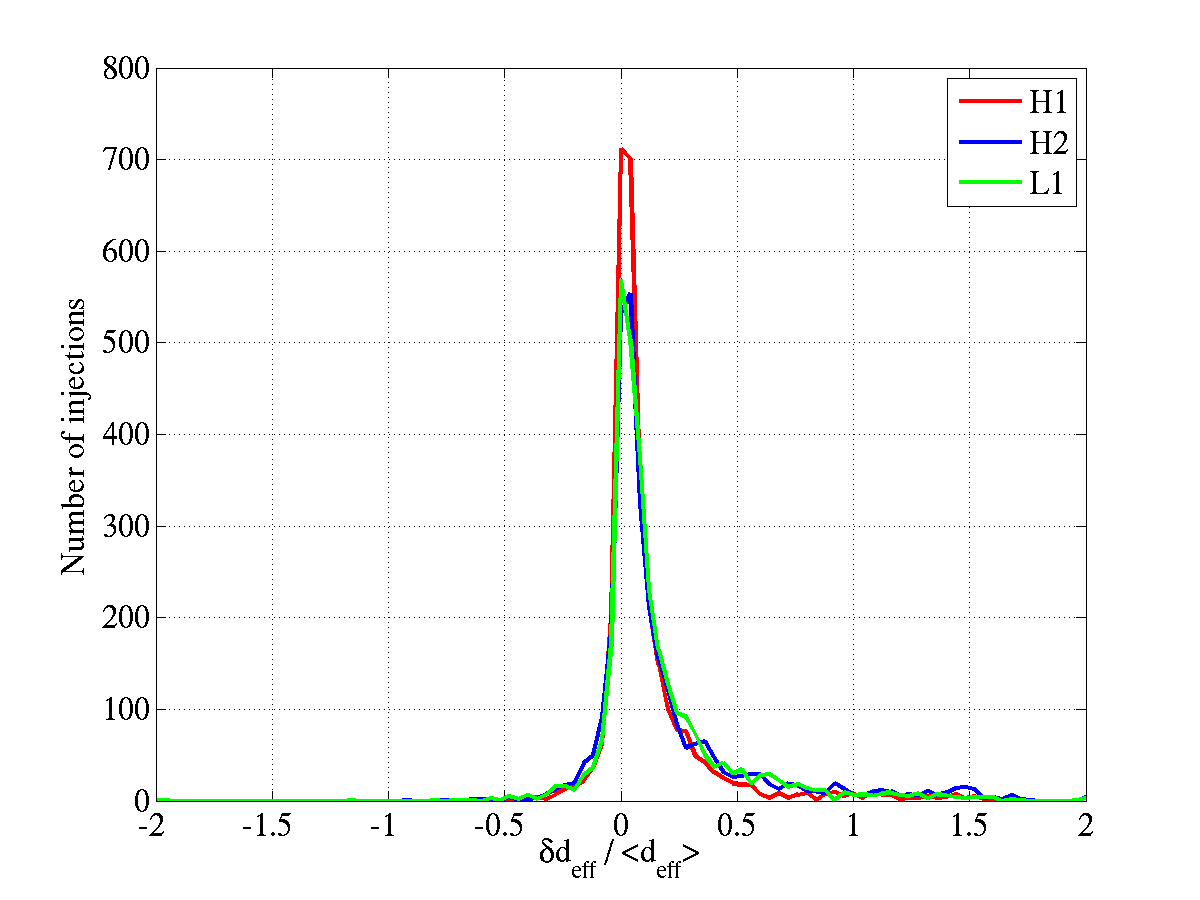}
\caption{Histogram of the fractional difference between detected and injected effective distance in H1, H2 and L1 for injections found in triple coincidence.}
\label{fig:histH1H2L1ddd}
\end{center}
\end{figure}

\begin{figure}[h]
\centering \begin{center}
\includegraphics[scale=0.6]{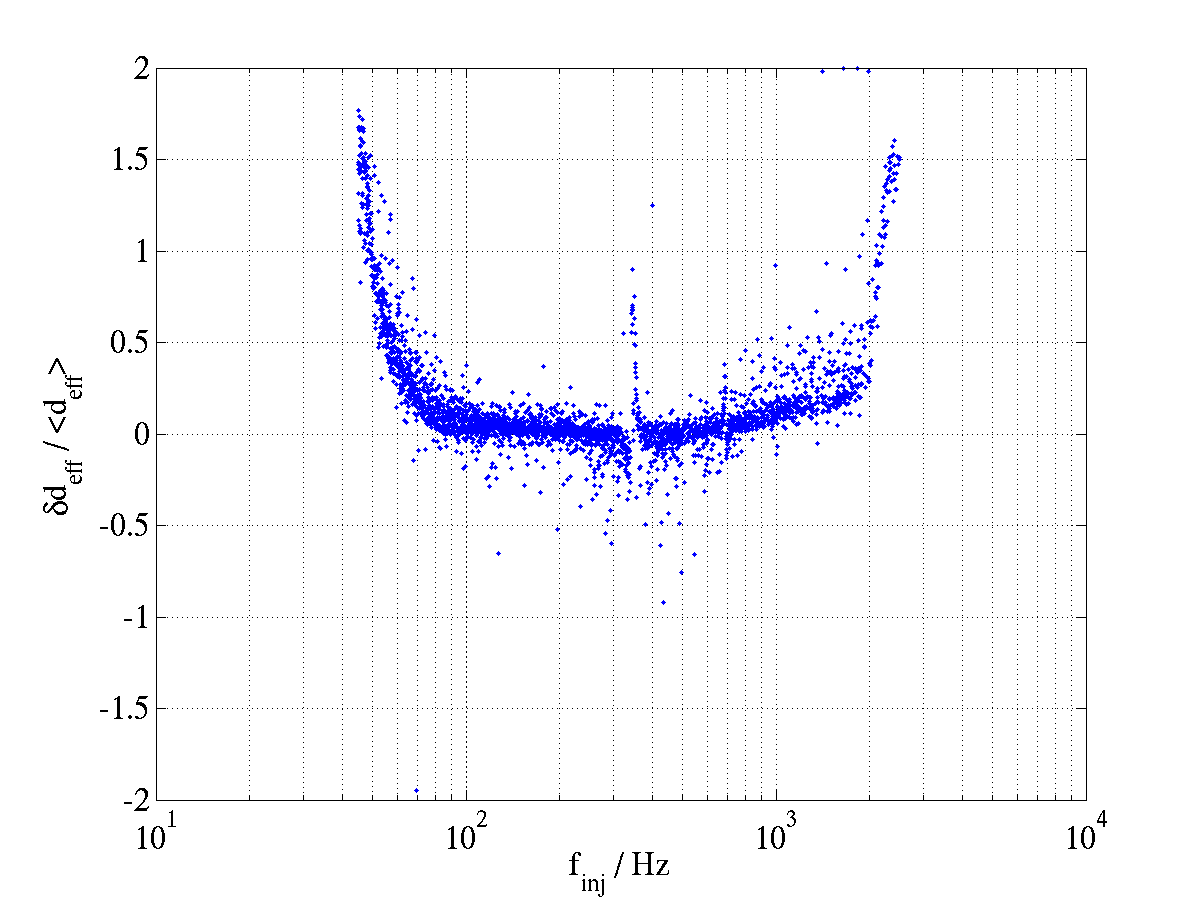}
\caption{Plot of the fractional difference between detected and injected effective distance versus frequency in H2 for injections found in triple coincidence.}
\label{fig:H2injdetdvf}
\end{center}
\end{figure}

\begin{figure}[h]
\centering \begin{center}
\includegraphics[scale=0.6]{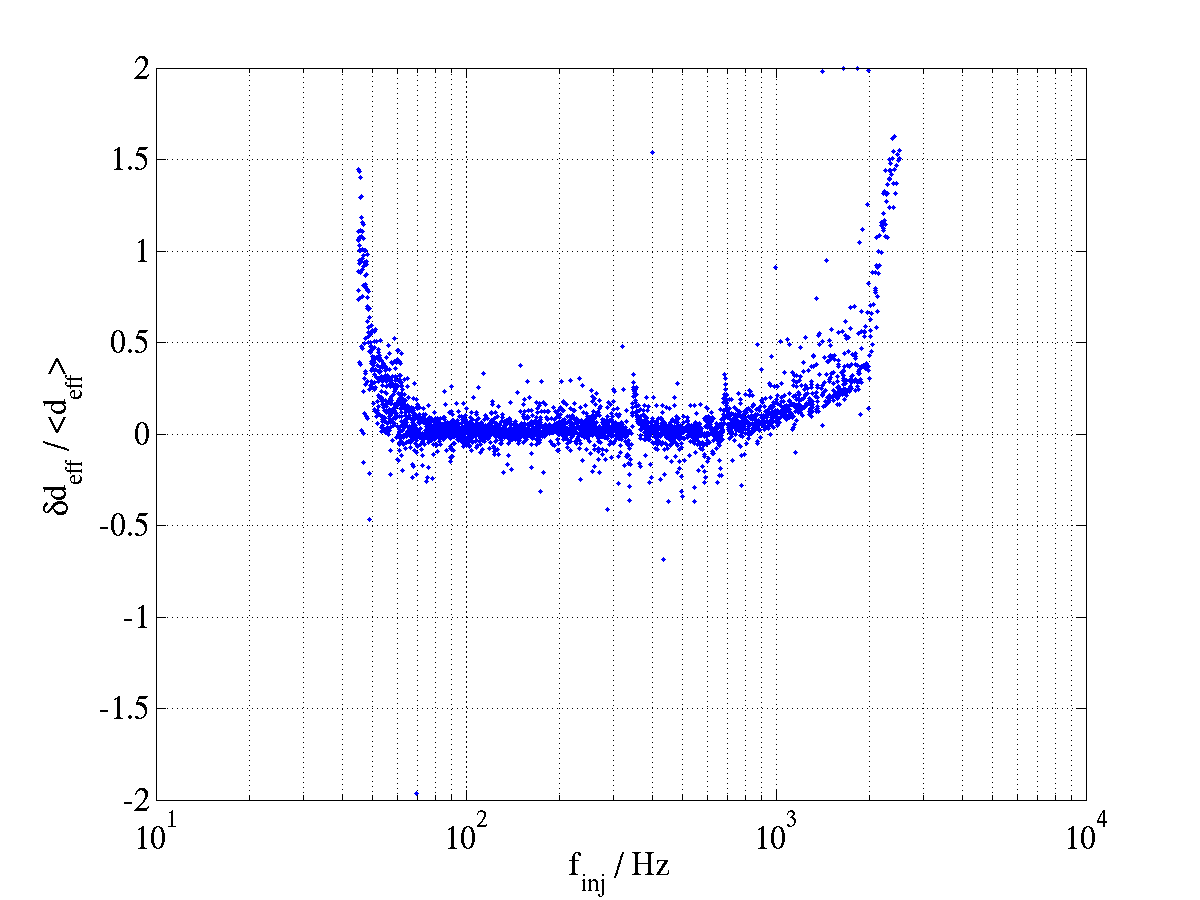}
\caption{Plot of the fractional difference between detected and injected
effective distance versus frequency in H1 for injections found in triple
coincidence.}
\label{fig:H1injdetdvf}
\end{center}
\end{figure}


\section{Comparison of Recovered Parameters Between Detectors}
\label{sec:paramacc}

Comparing the recovered parameters between detectors is also a useful exercise, as this gives us an idea how close we can expect the parameters of real signals to be. In chapter \ref{ch:tuning} we discussed the importance of such a comparison for tuning the coincidence test. Of course, these injections have survived the coincidence test and so we expect the parameters to be similar in different detectors. 

\subsection{Time} 
\label{sec:injdettime}
The difference between recovered times in H1 and H2, $\delta t$(H1-H2), shown in figure \ref{fig:H1H2histdt} is a sharply peaked distribution, as expected given their co-locality, whereas the difference between recovered times in the H1L1 $\delta t$(H1-L1) and H2L1 $\delta t$(H2-L1) pairs, shown in figure \ref{fig:H1L1H2L1histdt}, is a wide distribution where the light travel time of 10 ms can be clearly seen. 

\afterpage{\clearpage}

\begin{figure}[h]
\centering \begin{center}
\includegraphics[scale=0.6]{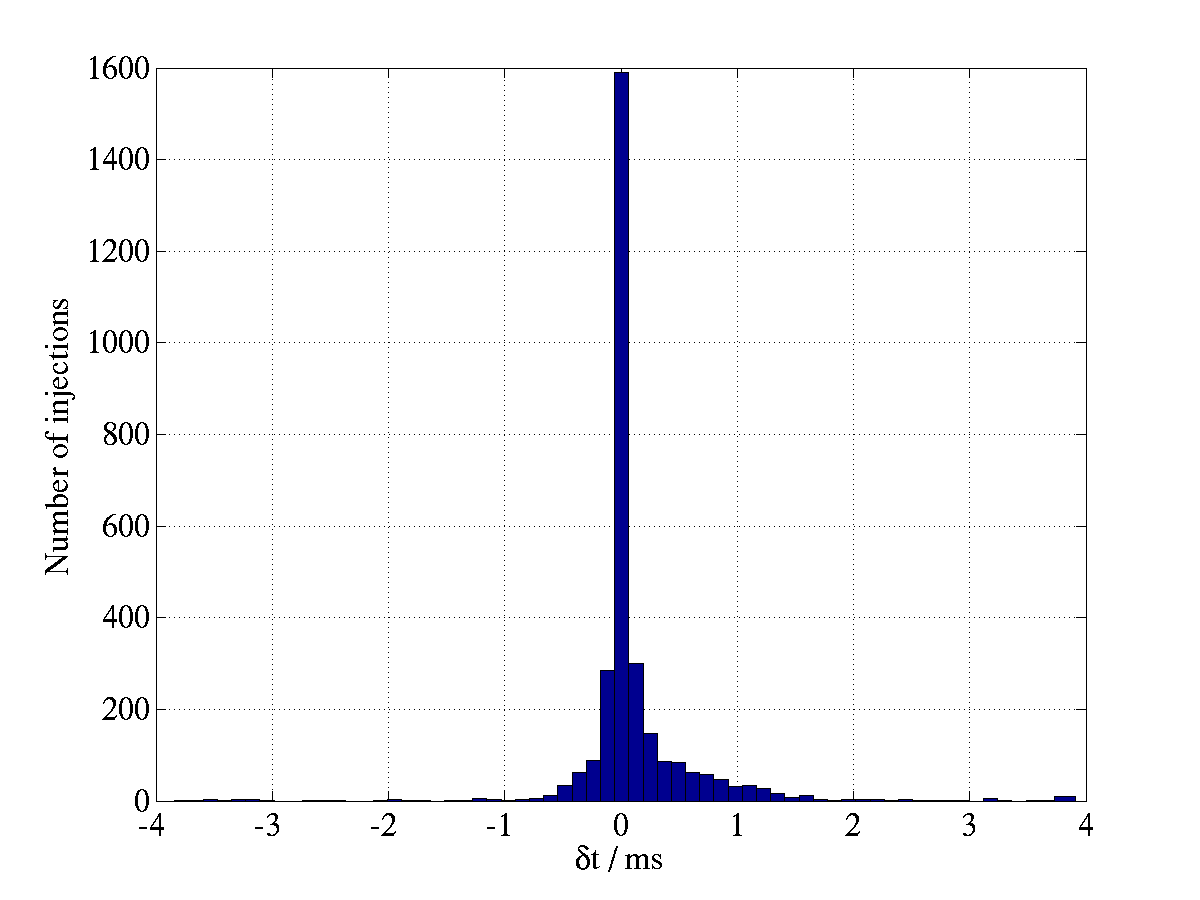}
\caption{Histogram of difference between H1 and H2 recovered times for injections found in triple coincidence.}
\label{fig:H1H2histdt}
\end{center}
\end{figure}

\begin{figure}[h]
\centering \begin{center}
\includegraphics[scale=0.6]{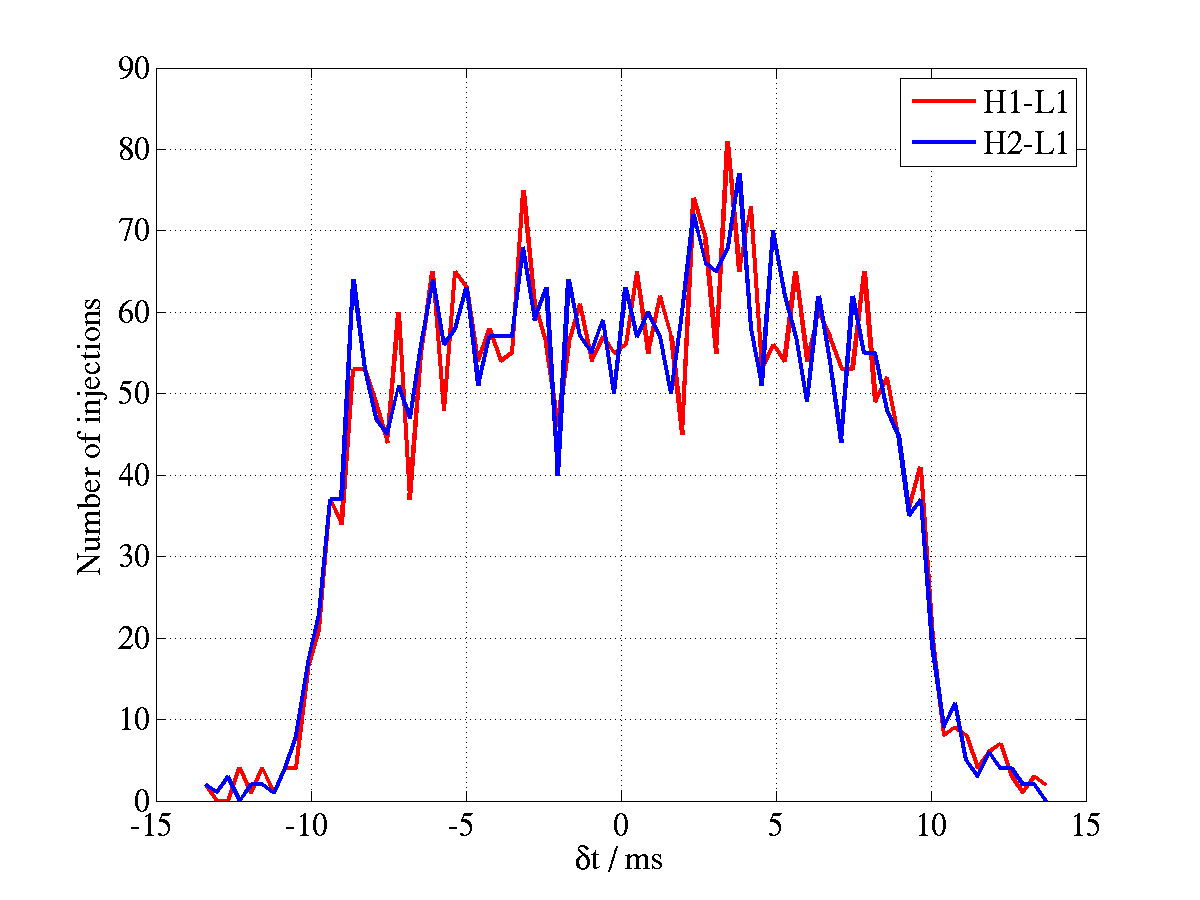}
\caption{Histogram of difference between H1L1 (red) and H2L1 (blue)  recovered times for injections found in triple coincidence.}
\label{fig:H1L1H2L1histdt}
\end{center}
\end{figure}

\begin{figure}[h]
\centering \begin{center}
\includegraphics[scale=0.6]{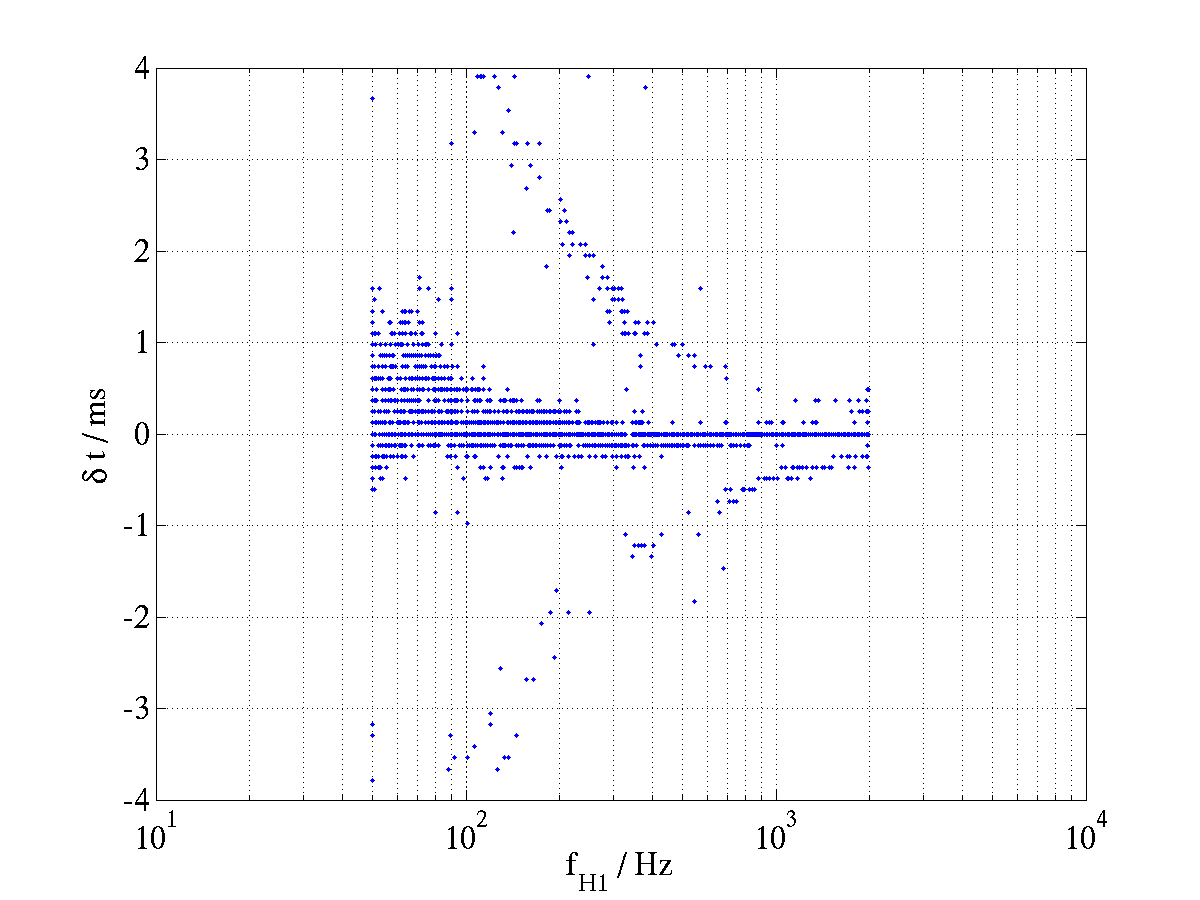}
\caption{Plot of difference between H1 and H2 recovered times as a function of
H1 recovered frequency for injections found in triple coincidence.}
\label{fig:H1H2ttvff}
\end{center}
\end{figure}

\begin{figure}[h]
\centering \begin{center}
\includegraphics[scale=0.6]{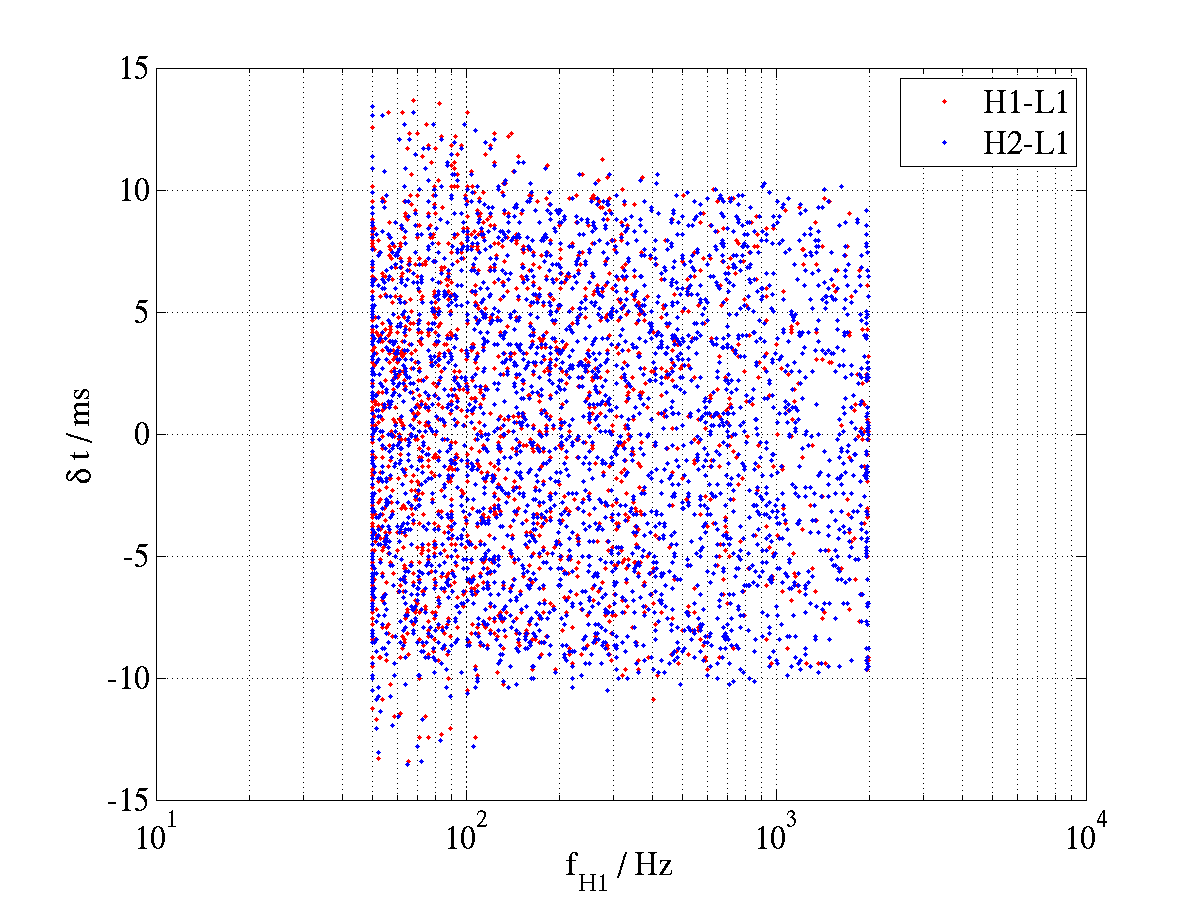}
\caption{Plot of difference between H1L1 (blue) and H2L1 (red)  recovered
times versus frequency for injections found in triple coincidence.}
\label{fig:H1L1H2L1ttvf}
\end{center}
\end{figure}

The scatter plot of $\delta t$(H1-H2) shown in figure \ref{fig:H1H2ttvff} displays interesting structure. The data points lie on equally separated discrete (horizontal) lines in $\delta t$; the lines are separated by the inverse of the sampling rate (8192)$^{-1}$s, showing just how well the time parameters are recovered for the majority of the injections. The next interesting structure is the overall shape of the population of points surrounding the $\delta t=0$ line, with the spread in $\delta t$ increasing as the frequency decreases. This is the same phenomenon discussed in section \ref{subsec:injdettime} for the comparison of detected and injected times. The third feature of note is the second population of points following the same shape, but separated from the main group by a larger $|\delta t|$. This population is actually the coincidence between two waves out of phase by $\pi$. In other words, the first peak of the injection in one detector was found in coincidence with the second peak in the other detector. For example, for a 300 Hz ringdown a half cycle is $\pm$1.7$\times$10$^{-3}$ s in duration, and as can be seen from the plot, this is approximately where the second population lies for this frequency.  This plot gives important information regarding the time coincidence requirement, implying that the width of this window should depend on the frequency (this will be discussed further in section \ref{sec:futcoinctestt}). Noting that a third population does not appear, we know that either that same peak or peaks separated by one half cycle are found in coincidence.

A scatter plot of the H1L1 time difference is shown in figure \ref{fig:H1L1H2L1ttvf}. We begin to see a spread in $\delta t$ below 200 Hz, but this is not as pronounced as in the H1H2 case.

\subsection{Metric Distance} 
\label{sec:detdetmetric}

The $ds^2$ histogram in figure \ref{fig:H1L1histds2} summarizes the mismatch between  the recovered parameters in H1 and L1. The bar at $ds^2=0$ indicates that approximately half of the injections were found with exactly the same template in both H1 and L1. The next bar at $ds^2=0.0075$ are the injections found with the same $Q$ and adjacent $f_0$ values. The third highest bar at $ds^2=0.015$ represent the injections found with the same $f_0$ but differing by one row of $Q$, and so on. The H1H2 and H2L1 plots show very similar structure. In all, 37\% of injections were found with the same template in all three detectors.  An interesting plot, shown in figure \ref{fig:H1L1fvdvds2}, is frequency versus distance for H1 with $ds^2$(H1,L1) as the colour axis. One might expect that as the effective distance increases the parameter accuracy would degrade, in particular close to the missed-found boundary seen in figure \ref{fig:mf_allvH}. The plot shows however that this was not so; even weak signals can be accurately recovered. Noise at the lowest and highest frequencies appear to have more of an effect as the plot shows an increased incidence of higher mismatched templates at those frequencies. Note that the coincidence window discussed in section \ref{sec:coinc} allows a window of $ds^2=0.05$ on both templates in a pair, allowing a total window of $ds^2=0.1$, as the range of the colour-bar shows. 

\afterpage{\clearpage}

\begin{figure}[h]
\centering \begin{center}
\includegraphics[scale=0.6]{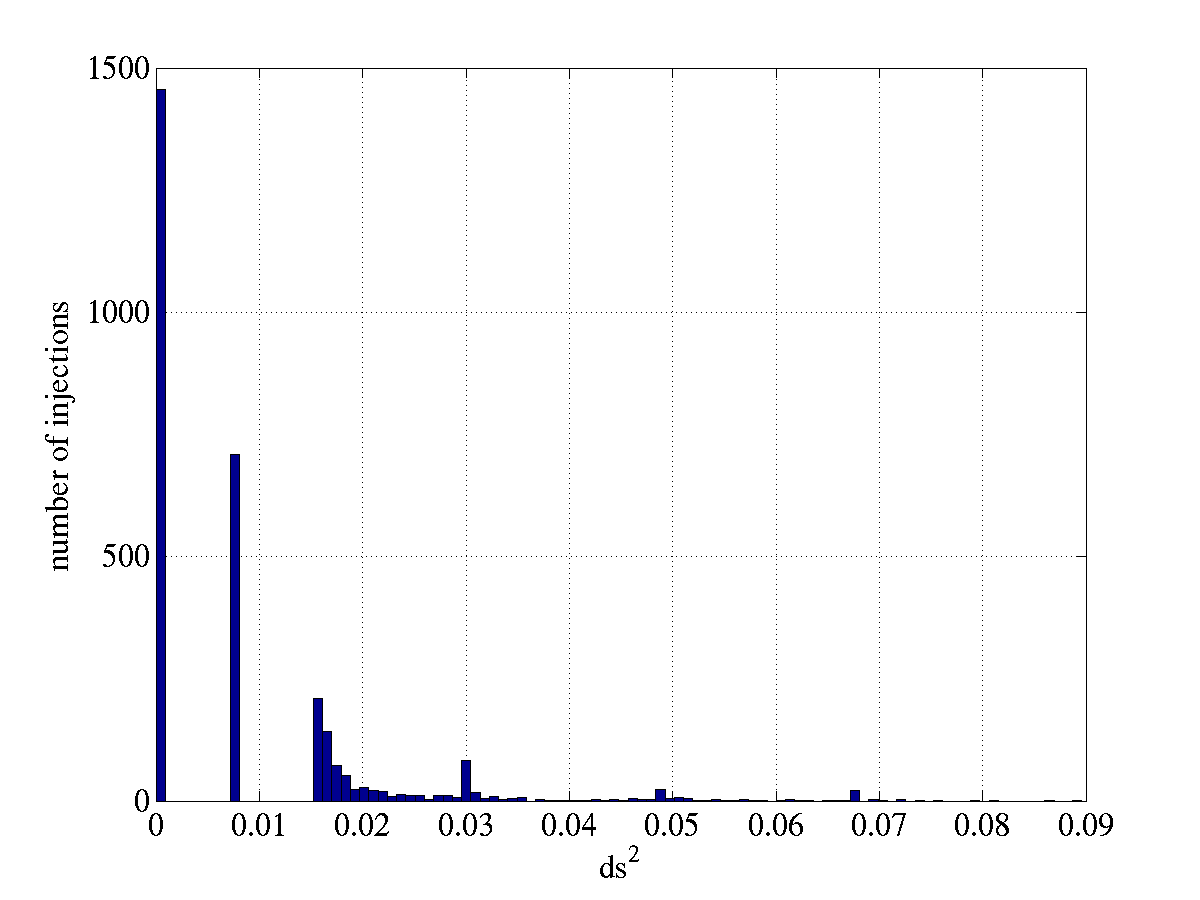}
\caption{Histogram of ds$^2$ between recovered parameters in H1 and L1 for injections found in triple coincidence.}
\label{fig:H1L1histds2}
\end{center}
\end{figure}

\begin{figure}[h]
\centering \begin{center}
\includegraphics[scale=0.6]{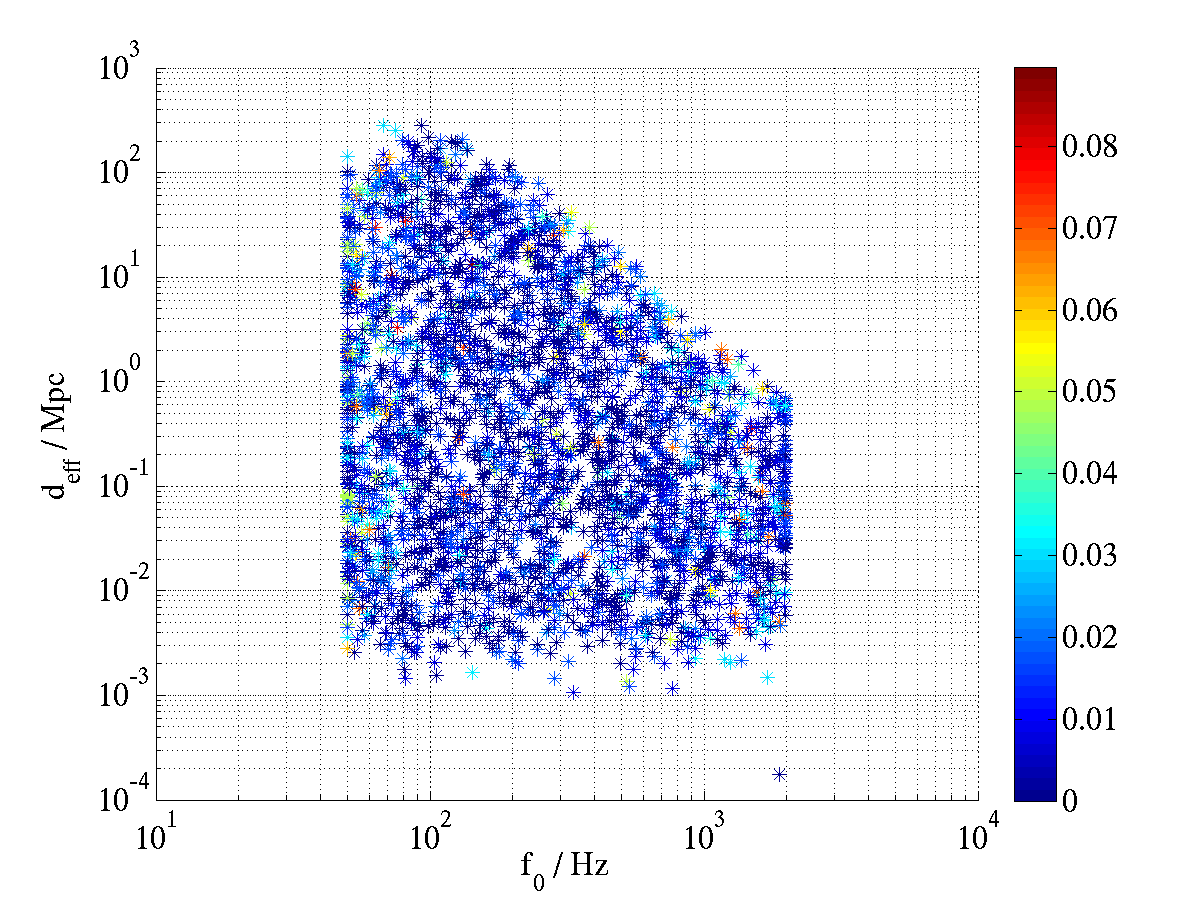}
\caption{Frequency versus effective distance for H1 with the colour-bar displaying $ds^2$(H1,L1), the metric distance between injections recovered in H1 and in L1.}
\label{fig:H1L1fvdvds2}
\end{center}
\end{figure}

\subsection{Effective Distance}
\label{sec:detdeteffd}

When comparing injected and detected quantities, effective distance was one of the parameters we looked at, as we expect to be able to reconstruct this quantity. However recall that effective distance is different for separated detectors as it depends on the sky location. Thus, in this section we should only expect to see a sharp narrow peak at zero in the H1H2 case, as co-located detectors should measure similar values of this quantity and this is the basis for the distance cut discussed in section \ref{sec:distancecut}. However, just for comparison, we compare the effective distance in all detector combinations. Once more we evaluate the fractional difference in effective distance defined in equation (\ref{eqn:fracdeltad})
\begin{equation}
\frac{\delta d_{eff}}{\left<d_{eff}\right>} =
\frac{ 2 \left[ d_{eff}(det_1)-d_{eff}(det_2)\right] }{
\left[d_{eff}(det_1)+d_{eff}(det_2)\right]},
\end{equation}
where $det_1$ and $det_2$ refer to detectors 1 and 2.
Figure \ref{fig:H1H2L1histdd} shows three distributions in $\delta d_{eff}/\left< d_{eff} \right>$. As expected, the H1H2 distribution is sharply peaked while the H1L1 and H2L1 distributions are much broader. The latter two distributions are symmetric about zero while the H1H2 shows a slight asymmetry. A scatter plot of the H1H2 distribution revealed a frequency dependence not seen in the other combinations. Figure \ref{fig:H1H2ddvf} shows that at low frequency $\delta d_{eff}/\left< d_{eff} \right>$ rapidly falls off to large negative values, indicating that the distance measured in H2 is greater than that measured in H1. This can be explained by referring back to figures \ref{fig:H2injdetdvf} and \ref{fig:H1injdetdvf}, which compare injected and recovered parameters. Here we see that in both plots $\delta d_{eff}/\left< d_{eff} \right>$ increases as the frequency reaches the low- and high-frequency ends of the template bank. However, in comparing these it is clear that in H2 $\delta d_{eff}/\left< d_{eff} \right>$ increases faster than in H1 at low frequencies; that is, H2 is overestimating the distance to a larger degree than H1 is. Thus when we take the difference between these two quantities we observe an asymmetric tail.

\afterpage{\clearpage}

\begin{figure}[h]
\centering \begin{center}
\includegraphics[scale=0.6]{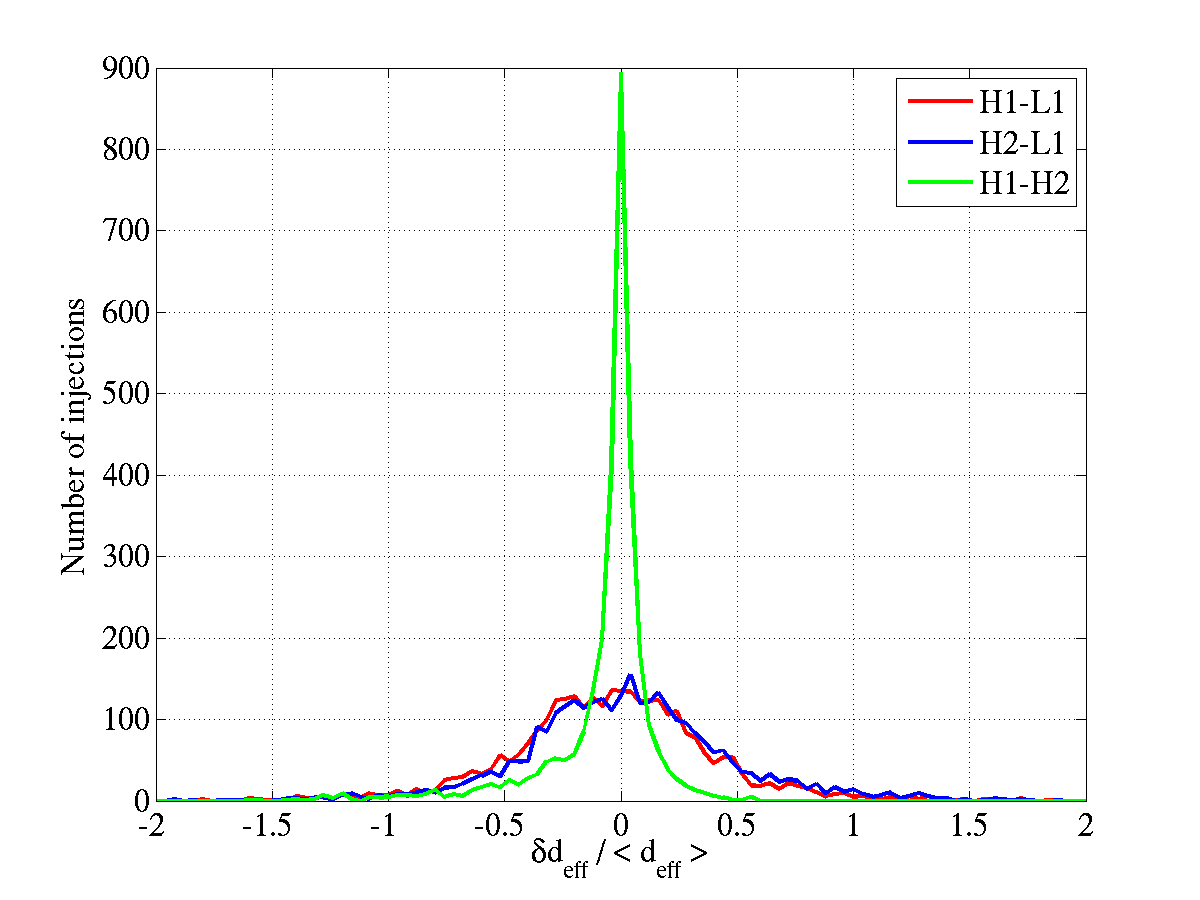}
\caption{Histogram of the fractional difference in H1H2 (green), H1L1 (red), and (H2L1) blue recovered effective distance for injections found in triple coincidence.}
\label{fig:H1H2L1histdd}
\end{center}
\end{figure}

\begin{figure}[h]
\centering \begin{center}
\includegraphics[scale=0.6]{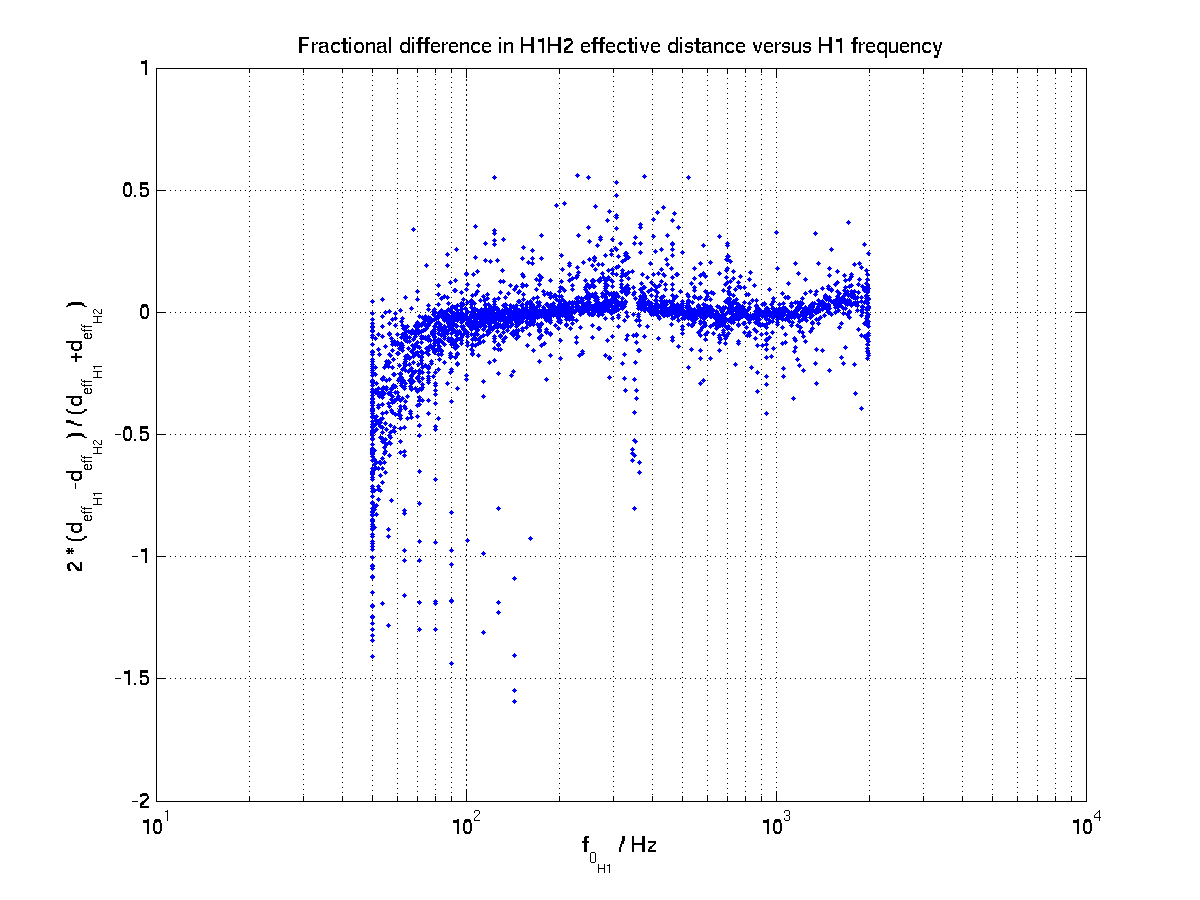}
\caption{Plot of the fractional difference in H1 and H2 recovered effective distance versus H1 frequency for injections found in triple coincidence.}
\label{fig:H1H2ddvf}
\end{center}
\end{figure}


\section{Comparison of the Background Estimation via Time-slides with the
Playground}
\label{sec:pgbg}
We discussed the estimation of the rate of accidental coincidences in section
\ref{sec:timeslides} and the definition of the playground data set in section
\ref{sec:playground}. In this section we compare the prediction of the background distribution of the final tuned pipeline with the playground.

\subsection{Triple Coincidences}

The number of background events found in triple coincidence was quite small;
from 100 timeslides there were just 14 events, making the triple-time false
alarm rate less than one event per run. Figure \ref{fig:tripH1L1_bgpg_snrsnr}
shows the projection of the SNR distribution of triple coincident background
triggers onto the H1L1 plane. The plot shows that the background events were found at low SNR in all three detectors. There were no events found in triple coincidence in the playground. This is consistent with the prediction of the background from the timeslide estimate.

\begin{figure}[htb]
\centering
\begin{center}
\includegraphics[scale=0.6]{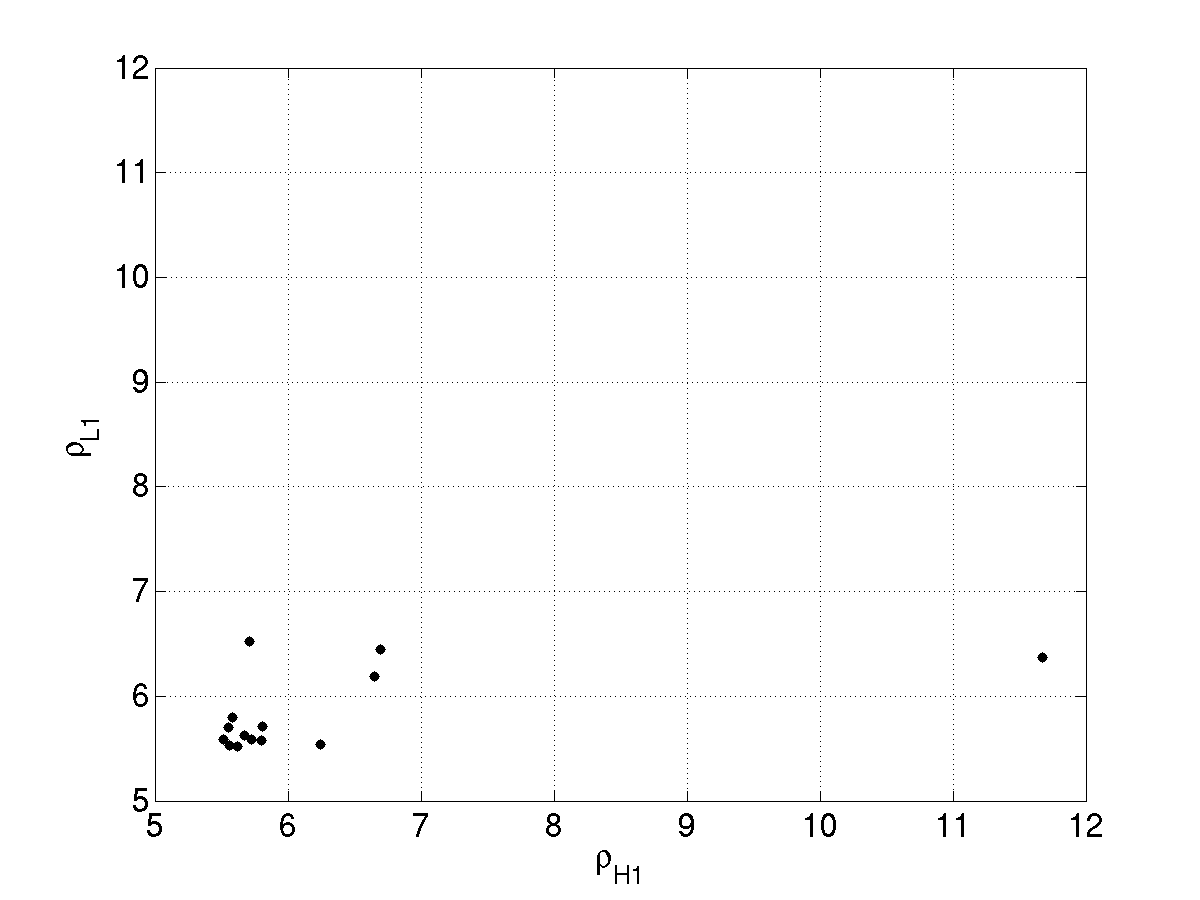}
\caption{L1 SNR versus H1 SNR for background triggers found in triple coincidence. }
\label{fig:tripH1L1_bgpg_snrsnr}
\end{center}
\end{figure}


\subsection{Double Coincidences}

In contrast, however, the number of background events found in double
coincidence was considerably higher. From 100 timeslides we found an average of 620, 150, and 800 background
events in H1L1, H1H2, and H2L1, respectively per timeslide. It should be noted that, as mentioned previously, the H1H2 false alarm rate is known to be underestimated.

There were 53, 21 and 69 double coincidences found in H1L1, H1H2 and H2L1 respectively in the playground data set. 

A plot of the SNR distribution for timeslides and playground found in double
coincidence in H1 and L1 is shown in figure \ref{fig:doubH1L1_bgpg_snrsnr},
for H1 and H2 in figure \ref{fig:doubH1H2_bgpg_snrsnr}, and for H2 and L1 in figure
\ref{fig:doubH2L1_bgpg_snrsnr}.  In addition to having a much larger false alarm rate
than triples, the double coincidence timeslide plots also show a different
type of SNR distribution; in addition to a central component close to the
diagonal, the plots show long tails extending to high SNR. Note that the H1H2
background distribution we see the effects of the  distance cut.
The playground triggers were all found with SNRs lower than 20.

We can compare the number of coincidences found in the playground
to the estimated background. Figures \ref{fig:doubH1L1pgbghist},
\ref{fig:doubH1H2pgbghist}, and \ref{fig:doubH2L1pgbghist} show histograms
of the number of double coincidences in each of the one hundred timeslides, along with
the number of double coincidences found in the playground scaled to the full data set
for H1L1, H1H2, and H2L1 respectively. If the
background estimate is an accurate measure of the false alarm rate, one
would expect that the scaled number
of playground events should be to be comparable to number of events in each timeslide.
We find that, although the scaled playground is lower than the average number
of timeslide double coincidences for H1L1 and H2L1, it is within the estimated
error. As we have discussed previously the timeslide method of background
estimation in H1H2 is flawed and we expect to see a greater number of
coincidences than predicted. Figure \ref{fig:doubH1H2pgbghist} shows that this
is indeed the case.
\afterpage{\clearpage}
\begin{figure}[htb]
\centering
\begin{center}
\includegraphics[scale=0.6]{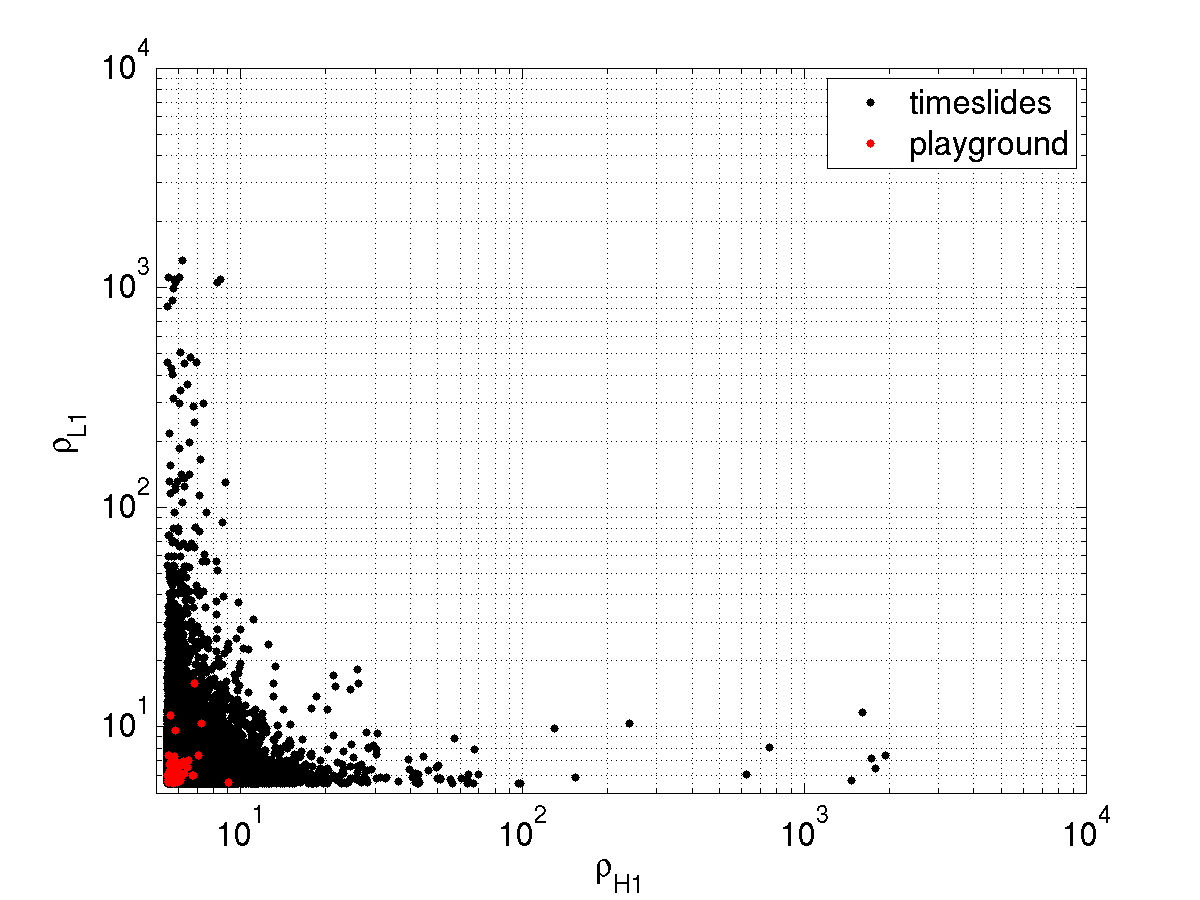}
\caption{The signal-to-noise distribution of playground and timeslides found
in double coincidence in L1 and H1.}
\label{fig:doubH1L1_bgpg_snrsnr}
\end{center}
\end{figure}

\begin{figure}[h]
\centering \begin{center}
\includegraphics[scale=0.6]{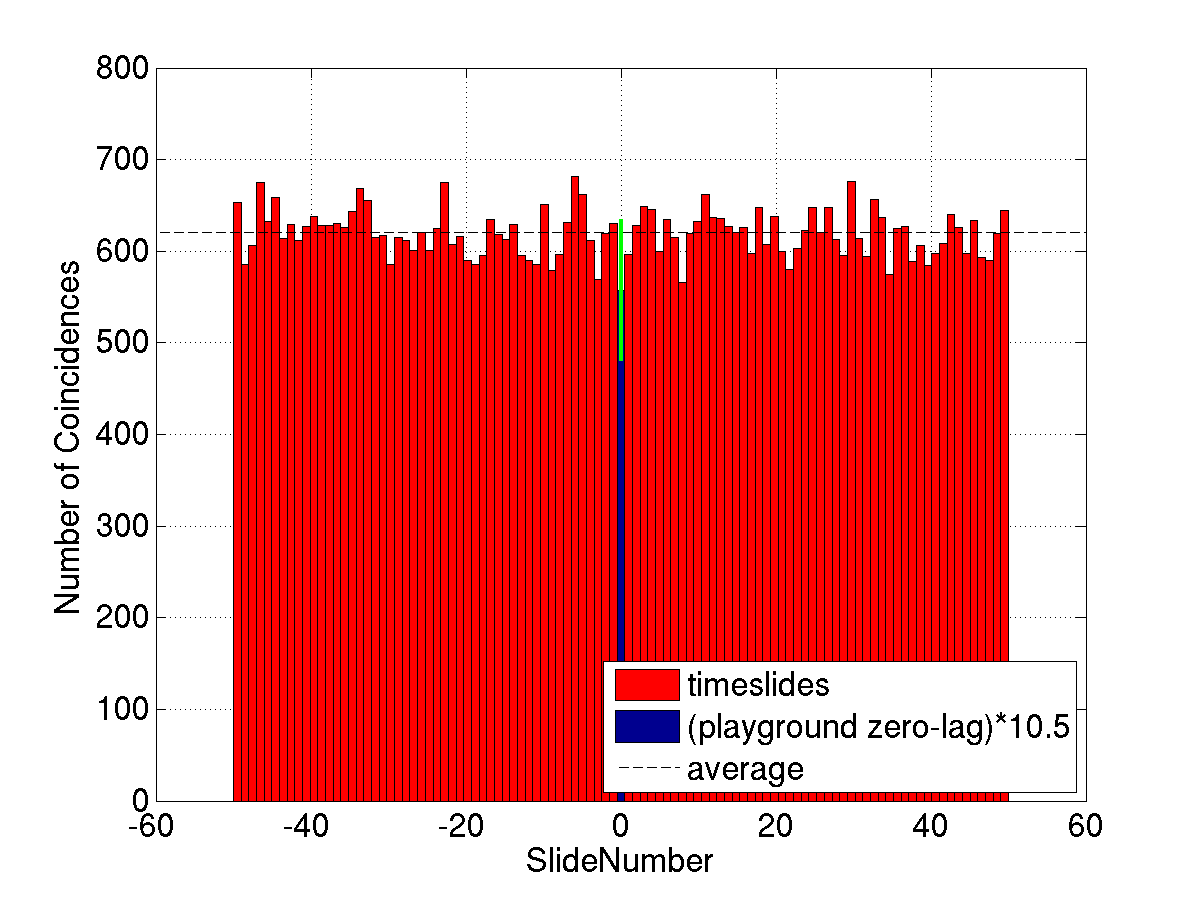}
\caption{Histogram of number of H1L1 double coincidences per timeslide and in
the playground scaled to the full data set.}
\label{fig:doubH1L1pgbghist}
\end{center}
\end{figure}

\afterpage{\clearpage}

\begin{figure}[htb]
\centering
\begin{center}
\includegraphics[scale=0.6]{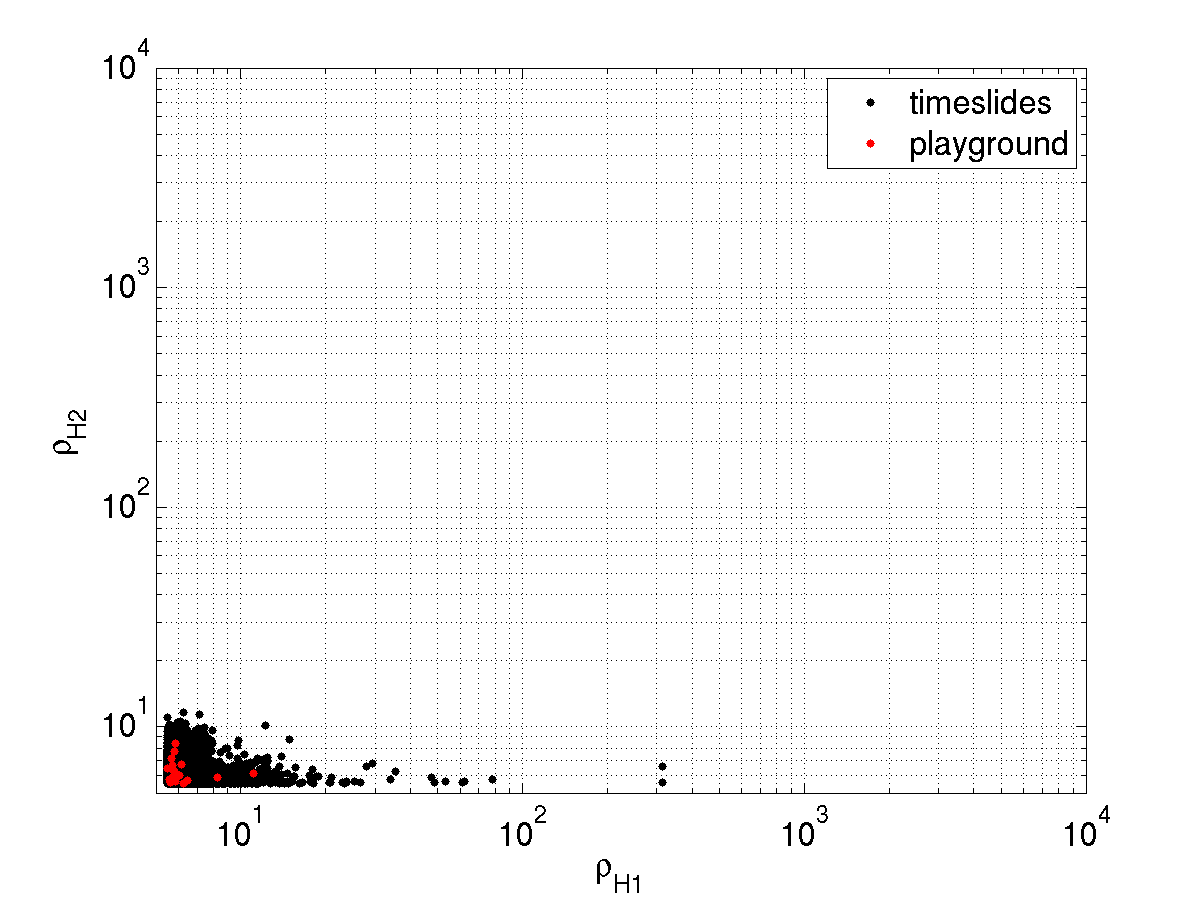}
\caption{The signal-to-noise distribution of playground and timeslides found
in double coincidence in H2 and H1 prior to the implementation of the distance
cut.}
\label{fig:doubH1H2_bgpg_snrsnr}
\end{center}
\end{figure}

\begin{figure}[h]
\centering \begin{center}
\includegraphics[scale=0.6]{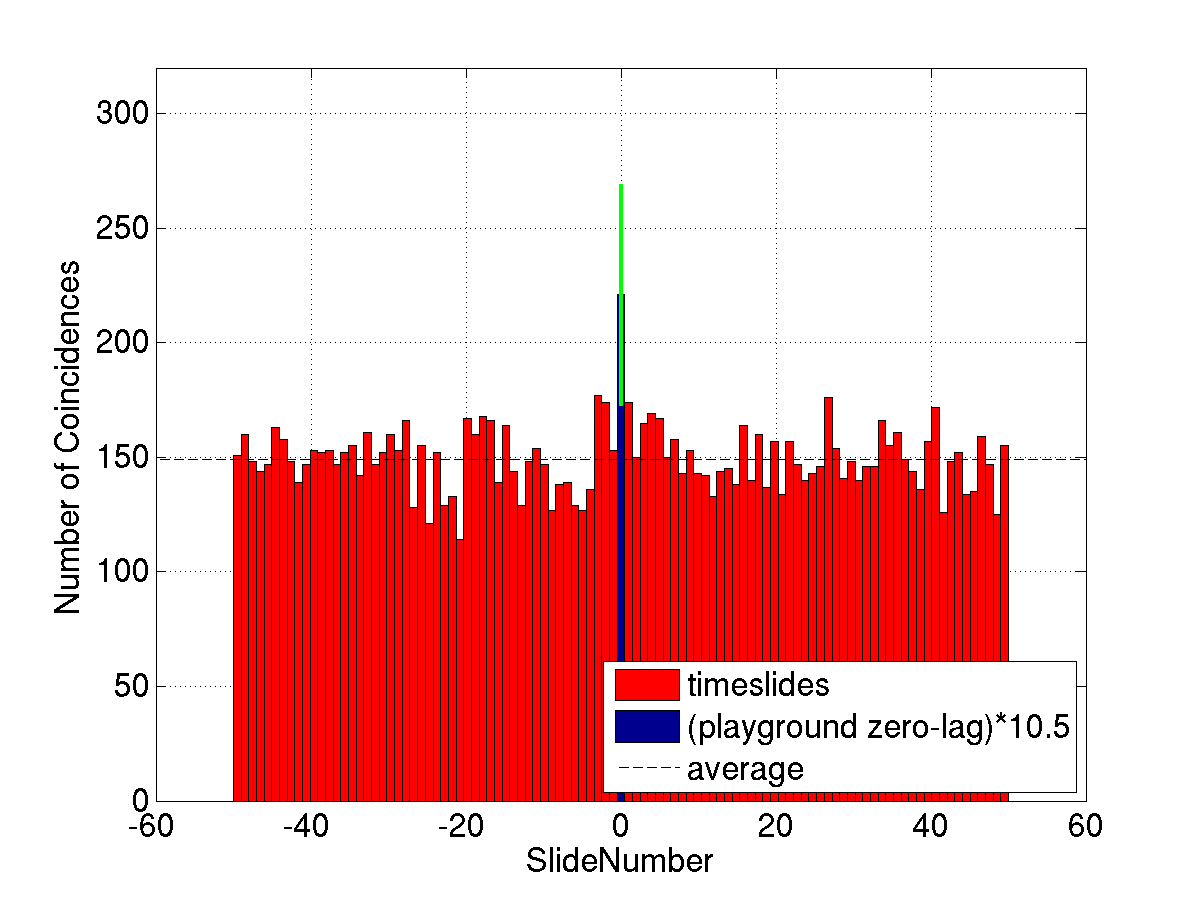}
\caption{Histogram of number of H1H2 double coincidences  per timeslide and
in
the playground scaled to the full data set.}
\label{fig:doubH1H2pgbghist}
\end{center}
\end{figure}

\begin{figure}[htb]
\centering
\begin{center}
\includegraphics[scale=0.6]{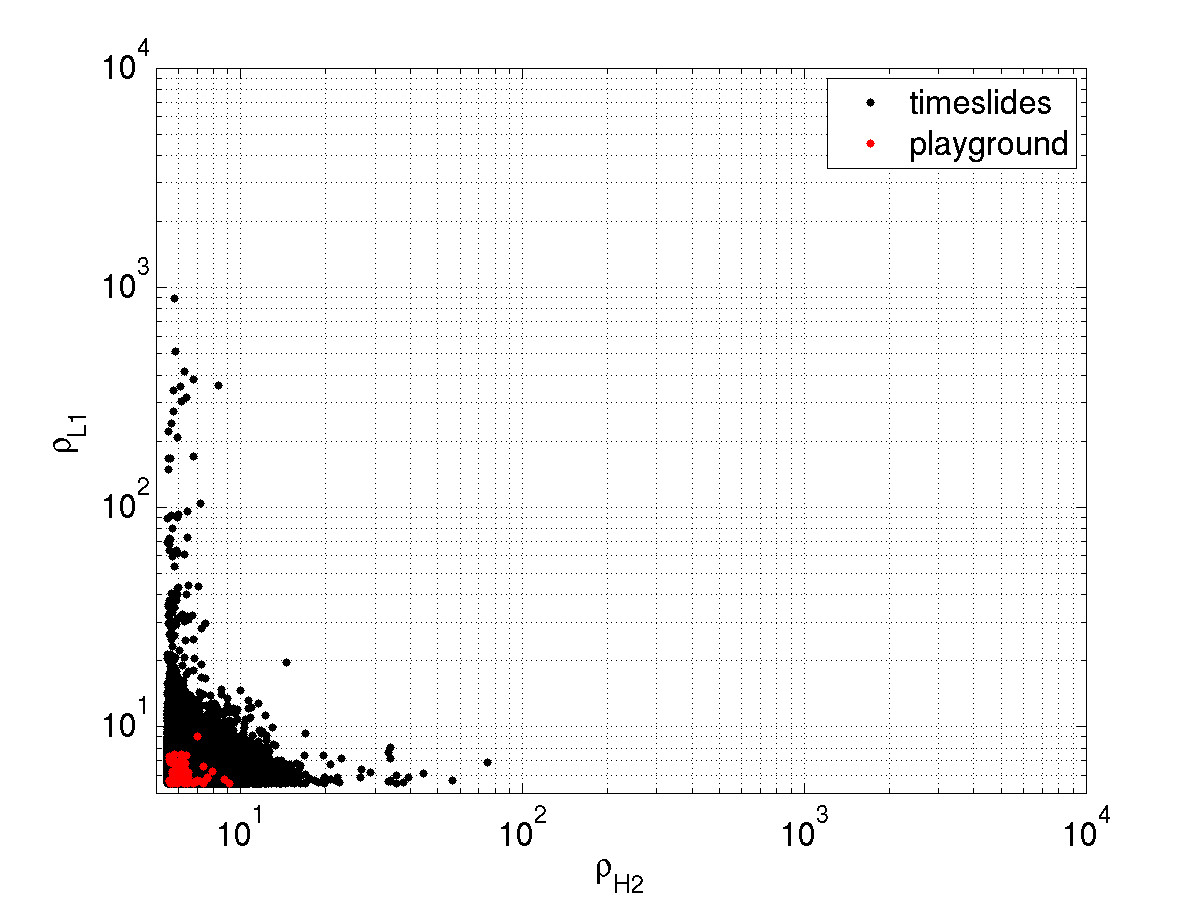}
\caption{The signal-to-noise distribution of playground and timeslides found
in double coincidence in L1 and H2.}
\label{fig:doubH2L1_bgpg_snrsnr}
\end{center}
\end{figure}

\begin{figure}[h]
\centering \begin{center}
\includegraphics[scale=0.6]{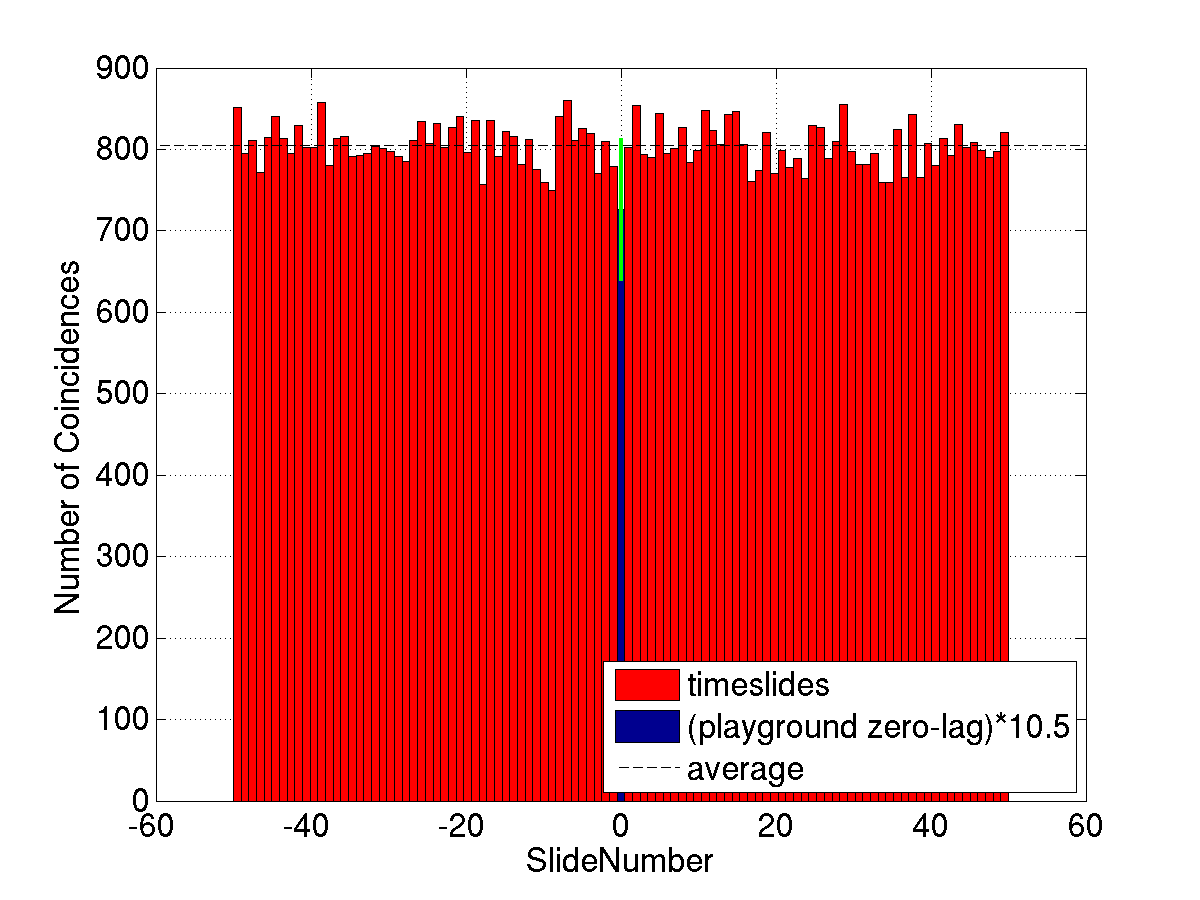}
\caption{Histogram of number of H2L1 double coincidences per timeslide and in
the playground scaled to the full data set.}
\label{fig:doubH2L1pgbghist}
\end{center}
\end{figure}


\section{Comparison of Recovered Injection Parameter Accuracy for Timeslides found in Double Coincidence}

We have just demonstrated that the expected rate of false alarms in double coincidence is very high. Here we investigate the possibility that by comparing the accuracy between timeslide components to the accuracy between injection components we could distinguish between a signal and a noise event. We look at the mismatch between pairs of detectors and the difference in recovered effective distance.

\subsection{Metric Distance}

First we compare how close in parameter space the background events were in comparison to the injections by plotting histograms of $ds^2$ normalized to the total number of events. (The injection distribution was already discussed in section \ref{sec:detdetmetric}.) Figure \ref{fig:H1L1bginjds2} shows a comparison between H1 and L1. Here again we see a series of peaks due to the fact that the same template bank is used throughout the search. We see that the fraction of timeslide coincidences is low, but not zero at $ds^2=0$, telling us that in double coincidence there is a non-negligible chance that two independent noise triggers could be found with the exact same parameters in widely separated sites. In fact the timeslide estimation of the H1L1 background predicts that in a typical S4 run we can expect 30 background events to be found in H1 and L1 with exactly the same waveform parameters! The fraction of injections found at higher values of the mismatch drops rapidly and the fraction of false alarms increases slightly, showing that a candidate event found with a large values of $ds^2$ is more likely to be background than a real signal. Almost identical distributions were seen for H1H2 and H2L1 combinations.

\begin{figure}[h]
\centering \begin{center}
\includegraphics[scale=0.6]{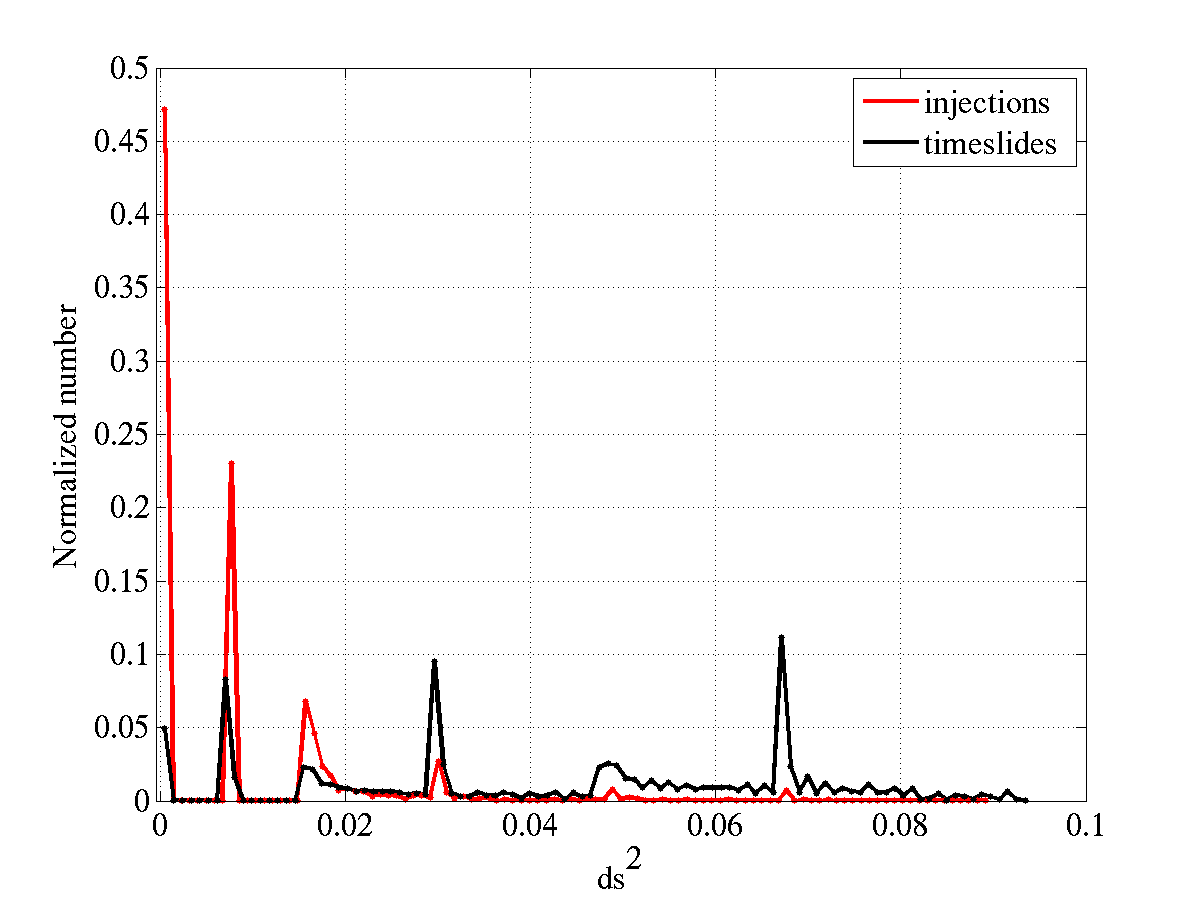}
\caption{The normalized distribution of the mismatch between H1 and L1 for
injections (red) and double coincident timeslides (blue).} 
\label{fig:H1L1bginjds2}
\end{center}
\end{figure}


\subsection{Effective Distance}

Recall from our discussion in section \ref{sec:detdetmetric} that we really
can only make use of the H1H2 comparison of effective distances as we expect
these to be very close for real signals. Figure \ref{fig:H1H2bginjdd} shows
the comparison of $\delta d_{eff}/\left< d_{eff} \right>$ between H1 and H2 for timeslides and injections. The plot shows that, whereas the injection distribution is peaked at $\delta d_{eff}/\left< d_{eff} \right>=0$, the timeslide distribution is peaked at $\delta d_{eff}/\left< d_{eff} \right>\sim 0.3$. We can also see the point where the distance cut was implemented, beyond where the injection distribution ended at $\delta d_{eff}/\left< d_{eff} \right>=0.67$. This shows the advantage of having a co-located half-length interferometer; if H2 was the same length as H1, the timeslide peak would also lie at $\delta d_{eff}/\left< d_{eff} \right>=0$.

The background distributions of $\delta d_{eff}/\left< d_{eff} \right>=0.67$ in H1L1 and H2L1 are interesting too. Firstly the plot shows that because the background distribution is broad and peaked close to zero just like the injection distribution, we cannot implement a distance cut. Both distributions display a second peak shifted from the first. This is because the L1 sensitivity was decreased for a period during the run, as the effective distance is proportional to the sensitivity, the effective distance of the background was also less. 

From this study we can conclude that it is impossible to distinguish between a
signal and background from the accuracy of parameters between pair of
detectors.

\begin{figure}[h]
\centering \begin{center}
\includegraphics[scale=0.6]{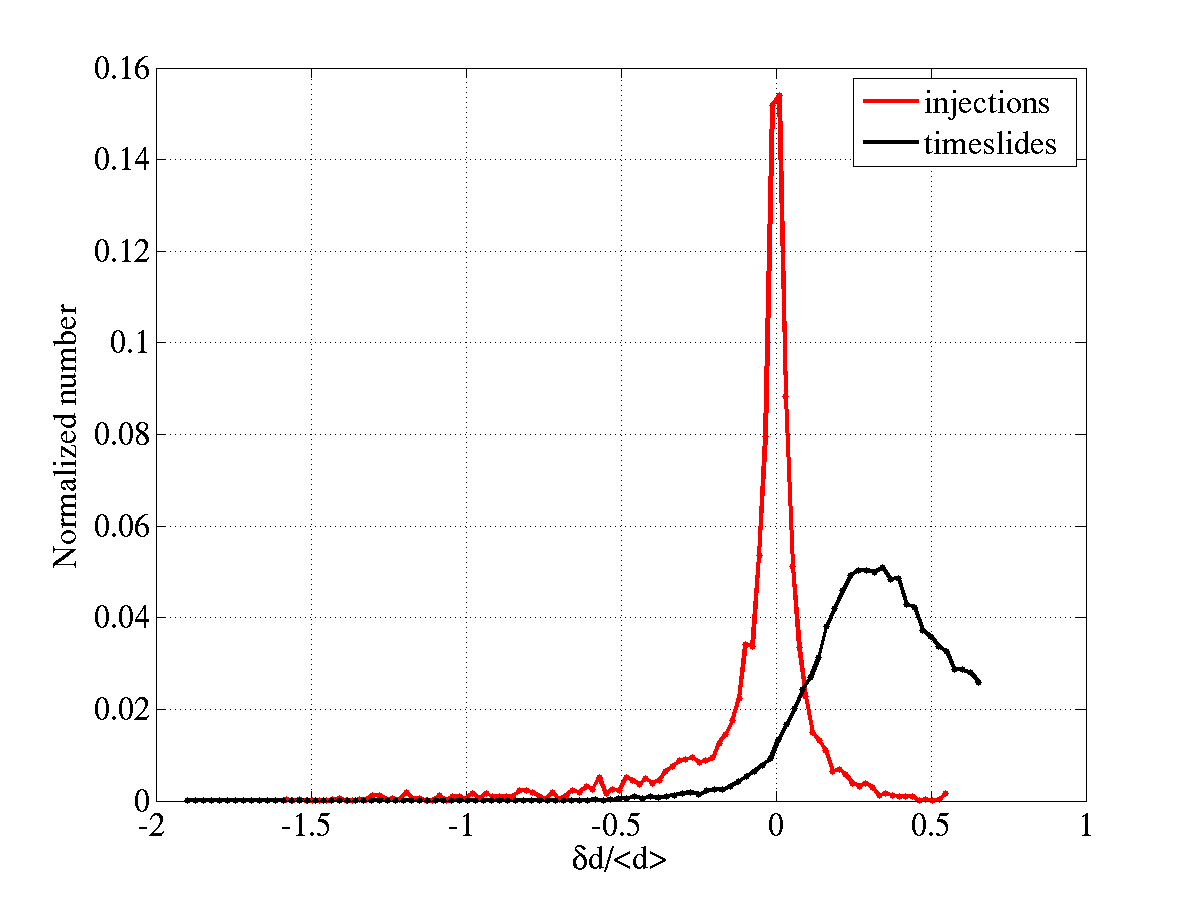}
\caption{The normalized distribution of the fractional difference in effective
distance between H1 and H2 for injections (red) and double coincident timeslides (blue).}
\label{fig:H1H2bginjdd}
\end{center}
\end{figure}

\afterpage{\clearpage} 
\begin{figure}[h]
\centering \begin{center}
\includegraphics[scale=0.6]{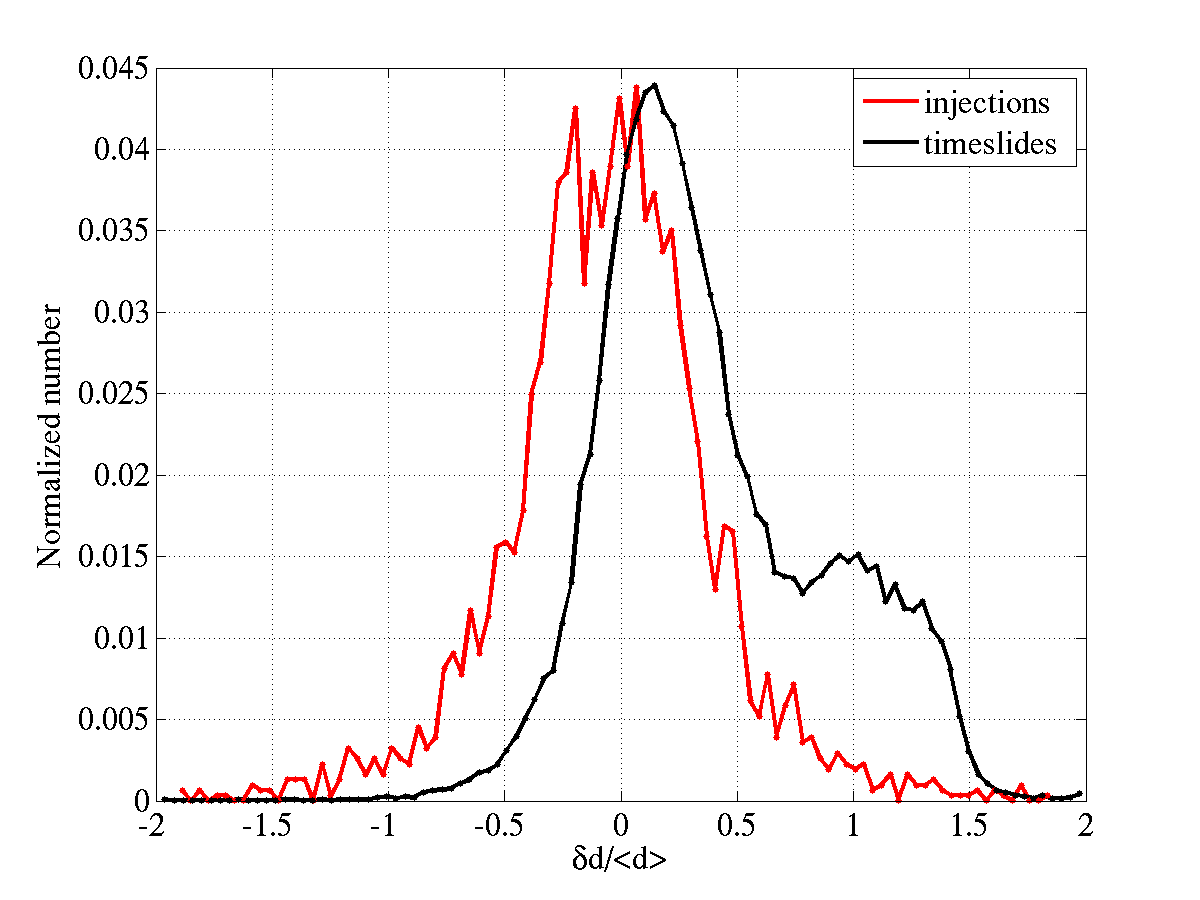}
\caption{The normalized distribution of the fractional difference in effective
distance between H1 and L1 for injections (red) and double coincident
timeslides (blue).}
\label{fig:H1L1bginjdd}
\end{center}
\end{figure}

\begin{figure}[h]
\centering \begin{center}
\includegraphics[scale=0.6]{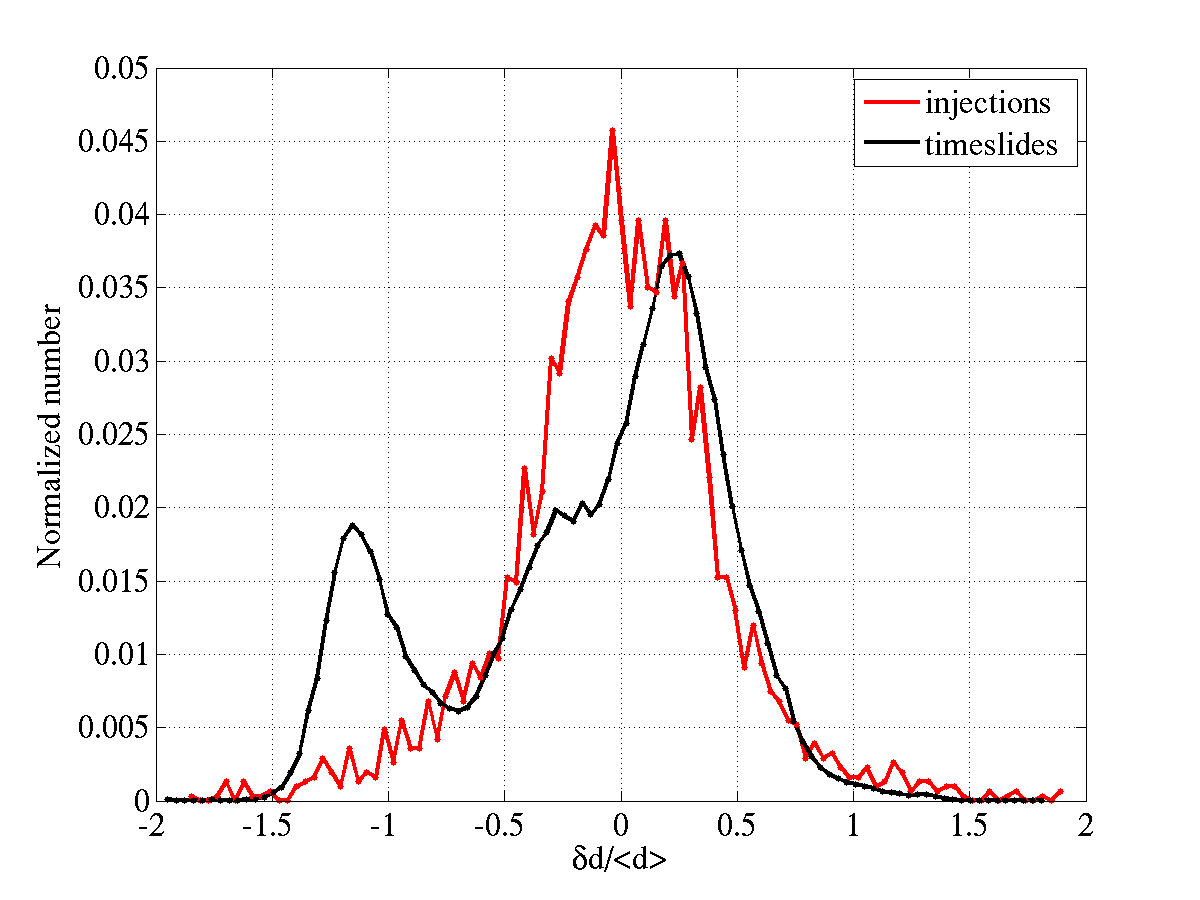}
\caption{The normalized distribution of the fractional difference in effective
distance between L1 and H2 for injections (red) and double coincident
timeslides (blue).}
\label{fig:L1H2bginjds2}
\end{center}
\end{figure}

\chapter{Results}
\label{ch:results}

\section{Introduction}
Once we were satisfied with the tuning of the search the pipeline was applied to the full data set and the search was unblinded. A cumulative histogram of the detection statistic of background events was compared to that of foreground events for each category of coincidences. A significant excess of the foreground distribution above the background distribution could indicate the presence of a gravitational wave. On opening the box we found that {\it no} triple coincidence signals survived to the end of the pipeline. As expected there were a number of foreground triggers found in double coincidence and while some of those categories showed a slight excess above the background we did not find sufficient evidence to claim a detection of gravitational waves. In section \ref{sec:openbox} we describe our findings on unblinding each category of coincidences in the search. In section \ref{sec:followup} the double coincidences with the highest detection statistic are followed up. In each case the presence of instrumental artifacts or environmental disturbances were sufficient to explain the coincident triggers. The upper limit on the rate of ringdowns is calculated in section \ref{sec:upperlimit}.


\section{Opening the Box} \label{sec:openbox}

In each of the cumulative histograms described in this section we plot the background distribution scaled down by a factor of 100 (as this many timeslides were performed) for comparison with the zero-lag events. The histogram is a function of the detection statistic, which for double coincidences is $\rho_{\textrm{DS}}=\textrm{min}\{\rho_1+\rho_2, 2\rho_1+2.2,2\rho_2+2.2\}$ where $\rho_1$ and $\rho_2$ refer to the signal-to-noise ratio (SNR of triggers from interferometers 1 and 2.  The error bars on the background distribution denote a one $\sigma$ error, where $\sigma=\sqrt{N/100}$ and $N$ is the number of background events per bin. In all of the scatter plots referred to in this section we plot all of the coincident events from 100 timeslides along with the zero-lag triggers. The inset in each plot zooms in on the low-SNR region containing most of the zero-lag triggers. Tables detailing the ten loudest candidate events in each category can be found in appendix \ref{app:topten}.

\subsection{Triples in Triple Time}
No triple coincident zero-lag events were found.

\subsection{H1L1 Doubles in Triple Time}
Figure \ref{fig:bgintdinttH1L1} shows the cumulative histogram of H1L1 doubles in triple time. The plot shows that the foreground was within the expected range of accidental coincidences. Figure \ref{fig:bgintdinttsnrH1L1} shows the SNR from the individual zero-lag and background triggers.

\subsection{H1H2 Doubles in Triple Time}
Figure \ref{fig:bgintdinttH1H2} shows the expected excess of zero-lag triggers due to our inability to predict the H1H2 false alarm rate. There was one trigger with an exceptionally high detection statistic,  $\rho_{DS} \sim 63$. Figure \ref{fig:bgintdinttsnrH1H2} shows the individual SNRs were $\rho_{\textrm{H1}}=1300$ and $\rho_{\textrm{H2}}=30$.

\subsection{H2L1 Doubles in Triple Time}
Figure \ref{fig:bgintdinttH2L1} shows that the H2L1 foreground was consistent with the background. A plot of $\rho_{\textrm{L1}}$ versus $\rho_{\textrm{H2}}$ is shown in figure \ref{fig:bgintdinttsnrH2L1}.

\subsection{H1L1 Doubles in Double Time}
For this category we see a deficit of foreground triggers compared to background in figure \ref{fig:bgintdindtH1L1}. Figure \ref{fig:bgintdindtsnrH1L1} shows the distribution of foreground and background triggers in SNR.

\subsection{H1H2 Doubles in Double Time}
Once more we see an excess in the H1H2 foreground due to our inability to estimate the background accurately. Figure \ref{fig:bgintdindtH1H2} displays the H1H2 cumulative histogram and figure \ref{fig:bgintdindtsnrH1H2} shows the $\rho_{\textrm{H2}}$ versus $\rho_{\textrm{H1}}$ scatter plot.

\subsection{H2L1 Doubles in Double Time}
Figure \ref{fig:bgintdindtH2L1} shows that the H2L1 double time foreground was consistent with background. Figure \ref{fig:bgintdindtsnrH2L1} displays the $\rho_{\textrm{L1}}$ versus $\rho_{\textrm{H2}}$ scatter plot.

\afterpage{\clearpage}

\begin{figure}[ht] 
\centering
\begin{center}
\includegraphics[scale=0.55]{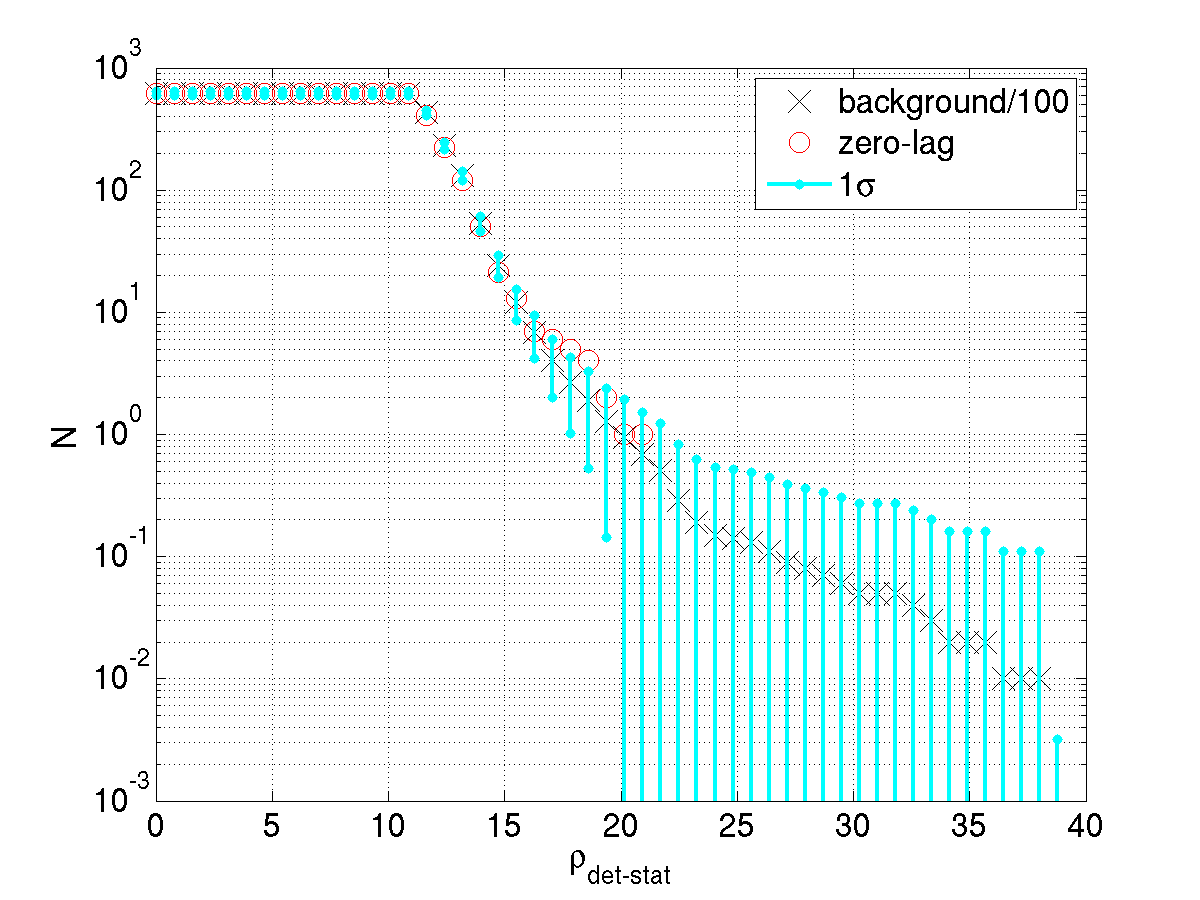}\\
\caption{Cumulative histogram of zero-lag and background coincidences: H1L1 doubles in triple time.}
\label{fig:bgintdinttH1L1}
\includegraphics[scale=0.55]{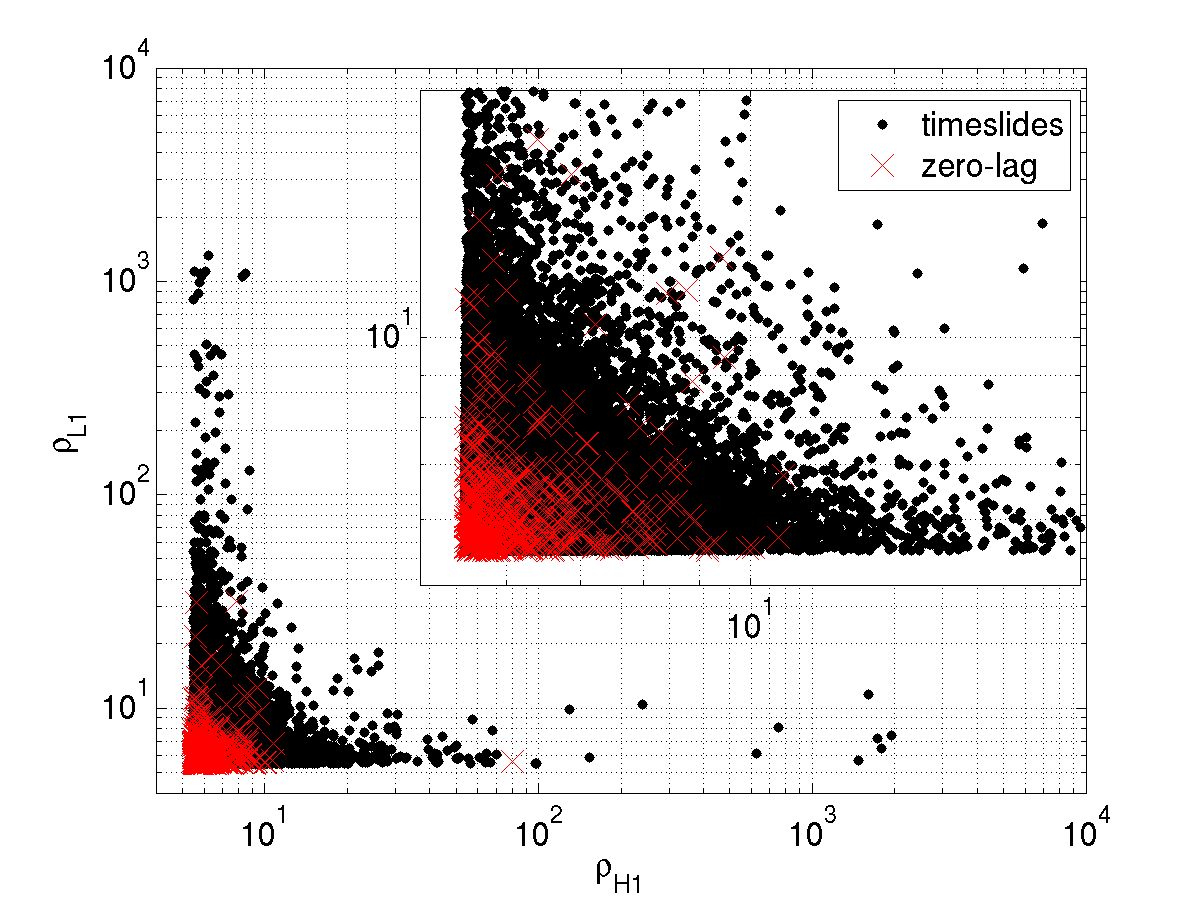}
\caption{Scatter plot of L1 versus H1 signal-to-noise ratio for double coincident zero-lag and background (100 timeslides) triggers in triple time. The inset is an enlargement of the region $\rho=5$ to 200.}
\label{fig:bgintdinttsnrH1L1}
\end{center}
\end{figure}

\begin{figure}[ht] 
\centering
\begin{center}
\includegraphics[scale=0.55]{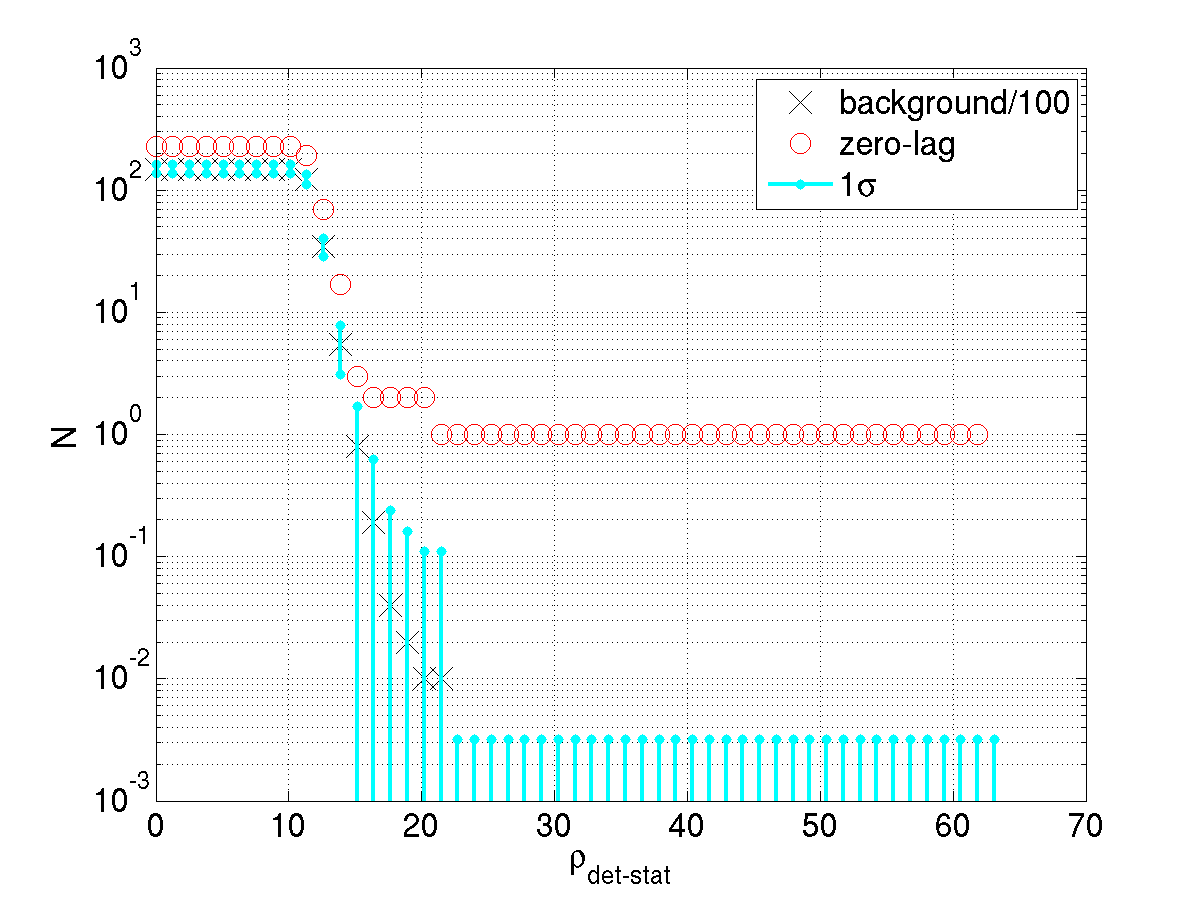}\\
\caption{Cumulative histogram of zero-lag and background coincidences, H1H2 doubles in triple time.}
\label{fig:bgintdinttH1H2}
\includegraphics[scale=0.55]{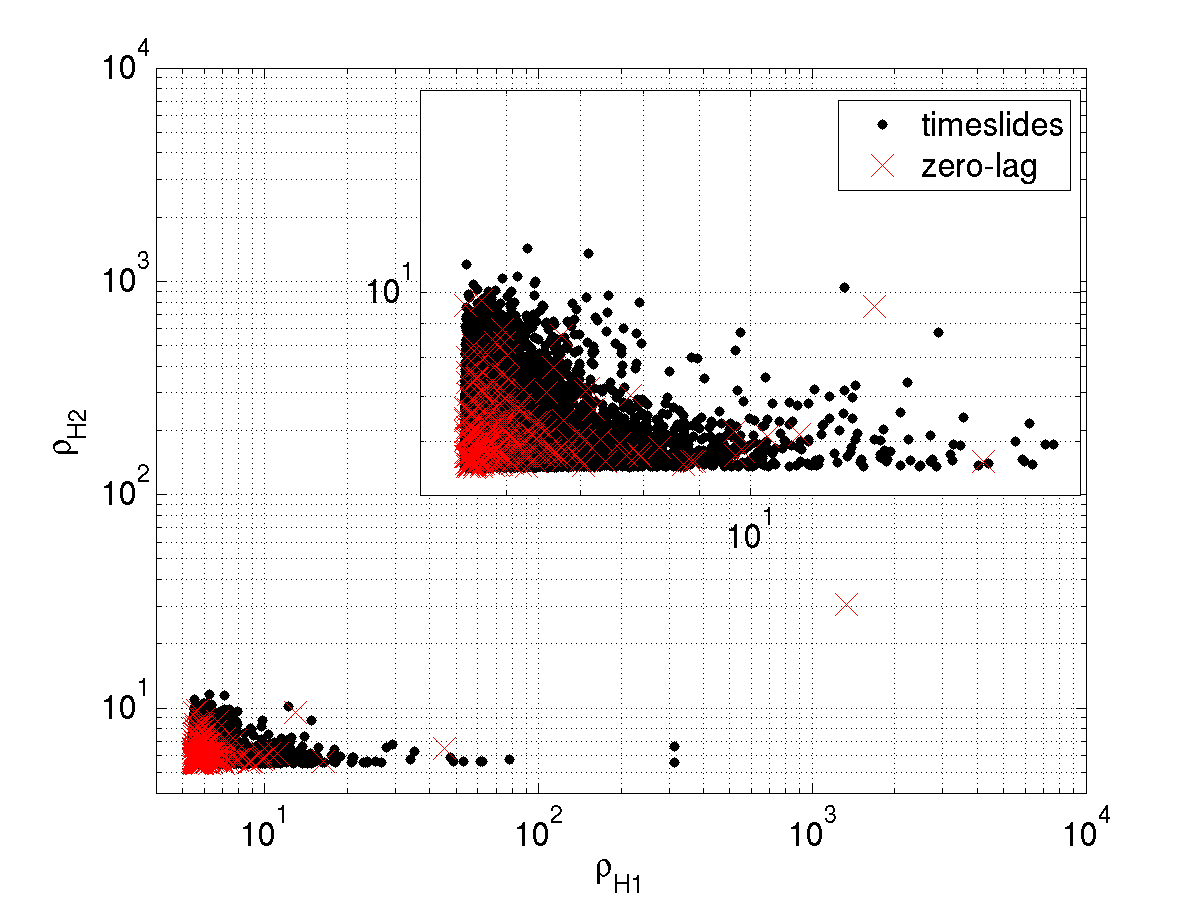}
\caption{Scatter plot of H2 versus H1 signal-to-noise ratio for double coincident zero-lag and background (100 timeslides) triggers in triple time. The inset is an enlargement of the region $\rho=5$ to 200.}
\label{fig:bgintdinttsnrH1H2}
\end{center}
\end{figure}

\begin{figure}[ht] 
\centering
\begin{center}
\includegraphics[scale=0.55]{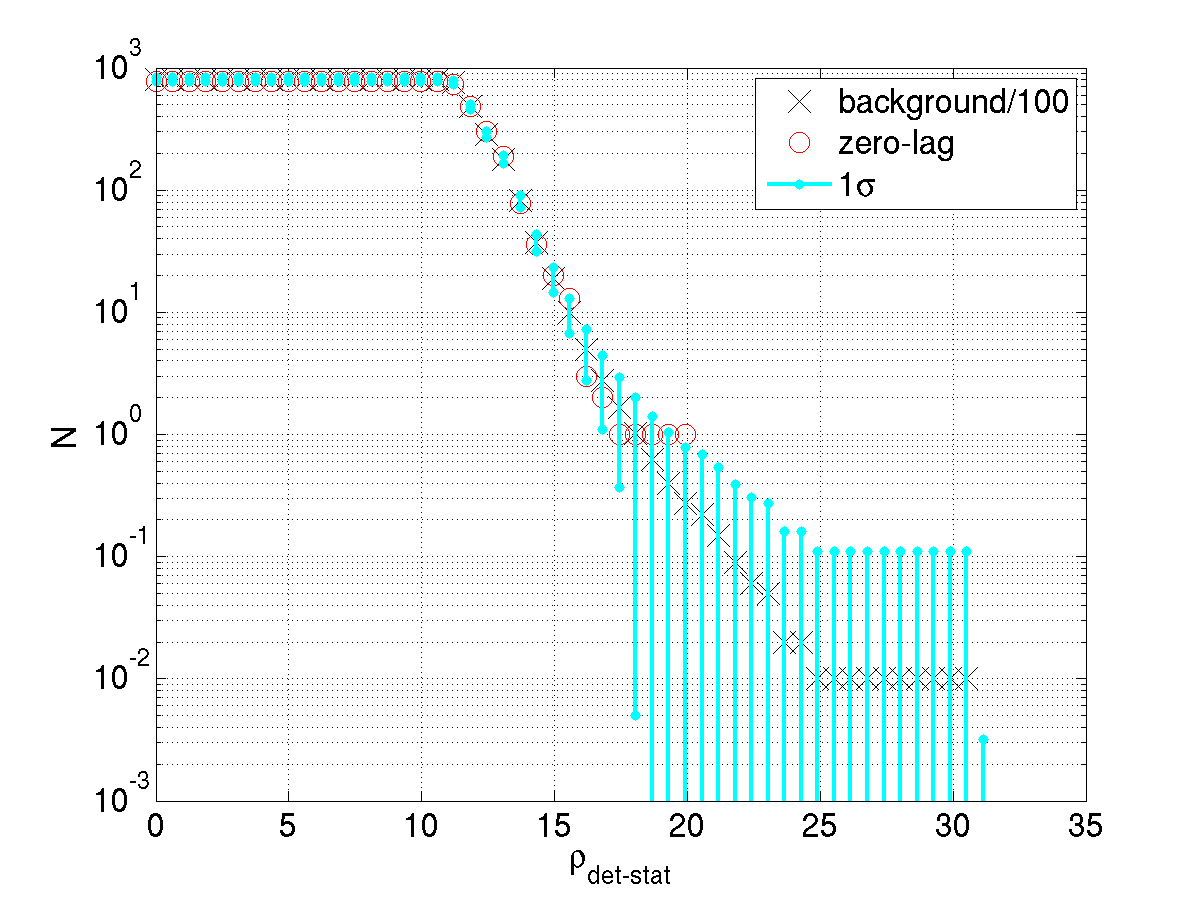} \\
\caption{Cumulative histogram of zero-lag and background coincidences, H2L1 doubles in triple time.}
\label{fig:bgintdinttH2L1}
\includegraphics[scale=0.55]{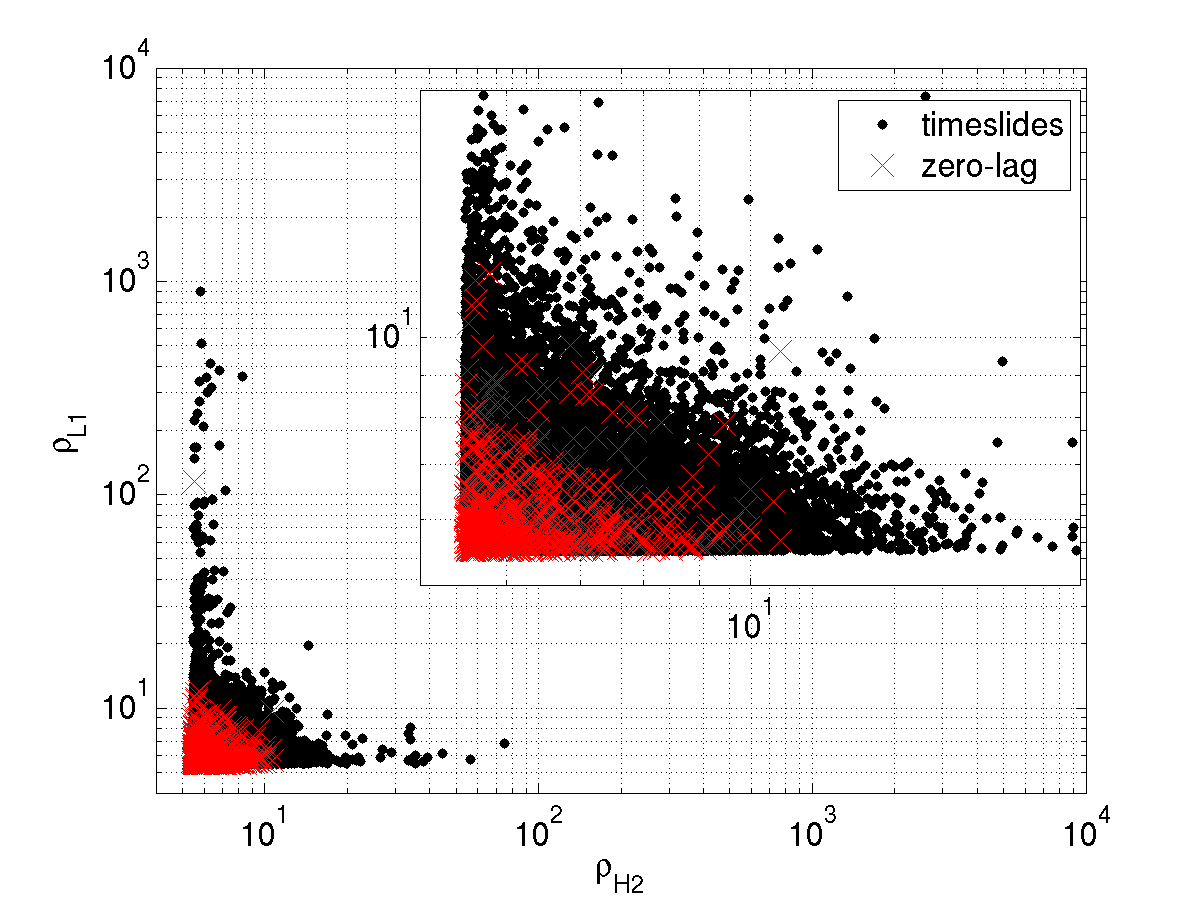}
\caption{Scatter plot of L1 versus H2 signal-to-noise ratio for double coincident zero-lag and background (100 timeslides) triggers in triple time. The inset is an enlargement of the region $\rho=5$ to 200.}
\label{fig:bgintdinttsnrH2L1}
\end{center}
\end{figure}

\begin{figure}[ht] 
\centering
\begin{center}
\includegraphics[scale=0.55]{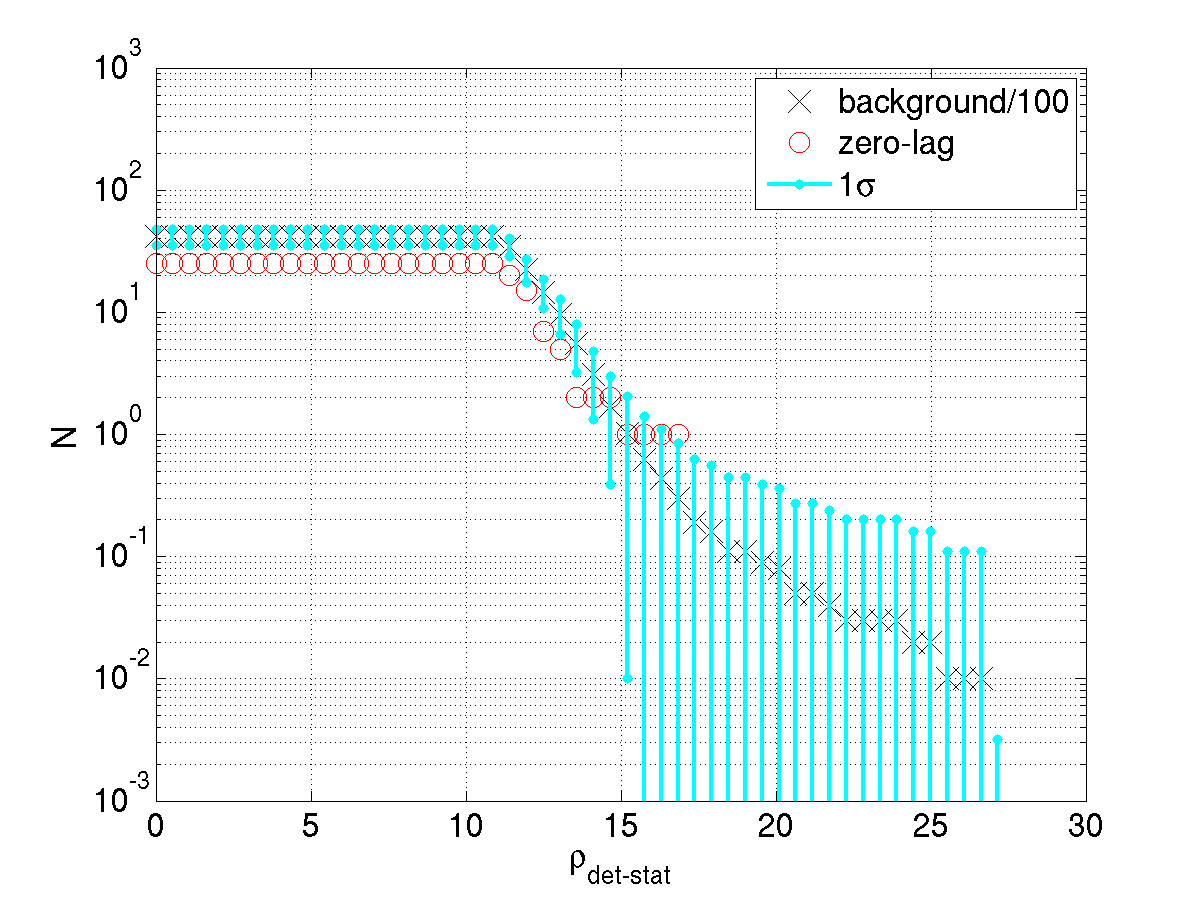} \\
\caption{Cumulative histogram of zero-lag and background coincidences, H1L1 doubles in double time.}
\label{fig:bgintdindtH1L1}
\includegraphics[scale=0.55]{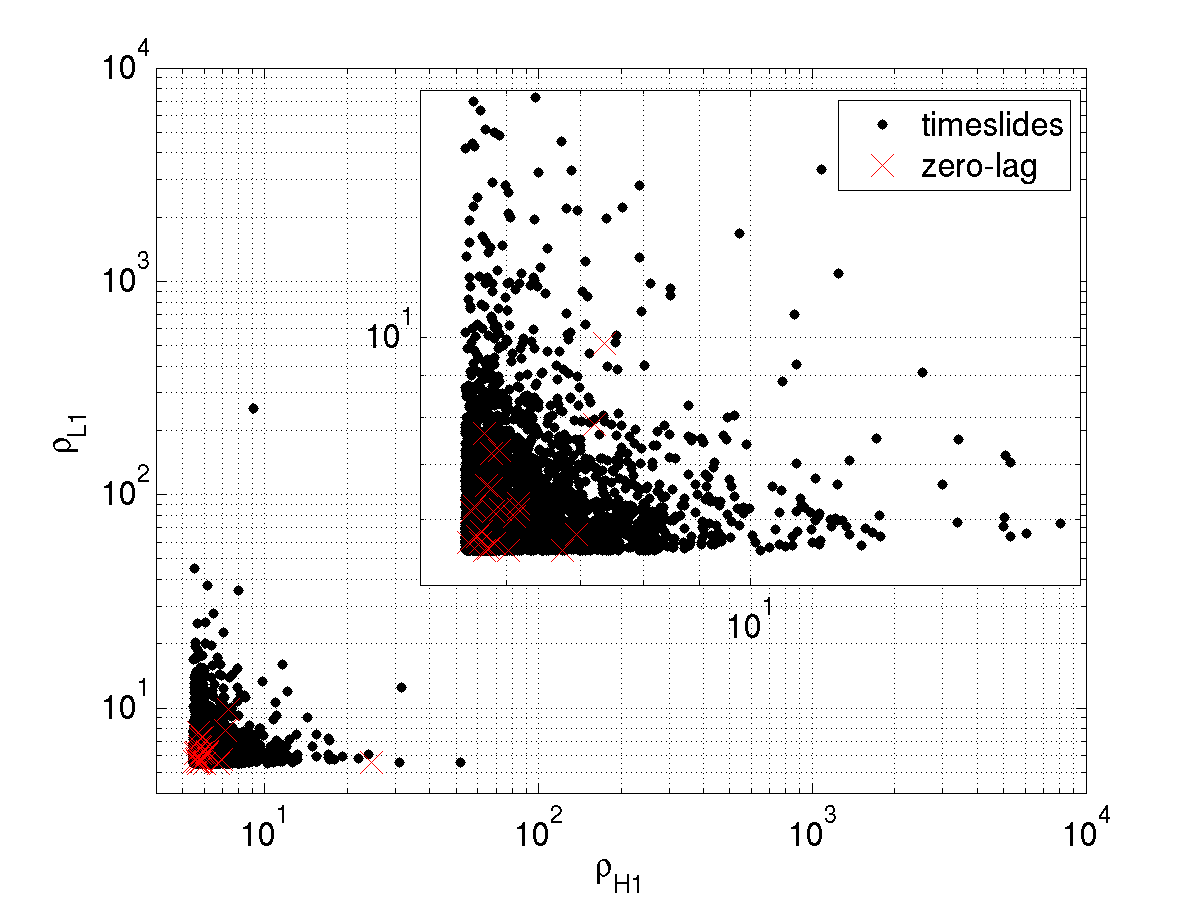}
\caption{Scatter plot of L1 versus H1 signal-to-noise ratio for double coincident zero-lag and background (100 timeslides) triggers in double time. The inset is an enlargement of the region $\rho=5$ to 200.}
\label{fig:bgintdindtsnrH1L1}
\end{center}
\end{figure}

\begin{figure}[ht] 
\centering
\begin{center}
\includegraphics[scale=0.55]{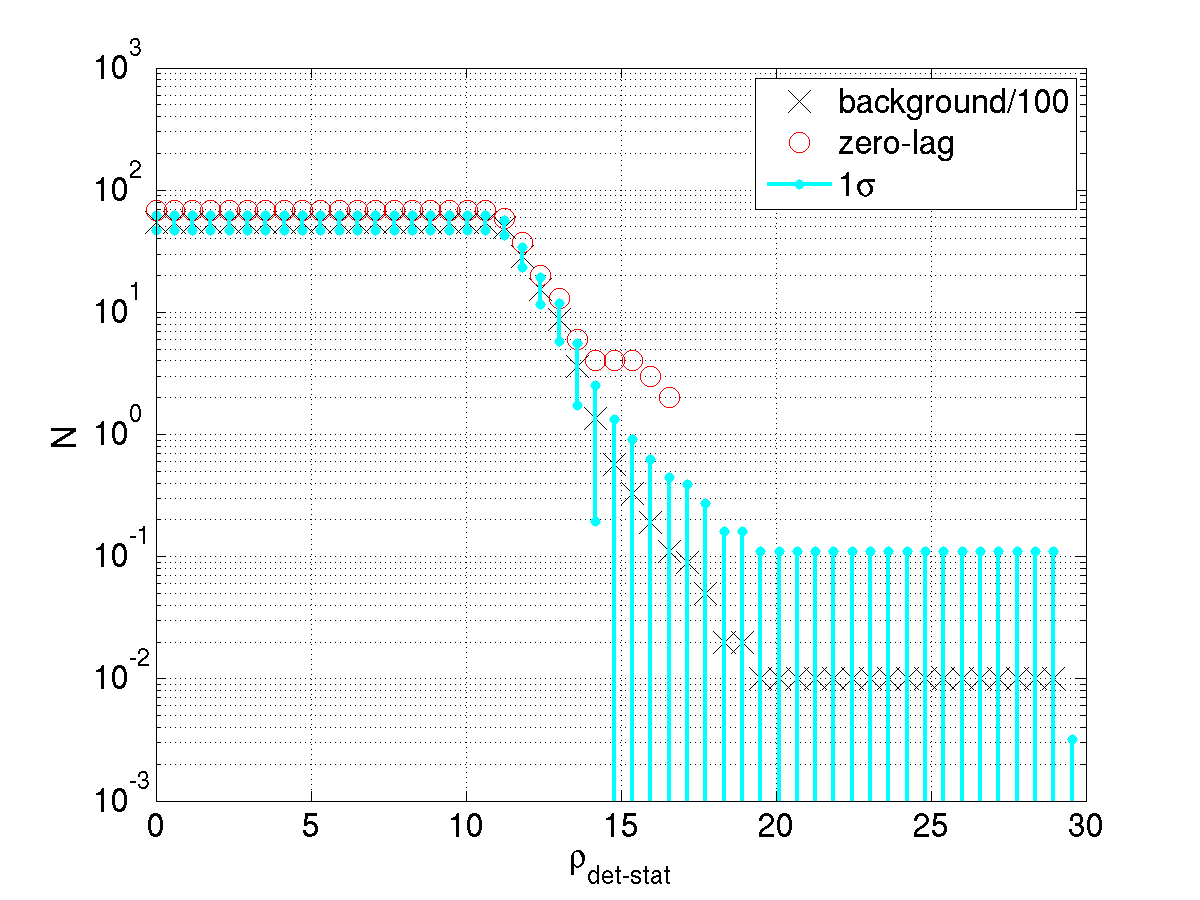} \\
\caption{Cumulative histogram of zero-lag and background coincidences, H1H2 doubles in double time.}
\label{fig:bgintdindtH1H2}
\includegraphics[scale=0.55]{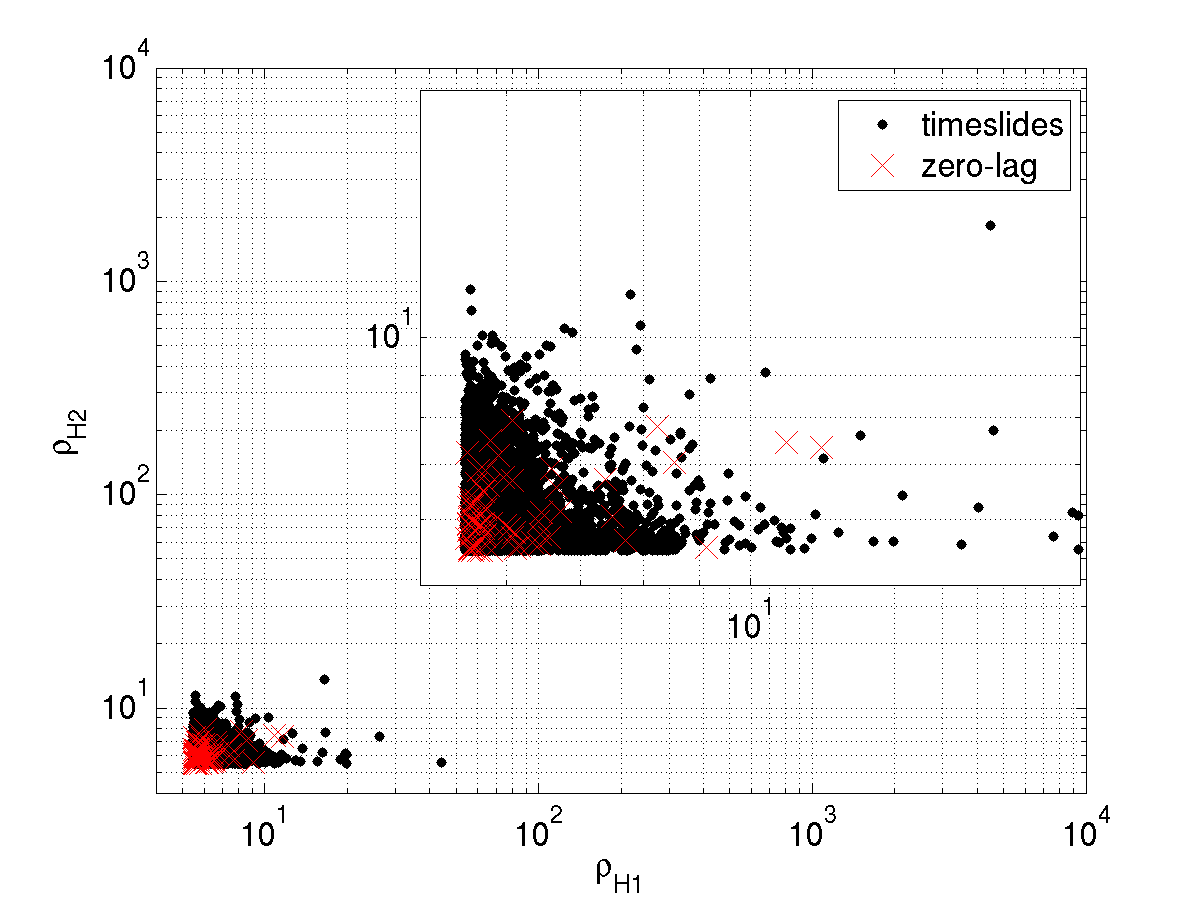}
\caption{Scatter plot of H2 versus H1 signal-to-noise ratio for double coincident zero-lag and background (100 timeslides) triggers in double time. The inset is an enlargement of the region $\rho=5$ to 200.}
\label{fig:bgintdindtsnrH1H2}
\end{center}
\end{figure}

\begin{figure}[ht] 
\centering
\begin{center}
\includegraphics[scale=0.55]{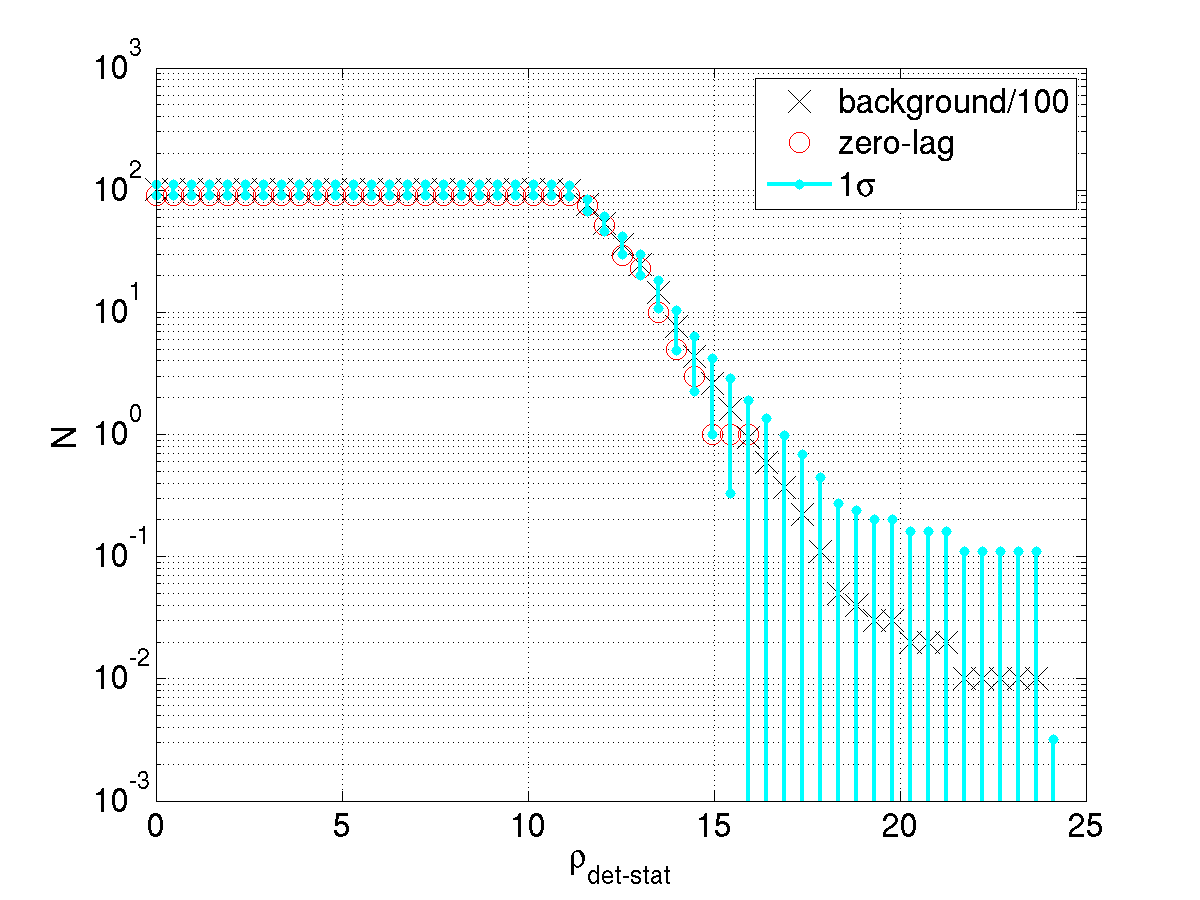}\\
\caption{Cumulative histogram of zero-lag and background coincidences, H2L1 doubles in double time.}
\label{fig:bgintdindtH2L1}
\includegraphics[scale=0.55]{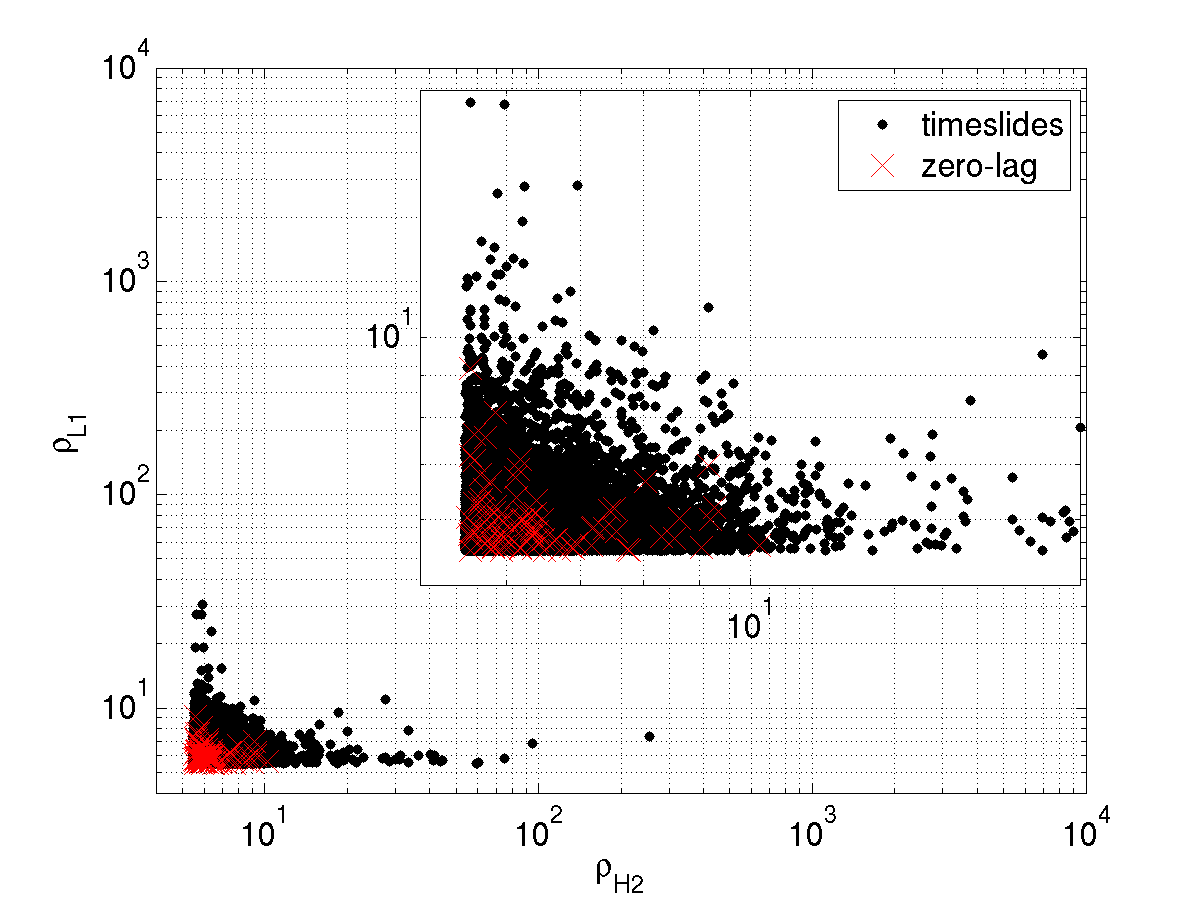}
\caption{Scatter plot of L1 versus H2 signal-to-noise ratio for double coincident zero-lag and background (100 timeslides) triggers in double time. The inset is an enlargement of the region $\rho=5$ to 200.}
\label{fig:bgintdindtsnrH2L1}
\end{center}
\end{figure}


\section{Following Up on the Loudest Candidates} \label{sec:followup}

Here we list a number of checks that can be applied to candidate events to increase or decrease our confidence in each as being caused by a gravitational wave. The histograms in the previous section showed that there were no events standing out above the background and thus we do not believe that we have detected a gravitational wave in this data set. However, as an exercise we apply this checklist to the three loudest candidates in each of the triple time categories and the loudest candidate in each of the double time categories and demonstrate how these candidates fail several of the tests. 

\subsection{Follow-Up Procedures}

\begin{itemize}

\item{Check what data quality flags (if any) were in place at the time of the candidate events. Candidate events occurring at the same time as category 4 data quality flags were generally downgraded in significance.}

\item{Run a qscan\footnote{Qscan is a script that creates time frequency maps of  selected channels around a desired time. It is a useful tool for obtaining an overview of excess power in the gravitational wave channel, excitation channels, auxiliary channels, and environmental channels.} \cite{Chat05} on the time under investigation. This has a twofold purpose: the first is to check if the signal in the gravitational wave channel is consistent with what we expect for a ringdown (a short signal at a single frequency) or an inspiral (a signal whose frequency increases with time), and the second is to see if there was excess power in any of the auxiliary channels which could explain the coupling of a non-gravitational wave source to the gravitational wave channel. For an example of what we expect in a qscan of the gravitational wave channel in the presence of a gravitational wave see the qscan of an inspiral-merger-ringdown hardware injection in figure \ref{fig:blindinj}. }

\item{Plot quantities such as the SNR time series, SNR versus frequency, and frequency time series around the time of the candidate event using the trigger files output after the filtering stage and compare with similar plots of injections and background events. 
}

\item{If a double coincident candidate event occurred during triple time, try to understand why it was not found in the third interferometer.}

\item{See if the event was also found in other searches, such as the S4 binary
black hole search (S4 BBH) \cite{S3S4} or the S4 burst search \cite{BurstS4}.
}

\end{itemize}

\begin{figure}[ht]
\centering
\begin{center}
\includegraphics[scale=0.6]{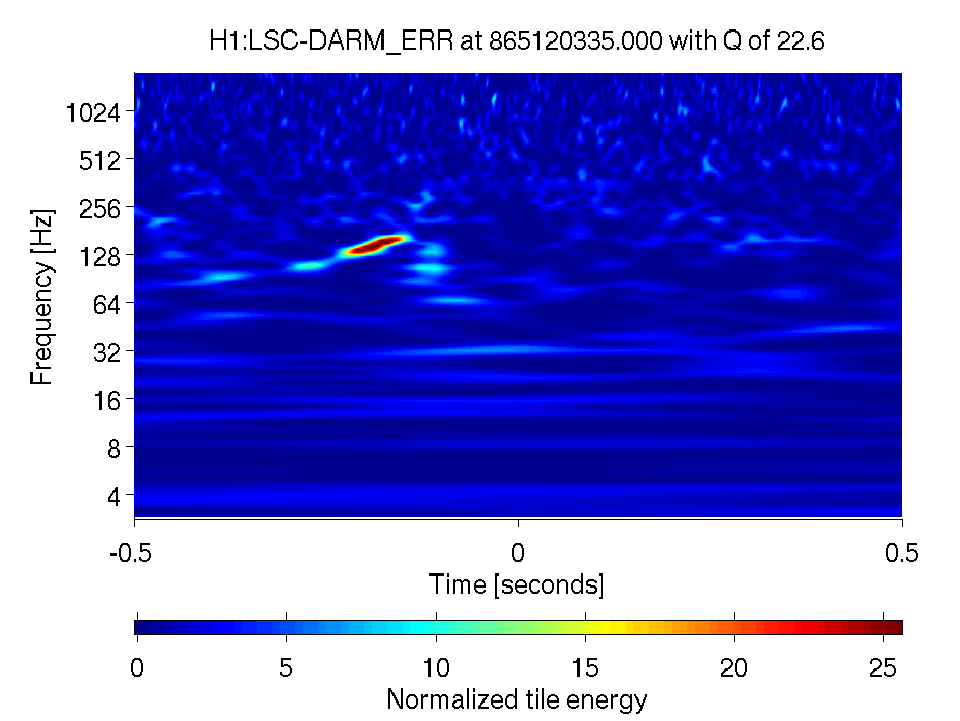}
\caption{The whitened spectrogram of the H1 gravitational wave channel showing a hardware injection of an inspiral-merger-ringdown signal during S5.}
\label{fig:blindinj}
\end{center}
\end{figure}


\subsection{H1L1 Doubles in Triple Time}

Candidate number 1: 

\begin{itemize}
\item{$t=794949585$, $\rho_{\textrm{DS}}=21$, $\rho_{\textrm{H1}}=9.4$, $\rho_{\textrm{L1}}=12.5$. }

\item{Examination of the data quality flag database shows that no category 4 data quality flags were on during the time of this candidate event.}

\item{If this was a real signal we would expect that $\rho_{\textrm{H2}} \approx 4.7$. However this is below the threshold for H2 and therefore if this was a gravitational wave it could not have been found as a triple coincidence.}

\item{This candidate was not in the top ten loudest candidates in the binary
black hole search.}

\item{A qscan of the H1 and L1 gravitational wave channels at this time (shown in figure \ref{fig:H1L1dintt1_794949585}) revealed a short broadband signal in each interferometer in an otherwise quiet time window. These characteristics indicate that the signal is not a ringdown (or an inspiral).}

\begin{figure} 
\centering
\subfigure[H1] 
{
\label{fig:H1L1dintt1} \includegraphics[width=6.7cm]{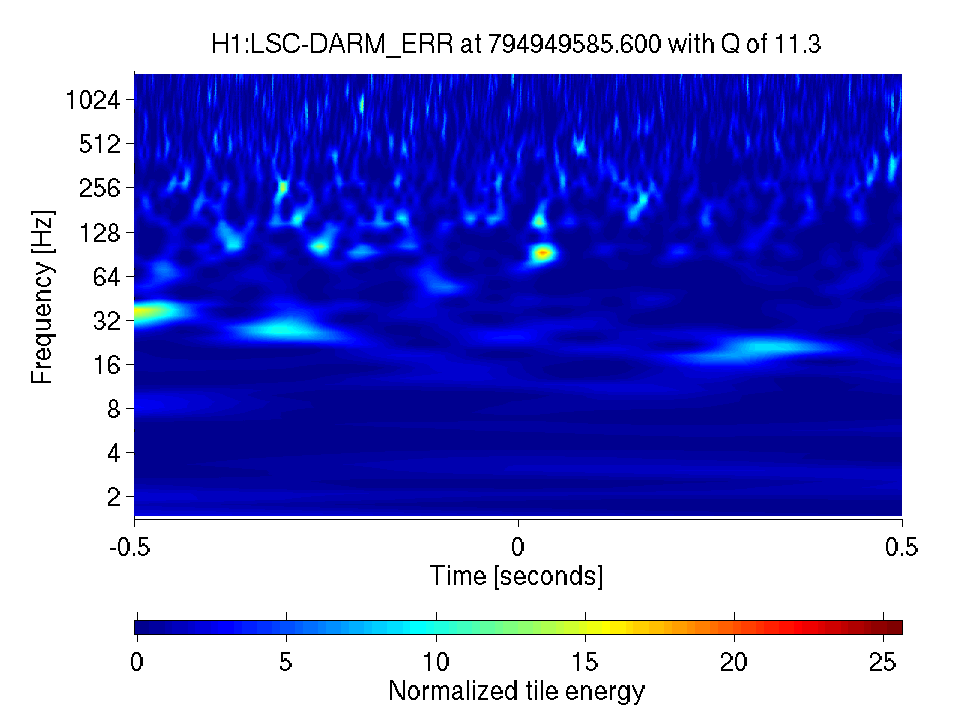} } \hspace{1cm}
\subfigure[L1] 
{ \label{fig:L1H1dintt1} \includegraphics[width=6.7cm]{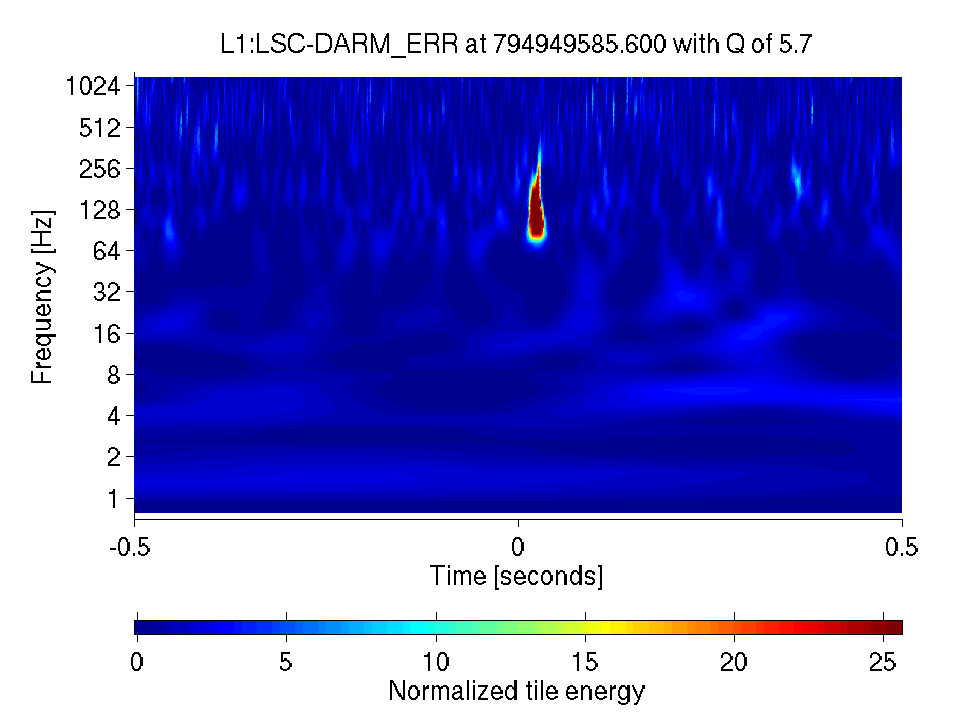} }
\caption{A qscan of the gravitational wave channel at GPS time 794949585, the loudest H1L1 candidate event in triple time.}
\label{fig:H1L1dintt1_794949585} 
\end{figure}

\end{itemize}

\noindent
Candidate number 2: 

\begin{itemize}
\item{$t=793829533$, $\rho_{\textrm{DS}}=20$, $\rho_{\textrm{H1}}=9$, $\rho_{\textrm{L1}}=11$.}

\item{The category 4 data quality flag L1:ASDC\_LOW\_THRESH was on at this time. This flag represents times when the amount of light in the dark port exceeds a certain threshold. This is usually indicates problems with alignment and raises our suspicions about the validity of this candidate.}

\item{The SNR of H1 indicates that this candidate was below the threshold for H2.}

\item{This candidate was not among the BBH search loudest triggers.}

\item{The qscan immediately eliminates this candidate. Figure \ref{fig:H1L1dintt2_793829533} shows a long broadband disturbance in L1.}

\begin{figure}
\centering
\subfigure[H1] 
{
\label{fig:H1L1dintt2}
\includegraphics[width=6.7cm]{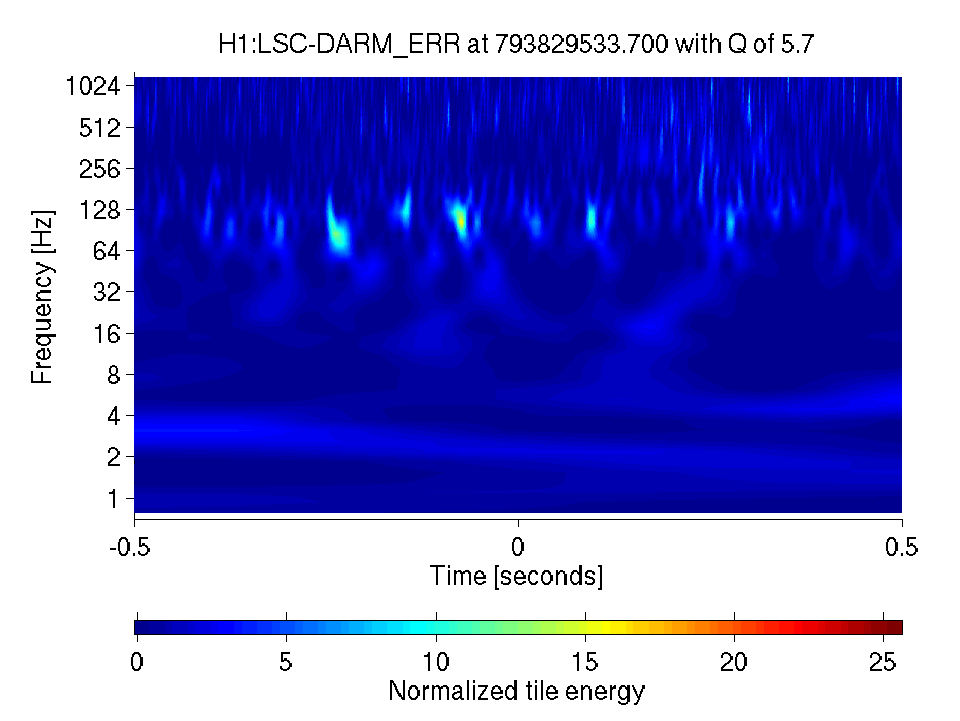} } \hspace{1cm}
\subfigure[L1] 
{ \label{fig:L1H1dintt2}
\includegraphics[width=6.7cm]{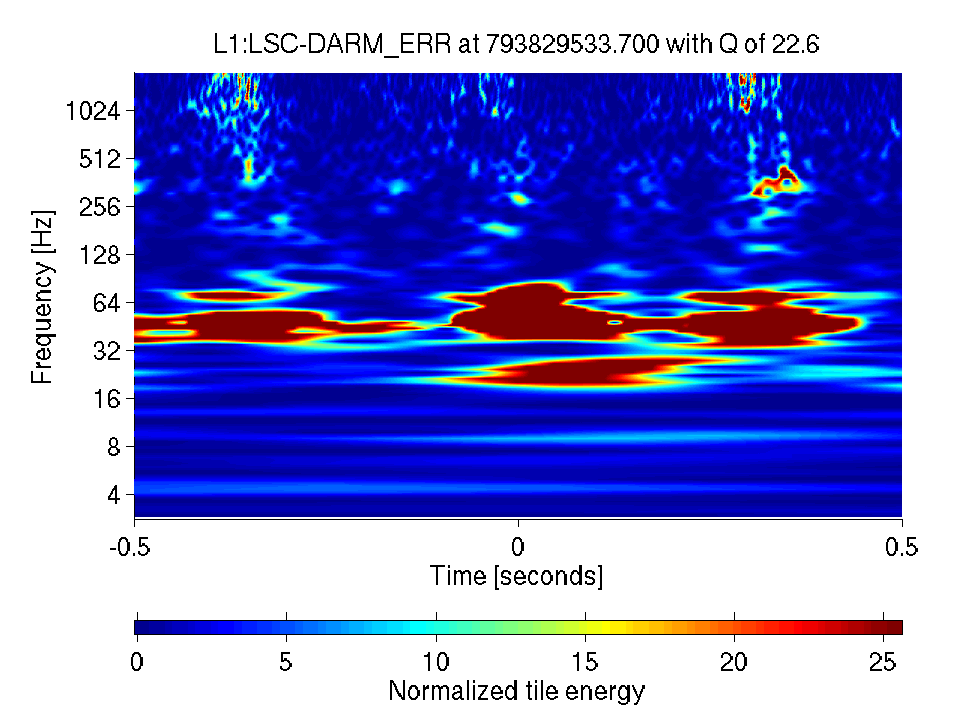} }
\caption{A qscan of the gravitational wave channel at GPS time 793829533, the second
loudest H1L1 candidate event in triple time.}
\label{fig:H1L1dintt2_793829533} 
\end{figure}

\end{itemize}

\noindent
Candidate number 3: 

\begin{itemize}
\item{$t=794291462$, $\rho_{\textrm{DS}}=19$, $\rho_{\textrm{H1}}=8$, $\rho_{\textrm{L1}}=11$.}

\item{Like the previous two candidates this also was too quiet to have been seen in H2.}

\item{The qscan, figure \ref{fig:H1L1dintt3_794291462}, once more indicates a noisy time in L1 during which the detection of gravitational waves would be unlikely.}

\begin{figure}
\centering
\subfigure[H1] 
{
\label{fig:H1L1dintt3}
\includegraphics[width=6.7cm]{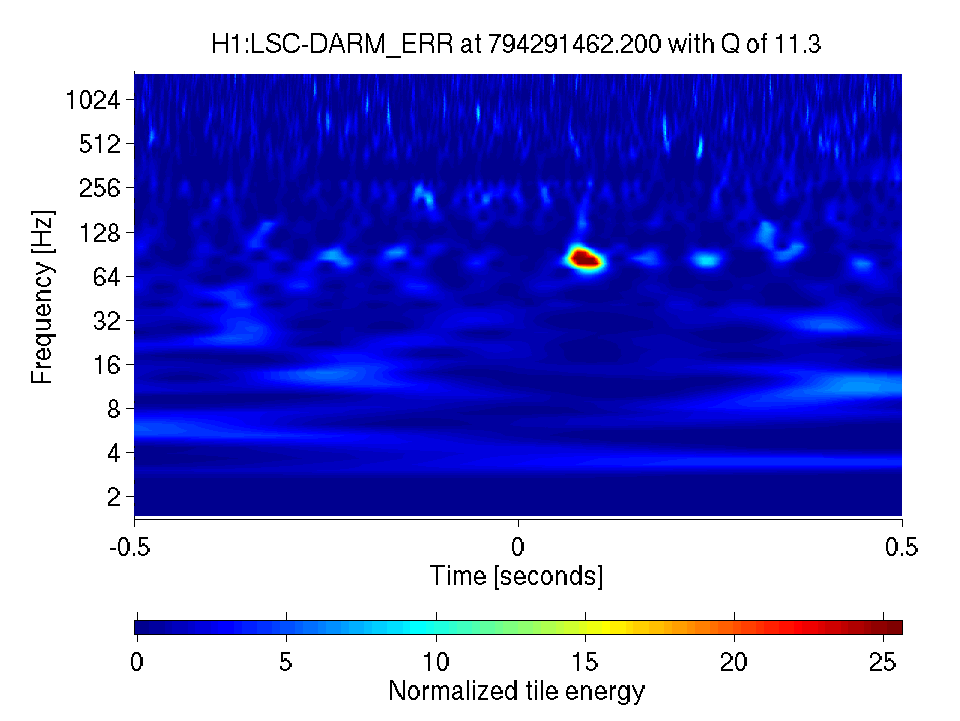} } \hspace{1cm}
\subfigure[L1] 
{ \label{fig:L1H1dintt3}
\includegraphics[width=6.7cm]{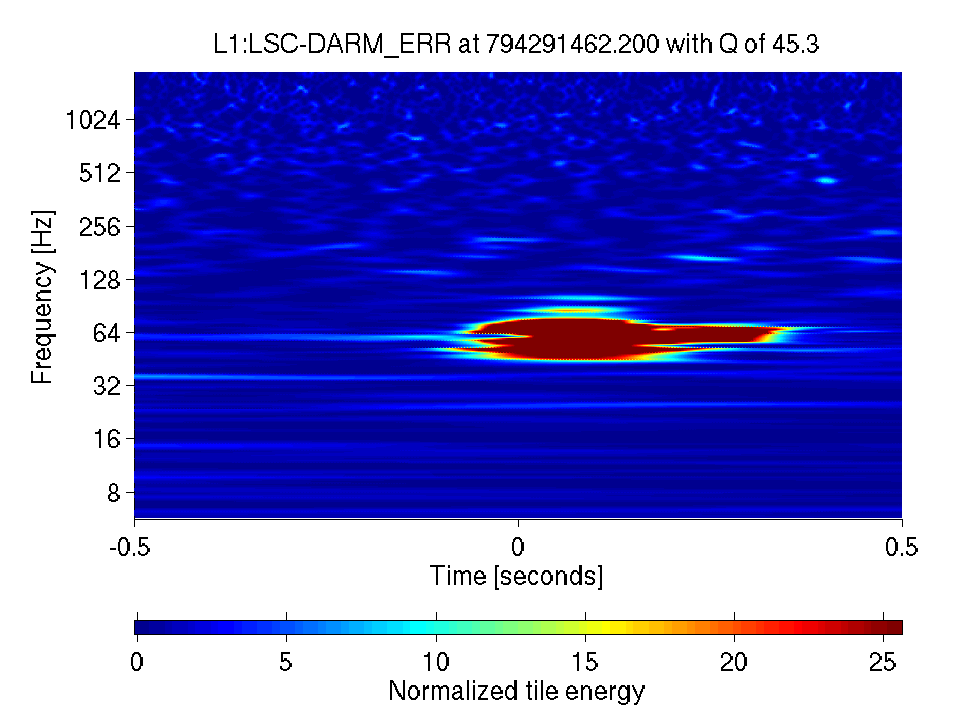} }
\caption{A qscan of the gravitational wave channel at GPS time 794291462, the third
loudest H1L1 candidate event in triple time.}
\label{fig:H1L1dintt3_794291462} 
\end{figure}

\end{itemize}


\subsection{H1H2 Doubles in Triple Time}

Candidate number 1: 

\begin{itemize}
\item{$t=793253792$, $\rho_{\textrm{DS}}=63$, $\rho_{\textrm{H1}}=1300$, $\rho_{\textrm{H2}}=30$.}

\item{The data quality flag H1H2\_COHERENCE was on at the time, indicating times of strongly coherent noise between H1 and H2. This decreases our confidence in this coincidence as a candidate event.}

\item{Looking at the qscan for this candidate, figure \ref{fig:H1H2dintt1_793253792}, we see a large broadband glitch in both the H1 and H2 gravitational wave channels. While we can not say with certainty what the cause of this glitch is, we can conclude that it does not have the characteristics of a gravitational wave ringdown. It does appear to be coincident in time with a glitch in a magnetometer channel, however a correlation between these channels has not yet been firmly established.}

\begin{figure}
\centering
\subfigure[H1] 
{
\label{fig:H1H2dintt1}
\includegraphics[width=6.7cm]{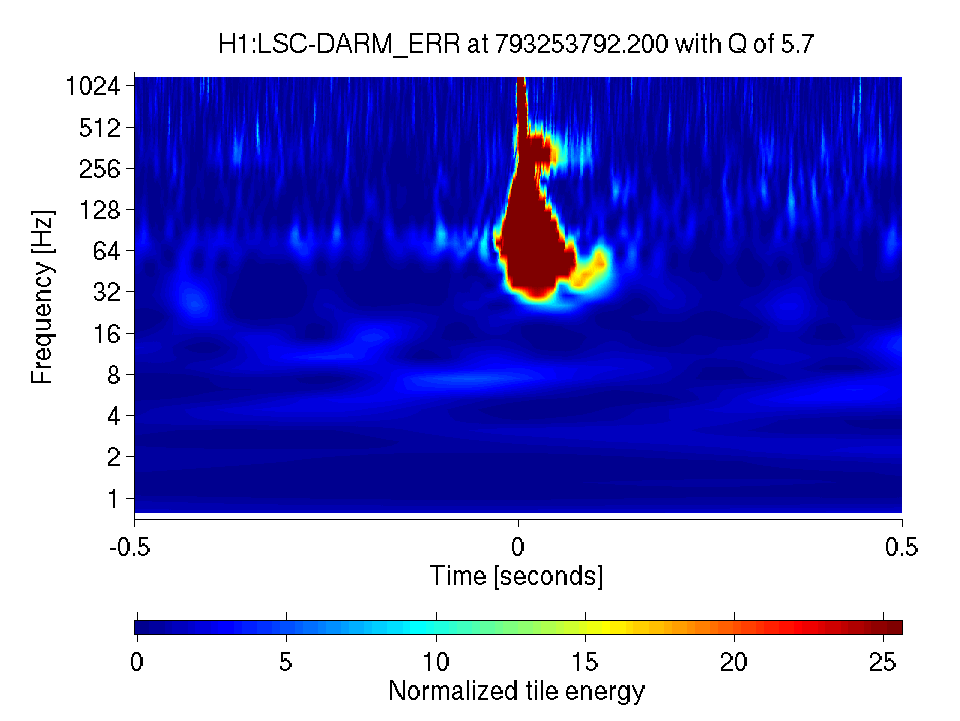} } \hspace{1cm}
\subfigure[H2] 
{ \label{fig:H2H1dintt1}
\includegraphics[width=6.7cm]{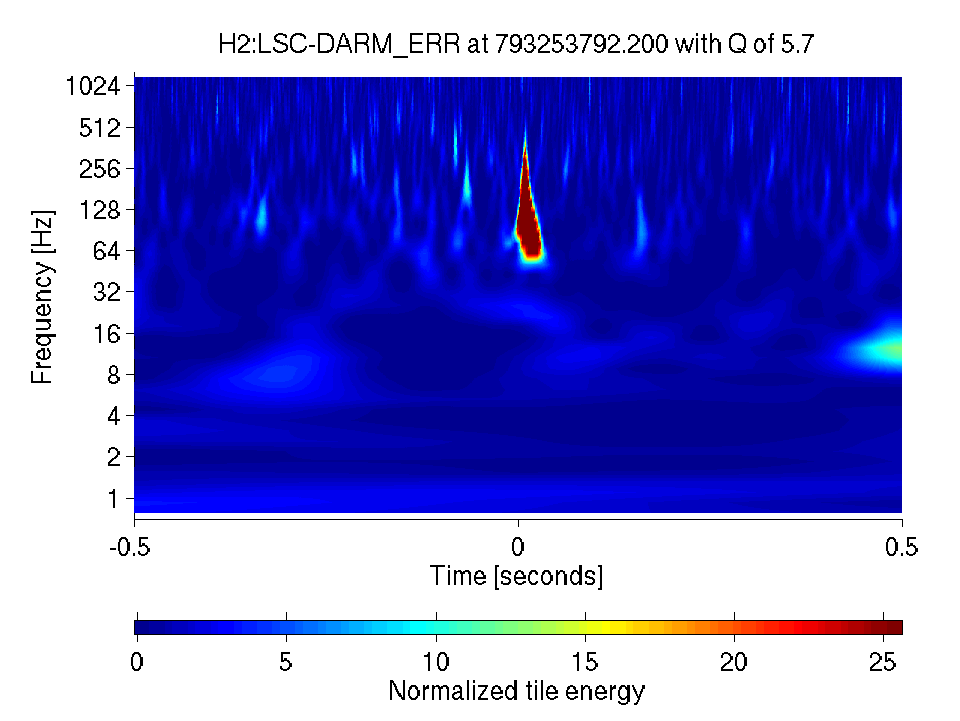} }
\caption{A qscan of gravitational wave channel at GPS time 793253792, the
loudest H1H2 candidate event in triple time.}
\label{fig:H1H2dintt1_793253792} 
\end{figure}

\end{itemize}
 
\noindent Candidate number 2: 

\begin{itemize}
\item{$t=794654729$, $\rho_{\textrm{DS}}=21$, $\rho_{\textrm{H1}}=13$, $\rho_{\textrm{H2}}=10$.}

\item{Figure \ref{fig:H1H2dintt2_794654729}, the qscan of the gravitational wave channel revealed a $\sim 0.5$ s long narrow-band signal in both H1 and H2.  In addition, a large number of environmental channels showed a significant disturbance. In the magnetometer channels this manifested itself as a $\sim 7$ s long broadband disturbance, as can be seen in figure \ref{fig:H1H2dintt1mag}. All of the accelerometer channels that were triggered displayed a shorter ($<$ 0.5 s) glitch at $\sim 128$ Hz; this can been seen in figure \ref{fig:H2H1dintt1acc}. This also appeared in the microphone channels with the same frequency and in some voltmeter channels at $\sim 292$ Hz. Some of the magnetometer channels also showed this line feature at $\sim 192$ Hz. This candidate is clearly not a gravitational wave.}

\begin{figure}
\centering
\subfigure[H1] 
{
\label{fig:H1H2dintt2}
\includegraphics[width=6.7cm]{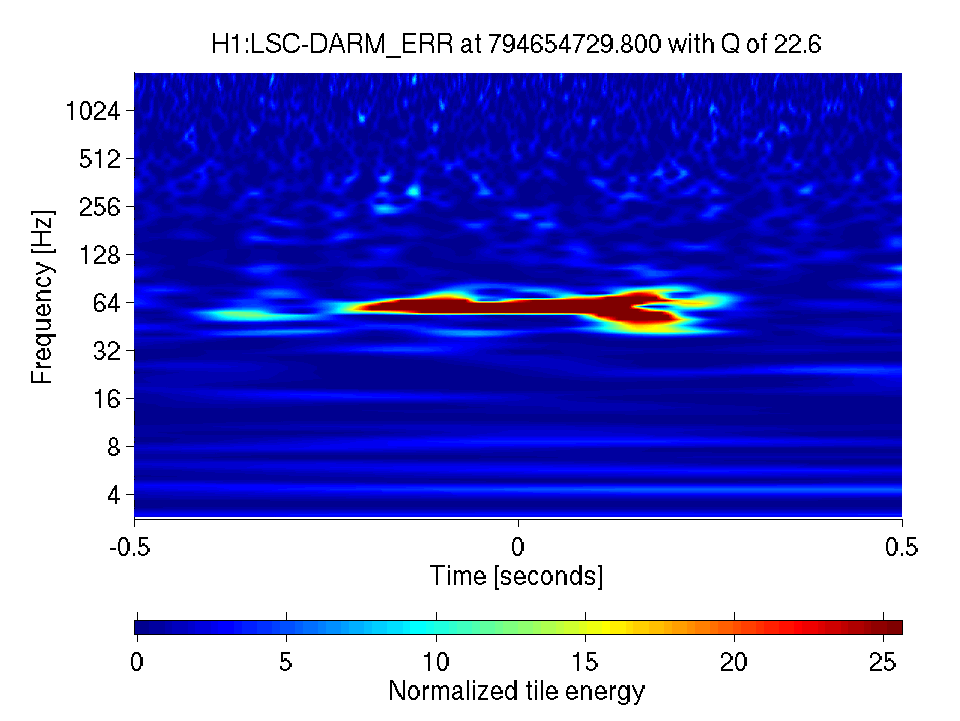} } \hspace{1cm}
\subfigure[H2] 
{ \label{fig:H2H1dintt2}
\includegraphics[width=6.7cm]{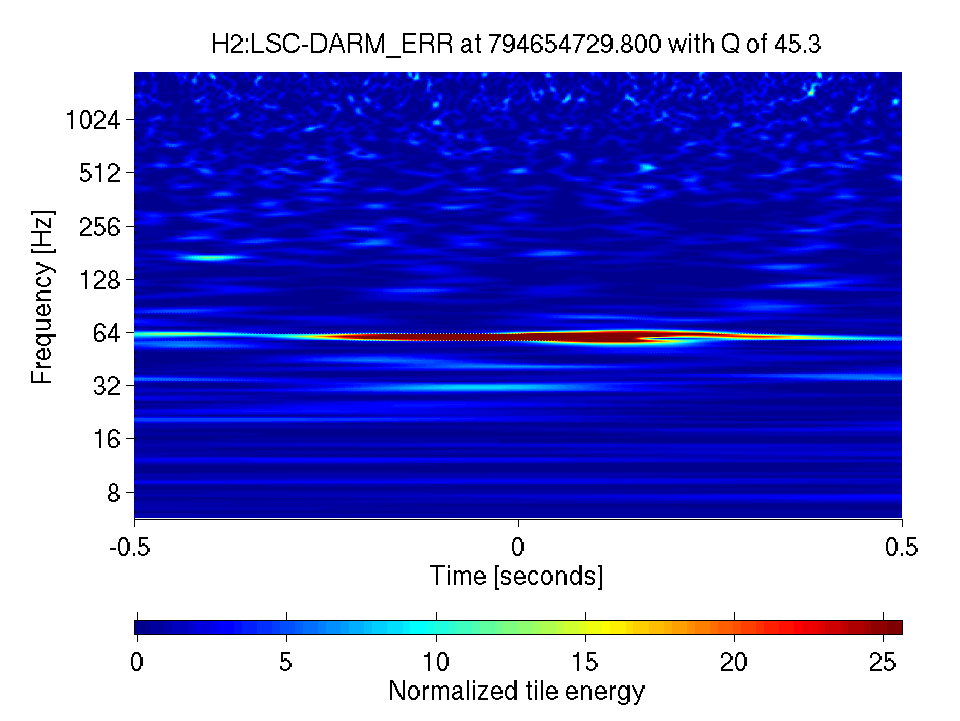} }
\caption{A qscan of the gravitational wave channel at GPS time 794654729, the
second loudest H1H2 candidate event in triple time.}
\label{fig:H1H2dintt2_794654729} 
\end{figure}

\begin{figure}
\centering
\subfigure[The magnetometer channel H0:PEM-LVEA\_MAGX.] 
{
\label{fig:H1H2dintt1mag}
\includegraphics[width=6.7cm]{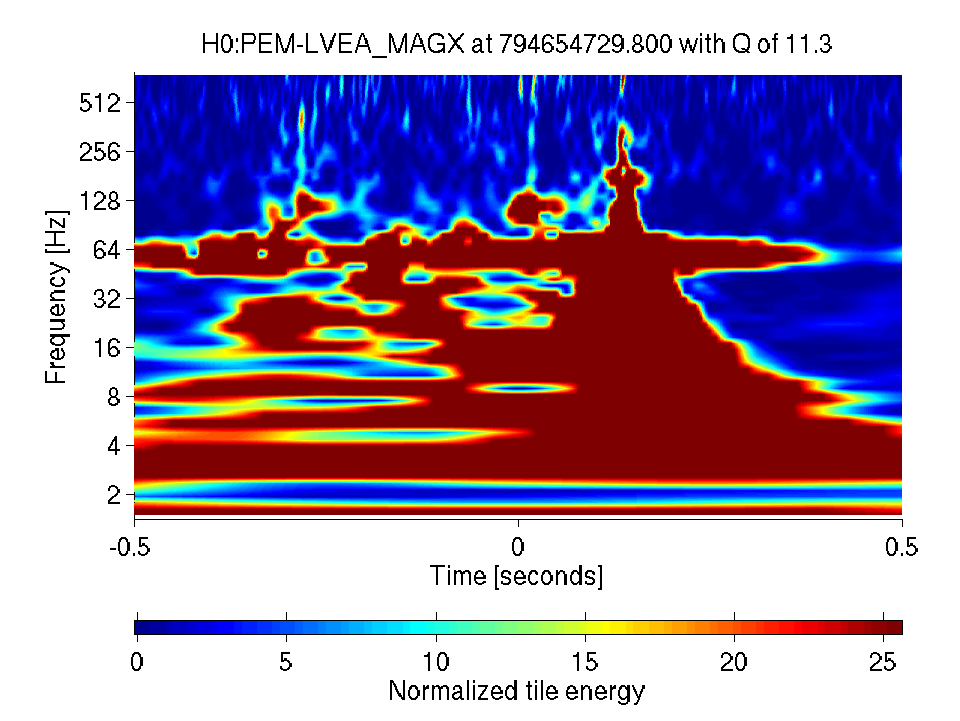} } \hspace{1cm}
\subfigure[The accelerometer channel H0:PEM-ISCT4\_ACCX.] 
{ \label{fig:H2H1dintt1acc}
\includegraphics[width=6.7cm]{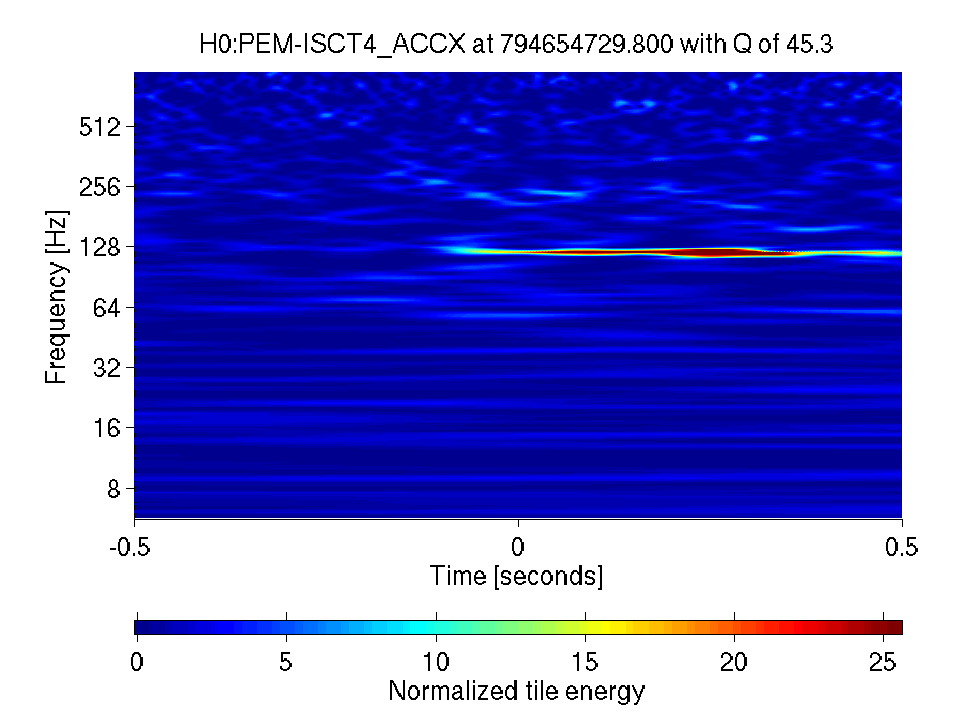} }
\caption{A qscan of two environmental channels triggered at 794654729.}
\label{fig:H1H2dintt1_794654729} 
\end{figure}

\end{itemize}

\noindent
Candidate number 3: 

\begin{itemize}
\item{$t=795398069$, $\rho_{\textrm{DS}}=15$, $\rho_{\textrm{H1}}=6$, $\rho_{\textrm{H2}}=8$.}

\item{The H1 category 4 data quality flag HIGH\_PIXEL\_FRACTION\_1KHZ was on at the time of the candidate, indicating a large deviation from Gaussianity, making this coincidence more likely due to noise than a gravitational wave.}

\item{A plot of SNR versus frequency shows a large portion of the both banks rang off over a small SNR range.
}

\item{For this candidate the gravitational wave channel qscan, figure \ref{fig:H1H2dintt3_795398069}, shows a series of small glitches at various frequencies indicating a noisy period of time in the detector.}

\begin{figure}
\centering
\subfigure[H1] 
{
\label{fig:H1H2dintt3}
\includegraphics[width=6.7cm]{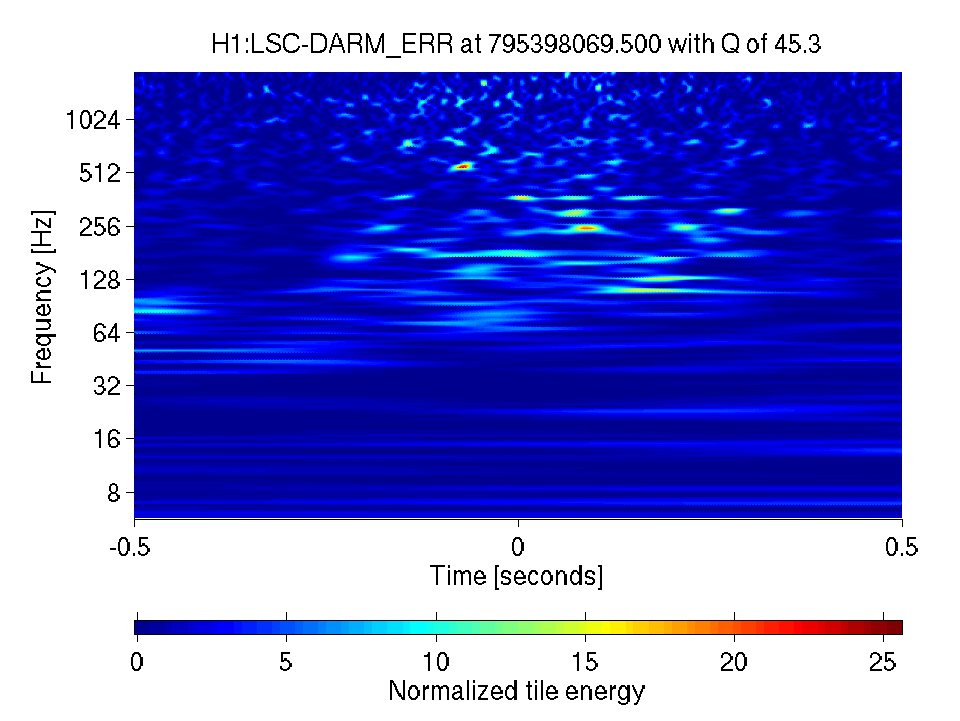} } \hspace{1cm}
\subfigure[H2] 
{ \label{fig:H2H1dintt3}
\includegraphics[width=6.7cm]{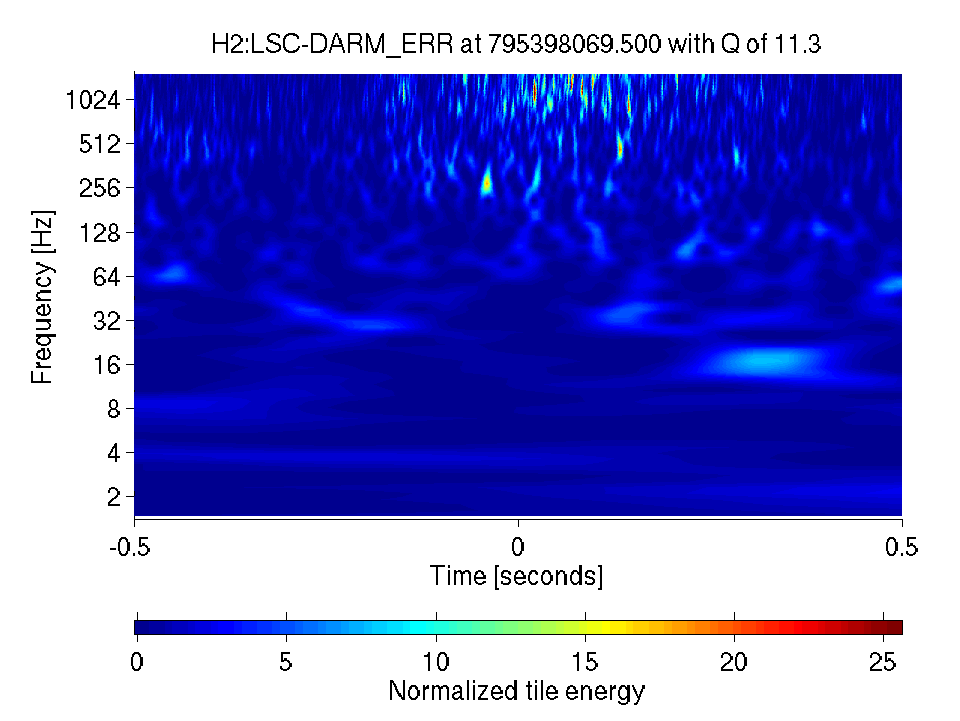} }
\caption{A qscan of the gravitational wave channel at GPS time 795398069, the third loudest H1H2 candidate event in triple time.}
\label{fig:H1H2dintt3_795398069} 
\end{figure}

\end{itemize}


\subsection{H2L1 Doubles in Triple Time}

Because of the similarity of the loudest three candidates in this category we discuss them collectively.

\begin{itemize}
\item{Candidate number 1: $t=794966223$, $\rho_{\textrm{DS}}=20$, $\rho_{\textrm{H2}}=10$, $\rho_{\textrm{L1}}=10$. \\
Candidate number 2: $t=794490884$, $\rho_{\textrm{DS}}=17$, $\rho_{\textrm{H2}}=9$, $\rho_{\textrm{L1}}=8$. \\
Candidate number 3: $t=793181575$, $\rho_{\textrm{DS}}=16$, $\rho_{\textrm{H2}}=9$, $\rho_{\textrm{L1}}=7$.}

\item{The qscans of the gravitational wave channel for candidates 1, 2, and 3 are shown in figures \ref{fig:H2L1dintt1_794966223}, \ref{fig:H2L1dintt2_794490884}, and \ref{fig:H2L1dintt3_793181575} respectively. Each demonstrates that the coincident triggers occurred during noisy times and are most likely false alarms.}

\begin{figure}
\centering
\subfigure[H2] 
{
\label{fig:H2L1dintt1}
\includegraphics[width=6.7cm]{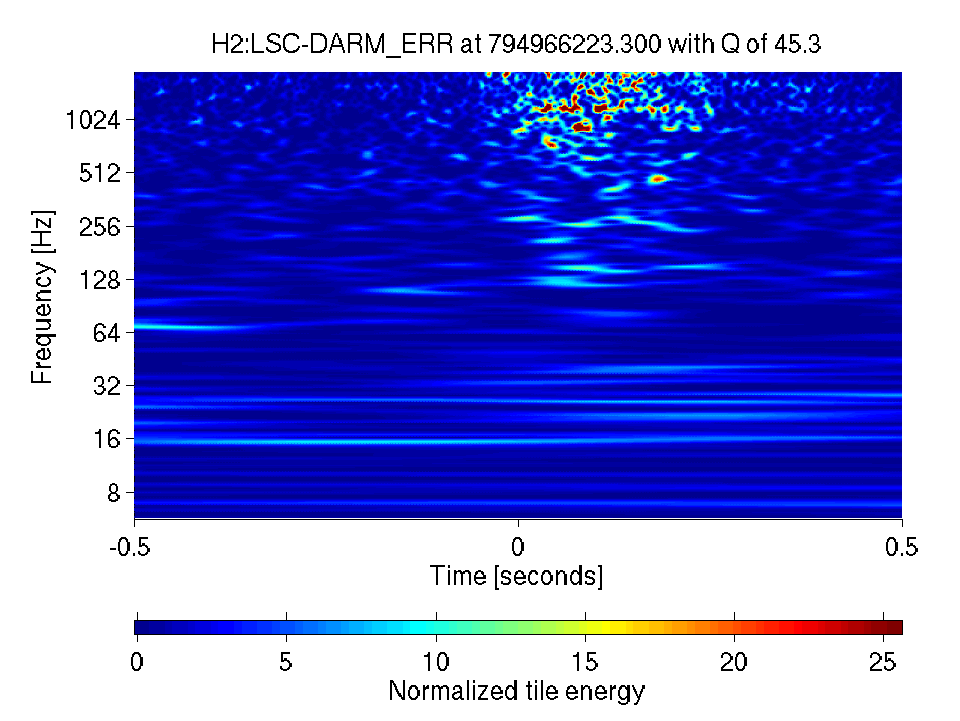} } \hspace{1cm}
\subfigure[L1] 
{ \label{fig:L1H2dintt1}
\includegraphics[width=6.7cm]{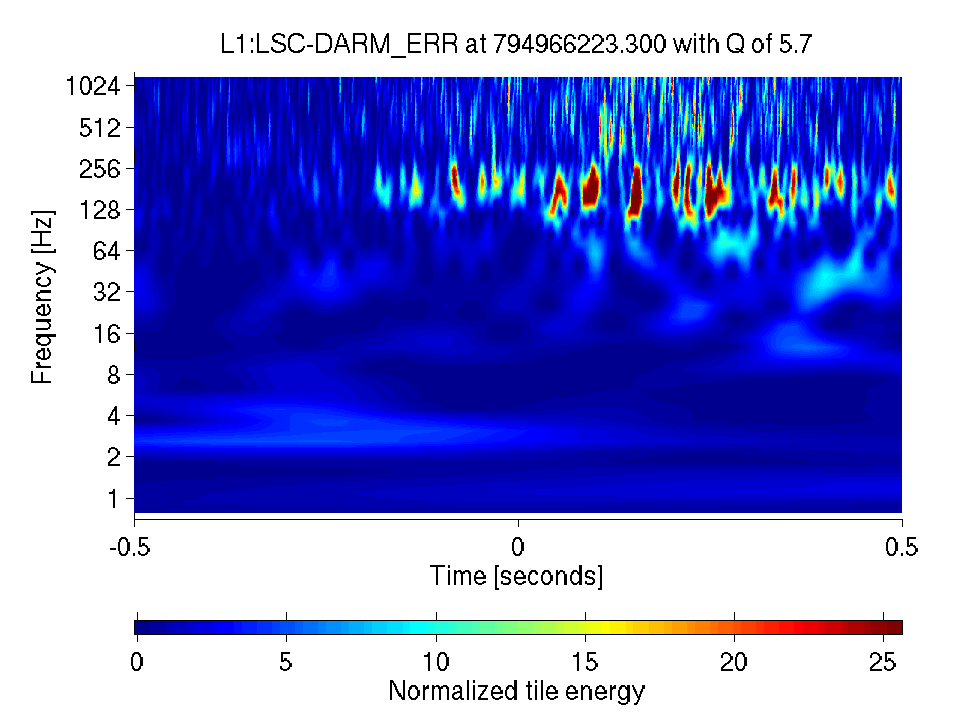} }
\caption{A qscan of the gravitational wave channel at GPS time 794966223, the loudest H2L1 candidate event in triple time.}
\label{fig:H2L1dintt1_794966223} 
\end{figure}

\begin{figure}
\centering
\subfigure[H2] 
{
\label{fig:H2L1dintt2}
\includegraphics[width=6.7cm]{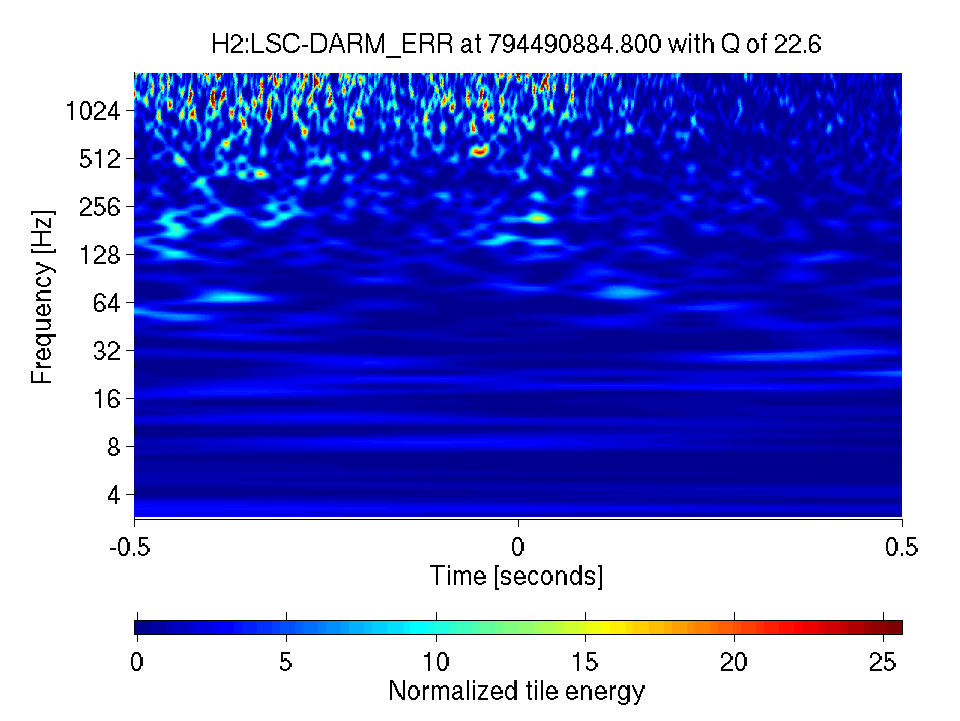} } \hspace{1cm}
\subfigure[L1] 
{ \label{fig:L1H2dintt2}
\includegraphics[width=6.7cm]{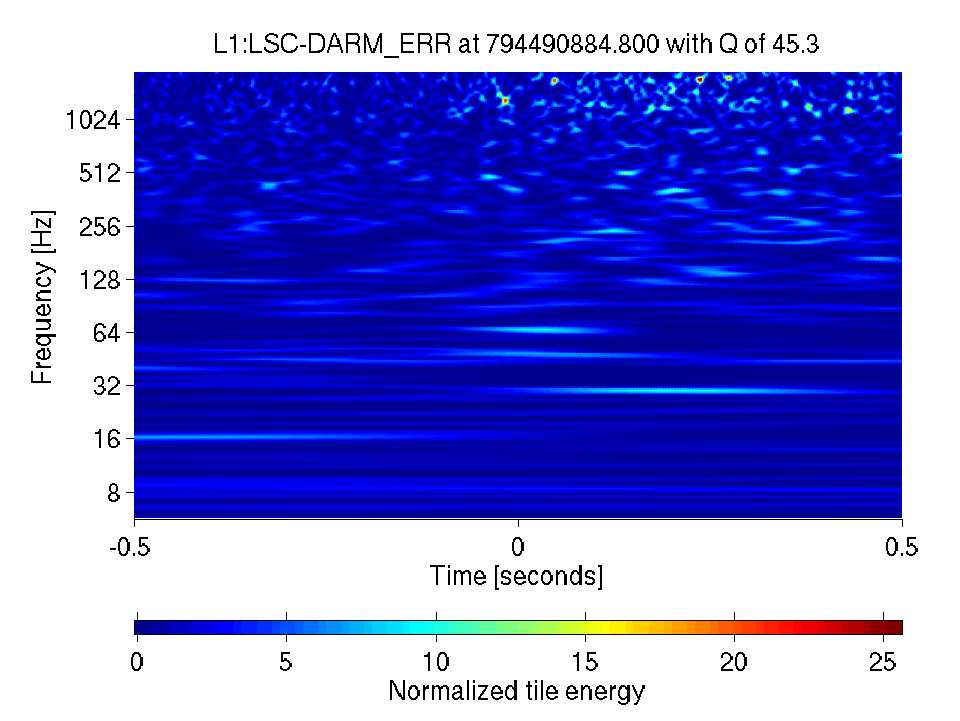} }
\caption{A qscan of the gravitational wave channel at GPS time 794490884, the
second loudest H2L1 candidate event in triple time.}
\label{fig:H2L1dintt2_794490884} 
\end{figure}

\begin{figure}
\centering
\subfigure[H2] 
{
\label{fig:H2L1dintt3}
\includegraphics[width=6.7cm]{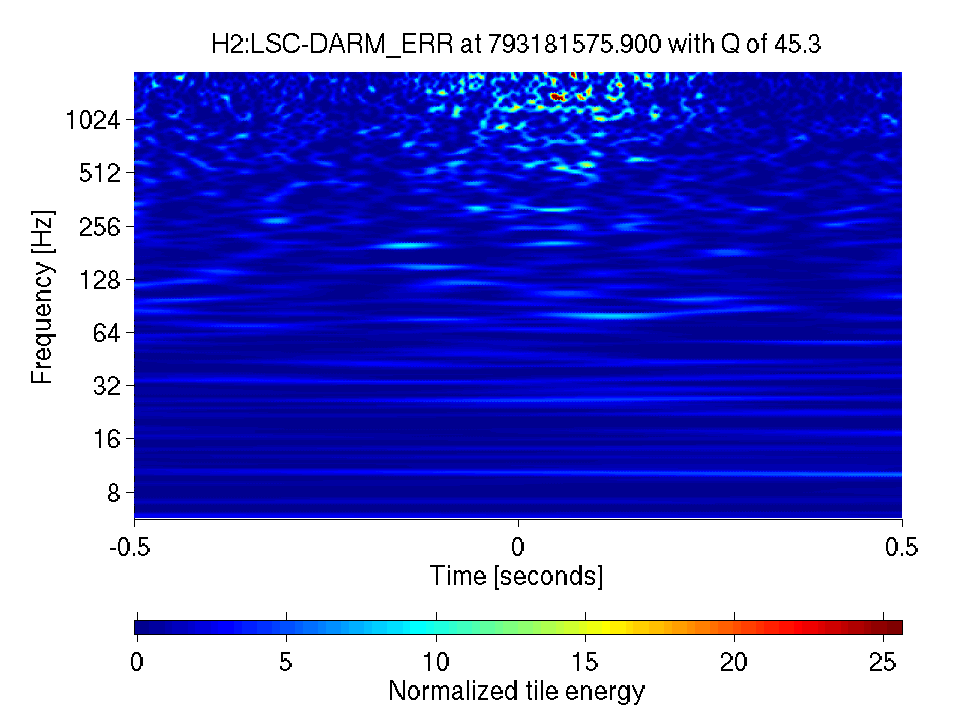} } \hspace{1cm}
\subfigure[L1] 
{ \label{fig:L1H2dintt3}
\includegraphics[width=6.7cm]{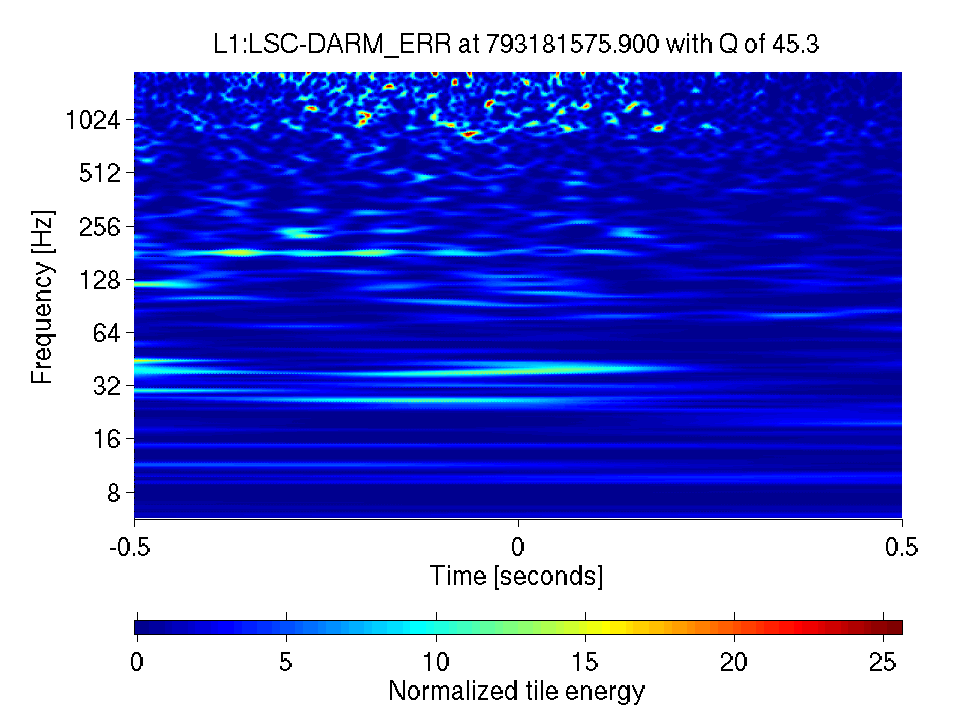} }
\caption{A qscan of gravitational wave channel at GPS time 794490884, the third loudest H2L1 candidate event in triple time.}
\label{fig:H2L1dintt3_793181575} 
\end{figure}

\item{For each of the candidates H1 was in science mode and if the H2 signal was due to a gravitational wave then, given that H1 is approximately a factor of 2 more sensitive, the event would also have been seen in H1.}

\item{Several data quality flags were on for candidates 1 and 3, some of which were category 4 flags reinforcing the false alarm claim. At the time of candidate 1 the H2 flag DUST\_ELEVATED indicating an increased particle count in the dark port (which could be responsible for glitches) and the L1 category 4 flags ACOUSTIC\_ELEVATED (indicating elevated acoustic noise in the 62--188 Hz band) and ASDC\_LOW\_THRESH were on.}

\end{itemize}


\subsection{H1L1 Doubles in Double Time}

Candidate number 1:
\begin{itemize}
\item{$t=793258551$, $\rho_{\textrm{DS}}=17$, $\rho_{\textrm{H1}}=7$, $\rho_{\textrm{H2}}=10$.}
\item{A qscan of the gravitatational wave channel is shown in figure
\ref{fig:H1L1dindt1_793258551}.}

\begin{figure}
\centering
\subfigure[H1] 
{
\label{fig:H1L1dindt1}
\includegraphics[width=6.7cm]{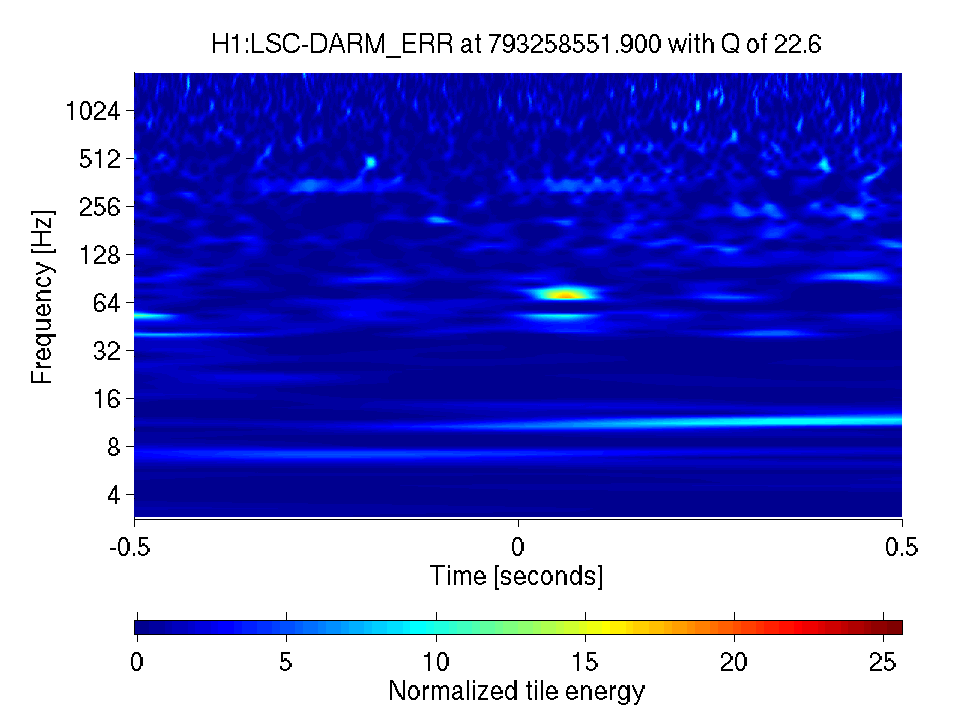} } \hspace{1cm}
\subfigure[L1] 
{ \label{fig:L1H1dindt1}
\includegraphics[width=6.7cm]{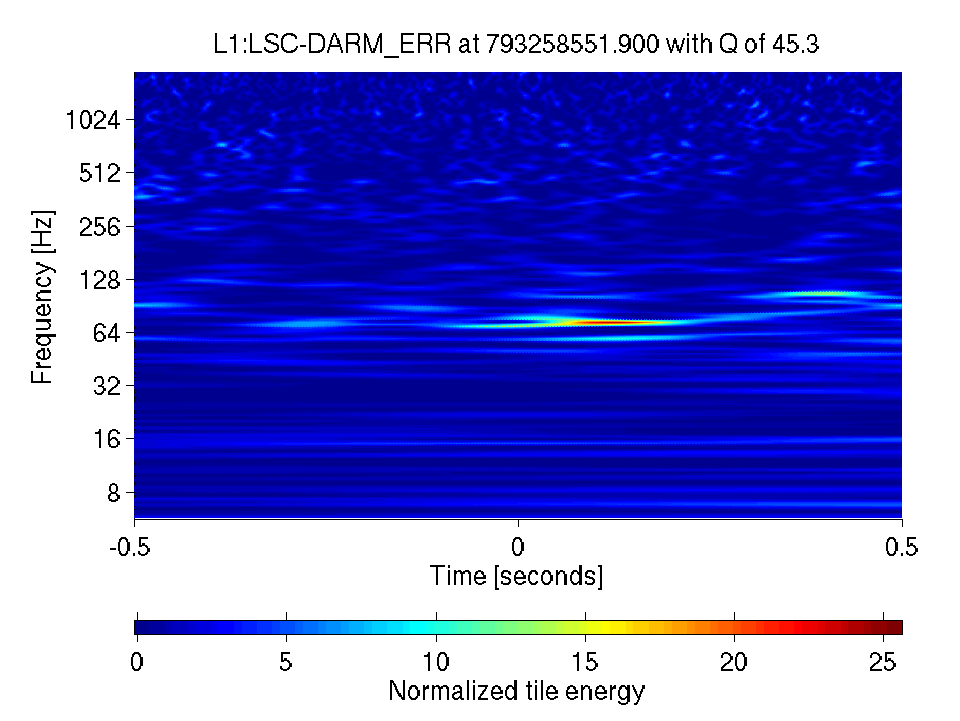} }
\caption{A qscan of the gravitational wave channel at GPS time 793258551, the
loudest H1L1 candidate event in triple time.}
\label{fig:H1L1dindt1_793258551} 
\end{figure}

\end{itemize}


\subsection{H1H2 Doubles in Double Time}

Candidate number 1:
\begin{itemize}
\item{$t=793589170$, $\rho_{\textrm{DS}}=17$, $\rho_{\textrm{H1}}=11$, $\rho_{\textrm{H2}}=8$.}

\item{The category 4 data quality flag SEISMIC\_0D8\_2D0 was on during this time indicating an excess of seismic activity between 0.8 and 2.0 Hz. Such a disturbance is likely to couple to both H1 and H2, producing the coincidence that we see.}

\item{A qscan of the gravitational wave channel at the time of this candidate event is shown in figure \ref{fig:H1H2dindt1_793589170}.}

\begin{figure}
\centering
\subfigure[H1] 
{
\label{fig:H1H2dindt1}
\includegraphics[width=6.7cm]{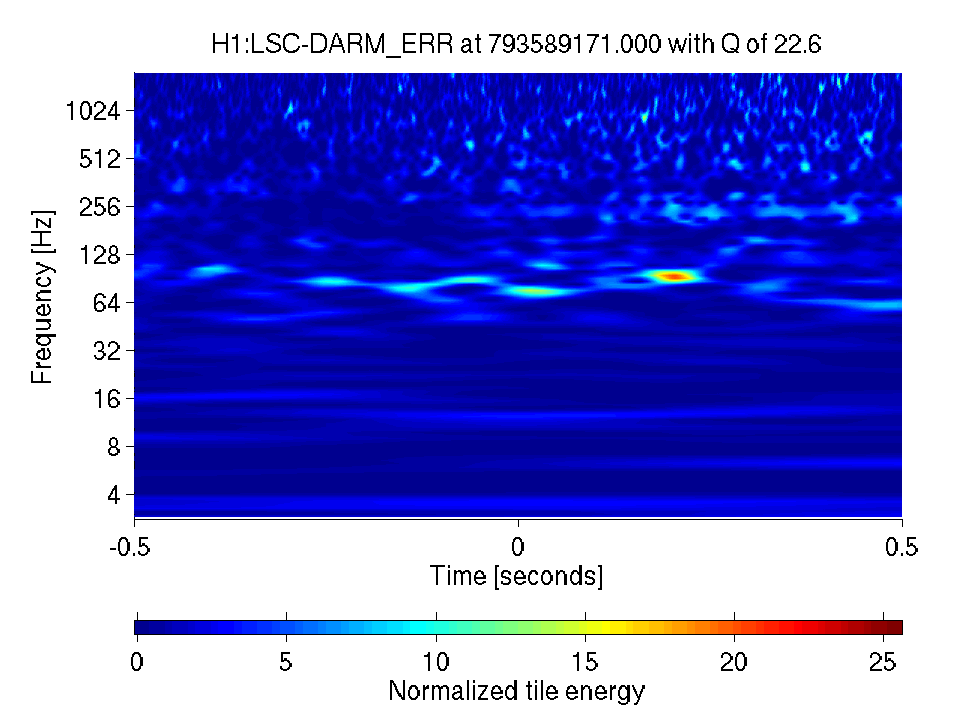} } \hspace{1cm}
\subfigure[L1] 
{ \label{fig:H2H1dindt1}
\includegraphics[width=6.7cm]{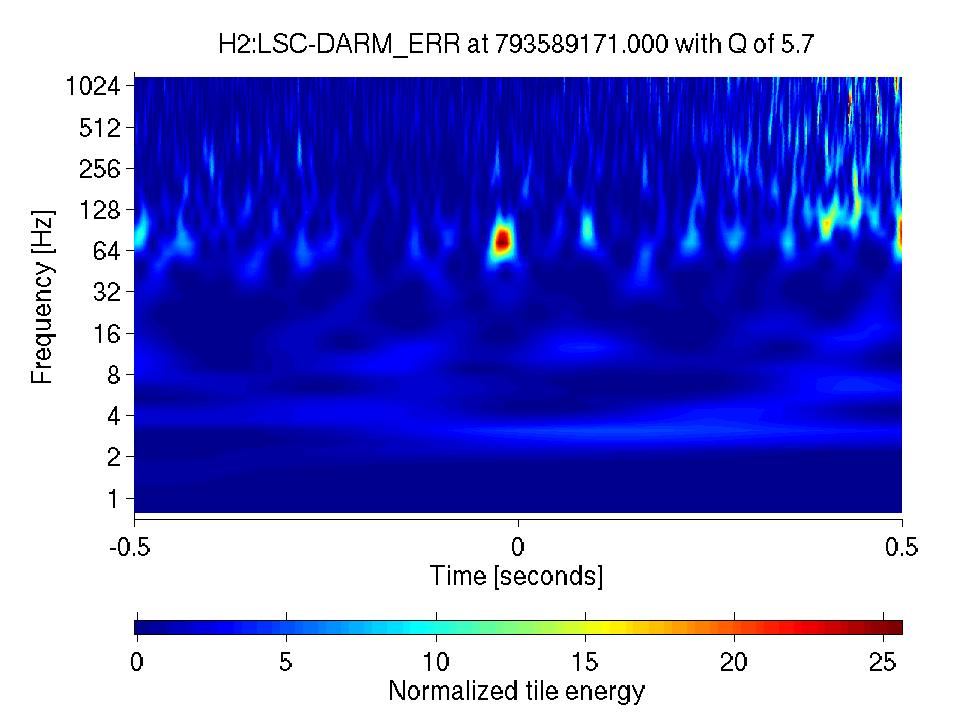} }
\caption{A qscan of the gravitational wave channel at GPS time 793589170, the
loudest H1H2 candidate event in triple time.}
\label{fig:H1H2dindt1_793589170} 
\end{figure}

\item{The category 4 data quality flag SEISMIC\_0D8\_2D0 was on during this time.}

\end{itemize}


\subsection{H2L1 Doubles in Double Time}

Candidate number 1:
\begin{itemize}
\item{$t=794432410$, $\rho_{\textrm{DS}}=16$, $\rho_{\textrm{H1}}=9$, $\rho_{\textrm{H2}}=7$.}

\item{The qscan of this candidate, figure \ref{fig:L1H2dindt1_794432410}, is quite similar to the H2L1 double coincident candidates in triple time; once more we find that the coincidence occurred during a noisy time in the detector.}

\begin{figure}
\centering
\subfigure[H2] 
{
\label{fig:H2L1dindt1}
\includegraphics[width=6.7cm]{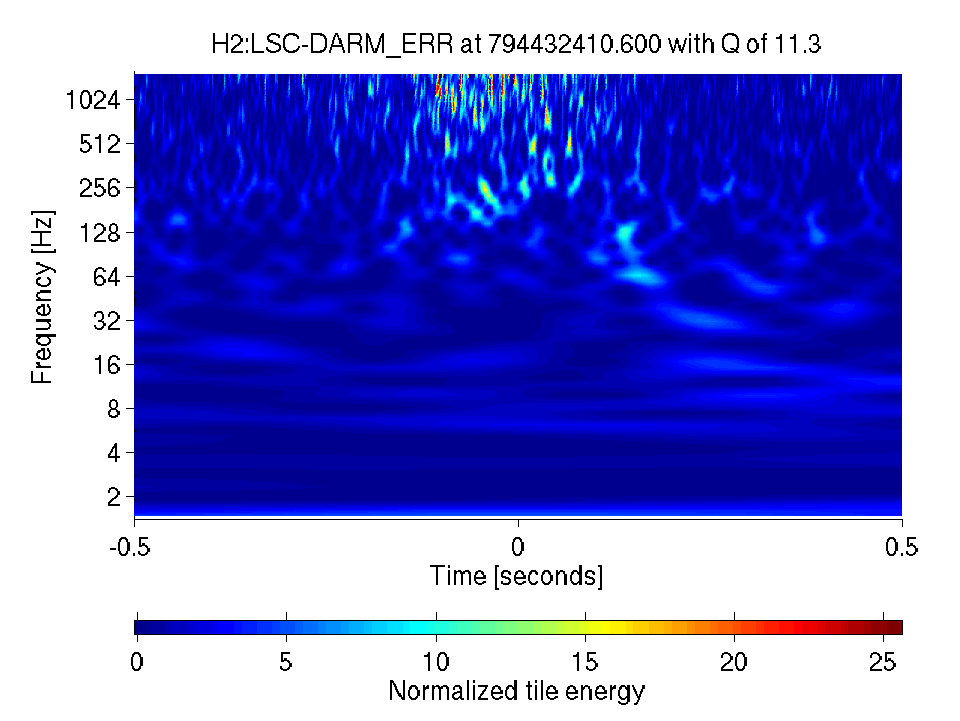} } \hspace{1cm}
\subfigure[L1] 
{ \label{fig:L1H2dindt1}
\includegraphics[width=6.7cm]{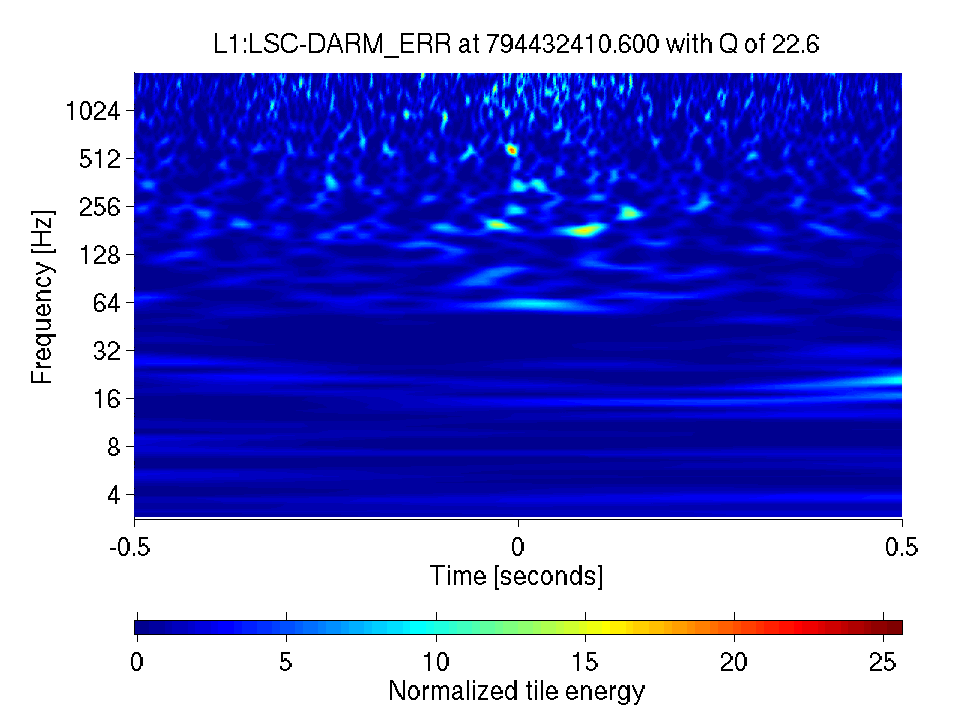} }
\caption{A qscan of the gravitational wave channel at GPS time 794432410, the
loudest H2L1 candidate event in triple time.}
\label{fig:L1H2dindt1_794432410} 
\end{figure}

\end{itemize}


\section{The Upper Limit}
\label{sec:upperlimit}

This goal of this search is to detect gravitational wave ringdowns. However if a gravitation wave is not detected, as is the situation here, we can place an upper limit on the rate of black hole ringdowns in a volume of space. In doing this we only consider times when it was possible to detect a signal in all three detectors. As we have seen in previous chapters, the level of false alarms for two-detector coincidence was very high, making detection confidence in those times very low. As we will demonstrate at the end of this chapter, confining ourselves to triple time does not make a significant difference to the upper limit we calculate. 

Calculating an upper limit is achieved by taking the astrophysical population of black holes and our ability to detect ringdowns into account. For now we express this as the cumulative luminosity $C_L$ and explain this quantity in more detail in the next section. We use the detection statistic of the loudest triple coincident event $\rho_{max}$ detected as a threshold above which we evaluate $C_L$ and calculate an upper limit. In this search there were no triple coincident events in the zero-lag data and so we set the $\rho_{max}$ to the search threshold.

\subsection{Bayesian and Frequentist Approaches to the Upper Limit}

Two schools of thought exist on how an upper limit should be calculated; one uses Bayesian probabilities while the other employs frequentist probabilities. The primary difference between the two methods is that the Bayesian method uses prior knowledge about the rate in calculating probabilities whereas the frequentist method does not. A nice derivation and discussion of Bayesian and frequentist upper limits in gravitational wave searches can be found in \cite{Brad04}, and \cite{Fair07} is a detailed reference on Bayesian upper limits for LIGO searches. Here we summarize the main points as they apply to the ringdown search.

\subsubsection{Frequentist}
The probability that there were no gravitational wave events with SNR greater
than $\rho$ is given by
\begin{equation}
P(\rho|R,T)=e^{-RTC_L(\rho)}
\label{eqn:PrhoRT}
\end{equation}
where $R$ is the rate of ringdown events per unit cumulative luminoisty and $T$ is the observation time. If there was a background $B$ present then the probability that no background events were present with SNR greater than $\rho$, $P_B(\rho)$, may be taken into account giving an overall probability of
\begin{equation}
P(\rho|R,T,B)=P_Be^{-RTC_L(\rho)}. \label{eqn:Pofrho}
\end{equation}
We can choose a confidence level $\alpha$ at which we wish to evaluate the rate of events above an SNR of $\rho_{max}$ and solve the equation $1-\alpha=P(\rho_{max}|RT,B)$ for $R$. This gives
\begin{equation}
R_\alpha=-\frac{ \ln(1-\alpha)-\ln(P_B(\rho_{max})) }{ T C_L(\rho_{max})}.
\end{equation}
In the ringdown analysis there were no events found in triple coincidence, and so $C_L$ was evaluated at threshold. The level of foreground was perfectly consistent with background (we had less than one event in both cases) and so the probability that a foreground event was associated with noise is 1. Thus $P_B=1$, $\ln(P_B)=0$, and the 90\% confidence frequentist upper limit on the rate is 
\begin{equation}
R_{90\%}=\frac{2.303}{TC_L(\rho_{max})}.
\end{equation}

\subsubsection{Bayesian}
A Bayesian upper limit is calculated from the posterior probability $P(R<R_\alpha , T|\rho_{max})$
\begin{equation}
P(R<R_\alpha, T|\rho_{max}) =\mathcal{N}^{-1} \int_0^{R_\alpha}  p(R,T) \; p(\rho_{max}|R,T) \textrm{ d}R
\end{equation}
where $p(R,T)$ is the prior distribution on the event rate, $\mathcal{N}$ is a normalization constant and the likelihood is
\begin{equation}
p(\rho |R,T,B) =\frac{ \textrm{d}P(\rho|R,T,B) }{ \textrm{d} \rho}= -P_BRT \frac{\textrm{d}C_L(\rho) }{ \textrm{d}\rho } e^{-RT C_L(\rho)}
\end{equation}
where the second equality makes use of equation (\ref{eqn:Pofrho}).
The upper limit is determined by solving $P(R<R_\alpha, T|\rho_{max})=\alpha$ for $R_\alpha$. In this search we do not have any information on the rate of ringdowns and so we choose a uniform prior, $p(R,T)=1$, and obtain
\begin{equation}
1-\alpha=e^{-R_\alpha T C_L(\rho_{max})} \left[ 1+\xi\; R\; T\; C_L(\rho_{max})\right]
\end{equation}
where
\begin{equation}
\xi=\Bigg[1-\frac{\mathrm{d} \ln(P_B)}{\mathrm{d} \ln(C_L)}\Bigg|_{C_L(\rho_{max})} \Bigg] ^{-1}.
\end{equation}
In the ringdown search the loudest event was consistent with background, $\xi=0$, and thus the 90\% rate upper limit is given by
\begin{equation}
R_{90\%}=\frac{2.303}{TC_L(\rho_{max})}.
\end{equation}
Thus, in this particular case both the frequentist and Bayesian approaches give the same upper limit.

\subsection{Cumulative Luminosity}
In the previous section we introduced the quantity $C_L$, the cumulative blue light luminosity, and stated that this was a measure of our ability to detect gravitational wave ringdowns from a given population of sources. The sources of interest are black holes in the high end of the stellar mass ranges and in the lower end of the intermediate mass range $6\ M_\odot<M<600\ M_\odot$. As described in chapter \ref{ch:astro}, we know very little about the population of stellar mass black holes and even less about intermediate-mass black holes, --- indeed there has been no strong evidence to date for their existence. However we do know that the formation of stars in general scales with the blue-light luminosity emitted by galaxies, and as it is expected that the rate of binary coalescence follows the rate of star formation, we work under the assumption that the rate of binary coalescence also scales with blue light luminosity. In an effort to interpret the results of LIGO binary coalescence searches a catalog of nearby galaxies that could host compact binary systems was compiled \cite{Kopp08}. 
Beyond $\sim30$ Mpc the cumulative luminosity scales as the cube of distance; this is the regime of interest for the current search. The relationship between cumulative luminosity and distance is given by
\begin{equation}
C_L=\rho_L V_{eff}
\end{equation}
where $\rho_L=(1.98 \pm 0.16)\times 10^{-2} \textrm{L}_{10} \textrm{Mpc}^{-3}$. $C_L$ has units of L$_{10}$ which is defined as L$_{10}=10^{10}L_{B,\odot}$, and $L_{B,\odot}=2.16 \times 10^{33}$ erg s$^{-1}$ is the solar blue light luminosity. The effective volume $V_{eff}$ is the volume of space we are sensitive to, which we quantify in terms of a detection efficiency expressed as a function of distance $\varepsilon(r)$
\begin{equation}
V_{eff}(r) = 4 \pi \int \varepsilon(r) r^2 dr. \label{eqn:Veffr}
\end{equation}
Thus, the rate of ringdowns is given in units of yr$^{-1}$ L$_{10}^{-1}$ by
\begin{equation}
R=\frac{2.303}{T \rho_L V_{eff}}.
\end{equation}

\subsubsection{Efficiency}
We evaluate the efficiency of detecting gravitational wave ringdowns from a hypothetical population of sources by injecting simulated signals into the data stream and searching for these signals, implementing the same pipeline used to detect a real ringdown in the noise. This population was discussed in section \ref{sec:injections}, and our ability to detect them was discussed in chapter \ref{ch:paramest}. The ratio of the number of injections found in triple coincidence compared to the number injected gives a measure of the efficiency for a given distance.

Evaluation of $V_{eff}$ is somewhat complicated by the fact that we evaluated
the efficiency as a function of logarithmic distance rather than linear
distance. Thus we need to make the substitution
\begin{equation}
u=\log_{10}(r)  = \frac{\ln(r)}{\ln(10)},
\end{equation}
which can be expressed as 
\begin{equation}
r=10^u.
\end{equation}
This gives
\begin{equation}
du=\frac{1}{\log_e(10)} \frac{1}{r}dr
\end{equation}
or in terms of $dr$
\begin{equation}
dr=\log_e(10) 10^u du.
\end{equation}
Equation (\ref{eqn:Veffr}) can be rewritten as
\begin{equation}
V_{eff}(u)=4 \pi \ln(10) \int \varepsilon(u) 10^{3u} du. 
\end{equation}
In practice $\varepsilon$ and $r$ are discrete quantities and so $V_{eff}$ is expressed as
\begin{equation}
V_{eff}=4 \pi \ln(10) \sum_i \varepsilon(u_i) r_i^3 \Delta u.
\end{equation}

\subsection{Calculating the Upper Limit}
As we saw from figure \ref{fig:mf_allvH} the efficiency varies dramatically with frequency, and therefore, in the calculation of the upper limit we divide the frequency space into separate bands, based roughly on the different levels of sensitivity: 45--100 Hz, 100--200 Hz, 200--500 Hz 500--1000 Hz, 1--2 kHz. For each band we calculate the efficiency as a function of effective distance shown in figures \ref{fig:effvd45100} to \ref{fig:effvd10002500}, the effective volume $V_{eff}$ and corresponding radius $r_{eff}=\left(V_{eff}/\frac{4\pi}{3} \right)^{1/3}$, a measure of the sensitivity $V_{eff}T$, and the 90\% rate upper limit $R_{90}$. These quantities are displayed in table \ref{tab:tinttul}. As mentioned above, the assumption of uniform density is only valid at large distances; in the frequency bands with lower sensitivity we are restricted to short distances, and so the calculation becomes invalid and thus we do not quote an upper limit for these. This is denoted by ``N/A'' in the table.
\afterpage{\clearpage}
\begin{figure}[ht] 
\centering
\begin{center}
\includegraphics[scale=0.6]{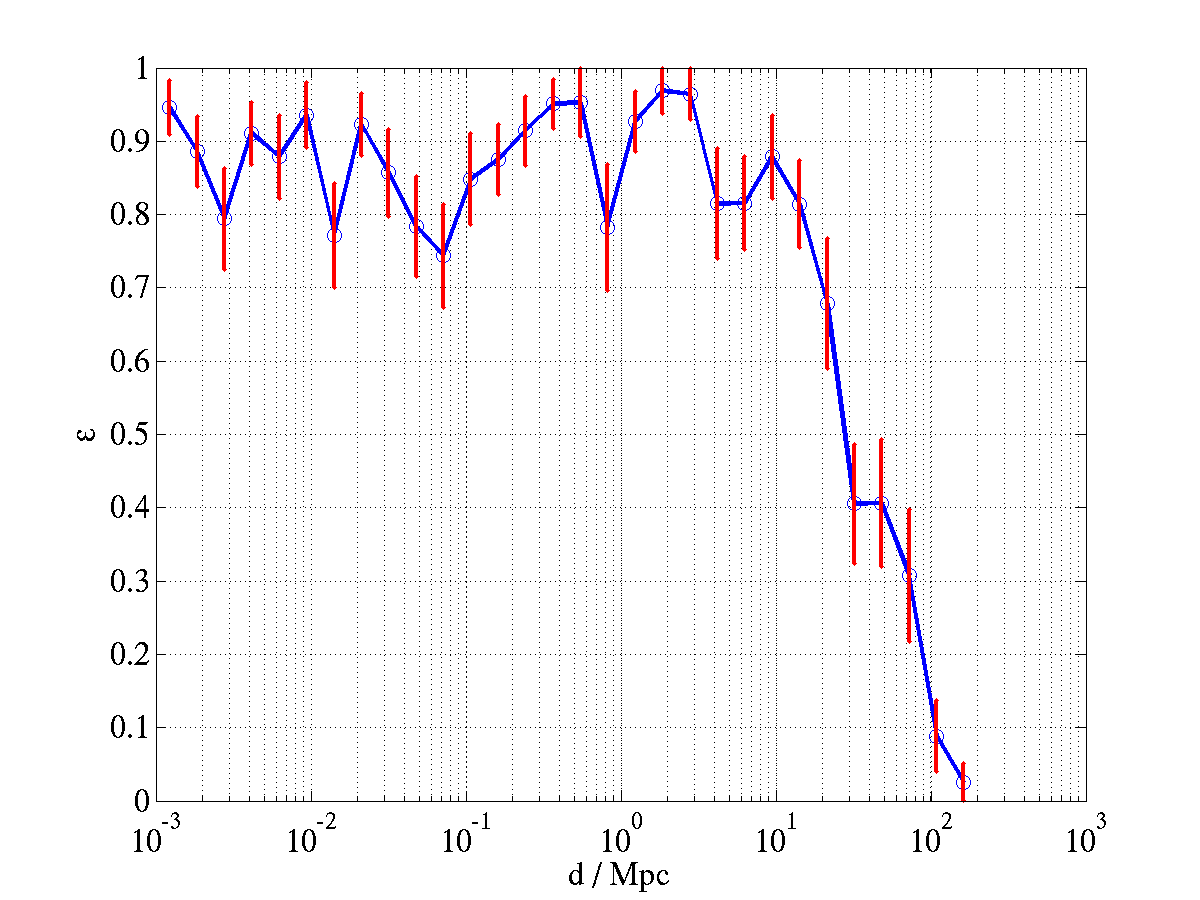}
\caption{The efficiency of detecting triples versus physical distance for injections made in the 45--100 Hz band.}
\label{fig:effvd45100}
\end{center}
\end{figure}

\afterpage{\clearpage}

\begin{figure}[ht]
\centering
\begin{center}
\includegraphics[scale=0.6]{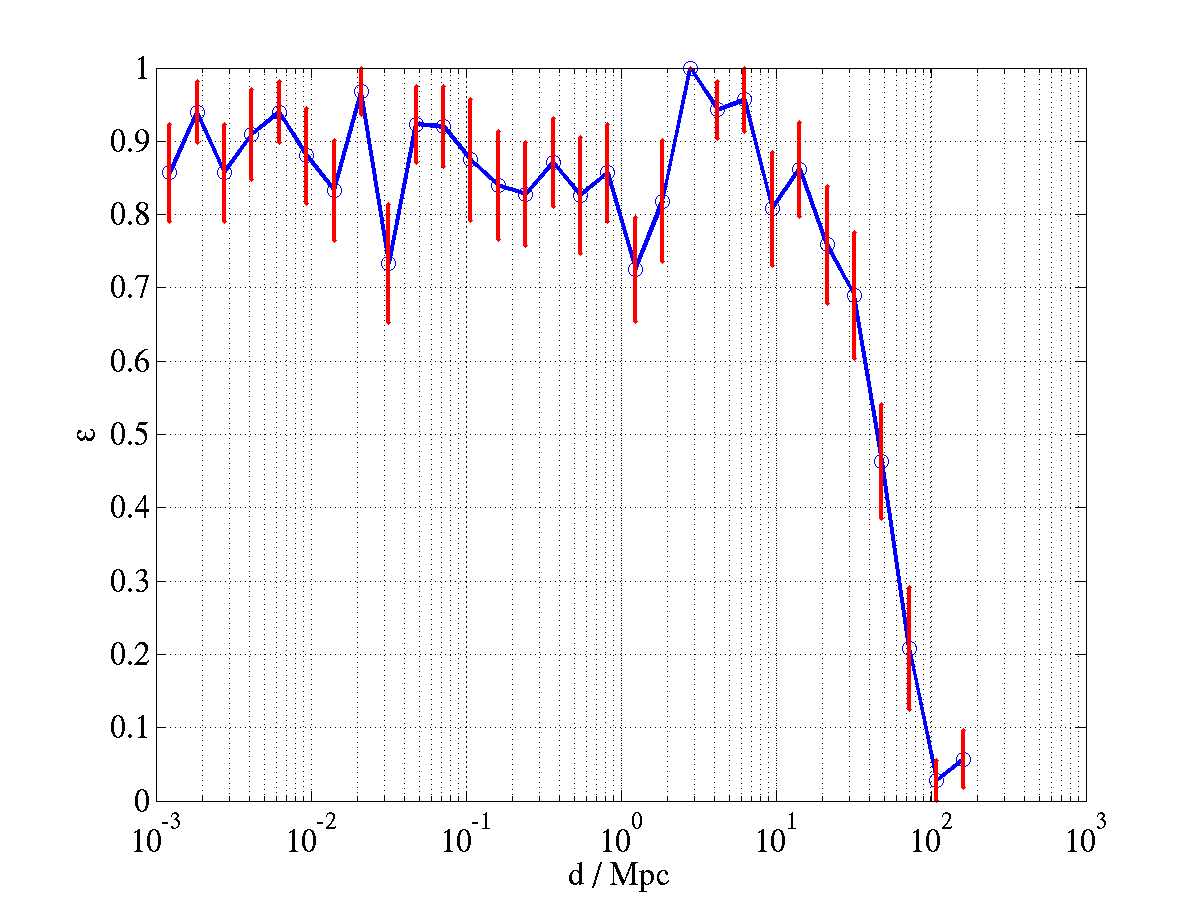}
\caption{The efficiency of detecting triples versus physical distance for injections made
in the 100--200 Hz band.}
\label{fig:effvd100200}
\end{center}
\end{figure}

\begin{figure}[ht]
\centering
\begin{center}
\includegraphics[scale=0.6]{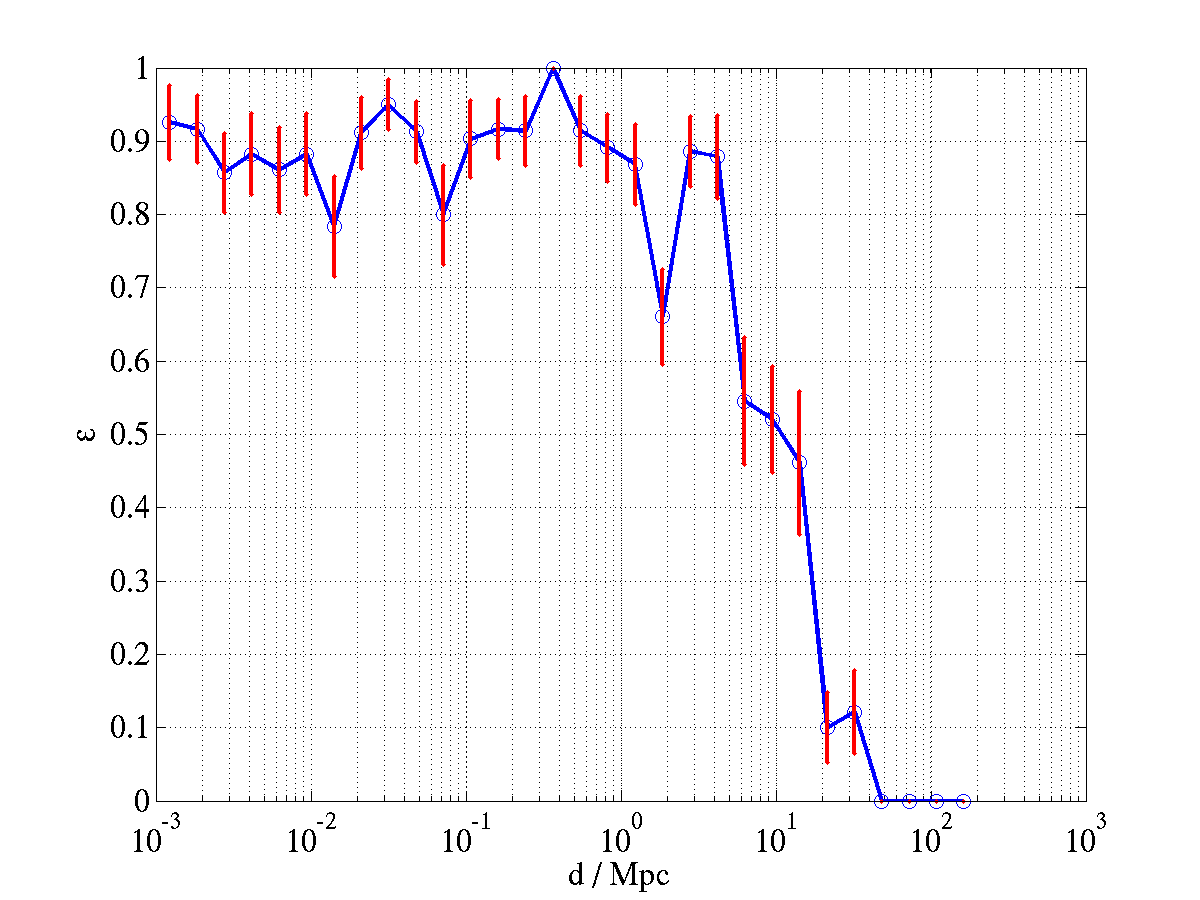}
\caption{The efficiency of detecting triples versus physical distance for injections made
in the 200--500 Hz band.}
\label{fig:effvd200500}
\end{center}
\end{figure}

\begin{figure}[ht]
\centering
\begin{center}
\includegraphics[scale=0.6]{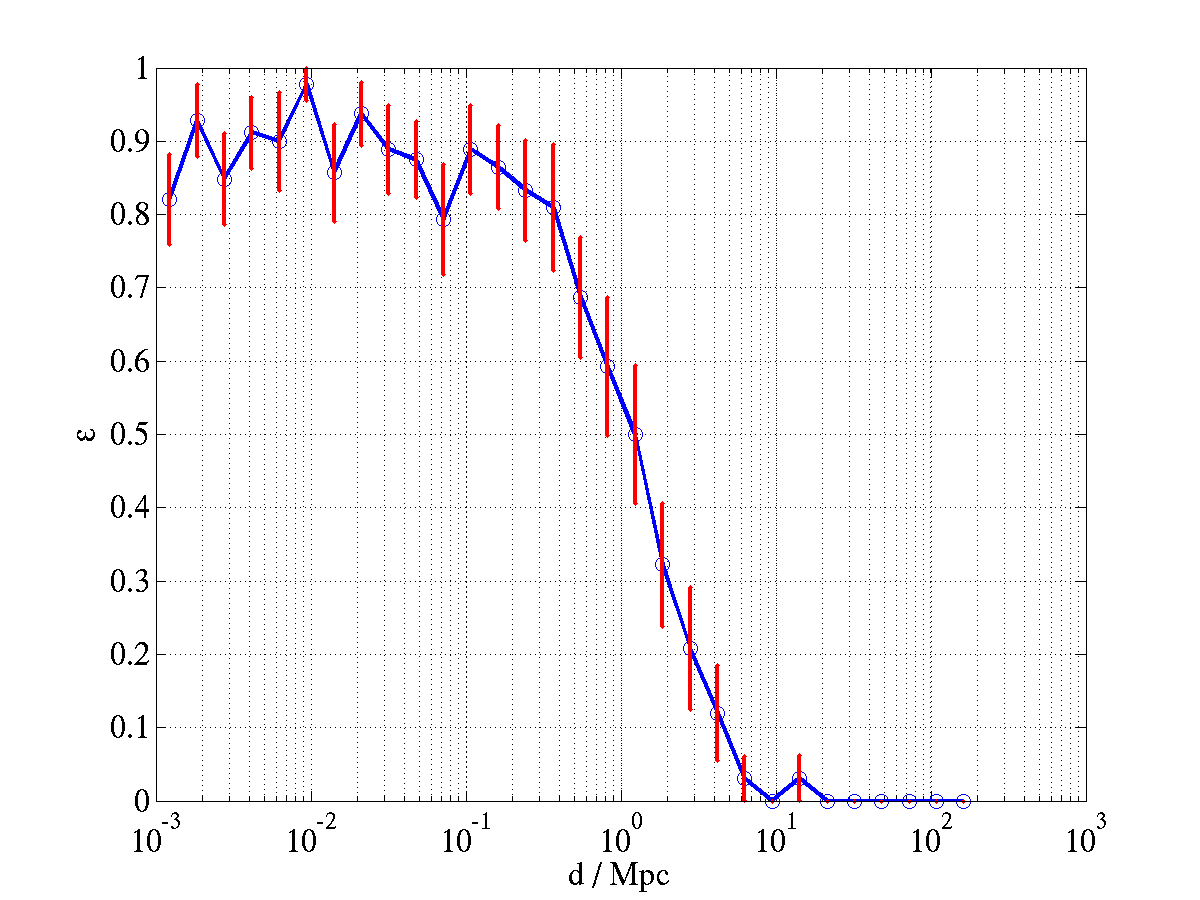}
\caption{The efficiency of detecting triples versus physical distance for injections made
in the 500--1000 Hz band.}
\label{fig:effvd5001000}
\end{center}
\end{figure}

\begin{figure}[ht]
\centering
\begin{center}
\includegraphics[scale=0.6]{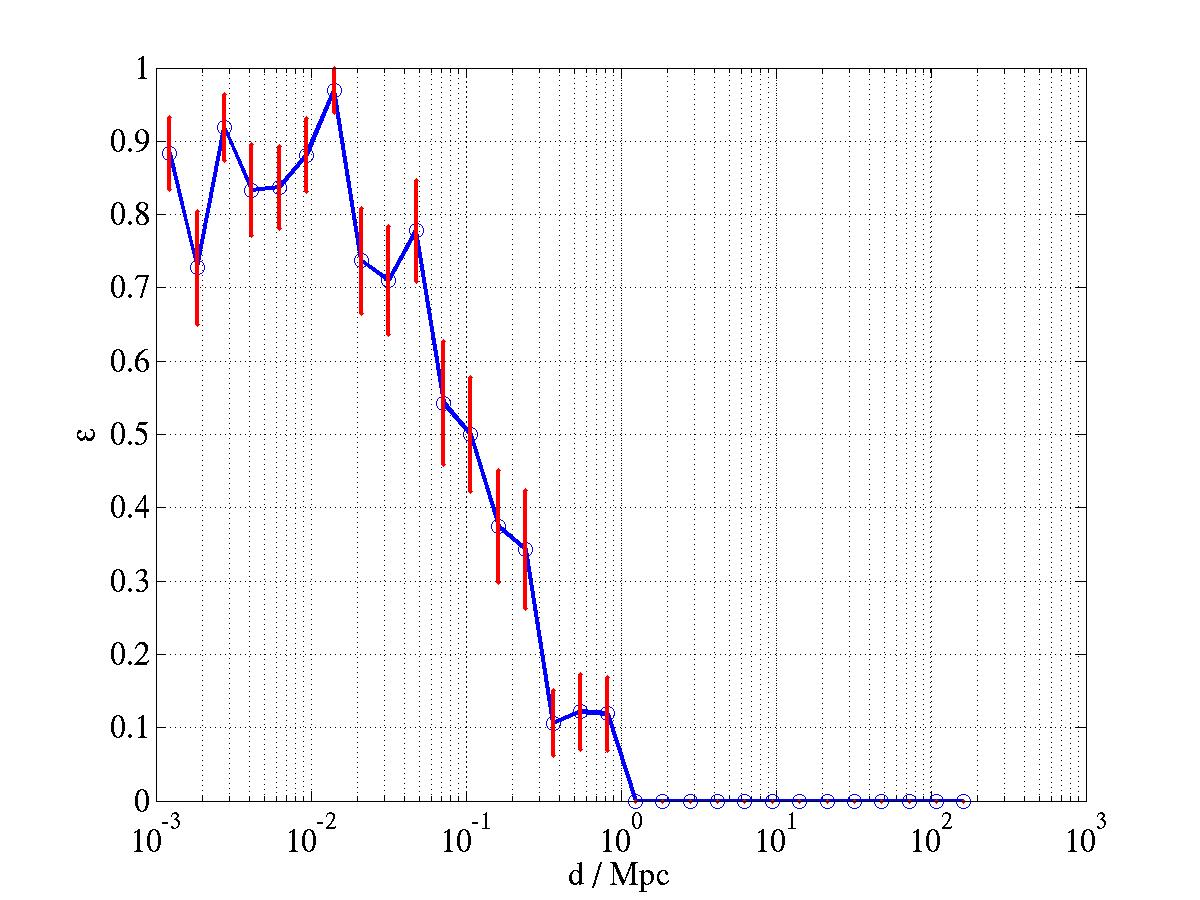}
\caption{The efficiency of detecting triples versus physical distance for injections made
in the 1000--2500 Hz band.}
\label{fig:effvd10002500}
\end{center}
\end{figure}

\begin{center}
\begin{small}
\begin{table}[htdp]
\centering
\caption{Upper limit for triples in triple time}
\begin{tabular}{c|lllll}
\hline \hline
f-band (Hz)& [45,100] & [100,200] & [200,500] & [500,1000] & [1000,2500]\\
\hline
M-band ($M_\odot$), $a=0$ & [260,120] & [120,60] & [60,24] & [24,12] & [12,5] \\
M-band ($M_\odot$), $a=0.994$ & [600,270] & [270,140] & [140,54] & [54,27] & [27,11] \\
\hline

$V_{eff}$ (Mpc$^3$) & 2.1$\times$10$^6$  & 2.3$\times$10$^6$ & 3.6$\times$10$^4$ 
                                         & 5.8$\times$10$^2$ & 5.1$\times$10$^{-1}$ \\
$r_{eff}$ (Mpc)     & 79                 & 82                & 20 
                                         & 5.2               & 0.49\\
$T$ (yr)            & 0.0375             & 0.0375            & 0.0375 
                                         & 0.0375            & 0.0375 \\
$V_{eff}T$ (Mpc$^3$ yr)  & 7.9$\times$10$^4$ & 8.6$\times$10$^4$  & 1.3$\times$10$^3$                
                                             & 2.2$\times$10$^1$  & 2$\times$10$^{-2}$\\
$R$ (yr$^{-1}$ L$_{10}^{-1}$) & 1.5$\times$10$^{-3}$ & 1.4$\times$10$^{-3}$ & N/A & N/A & N/A \\

\hline \hline
\end{tabular}
\label{tab:tinttul}
\end{table}
\end{small}
\end{center}

\subsection{Error Analysis}
Due to our lack of knowledge about the population of black holes we assign no systematic error to the astrophysical source population. Similarly, the waveform uncertainty is unquantifiable; the waveform we use is taken to be the definition of a black hole ringdown. We limit ourselves to evaluating systematic errors associated with the experimental apparatus and analysis method. The only systematic error associated with the former is calibration of the data. The only systematic error associated with the latter is with the limited number of Monte Carlo simulations to evaluate the efficiency. We consider only the 100--200 Hz band. 

As described in section (\ref{sec:calibration}) the response function is calculated from a reference loop gain function, a reference sensing function, and calibration coefficients recorded every minute. Errors in the calibration can cause the SNR of a signal to be incorrectly quantified, thus introducing inaccuracies in the distance. As the efficiency is a function of distance, care has to be taken to adjust the efficiency curve appropriately. The fractional uncertainty in amplitude $\delta$ was found from calibration studies \cite{Diet05} to be 5\%. Returning to equation (\ref{eqn:Veffr}), this means that we are actually evaluating $\varepsilon(r[1+\delta])$. Making the substitution of $u$ for $r[1+\delta]$,
\begin{equation}
\ V_{eff}=4\pi \int \frac{1}{(1+\delta)^3}  \varepsilon(u) u^2 du,
\end{equation}
shows we could be over- or under-estimating the effective volume by 15.67\%.
Thus the error in the volume in the 100--200 Hz band due to the calibration is $\delta V_{cal}=3.12\times 10^5$ Mpc$^{3}$.

The second source of error is due to the limited number of injections in our Monte Carlo (MC) simulations to evaluate the efficiency. Assuming binomial errors, the variance of the efficiency $\sigma^2_{\textrm{MC}}$ is
\begin{equation}
\sigma_{\textrm{MC}}^2=\frac{ \varepsilon (1-\varepsilon) }{N}
\end{equation}
where $N$ is the number of injections made. Thus the error in the effective volume, evaluated by multiplying $\sigma_{\textrm{MC}}^2$ by the square of the volume of each bin,
\begin{equation}
\sigma^2_{V_{\textrm{MC}}}=\sum_i \sigma^2_{\textrm{MC}_i} (dV_i)^2,
\end{equation}
was found to be $\sigma_{V_{MC}}=9.1\times10^5$ Mpc$^3$. Summing these errors in quadrature gives a total error of $\sigma_V=9.6\times 10^5$ Mpc$^3$. Multiplying by $1.64$ gives the 90\% confidence interval of $\delta V=1.57\times 10^6$ Mpc$^3$. To be conservative we apply a downward excursion to the effective volume, giving an upper limit of $R_{90\%}=4.3 \times 10^{-3}$ yr$^{-1}\textrm{L}_{10}^{-1}$.

\subsection{Including Doubles in the Upper Limit Calculation}
Before opening the box on the analysis the decision was made to examine all double and triple coincident triggers for a detection but consider only triples in the calculation of the upper limit. We knew that in the absence of signal-based vetoes the level of background was very high, and thus we would not gain very much by including them. As an exercise we quantify this. The efficiency of detecting injections in double coincidence in triple time was evaluated at the SNR of the loudest foreground event for H1L1, H1H2, and H2L1 pairs. The calculated rate, along with intermediate results of this calculation, are displayed in table \ref{tab:dinttul}. We did not perform injections in double time, however we do not expect the efficiency of doubles in double time to be significantly different from the efficiency of doubles in double time. We calculate the efficiency of each of the three pairs at the loudest event found for each in double time and using the double-time durations, calculated the upper limit. This is shown in table \ref{tab:dindtul}. Comparing with the 100--200 Hz band in table \ref{tab:tinttul}, the results show that doubles in triple time contribute an additional 7\% to the sensitivity, while doubles from double time contribute just 1\%.

\begin{center}
\begin{small}
\begin{table}[htdp]
\centering
\caption{Upper limit for doubles in triple time}
\label{tab:dinttul}
\begin{tabular}{c|lllll}
\hline \hline
f-band (Hz)& [45,100] & [100,200] & [200,500] & [500,1000] & [1000,2000]\\
\hline
M-band ($M_\odot$), $a=0$ & [260,120] & [120,60] & [60,24] & [24,12] & [12,5] \\
M-band ($M_\odot$), $a=0.994$ & [600,270] & [270,140] & [140,54] & [54,27] & [27,11] \\
\hline

$V_{eff}$ (Mpc$^3$)          & 4.1$\times$10$^5$ & 1.6$\times$10$^5$ & 2.5$\times$10$^1$ 
                                       & 3.2$\times$10$^2$ & 2.6$\times$10$^{-3}$ \\
$r_{eff}$ (Mpc)                   & 46                         & 34                         &1.8 
                                        & 0.91                       & 0.085\\
$T$ (yr)                                 & 0.0375                   & 0.0375                & 0.0375 
                                        & 0.0375                  & 0.0375 \\
$V_{eff}T$ (Mpc$^3$ yr)      & 1.5$\times$10$^4$ &5.9$\times$10$^3$ &9.4 $\times$10$^{-1}$                   
                                         &1.2$\times$10$^{-1}$  &9.6$\times$10$^{-5}$\\
$R$ (yr$^{-1}$ L$_{10}^{-1}$) &7.6$\times$10$^{-3}$ &2.0$\times$10$^{-2}$ &1.2$\times$10$^{2}$
                                          &9.8$\times$10$^3$    &1.2$\times$10$^{6}$ \\
\hline \hline
\end{tabular}
\end{table}
\end{small}
\end{center}

\begin{center}
\begin{small}
\begin{table}[htdp]
\centering
\caption{Upper limit for doubles in double time}
\label{tab:dindtul}
\begin{tabular}{c|lllll}
\hline \hline
f-band (Hz)& [45,100] & [100,200] & [200,500] & [500,1000] & [1000,2000]\\
\hline
M-band ($M_\odot$), $a=0$ & [260,120] & [120,60] & [60,24] & [24,12] & [12,5] \\
M-band ($M_\odot$), $a=0.994$ & [600,270] & [270,140] & [140,54] & [54,27] & [27,11] \\
\hline

$(V_{eff} T)_{\textrm{H1H2}}$ (Mpc$^3$ yr)      & 1.6$\times$10$^{-1}$ &2.2
                                         &2.8$\times$10$^{-2}$   &3.3$\times$10$^{-3}$                                
                                         &9.3$\times$10$^{-6}$\\
$(V_{eff} T)_{\textrm{H1L1}}$ (Mpc$^3$ yr)      & 2.2$\times$10$^3$ &8.3$\times$10$^2$ 
                                         &9.4$\times$10$^{-3}$   &6.4$\times$10$^{-3}$                                
                                         &1.2$\times$10$^{-6}$\\
$(V_{eff} T)_{\textrm{H2L1}}$ (Mpc$^3$ yr)      & 9.0                &6.8 
                                         &9.4$\times$10$^{-2}$   &7.6$\times$10$^{-3}$                                
                                         &7.4$\times$10$^{-6}$\\
$\sum_i (V_{eff} T)_i$ (Mpc$^3$ yr)      & 2.2$\times$10$^3$ &8.4$\times$10$^2$ 
                                         &1.3$\times$10$^{-1}$   &1.7$\times$10$^{-2}$                                
                                         &1.8$\times$10$^{-5}$\\
$R$ (yr$^{-1}$ L$_{10}^{-1}$) &5.4$\times$10$^{-2}$ &1.4$\times$10$^{-1}$ &8.9$\times$10$^{2}$
                                          &6.8$\times$10$^3$    &6.5$\times$10$^{6}$ \\
\hline \hline
\end{tabular}
\end{table}
\end{small}
\end{center}

\chapter{Search with Simulated Inspiral-Merger-Ringdown Signals}
\label{ch:imr}

\section{Introduction}
As described in previous chapters, the simulated signals on which the pipeline was tuned and the upper limit calculated consisted of isolated ringdown waveforms. However, an important source of gravitational wave ringdowns is expected to be binary black hole coalescences, in which case the ringdown will be preceded by an inspiral and merger. The inspiral and ringdown phases are well modeled, but analytic expressions of the merger waveform do not exist. However, recent breakthroughs in numerical relativity have given us a clearer picture of what to expect from the merger phase, and several groups are currently working on methods to utilize these results to provide analytic waveforms for use in coherent matched-filter searches.

To complete our investigation into the presence of ringdowns in S4 data it is necessary to check if the presence of an inspiral and merger would hamper or enhance our ability to detect and estimate the parameters of the ringdown using a ringdown-matched filter. To that end, we create inspiral-merger-ringdown (IMR) waveforms; as described in detail below, these waveforms consist of an inspiral waveform which is stitched in a continuous manner to a ringdown with the intervening signal representing the merger. The calculation of the ringdown parameters from the inspiral parameters is guided by the recent results of numerical relativity. Our simulations cover a much larger space than that for which numerical waveforms are currently available, and thus for now should be viewed as an approximation to the true waveform. In this analysis we inject both IMR and ringdown-only waveforms into the data and compare the outputs of both the single and coincident detector analyses. 

\section{Numerical Relativity}
Numerical relativity is a branch of computational physics concerned with solving Einstein's equation numerically. 2005 saw a major breakthrough in the field when for the first time an equal mass non-spinning binary black hole system was evolved from the last few orbits of inspiral through merger to ringdown, and the gravitational wave was extracted \cite{Pret05}. These and subsequent results \cite{Camp06}, \cite{Bake06}, indicate that for such a system the final spin of the black hole is expected to be close to 0.7. Studies of unequal mass non-spinning black holes suggest that the amplitude of the $l=m=2$ mode of the ringdown waveform decreases as the mass ratio of the binary components increases whereas the amplitude of the $l=m=3$ mode increases \cite{Herr07}. If the black holes are initially spinning, simulations show that the spin of the final black hole and the amount of radiation emitted also depend on the magnitude and inclination of the spins with respect to the orbital angular momentum \cite{Camp06b,Camp07,Camp07b,Bake07}.
 
Evolving a binary system from inspiral through to ringdown is computationally expensive; creating a bank of these waveforms for use in matched filter searches is simply not feasible. However efforts are currently underway within the LSC and elsewhere to use the results from numerical simulations of non-spinning black holes with mass ratios $1:1$ to $1:4$ in conjunction with the well-modeled inspiral waveform to create IMR waveforms. One method \cite{Buon07} extracts the mass and spin of the final black hole from the numerical waveforms, and from this calculates the fundamental quasi-normal mode and two overtones of the ringdown. To this an inspiral waveform (given by the effective one body model) is matched, giving a complete inspiral-merger-ringdown waveform. A second method ~\cite{Ajit07} constructs hybrid waveforms by matching a post-Newtonian inspiral to the merger and ringdown of numerical simulations, and from these proposes a family of phenomenological waveforms which closely match the hybrid waveform. Both these methods of creating accurate coalescence waveforms for use in LIGO analyses are at early stages of development but promise to give a larger range of sensitivity than templated searches for a single phase. This is illustrated in figure \ref{fig:AjithIMR} where, by using the initial LIGO sensitivity curve, a comparison is made between the horizon distance attainable by the inspiral and ringdown searches and the coherent IMR search described in \cite{Ajit07}. The figure shows that the coherent IMR search allows us to see to much larger distances. Model waveforms for spinning binaries and binaries with larger mass ratios are also under development.

\begin{figure}[h] \centering \begin{center}
\includegraphics[scale=0.6]{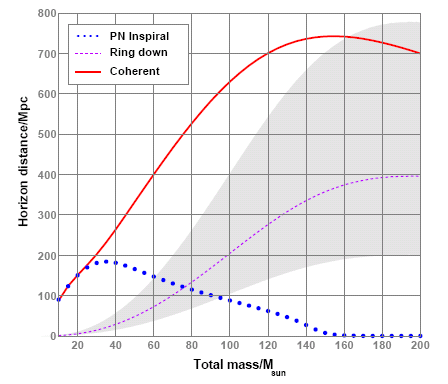}
\caption{A preliminary assessment of the performance of a phenomenological template bank (red line) for use in IMR searches compared to searches using only inspiral (blue dots) and only ringdown (red dashes) templates. The ringdown curve assumes $\epsilon=0.7\%$ and the shaded area represents $0.18\%\leq\epsilon\leq2.7\%$, and can be compared with figure \ref{fig:range} which displays the ringdown horizon distance for $\epsilon=1\%$. This figure is taken from \cite{Ajit07}.}
\label{fig:AjithIMR}
\end{center}
\end{figure}

\section{Creating the IMR Waveform}
At the present time numerical waveforms encompassing the range of ringdown injections made into S4 data are not available. We can, however, make an approximation using the well-modeled inspiral and ringdown waveforms currently used in those single-phase searches. Here we describe how the individual parts of the IMR waveform were created from a set of initial conditions and stitched together to form a coherent waveform for injection into the data.

As discussed in section \ref{sec:injections}, to inject a signal into the data (in software), the user supplies a list of input parameters that the appropriate waveform generation code uses to produce a waveform structure containing the plus and cross amplitudes, phase, and frequency for each time series data point in the waveform. For the IMR injections we start with the initial parameters of the inspiral phase. These include the component masses and spins, 
distance, and source position and orientation.  Using this information a spinning inspiral waveform\footnote{For more details about searches for gravitational waves from inspiraling spinning black holes see \cite{Jone08}.} is created up until the point where the binary separation is $6GM/c^2$. We denote the frequency at that point by $f_{6M}$. The next step is to estimate the parameters of the ringdown. The final spin of the black hole $\hat{a}$ can be estimated from the masses $m_1$ and $m_2$ and the spins $\hat{a}_1$ and $\hat{a}_2$ of the binary components using the results of numerical simulations as a guide \cite{Camp06b},
\begin{equation}
\hat{a}= \sqrt{ \frac{\hat{a}_1^2 m_1^2}{M_T^2} + \frac{\hat{a}_2^2 m_2^2}{M_T^2} + \frac{\eta}{0.25} 0.7 }
\end{equation}
where $M_T$ is the sum of the individual masses and the last term is the contribution of the orbital angular momentum with $\kappa$ representing the symmetric mass ratio $\eta=m_1m_2/\left(m_1+m_2\right)^2$. The estimation of final mass $M$ also uses results from numerical relativity \cite{Camp06b},
\begin{equation}
M=M_T \left[ 1-0.01\left(1+6\hat{a}^2\right) \right].
\end{equation}
Then, assuming that all the gravitational radiation is emitted in the $l=m=2$ mode, the ringdown waveform parameters $f_{r}$ and $Q$ may be calculated using equations (\ref{eqn:Echeverria_fofMa}) and (\ref{eqn:Echeverria_MoffQ}). 

Now that we know both $f_{6M}$ and $f_{r}$ the challenge is to match these in a smooth and continuous manner. This is implemented in two stages in the LAL \cite{LAL} code. First the evolution of the frequency from the inspiral stage is continued with $\dot{f}\sim f^{11/3}$ until 90$\%$ of the ringdown frequency $f_{0.9}$ is reached. From there the ringdown frequency is approached exponentially 
\begin{equation}
f(t)=f_{r} - K e^{-\lambda t}
\end{equation}
where $K=f_r-f_{0.9}$ and $\lambda= \dot{f}/A$.
In a similar manner the phase evolution from the inspiral stage is continued until $f$ reaches $f_{0.9}$ and then it is evolved as
\begin{equation}
\phi(t)=\phi_{0.9}+2\pi f t.
\end{equation}
The plus and cross amplitudes between the inspiral and ringdown are fit with a quadratic,
\begin{equation}
\mathcal{A} = \alpha_0 + \alpha_1 * t + \alpha_2 * t^2
\end{equation}
where $\alpha_0$ is the amplitude at the cut off point of the inspiral,
$\alpha_1=\dot{\alpha}_0$ (i.e.~, the rate of change of the amplitude
during the inspiral), and 
\begin{equation}
\alpha_2=\frac{ \left(\gamma -1\right) \left(\alpha_0+\tau \alpha_1 \right) }
{\left(1-\gamma\right)\tau^2+2\tau } -\tau,
\end{equation}
where $\gamma$ is a damping factor and $\tau$ is the length of time between
the end of the inspiral and the start of the ringdown. Finally the ringdown plus and cross amplitudes, frequency, and phase proceed as described in section \ref{sec:injections}. 

\afterpage{\clearpage}
\begin{figure}[h] 
\centering 
\begin{center}
\includegraphics[scale=0.55]{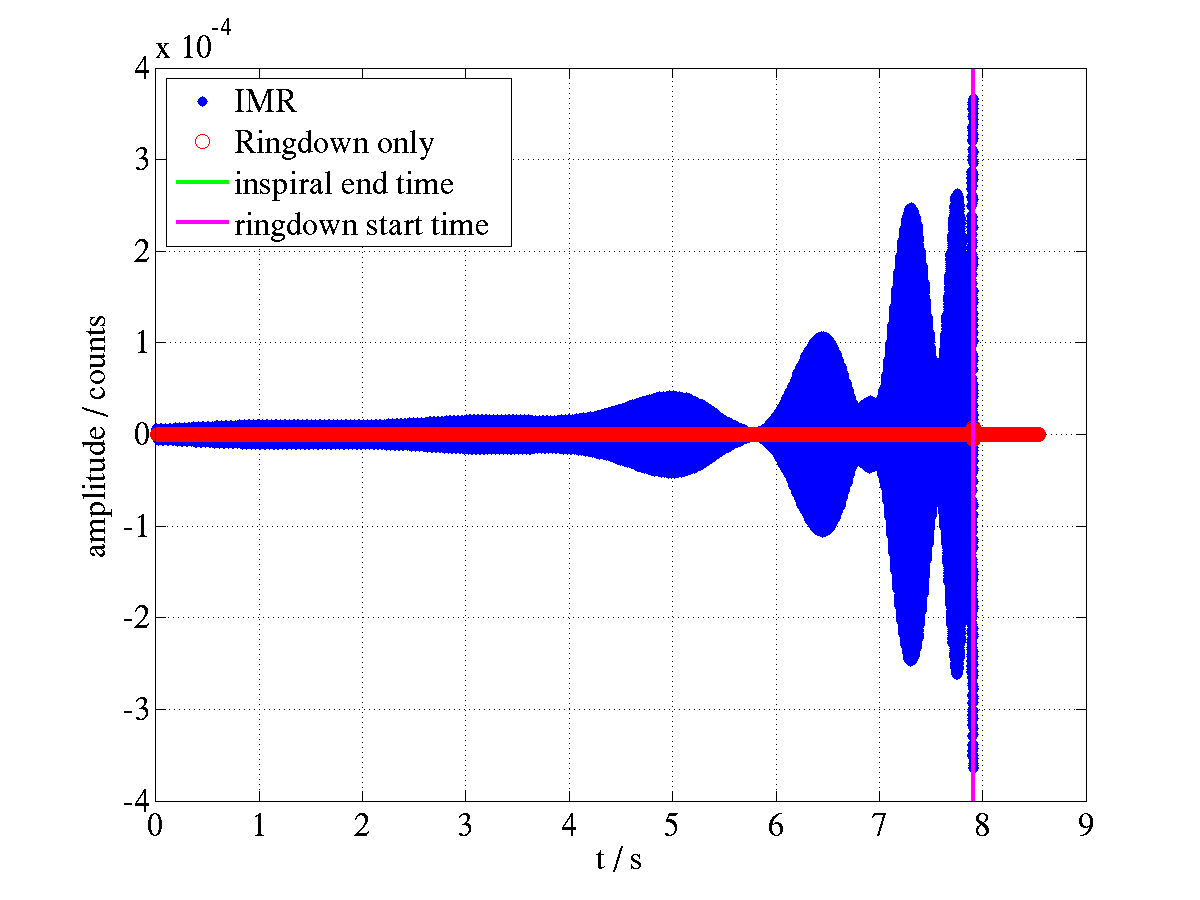}
\caption{The inspiral-merger-ringdown for a binary black hole system with component spins of 0.88 and 0.84 and masses of $8.9\ M_\odot$ and $6.3\ M_\odot$ (blue). Also shown is the ringdown-only waveform for the same system (red).} 
\label{fig:IMR_R}
\end{center}
\end{figure}

\begin{figure}[h] 
\centering
\begin{center}
\includegraphics[scale=0.55]{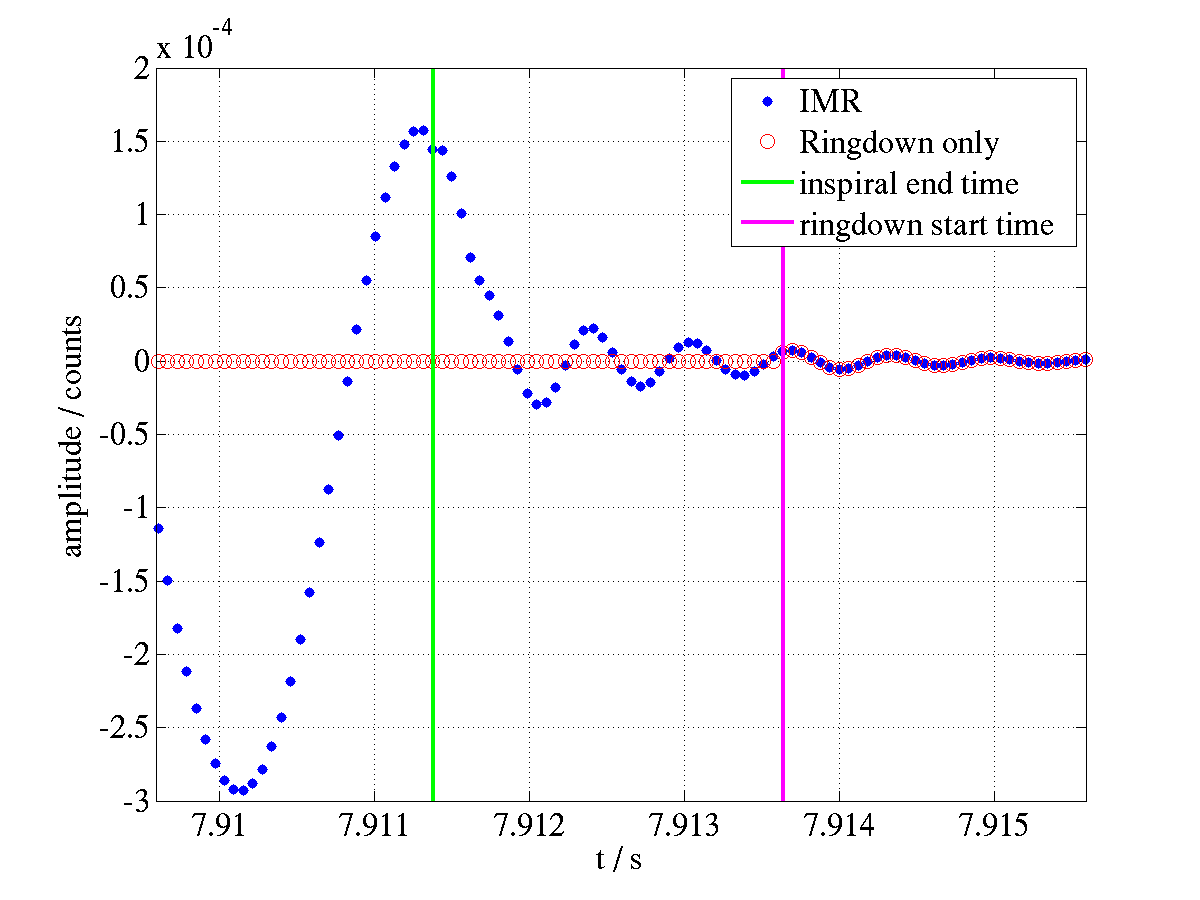}
\caption{The transition portion of the inspiral-merger-ringdown waveform (dots) and the ringdown-only waveform (circles) from figure \ref{fig:IMR_R}.}
\label{fig:IMR_Rz}
\end{center}
\end{figure}

Because we want to compare a simulated IMR signal with an isolated ringdown it is important that an identical ringdown is generated in each case. Thus, for a ringdown-only injection we create the full IMR waveform and set the inspiral and merger amplitudes to zero. To be consistent with the S4 ringdown analysis we define the ringdown to start at the point where the frequency becomes constant. Figure \ref{fig:IMR_R} shows one example of IMR and ringdown-only waveforms produced for a system with spins of 0.88 and 0.84 and masses of $8.9\ M_\odot$ and $6.3\ M_\odot$ respectively. The end time of the inspiral and the start time of the ringdown are marked, and it is only after the latter point that the amplitude of the ringdown-only injection becomes non-zero. This is best seen in figure \ref{fig:IMR_Rz} which zooms in on the transition between the inspiral and ringdown. After this point the amplitudes match exactly. Figure \ref{fig:fvtz} shows the frequency evolution of the same IMR waveform with the inset showing just the final 20 ms of the waveform. Here we see how the inspiral frequency increases rapidly until the constant ringdown frequency is reached. This marks the ringdown start time. In figure \ref{fig:MOvtz} we plot the same quantity in the dimensionless quantities. The final spin of the black hole is 0.9 and the ringdown $M\omega_R=0.68$. This is in agreement with figure \ref{fig:KerrREMOva} for the $l=m=2$ mode. We can also compare this to figure \ref{fig:Buonenergy}. 
\afterpage{\clearpage}
\begin{figure}[h] 
\centering
\begin{center}
\includegraphics[scale=0.55]{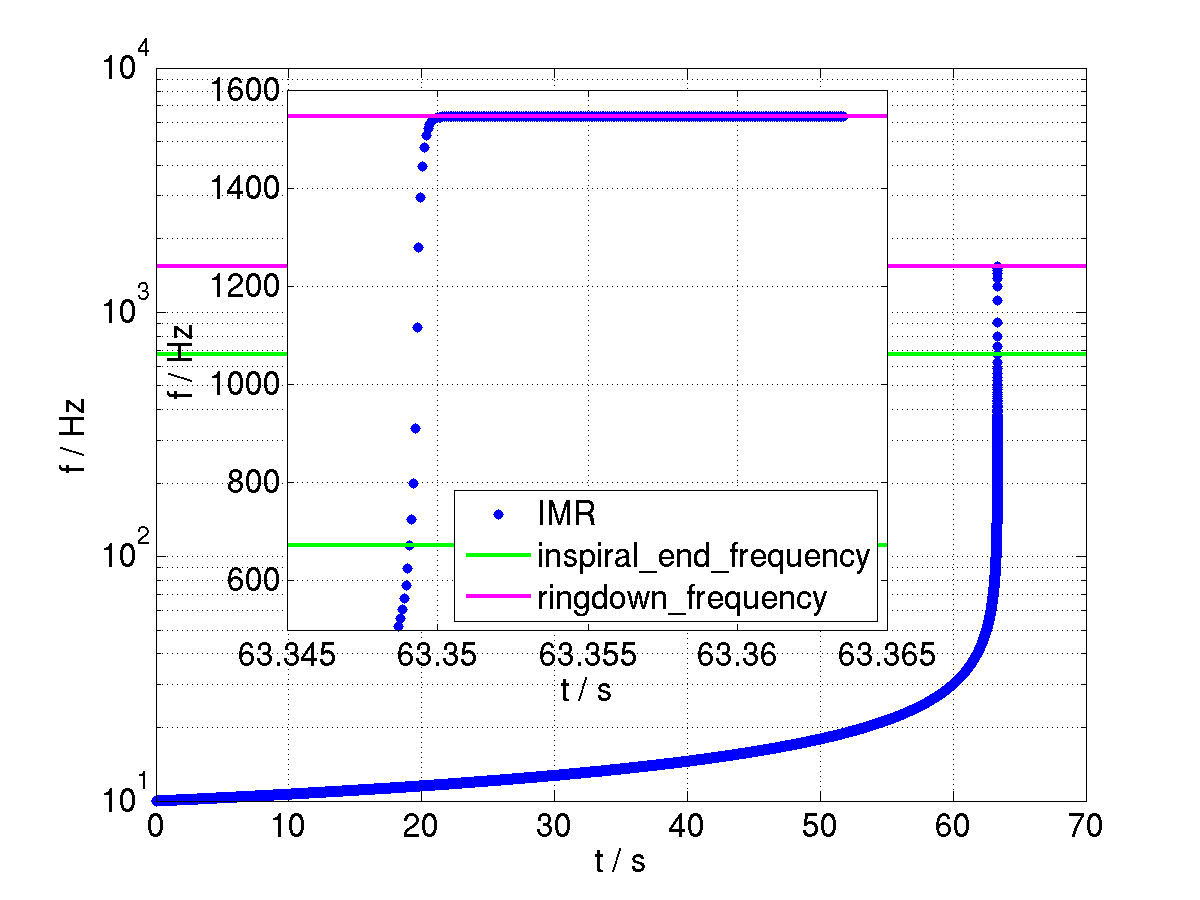}
\caption{The frequency time series of the coalescence. The inset zooms in on the inspiral-merger transition and ringdown phase.}
\label{fig:fvtz}
\end{center}
\end{figure}
\begin{figure}[h] 
\centering
\begin{center}
\includegraphics[scale=0.55]{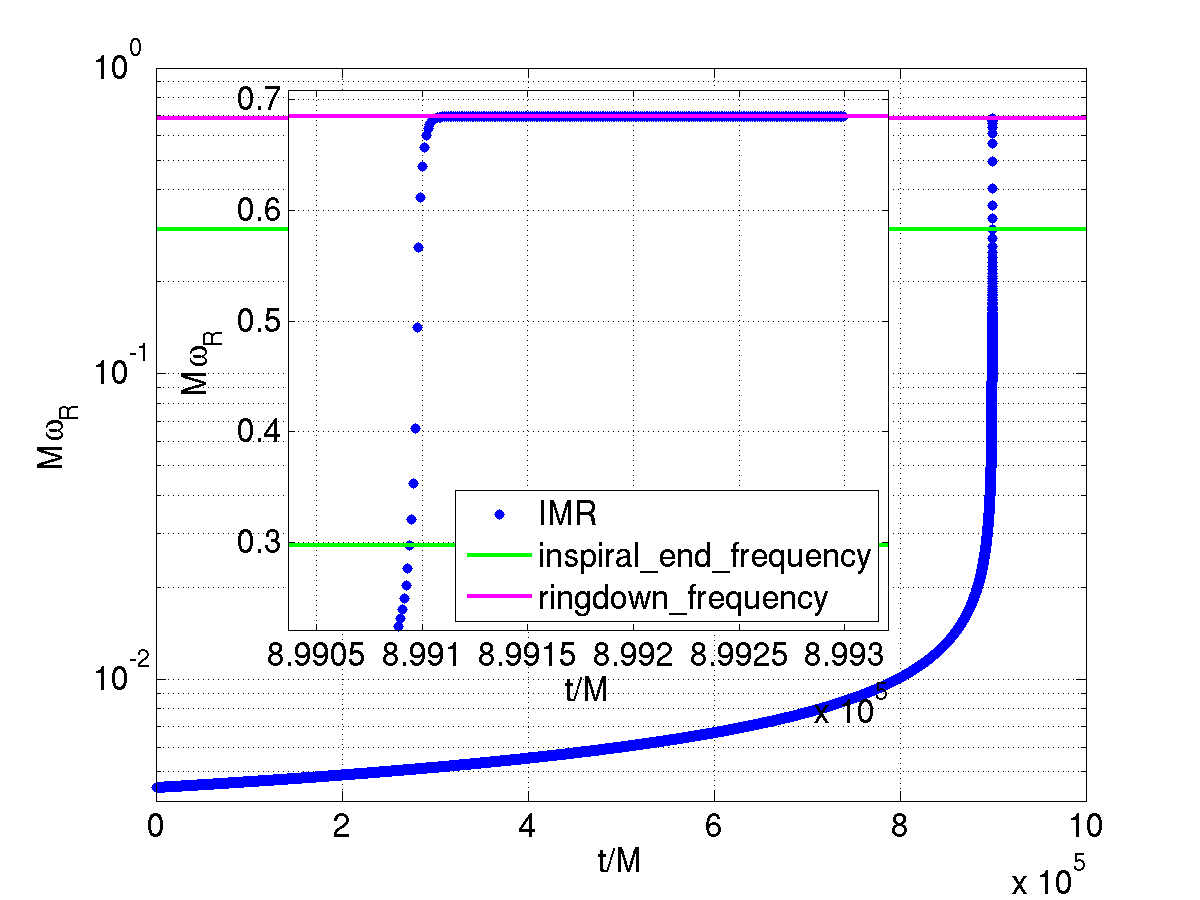}
\caption{$M\omega_R$ versus (unitless) time for the coalescence of a black hole binary. The final black hole has a spin of 0.9 and an $M\omega_R$ of 0.68, which agrees with figure \ref{fig:KerrREMOva} for the $l=m=2$ mode.}
\label{fig:MOvtz} 
\end{center} 
\end{figure}

If we want to evaluate how well we are recovering the injections it is necessary to calculate and record the ringdown parameters to compare with the output of the ringdown filter. Solving $a_+$ and $a_\times$ for $A$ and $\iota$ in equations (\ref{eqn:aplus}) and (\ref{eqn:across}) enables us to calculate the effective distance and the percentage of mass radiated as gravitational waves $\epsilon$. For every injection these parameters are written out to a  {\fontfamily{pcr}\selectfont sim\_ringdown table}.


\section{Single Detector Analysis}

The simulated signals were created uniformly in logarithmic component mass between $4\ M_\odot$ and $600\ M_\odot$, with a maximum total mass of $650\ M_\odot$. The distribution in distance was also logarithmic between 0.1 and 1 Mpc. Each component black hole had a spin $\hat{a}_1$, $\hat{a}_2$  whose magnitude was distributed uniformly in the range 0 and 1. This investigation was composed of two separate runs; in the first run we injected the full IMR waveform and in the second run only the ringdown was injected. The same initial conditions were used in both runs ensuring that an identical ringdown was injected each time.

\afterpage{\clearpage}

\begin{figure}[h] 
\centering
\begin{center}
\includegraphics[scale=0.55]{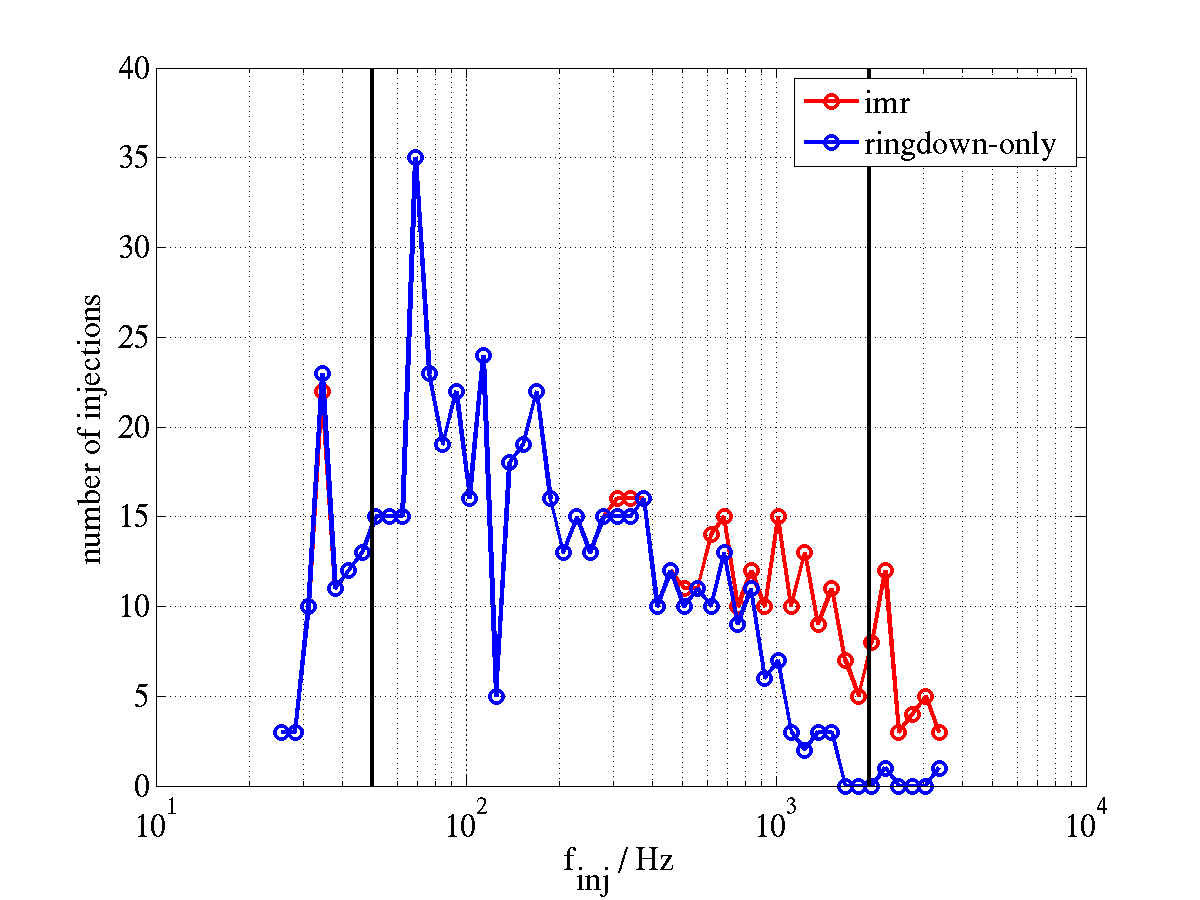} 
\caption{Number of IMR and ringdown-only injections recovered in H1 as a function of the frequency of the final ringdown. The vertical lines denote the template bank boundaries.}
\label{fig:histrimr}
\end{center}
\end{figure}
\begin{figure}[h]
\centering
\begin{center}
\includegraphics[scale=0.55]{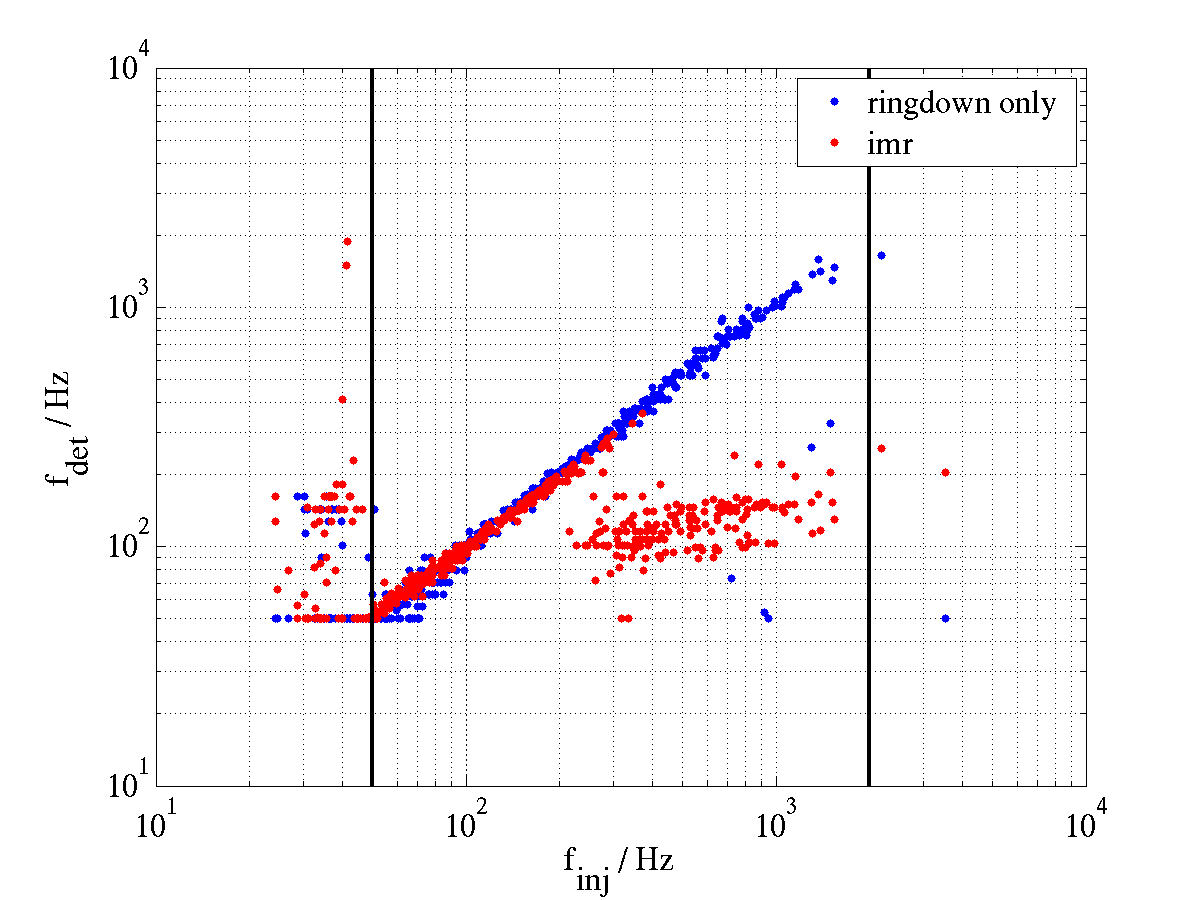}
\caption{Detected versus injected ringdown frequency for IMR and ringdown-only waveforms in H1.}
\label{fig:rimrfdetfinj}
\end{center}
\end{figure}

Comparing the trigger files output from the pipeline with the injection file allows us to count how many injections were found and how well the parameters were recovered. As shown in figure \ref{fig:histrimr}, we found that approximately 18\% more injections were found in the IMR run than in the ringdown-only run, with the excess appearing above an injected ringdown frequency $f_{inj}$ of $\sim 600$ Hz. Figure \ref{fig:rimrfdetfinj} shows that of the injections that were found in common by the two runs, the detected frequency $f_{det}$ of injections with $f_{inj}$ below about 200 Hz were fairly consistent between the two runs. Those injections with $f_{inj}>200$ Hz were not so consistent; the ringdown-only injections were found with templates close to the injected ringdown parameters, but the IMR injections were mostly found by templates in the 100--200 Hz band.

This can be explained as follows: for IMR injections with low ringdown frequency the inspiral part of the waveform is outside the LIGO band. As we increase the ringdown frequency, an increasing proportion of the inspiral and merger enters the band, matching an increasing number of templates. This is demonstrated in figures \ref{fig:ringfvt} and \ref{fig:imrfvt}, which show the templates that rang off during the 120 ms around a ringdown-only and an IMR injection, respectively, where the ringdown frequency was $f_{inj}\sim1500$ Hz.  In the ringdown-only case the only templates that ring are close in frequency to the frequency of the injection and do so right at the time of peak amplitude of the waveform, as indicated by the dashed lines. For the IMR case, however, most of the templates in the bank ring off. The templates ring off just as expected for the characteristic chirp frequency evolution of an inspiral; the inspiral enters the LIGO band at low frequency and its frequency increases until it reaches the ringdown. The template that rings up the loudest (and hence is the template associated with the injection) in this case is at $\sim 110$ Hz, as indicated in the plot by the red horizontal line. This is far from the ringdown frequency, denoted by the black horizontal line, but it is where the LIGO strain sensitivity is best (see figure \ref{fig:s4strain}).

\afterpage{\clearpage}

\begin{figure}[h]
\centering
\begin{center}
\includegraphics[scale=0.55]{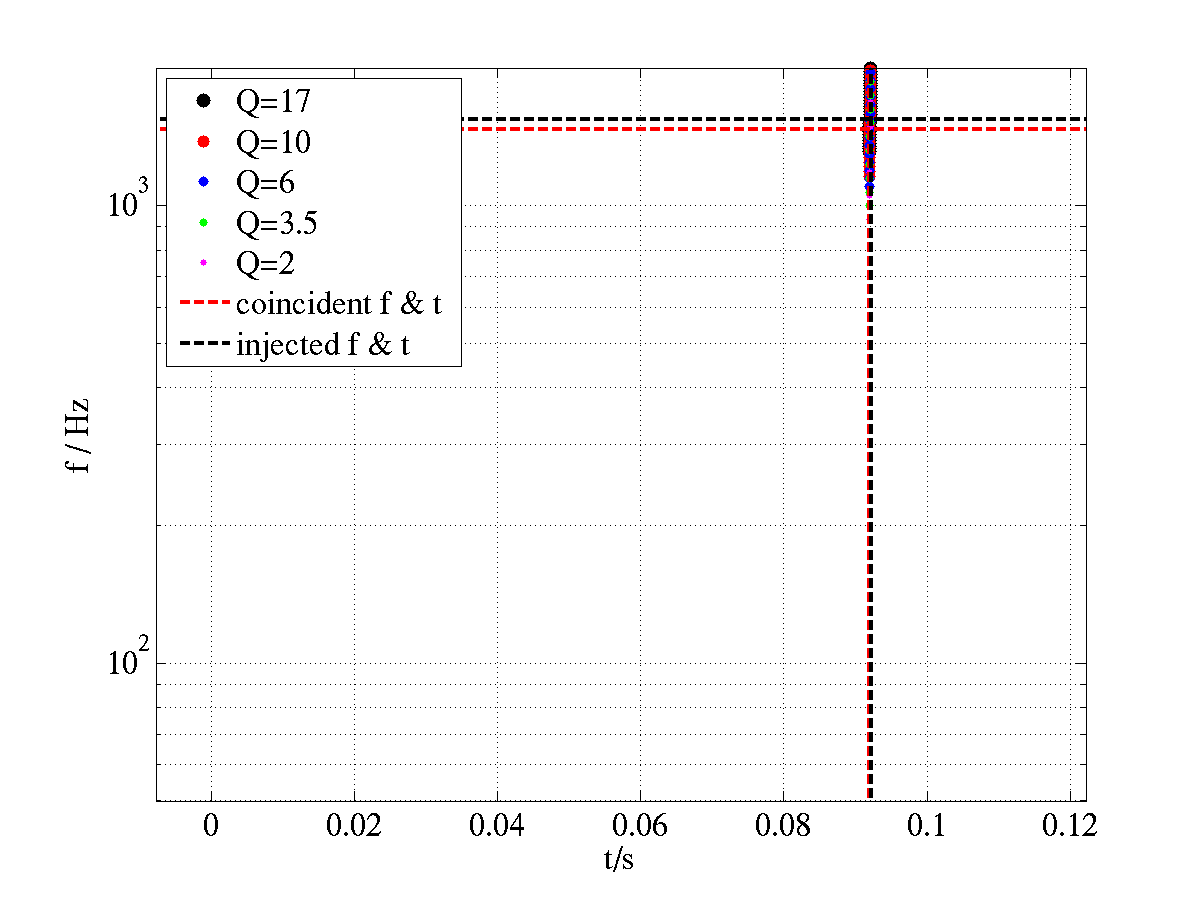} \\
\caption{Frequency versus time for the templates that rang up around the time of a
ringdown-only injection. The colour of the data points represents the quality
factor of the template. The black lines represent the frequency and time of the
injection, and the red lines represent the frequency and time of the template
with the largest signal-to-noise ratio.}
\label{fig:ringfvt}
\includegraphics[scale=0.55]{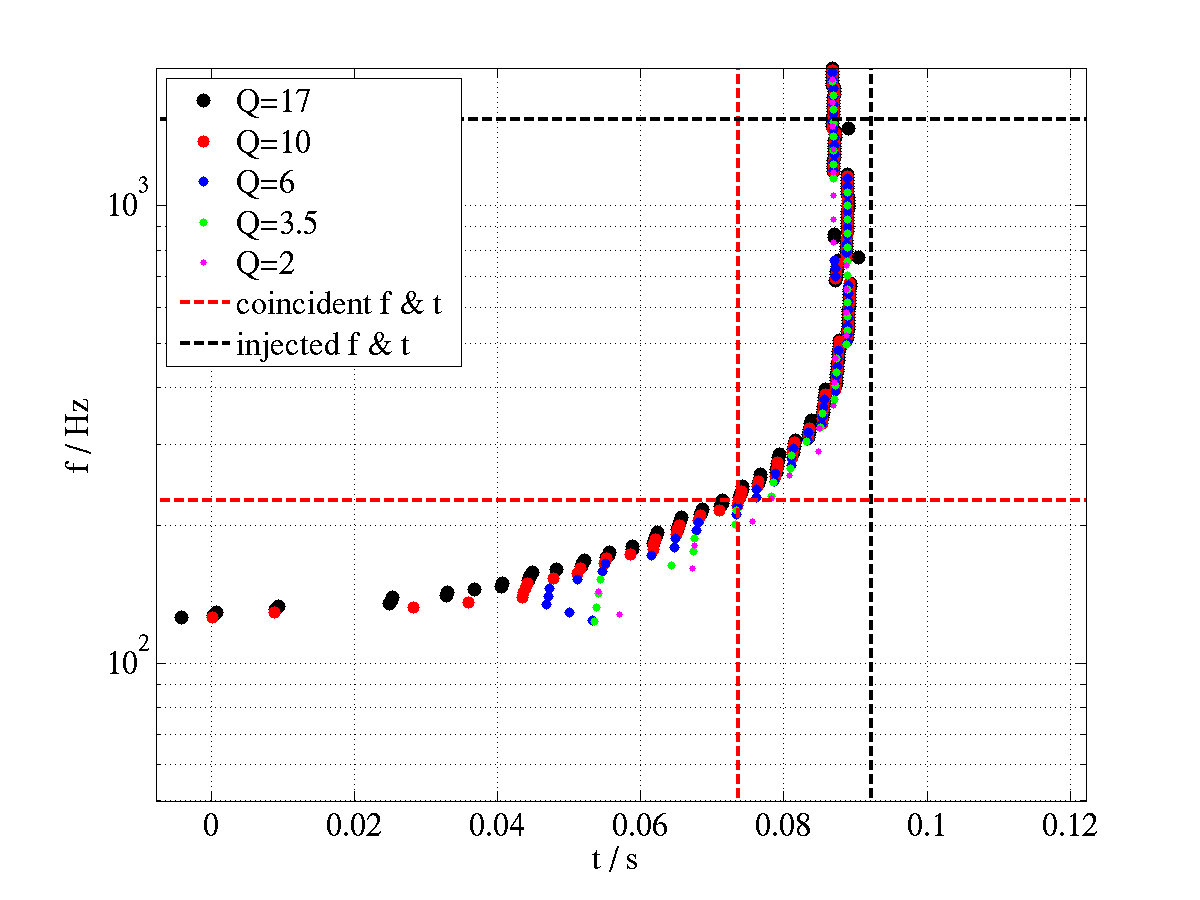} \\
\caption{Frequency versus time for the templates that rang up around the time of an IMR injection. The colour of the data points represents the quality factor of the template. The black lines represent the frequency and time of the injection, and the red lines represent the frequency and time of the template with the largest signal-to-noise ratio.}
\label{fig:imrfvt}
\end{center}
\end{figure}

In figure \ref{fig:imrwr} we plot the initial component masses of all the IMR injections with $f_{inj}>50$ Hz that were found at the correct ringdown frequency in red, and those that were found incorrectly in green. The plot shows that the majority of the injections that were found incorrectly by the ringdown search fall within the scope of the S4 BBH search, and thus that search should be able to find the signal and correctly identify the component masses. Most of those injections that were correctly identified lie outside that region, and so we can conclude that between the two searches the mass space is covered quite well.

Our results concur with a study by Baumgarte et al.~\cite{Baum} in which
two numerical waveforms --- one with ringdown frequency of $\sim 80$ Hz and the
other with $f\sim 280$ Hz ---  were filtered with ringdown templates. For the former
waveform the best-matched template triggered at the time of the ringdown,
whereas in the latter case, when the ringdown frequency was higher than LIGO's
most sensitive band, the best match occurred earlier, during the inspiral
phase.

\begin{figure}[h]
\centering
\begin{center}
\includegraphics[scale=0.6]{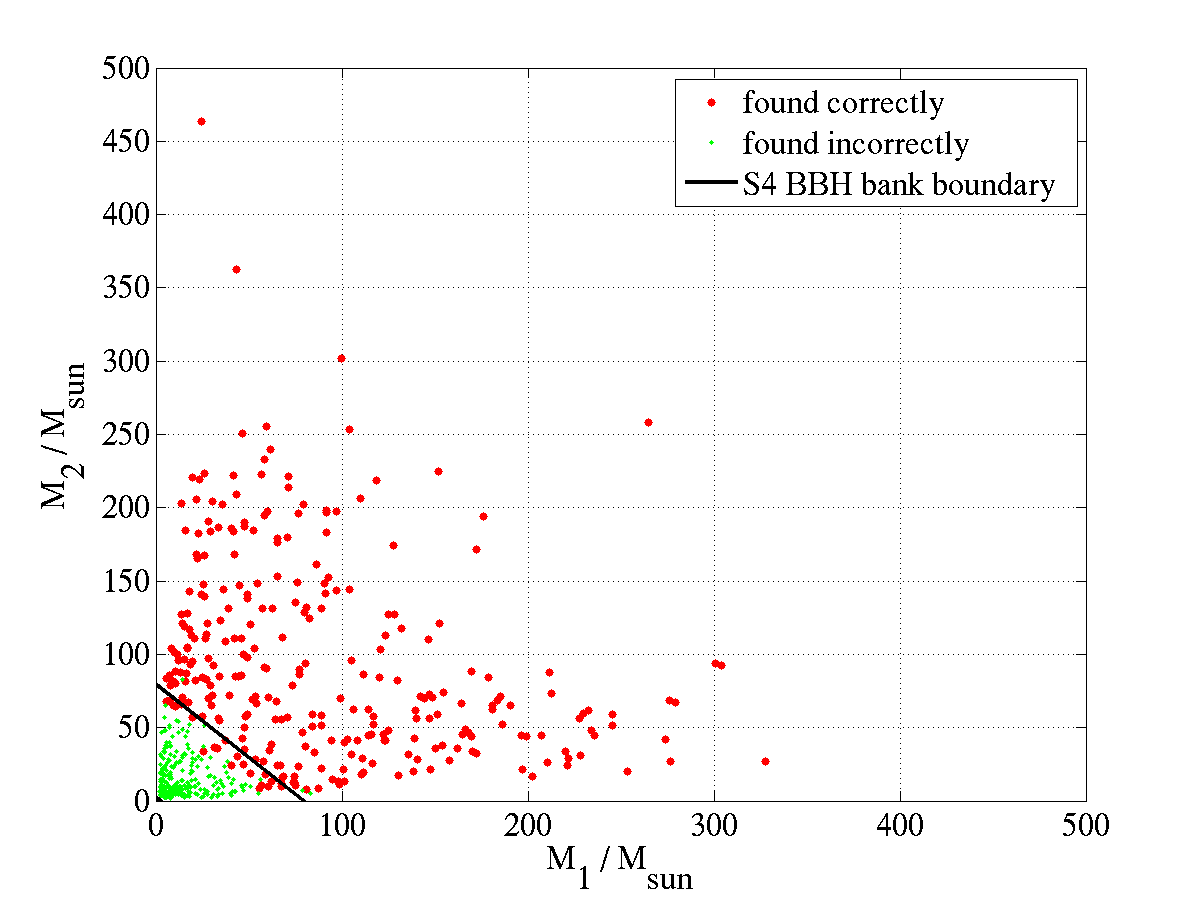}
\caption{Initial masses of the binary components for IMR injections found by
correct (red) and incorrect (green) ringdown templates. The black line
represents the upper limit to the mass range of the S4 binary black hole
inspiral search.}
\label{fig:imrwr}
\end{center}
\end{figure}


\section{Coincidence Analysis}
Thus far we have ascertained that although a ringdown may be detected with a lower frequency template, the presence of an inspiral and merger before the ringdown does not prevent it from being detected. The next important question is whether or not an IMR injection survives the coincidence test, as this is a key test in our pipeline. Recall from section \ref{sec:coinc} that for a given time window $\delta t$ triggers from different detectors are considered coincident if they lie within a specified parameter window $ds^2(f_0,Q)$ of each other. If multiple groups of coincidences (i.e., many triples and doubles) are found within $\delta t$, then the group with the loudest value of the detection statistic is chosen as the ``correct'' coincidence. 

\afterpage{\clearpage}

\begin{figure}[h] 
\centering
\begin{center}
\includegraphics[scale=0.55]{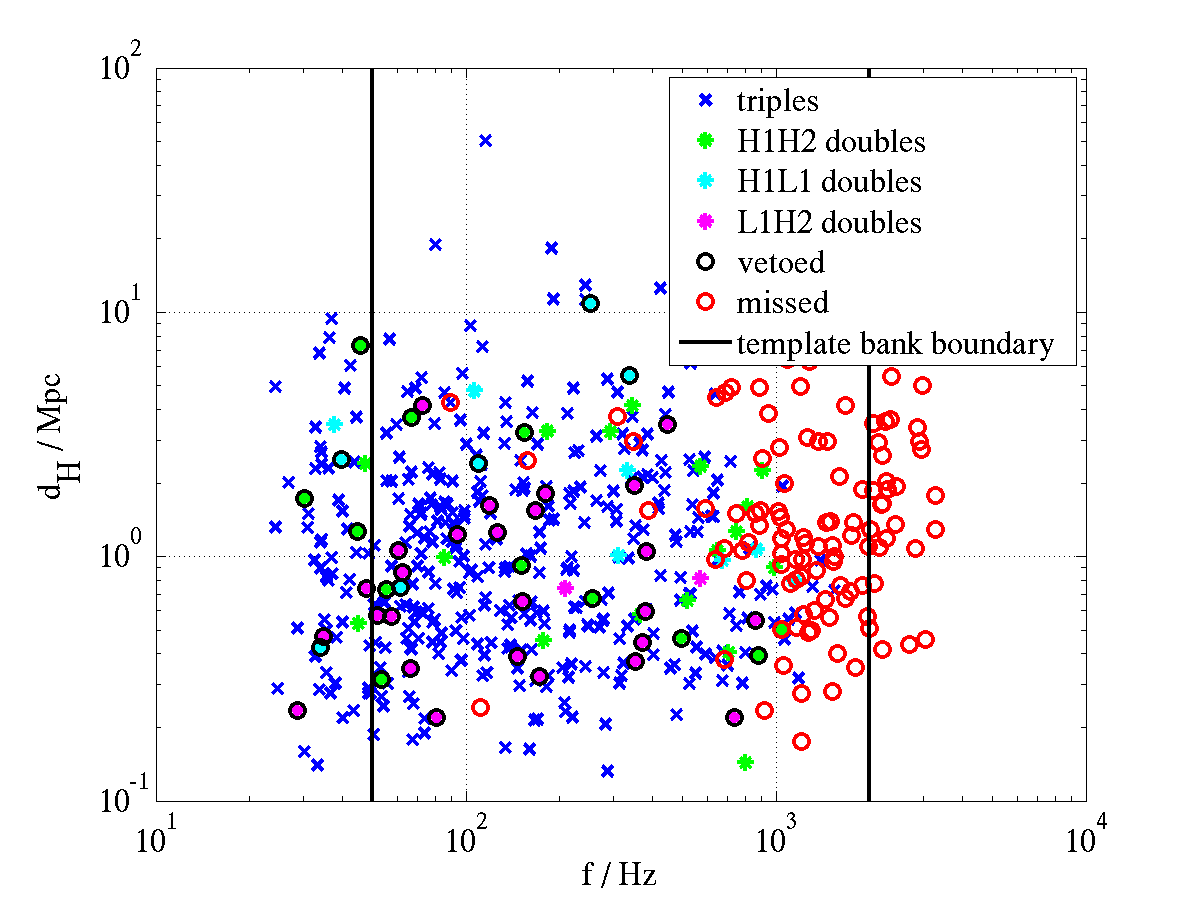}
\caption{Hanford effective distance versus injected ringdown frequency for
ringdown-only injections. The black vertical lines denote the template bank
boundaries.} \label{fig:ringmf}
\includegraphics[scale=0.55]{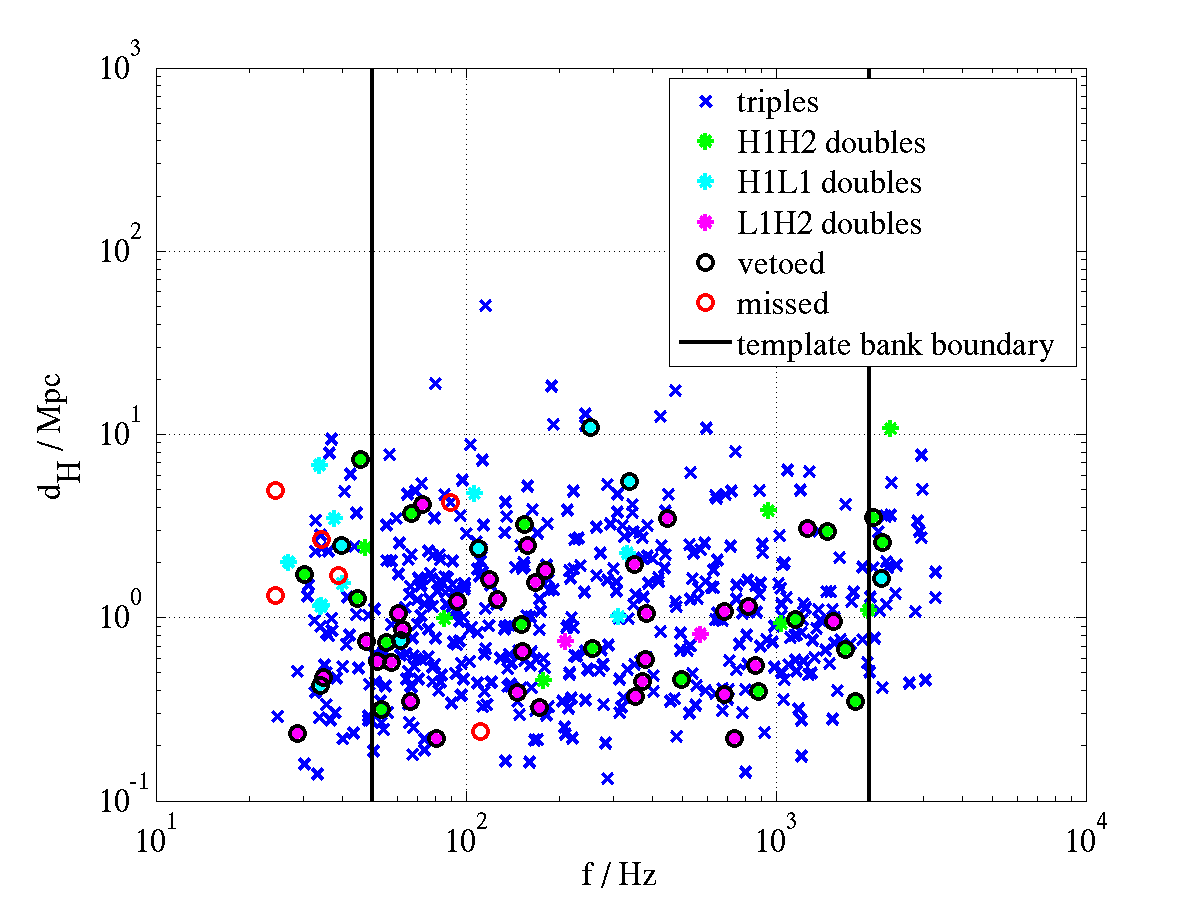} \\
\caption{Hanford effective distance versus injected ringdown frequency for IMR injections. The black vertical lines denote the template bank boundaries.}
\label{fig:imrmf}
\end{center}
\end{figure}

Ringdown-only and IMR injections were made into H1, H2, and L1 and the output coincidence files compared to the injection list; the results are shown in figures \ref{fig:ringmf} and \ref{fig:imrmf}, respectively, in plots of effective distance as measured at Hanford versus the injected ringdown frequency. These plots show the injections that were found in triple coincidence, those found in double coincidence (both because of vetoes and because they were missed in the third detector), and the missed injections. The ringdown-only results shown in figure \ref{fig:ringmf}, are, as expected, similar to those discussed in section \ref{sec:mfinjs}. What is interesting about the IMR coincidence, however, is that even though a large proportion of injections were found with the wrong ringdown parameters in all detectors, these were close enough to each other to allow the injection to pass the coincidence test! In fact, the efficiency of detection is higher in the IMR case than in the ringdown-only case (recall that parameter accuracy is not taken into account when calculating efficiencies). The high ringdown frequency injections that were missed in the ringdown-only case (because of high levels of noise above 1 kHz) were found in the IMR case because the inspiral part of the injection occurred at a less noisy (more sensitive) region of the template bank. This is a very encouraging result.


\section{Conclusion}
The calculation of the upper limit on the rate of ringdowns described in section \ref{sec:upperlimit} was based on our ability to recover injected signals. For that study we used isolated ringdown signals. The question here was how would this change if the ringdown was preceded by an inspiral and merger. This investigation has shown that the presence of an additional signal before the ringdown does not in any way hinder our ability to detect the signal. In fact, this model of an IMR injection improves our ability to detect coalescences with high ringdown frequencies i.e., low black hole masses. This increases our efficiency, which also positively impacts the upper limit we can set. From this we can conclude that not only is the upper limit for the S4 ringdown search presented in section \ref{sec:upperlimit} still valid, it may be regarded as a conservative upper limit.

What is impacted, however, is our ability to correctly recover the black hole's physical parameters; this study demonstrated that we can only correctly identify ringdown frequencies occurring below $\sim 200$ Hz. However, this lower limit on the accurate recovery of the mass of a black hole corresponds to the upper limit to the scope of the binary black hole inspiral search. Thus, a low-mass binary black hole coalescence  will be detected by both searches and correctly parameterized by the inspiral search, while high-mass coalescences should be detected and parameterized by the ringdown search (of course, only within the distance reach of the searches).

\chapter{The Future for Ringdown Searches}
\label{ch:future}
This thesis describes the first ringdown detection search in LIGO data and
has demonstrated that the pipeline is an effective method of searching
for triple coincident ringdown events. However this is just the beginning;
with every science run comes increased sensitivity and the possibility of
exploring a much larger population of astrophysical sources.

In the course of the analysis we have gained an understanding of the
character of ringdown waveforms in noisy data. 
In this chapter we list some of the unsolved issues, lessons learned, and
future recommendations for this particular search. We discuss some new ideas
for combining searches for the individual inspiral, merger and ringdown phases
of the binary coalescence and discuss the parameter space available to future
ringdown searches.


\section{Notes for Future Searches with the current Pipeline} 

\subsection{Searches for Triple Coincident Events}
We saw that in the S4 search the rate of false alarms in triple coincidence
was less than one event per run. Now that we have some understanding of the
characteristics of simulated ringdown waveforms in data we can tolerate a
somewhat higher level of background and use these known features to veto false
alarms. This gives us leeway to loosen some of the constraints on the search
and gain sensitivity.
We have demonstrated that the coincidence windows were sufficiently loose and that we did
not lose any injections because of clustering. However the search signal to
noise thresholds could be lowered further.
Given that a triple coincidence search is limited by H2 we recommend lowering
its SNR threshold. Decreasing the H2 threshold to 4 would allow triple
coincident signals to be seen with SNRs as low as 8 in H1 and L1 (as opposed
to 11 in the current search).
Given the rate of false alarms in double coincidence, attaining this
level of sensitivity without H2 is currently not possible.

\subsection{Searches for Double Coincident Events}
The results of the double coincidence analysis showed that the level of
background with the current pipeline was too high to detect gravitational
waves at threshold of 5.5. We are a long way from being able to claim a detection of gravitational
wave ringdown from co-located detectors however requiring two site coincidence
should in theory provide sufficiently strong evidence. We just discussed how
to increase our sensitivity to triples without changing the pipeline. However
increasing our sensitivity to doubles will require significant additions to
the pipeline.  We will need to work harder at reducing the level
of false coincidences.
One method of doing this is by implementing signal-based vetoes; vetoes based
on our knowledge of a signals shape in the time and frequency domains
\cite{Baba05}.
These have been implemented in inspiral searches and are effective in reducing
the false alarm rate.
Caution has to be exercised however when implementing these in the ringdown
search. The ringdown is a short duration single frequency waveform and is
likely (but not necessarily) to be preceded by an inspiral and thus any
signal-based vetoes must be tested on IMR waveforms to allow for this possibility.

\subsection{Coincidence Test}
\label{sec:futcoinctestt}
The coincidence test described in this analysis in which we use the metric to define coincident windows is an vast improvement on the traditional rectangular coincidence test. However the results of the injection (section \ref{sec:injdettime}) run revealed that the difference in the  time of arrival of the injected waveform was a strong function of frequency, particularly for the H1H2 pair (see figure \ref{fig:H1H2ttvff}). This plot showed that at high frequencies a much tighter time accuracy could be required. A frequency dependence time soincidence test should be considered for future searches as it is likely to reduce the false alarm rate considerably.


\subsection{Extending the Template Bank}
\label{sec:wings}
For the S4 search the region of frequency space searched over was from 50 Hz to 2 kHz. As mentioned in section \ref{sec:tunbank} we had hoped to extend the template bank to encompass a frequency range of 40Hz to 4kHz. However in the course of tuning the search and following up on missed injections a peculiar feature was observed. In plots of SNR versus frequency for injections, such as is shown in figure \ref{fig:snrwings}, the SNR falls off from the injection frequency as expected, but then begins to increase on both sides of the peak. 
This was observed in varying degrees of severity in every injection looked at regardless of the frequency of the injection. 
This feature became problematic when the templates far from the injection had higher SNR than those close to the injected frequency. When this occurred in one detector the injection failed the coincidence test and if it occurred in two or three detectors the injection was found at the wrong frequency. 
In the example shown in figure \ref{fig:snrwings} the injection was made at 200 Hz but was found close to 4 kHz. 
Weeks of investigations were dedicated to this problem but a solution was not
found and so the smaller bank was reinstated. These ``wings'' are still observable
with the smaller bank but the effect is small enough that they do not interfere with signal recovery.

This is an important problem to solve because the wider the frequency range we can search over, the larger the number of black hole ringdowns we are sensitive to. In particular, in increasing the upper frequency bound to 4 kHz we become
sensitive to gravitational waves from the entire mass range of non-spinning  stellar mass black holes. This would provide an excellent overlap with the binary black hole inspiral search.

\begin{figure}[htb]
\centering
\begin{center}
\includegraphics[scale=0.7]{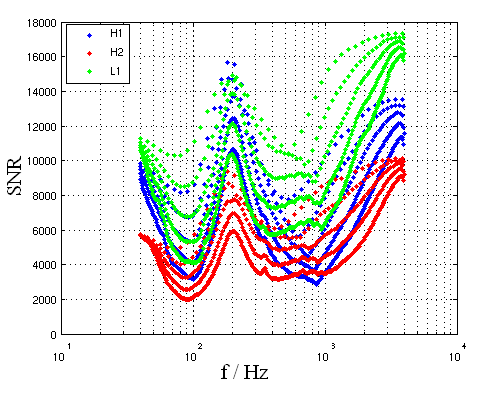}
\caption{A demonstration of the high SNR ``wings'' observed when the template bank
was extended to include frequenies between 40 Hz and 4 kHz. This plot shows the SNR versus frequency for a ringdown injection with central frequency of 200 Hz. }
\label{fig:snrwings}
\end{center}
\end{figure}


\section{Future Searches}

From an analysis point of view the hope for the future is to focus on IMR searches, bringing together the efforts of the three stand-alone pipelines: the binary black hole inspiral search, a burst search (which is sensitive to the merger phase) and the ringdown search. As we demonstrated in chapter \ref{ch:imr} there is a large degree of overlap between the searches. This combined effort could be implemented by running the filtering and coincidence steps separately and then comparing coincident triggers from the three searches. Another possibility is to combine the outputs of the filtering stage, and require coincidence between searches for each of the detectors and then look for coincidences between detectors.

The future holds exciting prospects for ringdown searches. This is best illustrated by figure \ref{fig:ieaLIGO}. Recall that the horizon distance is the distance to which we can detect a ringdown from an optimally oriented and located black hole of spin $\hat{a}$ with a signal to noise ratio of 8 in the detector. The figure shows horizon distance as a function of black hole mass for the predicted sensitivities of Initial LIGO (the blue curve), Enhanced LIGO (shown in green) and Advanced LIGO (shown in red)  assuming that 1\% of the mass is radiated as gravitational waves during the ringdown. The three curves in each group correspond to different spins, $\hat{a}=(0,0.49,0.9)$ for the curves from left to right, respectively. On the upper side of the horizontal axis the central ringdown frequency for a spin of 0.49 is also marked.

Comparing figures \ref{fig:range} and \ref{fig:ieaLIGO} demonstrates that we can expect to see $\sim 100$ Mpc further for Initial LIGO at design sensitivity than we did for S4. Preparations are currently underway to analyse data from the S5 run with the ringdown pipeline. S5 was the first science run at design sensitivity and includes one year of triple coincident data.

Enhanced LIGO is due to come on-line by the end of 2008. As figure
\ref{fig:ieaLIGO} shows, we can expect a factor two increase in sensitivity as
well as an extended mass range, with the lower mass limit extending further into the stellar mass black hole range.

The Advanced LIGO sensitivity curve used here is for the low power
configuration, (optimized for low frequency signals by reducing the radiation pressure quantum noise) and even at that the prediction is that our reach will be extended by an order of magnitude entering the regime of cosmological distances. At higher power this would be increased even further for low mass black hole ringdowns. The lower frequency limit in Advanced LIGO will be 12 Hz. This corresponds to a mass of 1000 $M_\odot$ for non-spinning black holes up to 2300 $M_\odot$ for rapidly spinning black holes. This makes the prospects for detection fro intermediate mass black holes very promising.

\begin{figure}[h]
\centering
\begin{center}
\includegraphics[scale=0.8]{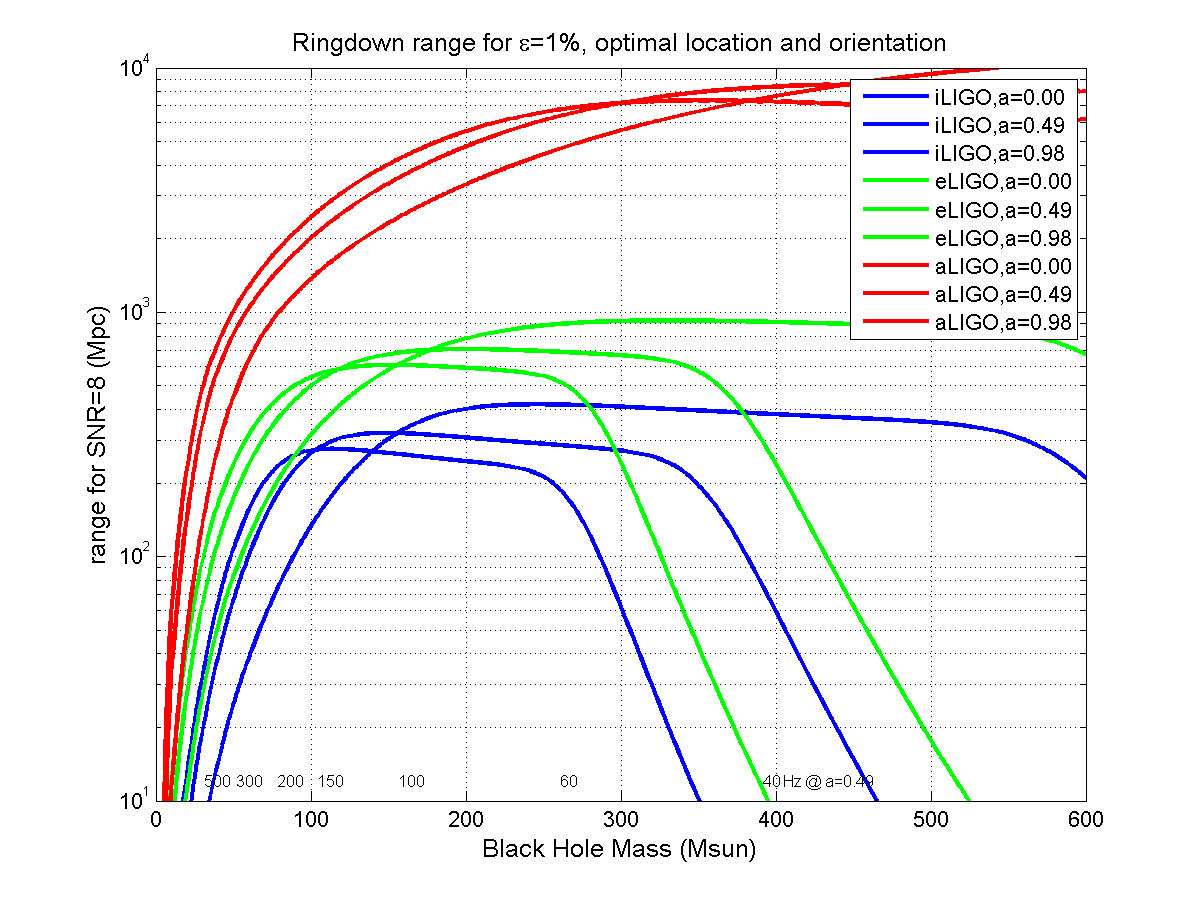}
\caption{Plot of horizon distance 
(distance to which a ringdown signal from an optimally oriented and located source will produce an SNR of 8 in the detector) 
versus mass for Initial LIGO (blue), Enhanced LIGO (green) and Advanced LIGO (red) in a low-power configuration. We have assumed that 1\% of the mass is radiated as gravitational waves. The curves in each group are for spins of 0, 0.49 and 0.98 going from left to right. The upper x-axis is the frequency for a spin of 0.49. (Plot from A. Weinstein.)}
\label{fig:ieaLIGO}
\end{center}
\end{figure}

\chapter{Summary and Concluding Remarks}
\label{ch:concl}
\section{Aim of the Search}

The gravitational radiation emitted by a perturbed black hole (e.g., the final stage of a compact binary coalescence (CBC))  is well modeled
and can be searched for using the method of matched filtering. 
However in the presence of non-Gaussian noise, optimal filtering alone is not sufficient to uncover a weak signal in the data. 
A powerful method of reducing the rate of false alarms is to require that a
trigger be seen at the same time in multiple detectors with similar parameters in order to be considered a candidate gravitational wave event.

In this study we have described the pipeline for a matched filter search with coincidence analysis and applied the pipeline to data from the fourth LIGO science run. The aim of the search was to detect gravitational waves from perturbed black holes and in the absence of a detection to place an upper limit on the rate of black hole ringdowns in the nearby universe.

\section{The Analysis Method}

We ran the search as a blind analysis to prevent any bias on the part of the
analyst from influencing the outcome of the search (for example setting cuts
based on triggers in the zero-lag data in order to get a better upper limit). We
tuned cuts, thresholds and coincidence windows with 
\begin{itemize}
\item{a coincidence analysis of simulated ringdown signals to gauge how well we can
expect to recover the parameters of a real signal,} 
\item{a coincidence analysis of time shifted data sets to estimate the rate of
false alarms,}
\item{and a representative subset of the data as a sanity check of the
analysis pipeline and a consistency check of the background estimation.}
\end{itemize}
Tuning is an iterative process with the outcome of the previous run
influencing the constraints on the subsequent run until the point is reached where
the maximum number of injected signals are recovered while the rate of
accidental coincidences is kept to a minimum.
Once the tuning was finalized a large scale Monte Carlo simulation of ringdown
waveforms was run to evaluate the efficiency of the search and to facilitate
parameter accuracy investigations.
A final background analysis was run to evaluate the false alarm rate.

It was decided in advance of unblinding the analysis to consider all double
and triple coincidences as possible gravitational wave candidates.
However in the absence of a detection only triple coincident signals would be used in
setting an upper limit.
Finally the tuned pipeline was applied to the data.

In a separate study we investigated the effects of
an inspiral and merger waveform preceding the ringdown on our ability to detect
gravitational waves with a ringdown filter and on our ability to recover the ringdown waveform parameters. We
injected inspiral-merger-ringdown waveforms into S4 data and ran with the
same pipeline and tuning described above. We then repeated the run injecting only the ringdown part of the pipeline and compared the results.

\section{Results}

\subsection{Opening the Box}
No candidate events were found in triple coincidence. A large number of
candidate events
were found in double coincidence, however the distribution was in agreement with the expected false alarm rate to within 1 sigma. Thus there was no evidence in the S4 data set of a detection of gravitational waves from perturbed black holes.

\subsection{Calculation of the Upper Limit}
Our Monte Carlo simulations revealed that the efficiency of detecting
gravitational wave ringdowns is highly frequency dependent. For that reason we
divided the simulations into five frequency bands and evaluated the
efficiency in each band.  The 100-200 Hz band was found to be the most sensitive. This
frequency range corresponds to a mass range of $60\ M_\odot \leq M \leq 120\ 
M_\odot$ for non spinning black holes and $150\ M_\odot\leq M \leq 300\ M_\odot$
for maximally spinning black holes assuming that the waveform is dominated by
the $l=m=2$ mode. The 90\% confidence upper limit of the rate of ringdowns in the $100-200$ Hz band was found to be $\mathcal{R}_{90\%}=4.3 \times 10^{-3} \textrm{ yr}^{-1} \textrm{ L}_{10}^{-1}$, where L$_{10}$ is a measure of the number of potential sources, equal to $10^{10}$ solar blue light luminosity.

\subsection{Parameter Accuracy}
We compared the injected and detected quantities and found that the accuracy with which the time of arrival of the injection could be determined was frequency dependent due to the uncertainty in the phase of the waveform. 
The waveform parameter accuracy was very high, with 70\% of the triple
coincident injections found with the correct template. We found that the match
between the injections and template decreased with the combination of high frequency and low quality factor values.

\subsection{Background Estimation}
Our background studies showed that with the chosen tuning the false alarm rate in triple coincidence was very low, less than one event per S4 run. In double coincidence the rate was much higher, more than 600 events per S4 run for two site coincidence. In fact the false alarm rate for signals found in detectors located in two widely separated sites with the exact same waveform parameters was 30 per S4 run. Clearly this is too high a rate to be able to confidently detect gravitational waves at the chosen threshold of $\rho_{DS}\sim12$. However at a higher threshold of $\sim 16$, less that 0.2 background events are expected in H1L1 and detection is possible.

\subsection{Inspiral-Merger-Ringdown}
We found that with the inclusion of the inspiral and merger waveforms our
efficiency of detecting simulated signals increased.
Isolated ringdowns at
high frequency are more difficult to recover than those at low frequency
because the level of noise above 500 Hz increases rapidly and our efficiency
decreases. However the inspiral and merger preceding the ringdown sweep
up through the LIGO band finding a significant match with ringdown
templates in the process. We found that the part of the signal in the most sensitive band of the detector, 100--200 Hz produced the loudest SNR and the ringdown was ``found'' in
this band. 
IMR injections with ringdown frequencies above 
200 Hz were generally detected with templates in the 100--200 Hz band.  
A consequence of this was that our ringdown parameter estimation was not accurate for IMRs with ringdown frequency above 200 Hz.

\section{Conclusions}
\begin{itemize}
\item{We have demonstrated using simulated signals that the pipeline presented in
this study is an effective means of detecting gravitational waves from
perturbed black holes in triple coincidence.}
\item{Our results verify that the timeslide method of determining the false alarm
rate and distribution with SNR is accurate for detectors at widely separate
sites, but not for co-located detectors.}
\item{We have found that the level of noise in double coincidence is too high to allow the detection of gravitational waves with the chosen SNR threshold.}
\item{We have shown that the accuracy with which the ringdown waveform parameters can be
recovered for ringdowns with frequency in excess of 200 Hz depends on whether or not an inspiral and merger preceded the ringdown. 
If our assumption that the radiation is dominated by the $l=m=2$ mode is correct we can estimate the mass of those ringdowns to a high degree of accuracy.}
\item{The next few years are very promising for IMR searches. Enhanced LIGO
will come online within a year increasing our sensitivity by a factor of 2. In
2014 Advanced LIGO will push the range in which we can search out
to cosmological distances in addition to extending the population of black
holes we are sensitive to beyond $10^3\ M_\odot$.}
\end{itemize}

\appendix
\chapter{Ringdown Search Configuration File}
\label{app:ini}

{\fontfamily{pcr}\selectfont
; ringdown pipeline configuration script. \\
; \\
; $Id$\\ 
; \\
; this is the configuration file for the inspiral DAG generation program \\
; lalapps\_inspiral\_hipe that creates a condor DAG to run the ringdown \\
; analysis pipeline. \\  \\
 
[condor] \\
; setup of condor universe and location of executables \\
universe    = standard \\
datafind    = /opt/lscsoft/glue/bin/LSCdataFind \\
tmpltbank   = /bin/false \\
inspiral    = /archive/home/lgoggin/bin/lalapps\_ring \\
inca        = /bin/false \\
thinca      = /archive/home/lgoggin/bin/lalapps\_rinca \\
trigtotmplt = /bin/false \\
sire        = /bin/false \\
cohbank     = /bin/false \\
chia        = /bin/false \\
inspinj     = /bin/false \\
frjoin      = /archive/home/lgoggin/bin/lalapps\_frjoin \\
coire       = /home/false \\ 

[pipeline] \\
; tagging information for the configure script \\
version = $Id$ \\
cvs-tag = $Name$ \\
; user-tag here can be overidden on the command line of \\ lalapps\_inspiral\_pipe \\
user-tag = \\
; data choice (playground\_only|exclude\_playground|all\_data) \\
playground-data-mask = all\_data \\  

[input] \\
; the segments file should be the output from segwizard with DQ flags applied
\\
; if no segment file if specified, assumed no data from that IFO. \\
h1-segments = H1triplesegs.txt \\
h2-segments = H2triplesegs.txt \\
l1-segments = L1triplesegs.txt \\
g1-segments = \\
ligo-channel = LSC-DARM\_ERR \\
geo-channel = \\
geo-bank = \\
geo-bank = \\
ligo-type = RDS\_R\_L3 \\
geo-type = \\
; injection file (if blank then no injections) \\
injection-file = HL-INJECTIONS\_8-793130413-2548800.xml \\
num-slides = \\ 
 
[calibration] \\
; location of the calibration cache and the cache files \\
path = /archive/home/lgoggin/projects/ringdown/s4/calibration \\
L1 = l1\_calibration.cache \\
H1 = h1\_calibration.cache \\
H2 = h2\_calibration.cache \\  
 
[datafind] \\
; type of data to use \\
type = RDS\_R\_L3 \\
url-type = file \\
match = localhost/archive \\  
 
[data] \\
; data conditioning parameters common to all ifos \\
pad-data = 8 \\
;segment-length = 1048576 \\
;number-of-segments = 16 \\
;sample-rate = 4096 \\
sample-rate = 8192 \\
block-duration = 2176 \\
segment-duration = 256 \\  
 
[ligo-data] \\
; data conditioning parameters for ligo data \\
highpass-frequency = 40 \\
cutoff-frequency = 45 \\
dynamic-range-factor = 1.0e+20 \\  
 
[geo-data] \\
; data conditioning parameters for geo data \\ 
 
[tmpltbank] \\
; not used in ringdown pipeline \\ 
 
[tmpltbank-1] \\
; not used in ringdown pipeline \\
 
[tmpltbank-2] \\
; not used in ringdown pipeline \\ 
 
[inspiral] \\
; analysis parameters -- added to all ring jobs \\
bank-max-mismatch = 0.03 \\
bank-min-frequency = 50 \\
bank-max-frequency = 2000 \\
bank-min-quality = 2.0 \\
bank-max-quality = 20.0 \\
bank-template-phase = 0 \\
maximize-duration = 1 \\
debug-level = 33 \\
;approximant = ringdown \\
;segment-overlap = 64 \\  
 
[no-veto-inspiral] \\
; not used in ringdown pipeline \\  
 
[veto-inspiral] \\
; not used in ringdown pipeline \\ 
 
[h1-inspiral] \\
; h1 specific inspiral paramters \\
threshold = 5.5 \\ 
 
[h2-inspiral] \\
; h2 specific inspiral parameters \\
threshold = 5.5 \\
 
[l1-inspiral] \\
; l1 specific inspiral parameters \\
threshold = 5.5 \\
 
[g1-inspiral] \\ 
 
[inspinj] \\
; not used in ringdown pipeline \\ 

[inca] \\
; not used in ringdown pipeline \\ 
 
[thinca] \\
; common coincidence parameters -- added to all thinca jobs \\
debug-level = 33 \\
multi-ifo-coinc = \\
maximization-interval = 1 \\
parameter-test = ds\_sq \\
h1-time-accuracy = 2 \\
h2-time-accuracy = 2 \\
l1-time-accuracy = 2 \\
;h1-freq-accuracy = 20 \\
;h2-freq-accuracy = 20 \\
;l1-freq-accuracy = 20 \\
;h1-quality-accuracy = 3 \\
;h2-quality-accuracy = 3 \\
;l1-quality-accuracy = 3 \\
h1-ds\_sq-accuracy = 0.05 \\
h2-ds\_sq-accuracy = 0.05 \\
l1-ds\_sq-accuracy = 0.05 \\
do-veto = \\
h1-veto-file = combinedVetoesH1-23.list \\
h2-veto-file = combinedVetoesH2-23.list \\
l1-veto-file = combinedVetoesL1-23.list \\ 
 
[thinca-1] \\
; not used in ringdown pipeline \\ 
 
[thinca-2] \\
; not used in ringdown pipeline \\ 
 
[thinca-slide] \\
; time slide parameters \\
h1-slide = 0 \\
h2-slide = 10 \\
l1-slide = 5 \\ 
 
[trigtotmplt] \\
; not used in ringdown pipeline \\ 
 
[sire] \\
; not used in ringdown pipeline \\ 
 
[sire-inj] \\
; not used in ringdown pipeline \\ 
 
[coire] \\
; not used in ringdown pipeline \\ 

[coire-inj] \\
; not used in ringdown pipeline \\ 

[cohbank] \\
; not used in ringdown pipeline \\ 

[coh-trig] \\
; not used in ringdown pipeline \\ 

[chia] \\
; not used in ringdown pipeline \\
} 

\chapter{Segmentation of the Data}
\label{app:segmentation}

Here we give a concrete example of how the data is segmented (segmentation
was initially discussed in chapter \ref{ch:search}).

In section \ref{sec:segmentation} we stated that the minimum length of an
analysis segment was 2176 s. In table \ref{tab:H1segs} we listed the first
five H1 science segments. The first science segment was 1020 s in duration and the
second was 240 s in duration; both of these are too short and have to be discarded.
Therefore, the first analysis segment comes from science segment number 3, beginning at GPS time
793166413 and ending at GPS time 793170673, with a total duration of 4260 s,
          and from this we create the first analysis segments.

As mentioned in section \ref{sec:readindata} an extra 8 s is read in before
the start of each analysis segment for data conditioning, and then removed
again before the filtering stage. However, for the first analysis segment in
a science segment, there is no previous analysis segment to ``borrow'' the 8 s
of data from and so the first 8 s of the science segment are effectively lost.
Thus, the first analysis segment begins at $793166413+8=793166421$ and ends
2176 s later at 793168597. An extra 8 s is also read in at the end of the
segment, and because the analysis segment is more than 2176 s long, this can
be taken from the next analysis segment.

For filtering purposes, the analysis segment is divided into sixteen 256 s
blocks, which overlap the previous block by 64 s and the next block by 64 s.
Then, for each block only triggers from the middle 128 s are recorded. As a
consequence, no triggers come from the first or last 64 s of the
analysis segment. Once again, for this particular analysis segment there is no segment preceding it and thus, any triggers from this time are lost. This is not so with the last 64 s of this first analysis segment; these are analysed in the
succeeding analysis segment. Therefore, for this analysis segment, the
first trigger will occur after $793166421+64=793166485$ and the last trigger
will appear before $793166485+16*128=793168533$, 64 s before the end time of
the analysis segment. 

The second segment begins 128 s before the first segment ended, at
$793168597-128 =793168469$ and ends 2176 s later, at 793170645. Now this time
the extra 8 s at the start of the segment can be read in from the previous
segment and therefore, no data is lost. Once more the segment is divided into
sixteen 256 s blocks each of which overlaps the previous and subsequent blocks
by 64 s, and the middle 128 s of each is analysed. As we started the second
analysis segment 128 s before the first segment ended, the first trigger from
the second segment can occur immediately after the last possible trigger in the
first segment. Therefore, no data is lost here. The last trigger in the second
segment can occur at $793168469+2048=793170581$.

Now the second segment ended 28 s before the end of the science segment; that
leaves $28+64=92$ s of unanalysed data. We analyse this by creating a segment
that starts $(2176+8)$ s before the end of the science segment, at 793168489, but
only analyse data after the last possible trigger in the second segment,
(i.e., at 793170581). As was the case at the start of the first analysis segment, the
last $8+64=72$ s are discarded, and so in this particular example, we get 20 s
of data from the last analysis segment in the science segment. The last
analysis segment ends at 793170665.

\begin{figure}[h]
\centering
\begin{center}
\includegraphics[scale=0.5]{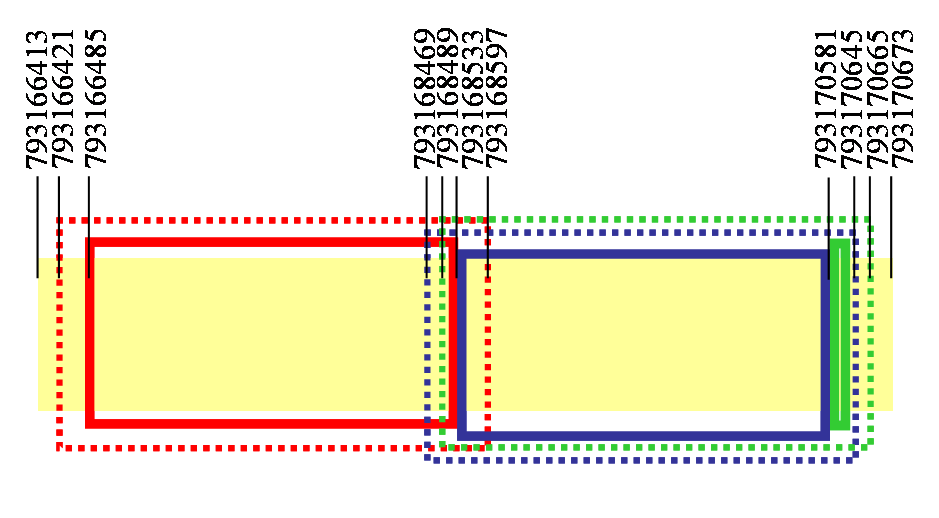}
\caption{An illustration of how science segment number 3 (yellow box) is divided into analysis segments (red, blue, and green dotted lines). The solid boxes denote the times that are analysed in each analysis segment.} 
\label{fig:segs}
\end{center}
\end{figure}

\chapter{Loudest Ten Events in Double Coincidence}
\label{app:topten}

\begin{center}
\begin{small}
\begin{table}[htdp]
\centering
\caption{The ten loudest candidate events from H1L1 doubles in triple time}
\label{tab:H1L1ttloud}
\begin{tabular}{c|rrrrrr}
\hline \hline
\#     &   $t$ (s)    &    $f$ (Hz)     &     $Q$     &  $d_{eff}$ (Mpc)    &    $\rho$    &     $\rho_{DS}$ \\
\hline
 1  &  794949585.625  &  107.56  &  3.46  &  186.60  &  9.42  &  21.05 \\
    &  794949585.614  &  103.84  &  10.04  &  149.07  &  12.50  &  \\
 2  &  793829533.702  &  81.39  &  10.04  &  303.72  &  8.76  &  19.72 \\
    &  793829533.693  &  79.43  &  10.04  &  84.72  &  11.41  &  \\
 3  &  794291462.267  &  73.83  &  10.04  &  313.25  &  8.41  &  19.03 \\
    &  794291462.254  &  81.99  &  5.91  &  70.34  &  11.32  &  \\
 4  &  794989787.461  &  87.56  &  10.04  &  288.37  &  9.48  &  18.93 \\
    &  794989787.458  &  81.99  &  5.91  &  190.72  &  9.45  &  \\
 5  &  794317560.631  &  65.36  &  10.04  &  338.90  &  7.86  &  17.93 \\
    &  794317560.629  &  68.63  &  17.01  &  19.13  &  31.51  &  \\
 6  &  794943595.760  &  153.33  &  10.04  &  151.11  &  8.87  &  17.71 \\
    &  794943595.771  &  153.65  &  17.01  &  118.65  &  8.84  &  \\
 7  &  794006110.807  &  132.55  &  3.46  &  216.83  &  7.23  &  16.67 \\
    &  794006110.795  &  127.60  &  2.00  &  46.28  &  10.38  &  \\
 8  &  793894630.682  &  63.86  &  17.01  &  318.93  &  7.75  &  16.03 \\
    &  793894630.673  &  64.79  &  17.01  &  78.58  &  8.29  &  \\
 9  &  795118144.597  &  85.44  &  5.91  &  272.38  &  8.31  &  15.94 \\
    &  795118144.604  &  79.43  &  10.04  &  276.42  &  7.63  &  \\
 10  &  794783103.777  &  50.00  &  5.91  &  323.03  &  6.87  &  15.93 \\
    &  794783103.776  &  50.00  &  5.91  &  105.96  &  15.78  &  \\
\hline \hline
\end{tabular}
\end{table}
\end{small}
\end{center}

\begin{center}
\begin{small}
\begin{table}[htdp]
\centering
\caption{The ten loudest candidate events from H1H2 doubles in triple time}
\label{tab:H1H2ttloud}
\begin{tabular}{c|rrrrrr}
\hline \hline
\#     &   $t$ (s)    &    $f$ (Hz)     &     $Q$     &    $d_{eff}$ (Mpc)    & $\rho$    &     $\rho_{DS}$ \\
\hline
 1  &  793253792.205  &  63.20  &  2.00  &  1.55  &  1336.49  &  63.10 \\
    &  793253792.207  &  50.00  &  2.00  &  44.09  &  30.45  &   \\
 2  &  794654729.865  &  54.51  &  17.01  &  164.21  &  12.98  &  21.26 \\
    &  794654729.864  &  54.51  &  17.01  &  84.23  &  9.53  &   \\
 3  &  795398069.574  &  1803.63  &  5.91  &  0.91  &  6.72  &  15.35 \\
    &  795398069.570  &  1700.01  &  17.01  &  0.52  &  8.63  &   \\
 4  &  793728359.581  &  52.21  &  17.01  &  26.55  &  45.28  &  15.14 \\
    &  793728359.584  &  54.51  &  17.01  &  135.21  &  6.47  &   \\
 5  &  793726436.436  &  1559.35  &  17.01  &  0.87  &  7.78  &  14.81 \\
    &  793726436.440  &  1493.45  &  17.01  &  0.96  &  7.03  &   \\
 6  &  795400260.902  &  160.99  &  10.04  &  111.05  &  9.67  &  14.64 \\
    &  795400260.899  &  151.46  &  17.01  &  135.64  &  6.22  &   \\
 7  &  795273330.160  &  82.75  &  17.01  &  252.29  &  11.09  &  14.51 \\
    &  795273330.158  &  83.95  &  17.01  &  214.01  &  6.16  &   \\
 8  &  795441060.564  &  70.63  &  17.01  &  183.24  &  10.37  &  14.42 \\
    &  795441060.560  &  68.63  &  17.01  &  162.54  &  6.11  &   \\
 9  &  794115186.367  &  1628.16  &  17.01  &  1.12  &  6.62  &  14.35 \\
    &  794115186.370  &  1675.71  &  17.01  &  0.70  &  7.73  &   \\
 10  &  795437362.744  &  1589.49  &  10.04  &  1.04  &  7.12  &  14.32 \\
    &  795437362.747  &  1493.45  &  17.01  &  0.84  &  7.20  &   \\
\hline \hline
\end{tabular}
\end{table}
\end{small}
\end{center}

\begin{center}
\begin{small}
\begin{table}[htdp]
\centering
\caption{The ten loudest candidate events from H2L1 doubles in triple time}
\label{tab:H2L1ttloud}
\begin{tabular}{c|rrrrrr}
\hline \hline
\#     &   $t$ (s)    &    $f$ (Hz)     &     $Q$     &    $d_{eff}$ (Mpc)    & $\rho$    &     $\rho_{DS}$ \\
\hline
 1  &  794966223.382  &  1795.38  &  10.04  &  0.45  &  10.65  &  20.27 \\
    &  794966223.393  &  1907.47  &  17.01  &  0.50  &  9.62  &  \\
 2  &  794490884.837  &  1660.93  &  5.91  &  0.60  &  9.48  &  17.32 \\
    &  794490884.844  &  1700.01  &  17.01  &  0.40  &  7.83  &  \\
 3  &  793181575.946  &  1991.64  &  17.01  &  0.36  &  9.17  & 16.38 \\
    &  793181575.951  &  1885.01  &  10.04  &  0.47  &  7.20  &  \\
 4  &  794934058.049  &  67.65  &  17.01  &  142.63  &  8.31  & 16.18 \\
    &  794934058.036  &  64.79  &  17.01  &  147.70  &  7.88 &  \\
 5  &  795117184.353  &  1088.11  &  17.01  &  2.15  & 7.16  &  16.17 \\
    &  795117184.352  &  1042.12  &  17.01  &  2.08  &  9.01  &  \\
 6  &  794477109.831  &  902.42  &  17.01  &  3.53  & 7.03  &  16.10 \\
    &  794477109.828  &  902.42  &  17.01  &  2.52 &  9.07  &  \\
 7  &  794925830.874  &  915.50  &  17.01  & 3.50  &  6.85  &  15.91 \\
    &  794925830.862  &  915.50  &  17.01  & 2.63  &  9.77  &  \\
 8  &  793676381.075  &  1593.87  &  5.91 &  0.78  &  7.87  &  15.89 \\
    &  793676381.068  &  1551.24  &  10.04 &  0.48  &  8.01  &  \\
 9  &  794484924.385  &  1203.45  & 17.01  &  1.52  &  7.75  &  15.86 \\
    &  794484924.373  &  1152.59  & 17.01  &  1.21  &  8.12  &  \\
 10  &  794939243.941  &  1148.01 &  3.46  &  1.76  &  7.18  & 15.71 \\
    &  794939243.940  &  931.57 &  3.46  &  2.54  &  8.53  &
 \\
 \hline \hline
\end{tabular}
\end{table}
\end{small}
\end{center}

\begin{center}
\begin{small}
\begin{table}[htdp]
\centering
\caption{The ten loudest candidate events from H1L1 doubles in double time}
\label{tab:H1L1dtloud}
\begin{tabular}{c|rrrrrr}
\hline \hline
\#     &   $t$ (s)    &    $f$ (Hz)     &     $Q$     &    $d_{eff}$ (Mpc)    & $\rho$    &     $\rho_{DS}$ \\
\hline
 1  &  793258551.938  &  66.98  &  10.04  &  370.92  &  7.37  &  16.93 \\
    &  793258551.941  &  70.63  &  17.01  &  176.60  &  9.82  &  \\
 2  &  794506571.018  &  50.00  &  5.91  &  295.90  &  7.21  &  15.04 \\
    &  794506571.007  &  51.46  &  17.01  &  212.73  &  7.84  &  \\
 3  &  794368837.863  &  181.31  &  2.00  &  112.87  &  5.73  &  13.37 \\
    &  794368837.859  &  257.63  &  2.00  &  12.00  &  7.65  &  \\
 4  &  794195114.960  &  125.61  &  17.01  &  88.24  &  24.44  &  13.36 \\
    &  794195114.972  &  118.81  &  5.91  &  139.28  &  5.58  &  \\
 5  &  794361114.970  &  60.76  &  10.04  &  365.64  &  5.91  &  13.20 \\
    &  794361114.960  &  65.36  &  10.04  &  94.98  &  7.29  &  \\
 6  &  794958144.714  &  59.43  &  17.01  &  287.89  &  5.81  &  13.03 \\
    &  794958144.724  &  56.92  &  17.01  &  172.74  &  7.23  &  \\
 7  &  795477196.874  &  79.87  &  2.00  &  255.50  &  6.95  &  12.70 \\
    &  795477196.885  &  92.78  &  5.91  &  362.71  &  5.75  &  \\
 8  &  795601342.002  &  53.73  &  17.01  &  327.60  &  6.15  &  12.44 \\
    &  795601342.000  &  54.51  &  17.01  &  225.26  &  6.29  &  \\
 9  &  794368869.630  &  1885.01  &  10.04  &  0.93  &  5.82  &  12.36 \\
    &  794368869.629  &  1885.01  &  10.04  &  0.14  &  6.55  &  \\
 10  &  794362732.279  &  1130.17  &  10.04  &  2.95  &  5.75  &  12.35 \\
    &  794362732.280  &  1012.90  &  5.91  &  0.83  &  6.61  &  \\
\hline \hline
\end{tabular}
\end{table}
\end{small}
\end{center}

\begin{center}
\begin{small}
\begin{table}[htdp]
\centering
\caption{The ten loudest candidate events from H1H2 doubles in double time}
\label{tab:H1H2dtloud}
\begin{tabular}{c|rrrrrr}
\hline \hline
\#     &   $t$ (s)    &    $f$ (Hz)     &     $Q$     &    $d_{eff}$ (Mpc)    & $\rho$    &     $\rho_{DS}$ \\
\hline
 1  &  793589170.965  &  71.65  &  17.01  &  223.57  &  10.80  &  17.10 \\
    &  793589170.967  &  72.05  &  10.04  &  161.15  &  7.45  &   \\
 2  &  795114450.689  &  78.68  &  5.91  &  213.69  &  11.61  &  16.88 \\
    &  795114450.691  &  75.65  &  10.04  &  173.89  &  7.34  &   \\
 3  &  794692945.319  &  1130.17  &  10.04  &  1.95  &  8.23  &  16.03 \\
    &  794692945.320  &  1169.30  &  17.01  &  1.40  &  7.80  &   \\
 4  &  795385587.580  &  648.09  &  17.01  &  6.64  &  8.53  &  15.58 \\
    &  795385587.576  &  614.69  &  10.04  &  7.55  &  7.04  &   \\
 5  &  795382679.975  &  568.84  &  5.91  &  9.57  &  7.39  &  14.10 \\
    &  795382679.972  &  585.98  &  17.01  &  9.28  &  6.71  &   \\
 6  &  793694651.919  &  1991.64  &  17.01  &  0.74  &  6.07  &  13.99 \\
    &  793694651.916  &  1991.64  &  17.01  &  0.43  &  7.93  &   \\
 7  &  795384035.598  &  1072.56  &  17.01  &  2.24  &  7.49  &  13.54 \\
    &  795384035.599  &  1119.89  &  17.01  &  2.14  &  6.05  &   \\
 8  &  795382177.870  &  1136.12  &  17.01  &  2.21  &  6.59  &  13.51 \\
    &  795382177.868  &  1186.25  &  17.01  &  1.48  &  6.92  &   \\
 9  &  794725076.726  &  69.53  &  5.91  &  350.69  &  7.70  &  13.36 \\
    &  794725076.726  &  63.20  &  2.00  &  185.92  &  5.66  &   \\
 10  &  794158080.377  &  52.50  &  10.04  &  252.45  &  9.12  &  13.32 \\
    &  794158080.380  &  52.21  &  17.01  &  153.64  &  5.56  &   \\
\hline \hline
\end{tabular}
\end{table}
\end{small}
\end{center}

\begin{center}
\begin{small}
\begin{table}[htdp]
\centering
\caption{The ten loudest candidate events from H2L1 doubles in double time}
\label{tab:H2L1dtloud}
\begin{tabular}{c|rrrrrr}
\hline \hline
\#     &   $t$ (s)    &    $f$ (Hz)     &     $Q$     &    $d_{eff}$ (Mpc)    & $\rho$    &     $\rho_{DS}$ \\
\hline
 1  &  794432410.606  &  1724.65  &  17.01  &  0.50  &  9.16  &  16.15 \\
    &  794432410.597  &  1651.77  &  17.01  &  0.18  &  6.99  &  \\
 2  &  793572623.619  &  1724.65  &  17.01  &  0.61  &  8.03  &  14.72 \\
    &  793572623.628  &  1675.71  &  17.01  &  0.49  &  6.68  &  \\
 3  &  794448103.696  &  1072.56  &  17.01  &  1.64  &  9.27  &  14.64 \\
    &  794448103.699  &  1055.52  &  5.91  &  0.77  &  6.22  &  \\
 4  &  793242141.579  &  1350.32  &  17.01  &  0.99  &  8.63  &  14.26 \\
    &  793242141.586  &  1369.89  &  17.01  &  1.07  &  6.03 &  \\
 5  &  795389328.279  &  1331.02  &  17.01  &  0.87  &  9.20  &  14.20 \\
    &  795389328.291  &  1369.89  &  17.01  &  1.67  & 6.00  &  \\
 6  &  793674572.094  &  972.01  &  5.91  &  3.20  &  5.86  &  13.93 \\
    &  793674572.087  &  858.96  &  5.91  &  2.66  &  8.11  &  \\
 7  &  795069044.029  &  1493.45  &  17.01  &  0.87  &  7.52  &  13.71 \\
    &  795069044.020  &  1589.49  &  10.04  &  1.13  &  6.20  &  \\
 8  &  795407874.988  &  1515.11  &  17.01  &  0.70  &  8.31  & 13.64 \\
    &  795407874.994  &  1472.11  &  17.01  &  1.52  &  5.72  &  \\
 9  &  793735902.987  &  1931.50  &  10.04  &  0.51  &  7.45  & 13.61 \\
    &  793735902.984  &  1879.52  &  5.91  &  0.23  &  6.16  & \\
10  &  794877527.436  &  667.02  &  17.01  &  6.15  & 8.54  &  13.58 \\
    &  794877527.435  &  657.49  &  17.01  &  10.08  & 5.69  &  \\

\hline \hline
\end{tabular}
\end{table}
\end{small}
\end{center}

\chapter{Projects Undertaken at the 40 m Interferometer}

Prior to the construction of the LIGO 4 and 2 km interferometers the 40 m
LIGO prototype, an laser interferometer with 40 m long arms located on the Caltech campus served as a test-bed for LIGO technologies. Currently it is used for testing Advanced LIGO and future generation technologies.

I was part of the 40 m team during this upgrade period. My first project
involved modeling the interferometer optics \cite{SURF}. For a given laser wavelength and length constraints I used Matlab to trace the path
of the beam through the interferometer and made recommendations for the radii
of curvature and placement of various in-vacuum optics. These included the
mode cleaner, a triangular configuration of mirrors designed to isolate a
single mode of light to send into the interferometer, and mode-matching
telescopes, a pair of lenses to match the light from the mode cleaner to the
beam-splitter. 

Once the new laser and optics were installed we found be useful to see the
beam on various optics throughout the interferometer. Cameras were installed
and connected to an electronics rack linked to monitors in the control room.
Using EPICS (Experimental Physics and Industrial Control System) I created an
interface for the user in the control room to select a particular camera and a
monitor in which to display the image. A similar control system was employed
to activate or deactivate lamps illuminating the optics chambers.


\begin{thebibliography}{<50>}

\bibitem{ELIGO} R. Adhikari, P. Fritschel, and S. Waldman, LIGO-T060156-01-I (2006).


\bibitem{Hart03}  J. B. Hartle, {\it Gravity: an introduction to Einstein's general relativity}, Addison Welsey (2003).
\bibitem{MTW} C. W. Misner, K. S. Thorne, and J. A. Wheeler, {\it Gravitation},
W. H. Freeman and Company (1970).
\bibitem{Saul94} P. R. Saulson, {\it Fundamentals of Interferometric Gravitational Wave Detection}, World Scientific Publishing (1994).
\bibitem{Diet05} A. Dietz, J. Garofoli, G Gonzales, M. Landry, B. O'Reilly,
and M. Sung,  LIGO-T050262-01-D (2005).
\bibitem{Siem04} X. Siemens, B. Allen, J. Creighton, M. Hewitson, and M. Landry, {\it Class. Quantum Grav.} {\bf 21}, S1723 (2004).
\bibitem{Schu87} B. F. Schutz and M. Tinto, {\it MNRAS}, {\bf 224}, 131
(1987).
\bibitem{Sigg98} D. Sigg, LIGO-P980007-00-D (1998).
\bibitem{Huls75} R. A. Hulse and J. H. Taylor, {\it Ap. J.}, {\bf 195}, L51 (1975).
\bibitem{Thor87} K. S. Thorne, {\it 300 Years of Gravitation}, edited by S.W. Hawking and W. Israel, Cambridge University Press, Cambridge, England (1987).
\bibitem{S1BNS} B. Abbott et al., ``Analysis of LIGO data for gravitational waves from binary neuton stars'', {\it Phys. Rev. D}, {\bf 69}, 122001 (2004).
\bibitem{S2BNS} B. Abbott et al., ``Search for gravitational waves from galactic and extra-galactic binary neutron stars'', {\it Phys. Rev. D}, {\bf 72}, 082001 (2005).
\bibitem{S2PBH} B. Abbott et al., ``Search for gravitational waves from primordial black hole binary coalescences in the galactic halo'', {\it Phys. Rev. D}, {\bf 72}, 082002 (2005).
\bibitem{S2BBH} B. Abbott et al., ``Search for gravitational waves from binary black-hole inspirals in LIGO data'', {\it Phys. Rev. D}, {\bf 73}, 062001 (2006).
\bibitem{S3SBBH} B. Abbott et al., ``Search of S3 LIGO data for gravitational wave signals from spinning black hole and neutron star binary inspirals'', arXiv:0712.2050 (2007).
\bibitem{S3S4} B. Abbott et al., ``Search for gravitational waves from binary inspirals in S3 and S4 LIGO data'', arXiv:0704.3368 (2007).
\bibitem{GRB070201} B. Abbott et al., ``Implications for the Origin of
GRB 070201 from LIGO Observations'', arXiv:0711.1163 (2007).
\bibitem{BurstS1} B. Abbott et al., ``First upper limits from LIGO on gravitational-wave bursts'', {\it Phys. Rev. D}, {\bf 69}, 102001 (2004).
\bibitem{BurstS2} B. Abbott et al., ``Upper limits on gravitational-wave bursts in LIGO's second science run'', {\it Phys. Rev. D}, {\bf 72}, 062001 (2005).
\bibitem{BurstS3} B. Abbott et al., ``Search for gravitational-wave bursts in LIGO's third science run'', {\it Class. Quant. Grav.}, {\bf 23}, S29 (2006).
\bibitem{BurstS4} B. Abbott et al., ``Search for gravitational-wave bursts in LIGO data from the fourth science run'', {\it Class. Quant. Grav.}, {\bf 24}, 5343 (2007).
\bibitem{GRBS2S3S4} B. Abbott et al., ``Search for Gravitational Waves Associated with 39 Gamma-Ray Bursts Using data from the Second, Third, and Fourth LIGO Runs'', arXiv:0709.0766 (2007).
\bibitem{PULS1} B. Abbott et al., ``Setting upper limits on the strength of periodic gravitational waves from PSR J1939+2134 using the first science data from the GEO 600 and LIGO detectors'', {\it Phys. Rev. D}, {\bf 69}, 082004 (2004).
\bibitem{PULS2} B. Abbott et al., ``Upper limits from the LIGO and TAMA detectors on the rate of gravitational-wave bursts'', {\it Phys. Rev. D}, {\bf 72}, 102004 (2005).
\bibitem{PULS2b} B. Abbott et al., ``Limits on gravitational wave emission from selected pulsars using LIGO data'', {\it Phys. Rev. Lett.}, {\bf 94}, 181103 (2005).
\bibitem{PULS2c} B. Abbott et al., ``Coherent searches for periodic gravitational waves from unknown isolated sources and Scorpius X-1: results from the second LIGO science run'', {\it Phys. Rev. D}, {\bf 76}, 082001 (2007).
\bibitem{PULS3S4} B. Abbott et al., ``Upper Limits on Gravitational Wave Emission from 78 Radio Pulsars'', {\it Phys. Rev. D}, {\bf 76}, 042001 (2007).
\bibitem{PULS4} B. Abbott et al., ``All-sky search for periodic gravitational waves in LIGO S4 data'',  {\it Phys. Rev. D}, {\bf 77}, 022001 (2008).
\bibitem{StocS1} B. Abbott et al., ``Analysis of first LIGO science data for stochastic gravitational waves'', {\it Phys. Rev. D}, {\bf 69}, 122004 (2004).
\bibitem{StocS3} B. Abbott et al., ``Upper limits on a stochastic background of gravitational waves'', {\it Phys. Rev. Lett.}, {\bf 95}, 221101 (2005).
\bibitem{StocS4} B. Abbott et al., ``Searching for Stochastic Background
of Gravitational Waves with LIGO'', {\it Ap. J.}, {\bf 659}, 918 (2007).
\bibitem{StocS4b} B. Abbott et al., ``Upper limit map of a background of gravitational waves'', {\it Phys. Rev. D}, {\bf 76}, 082003 (2007).
\bibitem{StocS4c} B. Abbott et al., ``First Cross-Correlation Analysis of Interferometric and Resonant-Bar Gravitational-Wave Data for Stochastic Backgrounds'', {\it Phys. Rev. D}, {\bf 76}, 022001 (2007).


\bibitem{Michell} J. Michell, {\it Philos. Trans. R. Soc. London}, {\bf 74}, 35 (1784).
\bibitem{Schw16b} K. Schwarzschild, {\it Sitzungsber. Gottingen Math. Phys.
Kl.} 424 (in German), arXiv:physics/9912033v1 (in English) (1916).
\bibitem{Oppe39} J. R. Oppenheimer and H. Snyder, {\it Phys. Rev.}, {\bf 56},
455 (1939).
\bibitem{Whee68} J. A. Wheeler, {\it Am. Sci.}, {\bf 56}, 1 (1968).
\bibitem{Whee71} J. A. Wheeler and R. Ruffini, {\it Phys. Today}, {\bf 24}, 30
(1971).

\bibitem{Regg57} T. Regge and J. A. Wheeler, {\it Phys. Rev.},  {\bf 108}, 1063 (1957).
\bibitem{Leav85} E. W. Leaver, {\it Proc. R. Soc. Lond.}, {\bf 402}, 285
(1985).
\bibitem{Bert04} E. Berti, arXiv:gr-qc/0411025v1 (2004).
\bibitem{Bach93} A. Bachelot and A. Motet-Bachelot, {\it Ann. Inst. Henri Poincar\'e}, {\bf 59}, 3 (1993).
\bibitem{Teuk73} S. A. Teukolsky, {\it Ap. J.}, {\bf 185}, 635 (1973).
\bibitem{Bert05} E. Berti, V. Cardoso, and C. Will, {\it Phys Rev. D}, {\bf
73}, 064030 (2005).
\bibitem{Vish70} C. V. Vishveshwara, {\it Nature}, {\bf 227}, 936 (1970).
\bibitem{Zeri70} F. J. Zerilli, {\it Phys. Rev. D}, {\bf 2}, 2141 (1970).
\bibitem{Pres71} W. H. Press, {\it Ap. J.} {\bf 170}, L105 (1971).
\bibitem{Pric72} R. H. Price, {\it Phys. Rev. D}, {\bf 5}, 2439 (1972).
\bibitem{Detw75} S. Chandrasekhar and S. Detweiler, {\it Proc. R. Soc. Lond.}, {\bf 344}, 441 (1975).
\bibitem{Ferr84} V. Ferrari and B. Mashhoon, {\it Phys. Rev. D}, {\bf 30}, 295 (1984).
\bibitem{Kokk99} K. D. Kokkotas and B. G. Schmidt, {\it Living Rev. Relativity}, {\bf 2}, 2, http://www.livingreviews.org/lrr-1999-2 (1999).
\bibitem{Eche89} F. Echeverria, {\it Phys. Rev. D}, {\bf 40}, 3194 (1989).
\bibitem{Ande01} N. Andersson, K.D. Kokkotas, {\it IJMPD}, {\bf 10}, 381, arXiv:gr-qc/0010102v1 (2001).
\bibitem{Davi71} M. Davis, R. Ruffini, W. H. Press and R. H. Price, {\it Phys.
Rev. Lett.}, {\bf 27}, 1466 (1971).
\bibitem{Flan98}  \'E. \'E. Flanagan and S. A. Hughes, {\it Phys. Rev. D.},
{\bf 57}, 4566 (1998).
\bibitem{Buon06} A. Buonanno, G. B. Cook, and F. Pretorius, {\it Phys. Rev. D}, {\bf 75}, 124018 (2007).
\bibitem{Kalo96} V. Kalogera and G. Baym, {\it Ap. J. Lett}, {\bf 470} L61
(1996).
\bibitem{Ferr05} L. Ferrarese and H. Ford, {\it Space Sci. Rev.}, {\it 116}, 523 (2005).
\bibitem{Mill04} M. C. Miller and E. J. M. Colbert, {\it IJMPD}, {\bf 13}, 1 (2004).
\bibitem{Crei99} J. D. E. Creighton, {\it Phys. Rev. D}, {\bf 60}, 022001 (1999). 
\bibitem{Alle97} B. Allen, ''GRASP documentation'', http://www.lsc-group.phys.uwm.edu (1997).
\bibitem{Adhi04} R. Adhikari, ``Sensitivity and noise analysis of 4 km laser
interferometer gravitational wave antennae'', Ph.D. Thesis, Massachusetts Institute of
Technology (2004).
\bibitem{LAL} ''LIGO Algorithm Library'', LIGO Project, California Institute of Technology, Pasadena, CA, http://www.lscgroup.phys.uwm.edu/daswg/projects/lal.html. 
\bibitem{Tsun04} Y. Tsunesada and the TAMA Collaboration, {\it Class. Quant. Grav.}, {\bf 21}, S703 (2004).






\bibitem{Turi60} G. L. Turin, {\it IRE. Trans. Inform. Theory}, {\bf 6}, 311 (1960).
\bibitem{Thor08} K. S. Thorne and R. Blandford, {\it Applications of Classical
Physics}, (unpublished) 2008.
\bibitem{Kuma05} V. Kumar, A. Mahalanobis, and R. Juday, {\it Correlation
Pattern Recognition},  Cambridge University Press, New York (2005).
\bibitem{Owen96} B. J. Owen, {\it Phys. Rev. D} {\bf 53}, 6749 (1996).




\bibitem{Cond} http://www.cs.wisc.edu/condor.


\bibitem{tuning}  The LSC, ''Tuning Matched Filter Searches for Compact
Binary Coalescence'', LIGO-T070109-01-Z (2007).
\bibitem{Powe} R. Powell, http://www.atlasoftheuniverse.com. 
\bibitem{Kopp08} R. K. Kopparapu, C. R. Hanna, V. Kalogera, R. O'Shaughnessy,
G. Gonz\'alez, P. R. Brady, and S. Fairhurst, {\it Ap. J.}, {\bf 675}, 1459 (2008).
\bibitem{Phin91} E. S. Phinney, {\it Ap. J.}, {\bf 380}, L17 (1991).
\bibitem{Osha05} R. O'Shaughnessy, C. Kim, T. Fragos, V. Kalogera, and K.
Belczynski, {\it Ap. J.}, {\bf 633}, 1076 (2005).
\bibitem{Port99} S. T. Portegies Zwart, J. Makino, S. L. W. McMillan, and P.
Hut, {\it A\&A}, {\bf 348}, 117 (1999).
\bibitem{Gurk06} M. A. G\"urkan, J. M. Fregeau, and F. A. Rasio, {\it Ap. J.}, {\bf 640}, L39 (2006).
\bibitem{Freg06} J. M. Fregeau, S. L. Larson, M. C. Miller, R. O'Shaughnessy, and F. A. Rasio, {\it Ap. J.}, {\bf 646}, L135 (2006).


\bibitem{Chat05} S. Chatterji, ''The search for gravitational wave bursts in
data from the second LIGO science run'', PhD Thesis, Massachusetts Institute of
Technology, Dept. of Physics (2005).
\bibitem{Brad04} P. R. Brady, J. D. E. Creighton, and A. G. Wiseman, {\it Class. Quantum Grav.} {\bf 21}, S1775 (2004).
\bibitem{Fair07} S. Fairhurst and P. R. Brady,  arXiv:0707.2410v1 (2007).


\bibitem{Pret05} F. Pretorius, {\it Phys. Rev. Lett.}, {\bf 95}, 121101 (2005).
\bibitem{Camp06} M. Campanelli, C. O. Lousto, P. Marronetti, and Y. Zlochower, {\it Phys. Rev. Lett.}, {\bf 96}, 111101 (2006).
\bibitem{Bake06} J. G. Baker, J. Centrella, D. Choi, M. Koppotz, and J. van Meter, {\it Phys. Rev. Lett.}, {\bf 96}, 111102 (2006).
\bibitem{Herr07} F. Herrmann, I. Hinder, D. Shoemaker, and P. Laguna {\it Class. Quant. Grav.}, {\bf 24}, S33 (2007).
\bibitem{Camp06b} M. Campanelli, C. O. Lousto, and Y. Zlochower, {\it Phys. Rev. D}, {\bf 74}, 084023 (2006).
\bibitem{Camp07} M. Campanelli, C. O. Lousto, Y. Zlochower, B. Krishnan, and D. Merritt {\it Phys. Rev. D}, {\bf 75}, 064030 (2007).
\bibitem{Camp07b} M. Campanelli, C. O. Lousto, Y. Zlochower, and D. Merritt {\it Ap. J.}, {\bf 659}, L5 (2007).
\bibitem{Bake07} J. G. Baker, W. D. Boggs, J. Centrella, B. J. Kelly, S. T. McWilliams, M. C. Miller, and J. R. van Meter, {\it Ap. J}, {\bf 668}, 1140 (2007).
\bibitem{Buon07} A. Buonanno, Y, Pan, J. G. Baker, J. Centrella, B. J. Kelly, S. T. Mc Williams, and J. R. van Meter, {\it Phys. Rev. D}, {\bf 76}, 104049 (2007).
\bibitem{Ajit07} P. Ajith et al., {\it Class. Quant. Grav.}, {\bf 24}, S689 (2007).
\bibitem{Jone08} B. Abbott et al., ``Search of S3 LIGO data for gravitational wave signals from spinning black hole and neutron star binary inspirals'', arXiv:0712.2050v2 (2008).
\bibitem{Baum} T. Baumgarte, P. Brady, J. D. E. Creighton, L. Lehner, F. Pretorius, and R. DeVoe, arXiv:gr-qc/0612100v1 (2006) 


\bibitem{Baba05} S. Babak, H. Grote, M. Hewitson, H. L\"uck and K. A. Strain,
{\it Phys. Rev. D}, {\bf 72}, 022002 (2005).


\bibitem{SURF} L. Goggin and A. Weinstein, ''Study of the Optical Parameters
of the 40m LIGO Prototype'', LIGO-T000122-00-R (2000).

\end{thebibliography}
\end{document}